\newif\ifjournal \journalfalse 
\pdfoutput=1
\documentclass[a4paper,11pt]{article}
\makeatletter\AtBeginDocument{\makeatother}
\usepackage{expl3}
\usepackage[T1]{fontenc}
\usepackage[utf8]{inputenc}
\usepackage{array,booktabs,multirow,graphicx,xcolor,lmodern,microtype,csquotes,amsmath,amsfonts,amssymb,amsthm,mathtools,multicol,geometry,etoolbox}
\usepackage[smaller]{acronym}
\newcommand{\IfAcUsed}[1]{\begingroup\expandafter\ifx\csname AC@\AC@prefix#1\endcsname\AC@used\endgroup\expandafter\@firstoftwo\else\endgroup\expandafter\@secondoftwo\fi}

\usepackage[backend = bibtex, style = numeric-comp, bibstyle = numeric, maxnames = 10, giveninits=true, sorting = none, block = ragged]{biblatex}
\renewcommand{\bibfont}{\scriptsize}
\DeclareNameFormat{author}{%
  \usebibmacro{name:family}%
    {\namepartfamily}{\namepartgiveni}{\namepartprefix}{\namepartsuffix}%
  \usebibmacro{name:andothers}}
\newif\ifurldone
\DeclareFieldFormat{eprint}{\global\urldonetrue\href{https://arxiv.org/abs/#1}{\nolinkurl{#1}}}
\DeclareBibliographyDriver{article}{%
  \usebibmacro{bibindex}%
  \usebibmacro{begentry}%
  \usebibmacro{author/translator+others}%
  \setunit{\printdelim{nametitledelim}}
  \usebibmacro{title}%
  \newunit\newblock
  \global\urldonefalse
  \iftoggle{bbx:eprint}{\printfield{eprint}}{}%
  \ifurldone\else
    \iftoggle{bbx:doi}{\printfield{doi} }{}%
    \iftoggle{bbx:url}{\usebibmacro{url+urldate}}{}%
  \fi
  \unskip
  \usebibmacro{finentry}}
\DeclareBibliographyAlias{book}{article}
\DeclareBibliographyAlias{booklet}{article}
\DeclareBibliographyAlias{collection}{article}
\DeclareBibliographyAlias{inbook}{article}
\DeclareBibliographyAlias{incollection}{article}
\DeclareBibliographyAlias{inproceedings}{article}
\DeclareBibliographyAlias{misc}{article}
\DeclareBibliographyAlias{thesis}{article}
\ExplSyntaxOn
\cs_generate_variant:Nn \tl_replace_all:Nnn { Nx }
\DeclareFieldInputHandler{title}
  {
    \tl_set:No \NewValue { \use:n #1 }
    \tl_replace_all:Nnn \NewValue { ^ } { \my_sp: }
    \tl_replace_all:Nxn \NewValue
      { \char_generate:nn { `_ } { 8 } } { \my_sb: }
  }
\cs_new_protected:Npn \my_sp:
  { \mode_if_math:TF { \sp } { \textsuperscript } }
\cs_new_protected:Npn \my_sb:
  { \mode_if_math:TF { \sb } { \textsubscript } }
\ExplSyntaxOff
\DeclareUnicodeCharacter{00D7}{\ensuremath{\times}}
\DeclareUnicodeCharacter{03A9}{\ensuremath{\Omega}}
\DeclareUnicodeCharacter{03B5}{\ensuremath{\epsilon}}
\DeclareUnicodeCharacter{03C4}{\ensuremath{\tau}}
\DeclareUnicodeCharacter{2212}{\ensuremath{-}}

\let \oldOmega = \Omega
\renewcommand{\Omega}{\ensuremath{\oldOmega}}
\let \oldbeta = \beta
\renewcommand{\beta}{\ensuremath{\oldbeta}}
\let \oldepsilon = \epsilon
\renewcommand{\epsilon}{\ensuremath{\oldepsilon}}
\newcommand{\mathscr}{\mathcal}

\usepackage{tikz}
\newcommand{\mathtikz}[2][]{\ensuremath{\vcenter{\hbox{\begin{tikzpicture}[#1]#2\end{tikzpicture}}}}} 
\newcommand{\quiver}[2][]{\mathtikz[semithick,node distance=3em,#1]{#2}} 
\usetikzlibrary{automata, arrows, calc, decorations.markings, decorations.pathreplacing, intersections,
  patterns, positioning, shapes.geometric, shapes.misc,topaths}
\tikzset{->-/.style = {decoration = {markings, mark = at position #1 with {\arrow{>}}}, postaction = {decorate}}}
\tikzset{color-group/.style  = {shape = rounded rectangle, minimum size = 2.5ex, inner sep = .5ex,     draw}}
\tikzset{flavor-group/.style = {shape = rectangle, minimum size = 2.5ex, inner sep = .5ex,     draw}}
\tikzset{cf-group/.style     = {shape = rounded rectangle, rounded rectangle right arc = none, draw}}
\tikzset{fc-group/.style     = {shape = rounded rectangle, rounded rectangle left arc  = none,  draw}}
\tikzset{cross/.style        = {minimum width=1pt, path picture={
      \draw[black, very thick]
               (path picture bounding box.south east) -- (path picture bounding box.north west)
               (path picture bounding box.south west) -- (path picture bounding box.north east);
          }}}
\usepackage{enumitem,hyperref,bookmark,cleveref}
\hypersetup{
  unicode,
  bookmarksnumbered,
  linktoc = all,
  hidelinks}

\theoremstyle{plain}
\newtheoremstyle{exercisestyle}{3pt}{3pt}{\slshape}{}{\normalfont\bfseries}{.}{.5em}{}%
\theoremstyle{exercisestyle}
\newtheorem{exercise}{Exercise}[section]

\DeclareMathOperator{\Vir}{Vir}
\DeclareMathOperator{\AdS}{AdS}

\DeclareMathOperator{\Pfaff}{Pfaff}

\DeclareMathOperator{\Tr}{Tr}
\let\Re\relax\let\Im\relax
\DeclareMathOperator{\Re}{Re}
\DeclareMathOperator{\Im}{Im}
\DeclareMathOperator{\Pexp}{Pexp}

\DeclareMathOperator{\rank}{rank}
\DeclareMathOperator{\ad}{ad}

\DeclarePairedDelimiter{\abs}{\lvert}{\rvert}
\DeclarePairedDelimiter{\vev}{\langle}{\rangle}
\DeclarePairedDelimiter{\ket}{\lvert}{\rangle}
\DeclareMathOperator{\diag}{diag}

\newcommand{\Chat}{\widehat{C}}
\newcommand{\ghat}{\widehat{g}}
\newcommand{\Vhat}{\widehat{V}}
\newcommand{\ZZ}  {{\mathbb{Z}}}
\newcommand{\RR}  {{\mathbb{R}}}
\newcommand{\RP}  {{\mathbb{RP}}}
\newcommand{\CC}  {{\mathbb{C}}}
\newcommand{\LL}  {{\mathbb{L}}}
\newcommand{\fA}  {{\mathfrak{A}}}
\newcommand{\cA}  {{\mathcal{A}}}
\newcommand{\cAbar}  {\overline{\mathcal{A}}}
\newcommand{\cB}  {{\mathcal{B}}}
\newcommand{\cI}  {{\mathcal{I}}}
\newcommand{\cL}  {{\mathcal{L}}}
\newcommand{\cM}  {{\mathcal{M}}}
\newcommand{\Fcal}  {{\mathcal{F}}}
\newcommand{\Mcal}  {{\mathcal{M}}}
\newcommand{\Ocal}  {{\mathcal{O}}}
\newcommand{\cO}  {{\mathcal{O}}}
\newcommand{\cW}  {{\mathcal{W}}}
\newcommand{\Nsusy}{\texorpdfstring{{\mathcal{N}}}{N}}
\newcommand{\del}{\partial}
\newcommand{\delbar}{\bar{\partial}}
\newcommand{\Phibar}{\overline{\Phi}}
\newcommand{\lie}[1]{\texorpdfstring{\mathfrak{#1}}{#1}}
\newcommand{\Lie}[1]{\texorpdfstring{\mathrm{#1}}{#1}}
\newcommand{\SU}{\Lie{SU}}
\newcommand{\USp}{\Lie{USp}}
\newcommand{\SO}{\Lie{SO}}
\newcommand{\Spin}{\Lie{Spin}}
\newcommand{\rep}[1]{\texorpdfstring{\mathbf{#1}}{#1}}
\newcommand{\Rsymm}{{\textnormal{R}}}
\newcommand{\Lag}{{\mathcal{L}}}
\newcommand{\old}{{\textnormal{old}}}
\newcommand{\twist}{{\textnormal{twist}}}
\newcommand{\vect}{{\textnormal{vector}}}
\newcommand{\hyper}{{\textnormal{hyper}}}

\newcommand{\bosonic}{{\textnormal{bosonic}}}

\newcommand{\gauged}{{\textnormal{gauged}}}
\newcommand{\pert}{{\textnormal{pert}}}
\newcommand{\cl}{{\textnormal{cl}}}
\newcommand{\oneloop}{{\textnormal{one-loop}}}
\newcommand{\inst}{{\textnormal{inst}}}
\newcommand{\vort}{{\textnormal{vort}}}
\newcommand{\Ztop}{Z_{\textnormal{top}}}
\newcommand{\Liouville}{{\textnormal{Liouville}}}
\newcommand{\Toda}{{\textnormal{Toda}}}
\newcommand{\SYM}{{\textnormal{\acs{SYM}}}}
\newcommand{\IR}{\texorpdfstring{{\textnormal{\ac{IR}}}}{IR}}

\newcommand{\CP}{{\mathbb{CP}}}
\newcommand{\rmd}{{\mathrm{d}}}
\newcommand{\rmD}{{\mathrm{D}}}
\newcommand{\tauLag}{\tau_{\textnormal{Lag}}}
\newcommand{\qLag}{q_{\textnormal{Lag}}}
\newcommand{\tauIR}{\tau_{\IR}}
\newcommand{\gfourd}{g_{\textnormal{4d}}}
\newcommand{\gfived}{g_{\textnormal{5d}}}
\newcommand{\xibar}{\overline{\xi}}
\newcommand{\kbar}{\overline{k}}
\newcommand{\qbar}{\overline{q}}
\newcommand{\zbar}{\overline{z}}
\newcommand{\sigmabar}{\overline{\sigma}}
\newcommand{\Hcal}{\mathcal{H}}
\newcommand{\SHc}{\ensuremath{\mathrm{SH}^c}}
\newcommand{\fullC}{{\overline{C}}}
\newcommand{\Xg}{\mathcal{X}(\lie{g})}
\newcommand{\Xsuii}{\mathcal{X}(\lie{su}(2))}
\newcommand{\XsuN}{\mathcal{X}(\lie{su}(N))}
\newcommand{\Theory}{\mathrm{T}}
\newcommand{\TgCD}{\Theory(\lie{g},C,D)}

\newcommand{\mystarsec}[2]{#1*{\texorpdfstring{\phantomsection
  \addcontentsline{toc}{\csname @gobble\expandafter\endcsname\string#1}{#2}}{}#2}}

\ifjournal
\newenvironment{apartetable}[2]{\table\centering\caption{\label{#1}#2}\smallskip}{\endtable}
\newcommand\apartehere{}
\newcommand\startaparte[1][]{}
\newcommand\stopaparte{}
\else
\usepackage{tcolorbox,wrapfig}
\newtcolorbox[blend into=tables]{apartetableAux}[2][]{float=htb, halign=center, fonttitle=\bfseries\boldmath, title={#2}, every float=\centering, left=1mm, right=1mm, #1}
\newenvironment{apartetable}[2]{\apartetableAux[label=#1]{#2}\vspace*{-1ex}}{\vspace*{-1ex}\endapartetableAux}
\long\def\apartehere#1\startaparte#2\stopaparte{\begin{aparte}#2\end{aparte}#1}
\newenvironment{aparte}[1][]{%
  \setlist[itemize]{leftmargin=0pt,
    itemindent=10pt,labelsep=3pt,
    itemsep=.1ex}%
  \wrapfigure{r}{.7\linewidth}%
  \begin{tcolorbox}[fonttitle=\bfseries\boldmath, left=1mm, right=1mm, #1]%
}{\end{tcolorbox}\vspace*{-\intextsep}\endwrapfigure}
\newcommand{\startaparte}{\begin{aparte}}
\newcommand{\stopaparte}{\end{aparte}}
\fi

\numberwithin{equation}{section}

\hypersetup{
  pdftitle = {A slow review of the AGT correspondence},
  pdfauthor = {Bruno Le Floch}}
\title{A slow review of the \acs{AGT} correspondence}
\author{Bruno Le Floch\thanks{CNRS, Laboratoire de Physique Théorique et Hautes Energies, Sorbonne Universit\'e, Paris, France}}
\date{January 2022}

\begin{document}

\acused{BPS}

\maketitle

\begin{abstract}
  Starting with a gentle approach to the \ac{AGT} correspondence from its 6d origin, these notes provide a wide (albeit shallow) survey of the literature on numerous extensions of the correspondence up to early 2020.
  This is an extended writeup of the lectures given at the Winter School \href{https://indico.desy.de/indico/event/23820/}{``\acs{YRISW} 2020''} to appear in a special issue of JPhysA.

  Class~S is a wide class of 4d $\Nsusy=2$ supersymmetric gauge theories (ranging from super-QCD to non-Lagrangian theories) obtained by twisted compactification of 6d $\Nsusy=(2,0)$ superconformal theories on a Riemann surface~$C$.
  This 6d construction yields the Coulomb branch and \acl{SW} geometry of class~S theories, geometrizes S-duality,
  and leads to the \ac{AGT} correspondence, which states that many observables of class~S theories are equal to 2d \ac{CFT} correlators.
  For instance, the four-sphere partition function of a 4d $\Nsusy=2$ $\SU(2)$ superconformal quiver theory is equal to a Liouville \ac{CFT} correlator of primary operators.

  Extensions of the \ac{AGT} correspondence abound:
  asymptotically-free gauge theories and \acl{AD} theories correspond to irregular \ac{CFT} operators,
  quivers with higher-rank gauge groups and non-Lagrangian tinkertoys such as~$T_N$ correspond to Toda \ac{CFT} correlators,
  and nonlocal operators (Wilson--'t~Hooft loops, surface operators, domain walls) correspond to
  Verlinde networks, degenerate primary operators, braiding and fusion kernels, and Riemann surfaces with boundaries.
\end{abstract}

\clearpage
\setlength{\columnsep}{3em} 
\footnotesize
\begin{multicols}{2}[\section*{\contentsname}]
  \makeatletter
  \@starttoc {toc}
\end{multicols}
\normalsize
\setlength{\columnsep}{1em} 

\clearpage

\section{\label{sec:intro}Introduction and outline}

\Acp{QFT} arise from many different constructions, be it Lagrangian descriptions, dimensional reduction or geometric engineering.  The resulting building blocks can then be further deformed (e.g.\ partially Higgsed), coupled (e.g.\ by gauging symmetries), or reduced by decoupling a subsector.
Theories living in different dimensions can also be fruitfully coupled together.

We explore these constructions, and some computation techniques, in the world of 4d $\Nsusy=2$ supersymmetric theories, specifically class~S theories~\cite{0904.2715} which are dimensional reduction of a 6d theory (``S'' stands for ``Six'').
Class~S includes the most commonly studied 4d $\Nsusy=2$ Lagrangian gauge theories (\ac{SYM}, \ac{SQCD}, quiver gauge theories, $\Nsusy=4$ \ac{SYM} and its mass deformation) and non-Lagrangian ones such as \ac{AD} theories~\cite{hep-th/9505062}, but also a plethora of previously unknown ones that have considerably broadened the set of known 4d $\Nsusy=2$ theories.

To construct a class~S theory we start from a 6d $\Nsusy=(2,0)$ \ac{SCFT} denoted by~$\Xg$, which is characterized by a simply-laced\footnote{A simple Lie algebra~$\lie{g}$ is simply-laced if all its roots have the same length.  Such algebras have an ADE classification: concretely, $\lie{g}$ is one of $\lie{su}(N)$, $\lie{so}(2N)$, $\lie{e}_6$, $\lie{e}_7$, or~$\lie{e}_8$.} Lie algebra~$\lie{g}$, for instance $\lie{su}(N)$.
We then reduce $\Xg$ on a Riemann surface~$C$ called the \ac{UV} curve\footnote{The two-dimensional Riemann surface~$C$ is a complex curve: it has complex dimension~$1$.}, while preserving 4d $\Nsusy=2$ supersymmetry thanks to a procedure called partial topological twist.
The Riemann surface can have punctures (removed points, so that $C=\fullC\setminus\{z_1,\dots,z_n\}$ with $\fullC$~being compact) at which boundary conditions must be prescribed.
Each choice of punctured Riemann surface, and data~$D_i$ describing the boundary condition at~$z_i$, leads to one 4d $\Nsusy=2$ class~S theory $\TgCD$.

Due to their 6d origin, nonperturbative dynamics of class~S theories are encoded in the geometry of~$C$.  For example the \ac{SW} curve~\cite{hep-th/9407087,hep-th/9408099} of a theory, which determines the low-energy effective action in a given Coulomb branch vacuum, is a branched cover of~$C$.  Strikingly, this idea extends to many observables of the class~S theory.
The \ac{AGT} correspondence~\cite{0906.3219} concerns the four-sphere (and ellipsoid) partition function:
\begin{equation}\label{AGT-ZS4}
  Z_{S^4_b}(\TgCD) = \vev*{\Vhat_{D_1}(z_1)\dots \Vhat_{D_n}(z_n)}_{\fullC}^{\Toda(\lie{g})}
\end{equation}
where the right-hand side is a correlator of vertex operators in the Liouville \ac{CFT} (for $\lie{g}=\lie{su}(2)$) or its generalization, Toda \ac{CFT}.
The vertex operators are inserted at each puncture~$z_i$ and depend on the data~$D_i$ characterizing punctures.

The rest of the introduction summarizes this review quickly: the reader should feel free to \hyperref[sec:6d]{skip to the main text}.
Sections~\ref{sec:6d}, \ref{sec:reduce}, and~\ref{sec:Lag} (summarized in \autoref{ssec:intro-S}) describe the theories~$\TgCD$ and the puncture data~$D_i$.
Sections~\ref{sec:loc} and~\ref{sec:AGT} (summarized in \autoref{ssec:intro-AGT}) explain how to define and compute both sides of~\eqref{AGT-ZS4}, namely $Z_{S^4_b}$ and Liouville \ac{CFT} correlators.
Finally, sections~\ref{sec:gen}, \ref{sec:ops}, \ref{sec:dim}, and~\ref{sec:con} describe numerous extensions of the correspondence, with pointers to the literature.
In \autoref{ssec:intro-reviews} we present the preexisting reviews on topics related to \ac{AGT}.\ifjournal\else\footnote{References (out before January 31, 2020) and comments on more recent developments welcome.}\fi

\subsection{\label{ssec:intro-S}Class~S theories}

In the main text we study the 6d $(2,0)$ theory~$\Xg$ (\autoref{sec:6d}), its twisted dimensional reductions to class~S theories (\autoref{sec:reduce}), and Lagrangian descriptions of some of these 4d $\Nsusy=2$ theories (\autoref{sec:Lag}).  Here we only give some outcomes of these discussions.  We often reduce to $\lie{g}=\lie{su}(N)$ for simplicity, but extensions to $\lie{g}=\lie{so}(2N)$ are also well-understood~\cite{0905.4074,0906.0359}.

\paragraph{Building blocks for~\(\TgCD\).}
A Riemann surface $C$ of genus~$g$ with $n$ punctures can be cut into $2g-2+n$ three-punctured spheres, also called trinions or pairs of pants\footnote{That number is zero or negative for the sphere with $0$, $1$ or~$2$ punctures and the torus with no punctures: these Riemann surfaces cannot be cut into three-punctured spheres, and the class~S construction does not give a 4d theory, see~\cite{1110.2657}.} glued together by tubes that connect pairs of punctures.
Such a description is often called a pants decomposition of~$C$.
Correspondingly, the general class~S theory $\TgCD$ can be decomposed into class~S theories called \emph{tinkertoys} that correspond to each three-punctured sphere (tinkertoys range in complexity from free hypermultiplets to previously unknown non-Lagrangian isolated \acp{SCFT}).
Each puncture is associated to a flavour symmetry, and connecting two punctures by a tube amounts to identifying the two associated flavour symmetries and gauging them using the same 4d $\Nsusy=2$ vector multiplet.
For instance a four-punctured sphere can be split into two three-punctured spheres (for suitable groups $G_1,\dots,G_5$):
\begin{equation}\label{intro-split}
  \Theory\Biggl(
  \mathtikz[semithick]{
    \draw (0,-.2) -- (1.7,-.2);
    \draw (0,-.1) circle (.05 and .1) node [above] {$\scriptstyle G_1$};
    \draw (0,0) arc (-90:0:.6 and .8);
    \draw (.7,.8) circle (.1 and .05) node [left] {$\scriptstyle G_2$};
    \draw (1.7,0) -- (1.4,0) arc (270:180:.6 and .8);
    \draw (1.7,-.1) circle (.05 and .1);
    \draw (1.7,-.2) -- (3.4,-.2);
    \draw (1.7,0) -- (2,0) arc (-90:0:.6 and .8);
    \draw (2.7,.8) circle (.1 and .05) node [right] {$\scriptstyle G_3$};
    \draw (3.4,0) arc (270:180:.6 and .8);
    \draw (3.4,-.1) circle (.05 and .1) node [above] {$\scriptstyle G_4$};
  }
  \Biggr)
  =
  \Theory\Biggl(
  \mathtikz[semithick]{
    \draw (0,-.2) -- (1.4,-.2);
    \draw (0,-.1) circle (.05 and .1) node [above] {$\scriptstyle G_1$};
    \draw (0,0) arc (-90:0:.6 and .8);
    \draw (.7,.8) circle (.1 and .05) node [left] {$\scriptstyle G_2$};
    \draw (1.4,0) arc (270:180:.6 and .8);
    \draw (1.4,-.1) circle (.05 and .1) node [above] {$\scriptstyle G_5$};
  }
  \Biggr)
  \mathrel{\mathop{\otimes}\limits_{\textnormal{gauge }G_5}}
  \Theory\Biggl(
  \mathtikz[semithick]{
    \draw (2,-.1) circle (.05 and .1) node [above] {$\scriptstyle G_5$};
    \draw (2,-.2) -- (3.4,-.2);
    \draw (2,0) arc (-90:0:.6 and .8);
    \draw (2.7,.8) circle (.1 and .05) node [right] {$\scriptstyle G_3$};
    \draw (3.4,0) arc (270:180:.6 and .8);
    \draw (3.4,-.1) circle (.05 and .1) node [above] {$\scriptstyle G_4$};
  }
  \Biggr) .
\end{equation}
In simple cases where tinkertoys are collections of hypermultiplets, this results in gauge theories with an explicit Lagrangian made of hypermultiplets and vector multiplets.
Thanks to the partial topological twist, the 4d theory does not depend on the metric of~$C$~\cite{1109.3724,1110.2657} but only on the complex structure of~$C$, which can be described by the ``length'' and ``angle'' of each tube.
These two parameters control the complexified gauge coupling ($q=e^{2\pi i\tau}$ with $\tau=\frac{\theta}{2\pi}+\frac{4\pi i}{g^2}$) that combines the Yang--Mills coupling~$g$ with the theta angle~$\theta$ of the 4d vector multiplet corresponding to the tube.  Weak coupling $g\to 0$ corresponds to a very long tube.

Of course, $C$~can be decomposed in many ways into three-punctured spheres: correspondingly, $\TgCD$ has many equivalent dual descriptions involving completely different sets of fields and gauge groups.
The weak gauge coupling regime of these descriptions correspond to regimes where the complex structure on~$C$ is well-described by one pants decomposition where three-punctured spheres are joined by very long tubes.
These regimes, which are different cusps of the space of complex structures on~$C$, are continuously connected by varying the gauge couplings.
In this way, gauge theories at strong coupling in one description may admit a different weakly-coupled description.
This phenomenon~\cite{0904.2715} generalizes S-duality of the $\SU(2)$ $N_f=4$ theory and of $\Nsusy=4$ \ac{SYM}.  The 6d construction thus makes these S-dualities manifest through~$C$.

In the 6d construction, the punctures at $z_i\in\fullC$ are codimension~$2$ defects that wrap the 4d spacetime on which the class~S theory is defined.
To preserve supersymmetry of the 4d theory the defects should be half-\mbox{\phantomsection\label{acro:BPS}\ac{BPS}}, namely preserve half of the original supersymmetry.
One must classify such defects (typically by moving along the Coulomb branch), and then the tinkertoys corresponding to three-punctured spheres.
Incidentally, the 6d theory also admits interesting half-\ac{BPS} codimension~$4$ operators supported on 2d subspaces, which enrich the correspondence.

\paragraph{Coulomb branch and \acl{SW} curve.}
One way to get a handle on the theory $\TgCD$ is to describe its supersymmetric vacua, especially its Coulomb branch, and give the low-energy behaviour of the theory near each vacuum.
This branch is spanned by Coulomb branch operators, namely local operators annihilated by all 4d antichiral supercharges.

Semiclassically, vacua of the 6d $(2,0)$ theory~$\Xg$ are parametrized modulo gauge transformations by some (commuting, diagonalizable) adjoint-valued scalars~$\Phi_I$, where $I=6,\dots,10$ is an index for the $\lie{so}(5)$ R-symmetry\footnote{In the M-theory construction of the 6d theory, $\lie{so}(5)$ rotates coordinates $x^6,\dots,x^{10}$.}.  Alternatively they are parametrized by (consistent) values for gauge-invariant polynomials (Casimirs) such as $\Tr(\Phi_I\Phi_J)$.
Coulomb branch vacua of the 4d theory are then configurations of the~$\Phi_I$ (or rather of the invariant polynomials) allowed to vary along the curve~$C$.
More precisely, tracking down how 4d $\Nsusy=2$ antichiral supercharges embed into 6d $\Nsusy=(2,0)$, we find two restrictions: the Casimirs depend holomorphically on the coordinate $z\in C$, and among all the $\Phi_I$ only $\Phi_z\coloneqq\Phi_6+i\Phi_7$ is non-zero.
Because the partial topological twist mixes a subalgebra $\lie{so}(2)$ of R-symmetry (under which $\Phi_z$ is charged) into the rotation group on~$C$, $\Phi_z\rmd z$~is tensorial (specifically a one-form) on~$C$.
Roughly speaking, then, the 4d Coulomb branch is parametrized by the adjoint-valued holomorphic one-form $\Phi_z\rmd z$ on~$C$ modulo gauge transformations, and more invariantly by \acp{VEV} $\vev{P_k(\Phi_z)}\rmd z^k$ of Casimir polynomials.

In the $\lie{g}=\lie{su}(N)$ case we repackage them as $\phi_k(z)=u_k(z)\rmd z^k$, $k=2,\dots,N$, defined in local coordinates $z\in C$ by expanding
\begin{equation}\label{intro-uk}
  \vev{\det(x-\Phi_z)} = x^N + \sum_{k=2}^N u_k(z) x^{N-k} .
\end{equation}
It is then useful to consider the zeros of this determinant
\begin{equation}\label{intro-SW}
  \Sigma = \biggl\{(z,x)\biggm| x^N + \sum_{k=2}^N u_k(z) x^{N-k} = 0\biggr\} \subset T^*C ,
\end{equation}
where $z$ is a coordinate on~$C$ and $x$ parametrizes the fiber of the cotangent bundle~$T^*C$, the bundle of one-forms\footnote{This just means $x\rmd z$ transforms as a tensor under changing the coordinate $z$ on~$C$.} on~$C$.
The complex curve $\Sigma$ depends on the choice of vacuum (specified by $\phi_2,\dots,\phi_N$) and turns out to be the \ac{SW} curve of $\TgCD$, presented as an $N$-fold (ramified) cover of~$C$.
It is equipped with a natural one-form $\lambda=x\rmd z$, the \ac{SW} differential.\footnote{Equation~\eqref{intro-SW} is often reformulated as $\lambda^N + \sum_{k=2}^N \phi_k(z) \lambda^{N-k} = 0$.}
From the \ac{SW} curve and differential $(\Sigma,\lambda)$ of $\TgCD$ in a given Coulomb branch vacuum one can derive the infrared effective action (the prepotential).
Masses of \ac{BPS} particles can also be extracted as integrals of~$\lambda$ along closed contours.

\paragraph{Tame punctures and tinkertoys.}
A puncture at $z_i\in\fullC$ is described in this language as a singularity of the gauge-invariants~$\phi_k$.  An important example is the \emph{full tame puncture} which imposes a first order pole $\Phi_z\sim m_i(z-z_i)^{-1}\rmd z+O(1)$ at~$z_i$, up to conjugation, where the residue $m_i\in\lie{g}_{\CC}$ is a suitably generic element of the complexification~$\lie{g}_{\CC}$ of~$\lie{g}$.
This mass\footnote{For $\lie{su}(N)$, the $N$ eigenvalues of~$m_i$ give residues of~$\lambda$ at each of the $N$~points of~$\Sigma$ projecting to~$z_i$.  Integrating~$\lambda$ to compute masses of \ac{BPS} particles picks up such residues, which are thus mass parameters.} parameter~$m_i$ can be understood as a constant value for the background vector multiplet scalar that couples to the flavour symmetry~$\lie{g}$ corresponding to the puncture at~$z_i$.
In gauge-invariant terms this first order pole translates to
\begin{equation}\label{intro-tame}
  \vev{P_k(\Phi_z)} = \frac{P_k(m_i)}{(z-z_i)^k} + \dots ,
\end{equation}
or equivalently $\phi_k \propto \rmd z^k/(z-z_i)^k + \dots$ with a leading-order coefficient determined from~$m_i$, using~\eqref{intro-uk} in the $\lie{su}(N)$ case.

In fact, when $C$~gets pinched and split into two in the limit where a tube becomes infinitely thin, this type of singularity generically occurs.
The main building block of class~S theories is thus \emph{the tinkertoy~$T_{\lie{g}}$}, namely the theory associated to a sphere with three full tame punctures.
A frequent notation is $T_N\coloneqq T_{\lie{su}(N)}$.
By matching \ac{SW} curves and \ac{SW} differentials of $\Theory(\lie{su}(2),C,D)$ theories to previously known theories such as $\SU(2)$ $N_f=4$ \ac{SQCD}, one checks that $T_2$~is simply a collection of $4$~free hypermultiplets~\cite{0904.2715}.
In general, however, the theory $T_{\lie{g}}$ is a non-Lagrangian theory, with (at least) one flavour symmetry $\lie{g}$ for each puncture.
For instance, $T_{\lie{su}(3)}$~is the Minahan--Nemeschansky \ac{SCFT} with flavour symmetry $\lie{e}_6\supset\lie{su}(3)^3$.

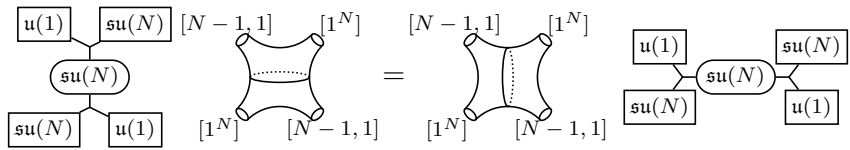
\begin{figure}\centering
  \begin{tikzpicture}[semithick]
    \begin{scope}[xshift=-4cm]
      \node(topL)[flavor-group] at (-.6,.7) {\scriptsize $\lie{u}(1)$};
      \node(topR)[flavor-group] at (.6,.7) {\scriptsize $\lie{su}(N)$};
      \node(mid)[color-group] at (0,0) {\scriptsize $\lie{su}(N)$};
      \node(botL)[flavor-group] at (-.6,-.7) {\scriptsize $\lie{su}(N)$};
      \node(botR)[flavor-group] at (.6,-.7) {\scriptsize $\lie{u}(1)$};
      \draw(topL)--(0,.4)--(topR);
      \draw(0,.4)--(mid)--(0,-.4);
      \draw(botL)--(0,-.4)--(botR);
    \end{scope}
    \begin{scope}[xshift=-1.5cm]
      \begin{scope}[rotate=45]
        \foreach\pm in{+,-}{
          \draw (\pm.7,0) circle (.05 and .1);
          \draw (0,\pm .7) circle (.1 and .05);
          \draw (-.7,-\pm.1) arc (\pm 90:0:.6 and .6);
          \draw (.7,-\pm.1) arc (\pm 90:\pm 180:.6 and .6);
        }
      \end{scope}
      \foreach\pm in {+,-}{
        \node at (\pm .8,\pm .7) {\scriptsize $[1^N]$};
        \node at (\pm .7,-\pm .7) {\scriptsize $[N-1,1]$};
      }
      \draw[densely dotted] (.39,0) arc (0:180:.39 and .07);
      \draw (.39,0) arc (0:-180:.39 and .07);
    \end{scope}
    \node {$=$};
    \begin{scope}[xshift=1.5cm]
      \begin{scope}[rotate=45]
        \foreach\pm in{+,-}{
          \draw (\pm.7,0) circle (.05 and .1);
          \draw (0,\pm .7) circle (.1 and .05);
          \draw (-.7,-\pm.1) arc (\pm 90:0:.6 and .6);
          \draw (.7,-\pm.1) arc (\pm 90:\pm 180:.6 and .6);
        }
      \end{scope}
      \foreach\pm in {+,-}{
        \node at (\pm .8,\pm .7) {\scriptsize $[1^N]$};
        \node at (\pm .7,-\pm .7) {\scriptsize $[N-1,1]$};
      }
      \draw[densely dotted] (0,.39) arc (90:-90:.07 and .39);
      \draw (0,.39) arc (90:270:.07 and .39);
    \end{scope}
    \begin{scope}[xshift=4.5cm]
      \node(lt)[flavor-group] at (-1,.4) {\scriptsize $\lie{u}(1)$};
      \node(lb)[flavor-group] at (-1,-.4) {\scriptsize $\lie{su}(N)$};
      \node(c)[color-group] at (0,0) {\scriptsize $\lie{su}(N)$};
      \node(rt)[flavor-group] at (1,.4) {\scriptsize $\lie{su}(N)$};
      \node(rb)[flavor-group] at (1,-.4) {\scriptsize $\lie{u}(1)$};
      \draw(lt)--(-.7,0)--(lb);
      \draw(-.7,0)--(c)--(.7,0);
      \draw(rt)--(.7,0)--(rb);
    \end{scope}
  \end{tikzpicture}
  \caption{\label{fig:SUN-SQCD}The $\lie{su}(N)$ class~S theory corresponding to a sphere with two full tame punctures (labelled $[1^N]$, flavour symmetry $\lie{su}(N)$) and two simple tame punctures (labelled $[N-1,1]$, symmetry $\lie{u}(1)$).  We depict two pants decompositions constructed from spheres with one simple and two full punctures, whose corresponding tinkertoy is a collection of hypermultiplets.  The decompositions lead to two S-dual Lagrangian descriptions of the theory as $\SU(N)$ \ac{SQCD} with $N_f=2N$.  The third pants decomposition (not depicted here) involves non-Lagrangian tinkertoys (for $N>2$).}
\end{figure}

There are more general \emph{tame punctures}, defined as points where one imposes a first order pole of~$\Phi_z$ with a residue~$m$ that may be non-generic.
The resulting tinkertoys amount to a \emph{partial Higgsing}: moving onto the Higgs branch of~$T_{\lie{g}}$ by turning on a nilpotent \ac{VEV} for (the moment map of) the symmetry carried by the puncture, thus reducing the symmetry.
For $\lie{su}(N)$ they are characterized by the pattern of equal eigenvalues of~$m$, encoded as a partition of~$N$, and they lead to lower-order poles for the~$\phi_k$.
The partition for a full tame puncture is $N=1+1+\dots+1$, also denoted by~$[1^N]$; it carries $\lie{su}(N)$ flavour symmetry (broken explicitly by the mass~$m$).
At the other extreme, the puncture corresponding to the partition~$[N]$ (all eigenvalues equal, hence vanishing) is a trivial absence of puncture since it is a pole with zero residue.
The next ``smallest'' puncture, called a \emph{simple tame puncture} corresponds to the partition $[N-1,1]$ so $m=\diag(m_1,\dots,m_1,-(N-1)m_1)$; it carries $\lie{u}(1)$ flavour symmetry, enhanced to $\lie{su}(2)$ for $N=2$ since in that case the simple and full punctures are identical.
Both the full and the simple tame punctures appear in the class~S description of $\SU(N)$ $N_f=2N$ \ac{SQCD}, as depicted in \autoref{fig:SUN-SQCD}.

While the gauge algebra carried by each tube is~$\lie{g}$ when all punctures are full tame punctures, more general tame punctures may lead to smaller gauge algebras.
For example, $\lie{su}(N)$ class~S includes \emph{linear quiver gauge theories} with gauge group $\prod_i \SU(N_i)$ (with $N_i\leq N$), one hypermultiplet in each bifundamental representation $N_i\otimes\overline{N_{i+1}}$, and $M_i\leq 2N_i-N_{i-1}-N_{i+1}$ hypermultiplets\footnote{When this bound is saturated the gauge coupling of that group does not run.  When it is obeyed but not saturated (so $M_i < 2N_i-N_{i-1}-N_{i+1}$) we get an asymptotically free gauge theory, which can be realized in class~S using wild punctures.  When the bound is violated instead, the theory is only an effective theory and does not have a class~S construction.} in fundamental representations $N_i$ of each~$\SU(N_i)$.  This is summarized in the quiver diagram
\begin{equation}\label{intro-SU-quiver}
  \quiver{
    \node (A) [color-group] at (0,0) {$\SU(N_1)$};
    \node (B) [color-group] at (1.8,0) {$\SU(N_2)$};
    \node (C) at (3.2,0) {${\cdots}$};
    \node (D) [color-group] at (4.6,0) {$\SU(N_p)$};
    \node (Am) [flavor-group] at (0,.7) {$M_1$};
    \node (Bm) [flavor-group] at (1.8,.7) {$M_2$};
    \node (Dm) [flavor-group] at (4.6,.7) {$M_p$};
    \draw (Am)--(A)--(B)--(C)--(D)--(Dm);
    \draw (B)--(Bm);
  }
\end{equation}

\subsection{\label{ssec:intro-AGT}Basic \acsfont{AGT} correspondence}

We summarize here two sections that build up to the full \ac{AGT} correspondence.
First, \autoref{sec:loc} describes how the (squashed) sphere partition function~$Z_{S^4_b}$ of quiver gauge theories is computed using supersymmetric localization, and especially the issue of instanton counting.
Then, \autoref{sec:AGT} explains basic aspects of Liouville \ac{CFT} and gives the precise statement of the \ac{AGT} correspondence for $\lie{g}=\lie{su}(2)$ generalized quivers.

\paragraph{Supersymmetric localization.}
In \autoref{sec:loc} we explain how to place class~S theories on the (squashed) four-sphere $S^4_b\coloneqq\{y_5^2+b^2(y_1^2+y_2^2)+b^{-2}(y_3^2+y_4^2)=r^2\}\subset\RR^5$ supersymmetrically, which incidentally requires masses to be purely imaginary.
We also explain how to evaluate the partition function on this ellipsoid using supersymmetric localization~\cite{0712.2824,1206.6359}.
This path integral technique applies to each 4d $\Nsusy=2$ Lagrangian description of $\TgCD$---if such a description exists.\footnote{Factorization properties of~$Z_{S^4_b}$ that we find upon cutting the Riemann surface also hold for non-Lagrangian class~S theories.  They are obtained by applying supersymmetric localization to the vector multiplets only, and not to the tinkertoys.}
Supersymmetric localization can reduce the infinite-dimensional path integral down to a finite-dimensional integral over supersymmetric configurations of the hypermultiplets and vector multiplets.
One finds configurations labeled by the (purely imaginary) constant value~$a$ of a vector multiplet scalar, which can be gauge-fixed to lie in the Cartan subalgebra of the gauge algebras.  These configurations are additionally dressed by point-like instantons at one pole ($y_5=r$) and anti-instantons at the other pole ($y_5=-r$) of~$S^4_b$.
The partition function then reads
\begin{equation}\label{ZS4-localized}
  Z_{S^4_b}(q,\qbar) = \int\rmd a\,Z_{\cl}(a,q,\qbar)\,Z_{\oneloop}(a) Z_{\inst}(a,q) Z_{\inst}(a,\qbar) ,
\end{equation}
where we omit the dependence on $\lie{g}$ and data~$D$ at the punctures but write explicitly the dependence on complex structure parameters~$q$ of the curve~$C$.
Here, $Z_{\cl}$~comes from the classical action of supersymmetric configurations; it depends non-holomorphically on the complex gauge couplings~$q$, but factorizes as $Z_{\cl}(a,q,\qbar)=Z_{\cl'}(a,q)Z_{\cl'}(a,\qbar)$.
Quadratic fluctuations around these configurations yield $Z_{\oneloop}(a)$, a straightforward product of special functions that is completely independent of the shape (complex structure) of~$C$.
Finally, (anti)\llap{-}instantons at each pole bring a factor of $Z_{\inst}$ that depends (anti)\llap{-}holomorphically on gauge couplings~$q$.

The factor $Z_{\inst}(a,q)=\sum_{k\geq 0}q^k Z_{\inst,k}(a)=1+O(q^1)$ is Nekrasov's instanton partition function~\cite{hep-th/0206161,hep-th/0306238} with parameters $\epsilon_1=b/r$, $\epsilon_2=1/(rb)$, computable in favourable cases.
The main difficulty is to compute each $k$-instanton contribution $Z_{\inst,k}(a)$,
which is an integral over the $k$-instanton moduli space.
This space is finite-dimensional but very singular, and its singularities are understood best for unitary gauge groups.
For linear quivers of unitary groups, which are obtained from~\eqref{intro-SU-quiver} by replacing all $\SU(N_i)$ gauge groups by $\Lie{U}(N_i)$, the Nekrasov partition function can be determined by equivariant localization or through IIA brane constructions.
The instanton partition function of the $\SU$ theories~\eqref{intro-SU-quiver} that we care about can then be derived by an appropriate decoupling of the $\Lie{U}(1)$ factors, which divides $Z_{\inst}(a,q)$ by simple factors such as powers of $(1-q)$~\cite{0906.3219}.
Various other methods have been devised, but there is as of yet no complete first principles derivation of~$Z_{\inst}$ for general class~S theories, and even when restricting to $\lie{g}=\lie{su}(2)$ with tame punctures.\footnote{I thank Jaewon Song for clarifications on this point.}

S-dual Lagrangian descriptions of the same theory, obtained by different pants decompositions of~$C$, should have the same partition function if S-duality is to hold.
The equality of explicit integral expressions~\eqref{ZS4-localized} is extremely challenging to prove, even for the $\SU(2)$ $N_f=4$ theory.
In fact the easiest way I know is to derive the \ac{AGT} correspondence in that case (e.g.~\cite{1012.1312}) and then rely on modularity properties on the 2d \ac{CFT} side shown in~\cite{hep-th/0104158,hep-th/0303150,0803.0919}.

\paragraph{Liouville \acs{CFT} correlators and basic \acs{AGT} correspondence.}

In \autoref{sec:AGT} we move on to the other side of the correspondence for $\lie{g}=\lie{su}(2)$, namely Liouville \ac{CFT} correlators.
Liouville \ac{CFT} depends on a ``background charge'' $Q=b+1/b\geq 2$ (the central charge is $c=1+6Q^2\geq 25$), which translates on the 4d side to a deformation parameter of $S^4$~into the ellipsoid~$S^4_b$.
As in any 2d \ac{CFT}, local operators organize into conformal families constructed by acting with the Virasoro algebra on primary operators.
In the Liouville \ac{CFT} these primaries are the vertex operators~$\Vhat_\alpha$, labeled by a continuous parameter $\alpha=Q/2+iP$ with $P\in\RR/\ZZ_2$ (called momentum), and they have equal holomorphic and antiholomorphic dimension $h(\alpha)=\alpha(Q-\alpha)=Q^2/4+P^2$.
In the $\lie{su}(2)$ case the data $D_j$ for each tame puncture reduces to specifying a mass $m_j\in i\RR/\ZZ_2$ (imaginary), naturally identified with a Liouville momentum (up to the sphere's radius~$r$): the \ac{AGT} correspondence then states
\begin{equation}\label{AGT-ZS4-Liouville}
  Z_{S^4_b}\bigl(\Theory(\lie{su}(2),C,m)\bigr) = \vev[\big]{\Vhat_{Q/2+rm_1}(z_1)\dots \Vhat_{Q/2+rm_n}(z_n)}_{\fullC}^{\Liouville} .
\end{equation}

As in any 2d \ac{CFT}, $n$-point functions of Virasoro primary operators on the Riemann surface~$\fullC$ have a useful expression for each pants decomposition of the punctured Riemann surface~$C$.
The idea is to insert a complete set of states along each cut in the decomposition, then use Virasoro symmetry to rewrite all resulting three-point functions in terms of those of primaries.  Schematically this gives
\begin{equation}\label{Liouville-into-blocks}
  \vev*{\Vhat_{\mu_1}(z_1,\bar{z}_1)\dots \Vhat_{\mu_n}(z_n,\bar{z}_n)}_{\fullC}^{\Liouville}
  = \int\rmd\alpha\,C(\mu,\alpha) \mathcal{F}(\mu,\alpha,q) \mathcal{F}(\mu,\alpha,\qbar) .
\end{equation}
Here we integrate over all internal momenta~$\alpha$ labelling the conformal family in each inserted complete set of states.
The factor $C(\mu,\alpha)$ is a combination of structure constants of Liouville \ac{CFT}.
The other two factors are conformal blocks, which are purely about representation theory of the Virasoro algebra, and which depend (anti)\llap{-}holomorphically on the complex structure parameters~$q$ of~$C$, including (cross-ratios of)~$z_i$.

Both sides of the \ac{AGT} correspondence admit the same kind of expressions \eqref{ZS4-localized} and~\eqref{Liouville-into-blocks} for each pants decomposition of~$C$, with one integration variable $a$ or~$\alpha$ for each tube, and a factorization of the dependence on~$q$ into holomorphic and antiholomorphic.
In fact these expressions match factor by factor: $Z_{\oneloop}(m,a) = C(\mu,\alpha)$ and $Z_{\cl'}(a,q)Z_{\inst}(m,a,q) = \mathcal{F}(\mu,\alpha,q)$.
An additional entry in the dictionary is that $\phi_2$ on the 4d side corresponds to the holomorphic stress-tensor $T(z)$ on the Liouville side in the classical limit $r\to\infty$: the leading term in an \ac{OPE} with $T(z)$ matches $r^2\phi_2(z)$,
\begin{equation}\label{intro-Ward}
  T(z)\Vhat_\mu(0) = \frac{h(\mu)\Vhat_\mu(0)}{z^2} + \dotsm
  \underset{r\to\infty}{\simeq} - \frac{r^2m^2}{z^2}\Vhat_\mu(0) + \dotsm
  \simeq - r^2 \phi_2(z) \Vhat_\mu(0) + \dots .
\end{equation}

We end \autoref{sec:AGT} by outlining the technical derivation of how Liouville \ac{CFT} appears upon reducing the 6d theory on~$S^4$~\cite{1605.03997}.

\paragraph{Extensions of the \ac{AGT} correspondence.}

The \ac{AGT} correspondence is generalized in two ways in \autoref{sec:gen}.
First, asymptotically free theories and \ac{AD}~theories are described by replacing tame punctures by wild punctures, which replaces primary vertex operators by irregular ones on the \ac{CFT} side.
Second, $\lie{su}(2)$ is replaced by $\lie{g}=\lie{su}(N)$: hypermultiplets are then replaced by non-Lagrangian building blocks~$T_N$ and Liouville \ac{CFT} by Toda \ac{CFT}.

In \autoref{sec:ops} we investigate how to include in the \ac{AGT} correspondence various gauge theory operators (local operators, Wilson--'t~Hooft loops, \dots).  The \ac{CFT} side features Verlinde loops, degenerate vertex operators, fusion and braiding kernels, and Riemann surfaces with boundaries.
The dictionary and references are summarized in \autoref{tab:ops}.

\ifjournal\else\medskip\fi

\begin{apartetable}{tab:ops}{\ac{AGT} correspondence for extended operators, sorted by codimension of the 6d operator or orbifold that yields them, and sorted by dimension on the 4d side.  Most entries are hyperlinked to the main text.}
  \begin{tabular}{c@{\;}c@{\;}m{13em}@{\ \ }m{12.5em}@{\ \ }m{6em}}
    &
    & Operator in class~S theory
    & \hyperref[ssec:AGT-Liou]{Liouville}/\hyperref[ssec:gen-rank]{Toda} \ac{CFT}
    & References
    \\\midrule
    \ \clap{\smash{\rotatebox[x=7.5em,y=\baselineskip]{90}{\begin{tikzpicture}
        \node at (0,0){Codimension~$4$};
        \node at (0,-.3) {$\overbrace{\hspace{8em}}$};
      \end{tikzpicture}}}}\ \
    & \multirow{2}[2]{*}{\hyperref[ssec:local]{0d}}
    & \hyperref[par:local-Coulomb]{Coulomb branch operator}
    & \hyperref[par:local-Coulomb]{Integrated current}
    & \cite{0909.4031}
    \\\cmidrule{3-5}
    &
    & \hyperref[par:local-orbifold]{Orbifold $\CC^2/\ZZ_M$}
    & \hyperref[par:local-orbifold]{Change \ac{CFT} to coset}
    & \cite{1105.5800,1105.6091,1106.1172,1106.2505,1106.4001,1107.4609,1109.4264,1110.2176,1110.5628,1111.2803,1205.0784,1208.0790,1210.7454,1211.2788,1306.3938,1308.2068,1409.3465,1705.03628}
    \\\cmidrule{2-5}
    & \hyperref[ssec:line]{1d}
    & \hyperref[par:line-dyon]{Dyonic loop:\newline\mbox{}\quad Wilson loop/'t~Hooft loop}
    & \hyperref[par:line-dyon]{Degenerate Verlinde loop:\newline\mbox{}\quad around a tube/transverse}
    & \raggedright\arraybackslash
      $\lie{su}(2)$~\cite{0907.2593,0909.0945,0909.1105,0911.1922,0912.5535,1003.1112,1404.0332},
      $\lie{g}$~\cite{1003.1151,1008.4139,1108.0242,1110.3354,1206.6896,1304.2390,1312.5001,1504.00121,1603.02939,1701.04090}
    \\\cmidrule{2-5}
    & \multirow{3}{*}{\hyperref[ssec:surf]{2d}}
    & \hyperref[par:surf-vort]{Vortex string operator}
    & \hyperref[par:surf-vort]{Degenerate vertex operator}
    & \cite{0909.0945,0911.1316,1004.2025,1006.4505,1007.2524,1008.0574,1011.4491,1102.0184,1107.2787,1111.7095,1303.2626,1303.4460,1307.6612,1403.3657,1407.1852,1610.03501,1710.06970,1711.11582}
    \\\cmidrule{3-5}
    \ \clap{\smash{\rotatebox[x=8em,y=\baselineskip]{90}{\begin{tikzpicture}
        \node at (0,0) {Codimension~$2$};
        \node at (0,-.3) {$\overbrace{\hspace{9em}}$};
      \end{tikzpicture}}}}\ \
    &
    & \hyperref[par:surf-mono]{\acl{GW} surface defect\newline or orbifold $\CC\times(\CC/\ZZ_M)$}
    & \hyperref[par:surf-mono]{Change \ac{CFT} by\newline Drinfeld--Sokolov reduction}
    & \cite{1005.4469,1008.1412,1011.0289,1012.1355,1102.0076,1105.0357,1203.1427,1205.3091,1209.2992,1301.0940,1405.6992,1408.4132,1509.07516,1512.01084}
    \\\cmidrule{2-5}
    &
    & \hyperref[par:wall-symm]{Symmetry-breaking wall} & \hyperref[par:wall-symm]{Verlinde loop}
    & \cite{1003.1112}
    \\\cmidrule{3-5}
    & \hyperref[ssec:wall]{3d}
    & \hyperref[par:wall-dual]{S-duality domain wall} & \hyperref[par:wall-dual]{Modular kernel}
    & \cite{1009.0340,1103.5748,1202.4698,1304.6721,1512.09128}
    \\\cmidrule{3-5}
    &
    & \hyperref[par:wall-BCFT]{Boundary} & \hyperref[par:wall-BCFT]{Boundary \ac{CFT}}
    & \cite{1708.04631,1710.06283}
    \\\cmidrule{2-5}
    & 4d
    & Coupling to a tinkertoy & Vertex operator
    & \cite{0906.3219,0907.2189,0908.0307,1008.5203,1012.4468,1106.5410,1107.0973,1111.5624,1203.2930,1212.3952,1309.2299,1403.4604,1412.8129,1501.00357,1601.02077,1704.07890,1711.04727,1802.09626}
  \end{tabular}
\end{apartetable}

We discuss some offshoots of the \ac{AGT} correspondence in \autoref{sec:dim}.
Placing the 6d theory onto other product spaces $M\times C$ (with some twist) leads to interesting relations between theories on~$M$ and on~$C$: the index/$q$\ac{YM} correspondence~\cite{1104.3850}, the 3d/3d correspondence~\cite{1108.4389}, the 2d/4d correspondence~\cite{1306.4320}.
In another direction, some class~S theories (especially linear quiver gauge theories) can be realized as reductions of 5d $\Nsusy=1$ theories.  Instanton partition functions have direct analogues in 5d as certain $q$-deformations of the 4d results.  This leads to a $q$-deformed \ac{AGT} correspondence~\cite{0910.4431} equating these 5d instanton partition functions to chiral correlators (``conformal blocks'') of $q$-deformed Virasoro or $W_N$~algebras.
The $S^5$ partition function involves \emph{three} instanton partition functions, and its proper translation to non-chiral correlators of a complete $q$-Toda theory is still under investigation~\cite{1710.07170}.
We end in \autoref{sec:con} with a quick outline of many topics omitted in this review, such as matrix models, topological strings, quantum integrable systems, etc.

\subsection{\label{ssec:intro-reviews}Earlier reviews}

There have been many good reviews related to the \ac{AGT} correspondence, including in several PhD theses.
I particularly recommend Tachikawa's very clear collection of reviews~\cite{1312.2684,1412.7121,1504.01481,1608.02964}.
\begin{itemize}
\item \textbf{6d $(2,0)$ \acp{SCFT}.}
  These theories, and more generally 6d $(1,0)$ \ac{SCFT}, are reviewed in~\cite{1805.06467} from an F-theory perspective.  For codimension~$2$ defects, which are central in the \ac{AGT} correspondence, see~\cite{Balasubramanian-2014dia}.

\item \textbf{4d $\Nsusy=2$ and \acl{SW}.}
  While there are nice introductions from the late 1990's~\cite{hep-th/9601007,argyres2000non} to the \ac{SW} solution of 4d $\Nsusy=2$ theories, I recommend more modern explanations such as Martone's notes in this school~\cite{2006.14038}, and the well-known review ``for pedestrians''~\cite{1312.2684} which covers a lot of ground, including how \ac{AD} theories arise from limits of \ac{SQCD}.  The book~\cite{1412.7145} discusses many modern relations between 4d $\Nsusy=2$ theories and other topics.  The review~\cite{1504.01481} is focussed on the very important non-Lagrangian 4d $\Nsusy=2$ theory~$T_N$.

\item \textbf{Localization and instanton counting.}  Supersymmetric localization is reviewed in the book~\cite{1608.02952}, and in particular the squashed four-sphere partition function in~\cite{1608.02962}.  Its expression involves Nekrasov's instanton partition function, for which a good starting point is~\cite{1412.7121}, followed by~\cite{Song:2012kgc} which discusses all gauge groups, subtleties regarding the $\Lie{U}(1)$ factor, and the choice of renormalization scheme.

\item \textbf{Toda \ac{CFT} and W-algebras.}\footnote{I thank Ioana Coman for pointers.}
  Liouville \ac{CFT} is reviewed in~\cite{1406.4290,1609.09523} among many others, and it is worth reading~\cite{1108.4417} for some subtleties.  There are no recent reviews on Toda \ac{CFT} or on W-algebras.
For W-algebras see the old~\cite{hep-th/9210010,hep-th/9302006} (and possibly~\cite{Bouwknegt:1994hmh}) or the truncations of~$W_{1+\infty}$ in~\cite{1411.7697,1910.00041}.
For Toda \ac{CFT} perhaps the early article~\cite{0709.3806} or my thesis~\cite{LeFloch-2015rpq}\footnote{Better reference very welcome: only the $A_{N-1}$ case is considered there, and only full, simple, and degenerate punctures rather than general tame punctured labeled by partitions of~$N$.}.

\item \textbf{\ac{AGT} for physicists.}  See~\cite{1608.02964} (or perhaps~\cite{1108.5632}, in Japanese) for a brief review, and the longer~\cite{1309.5299} ranging from \ac{SW} basics to \ac{AD} theories arising from degenerations of \ac{SQCD}.  The matrix model approach to \ac{AGT} is reviewed in~\cite{1412.7124,1507.00260}.

\item \textbf{\ac{AGT} for mathematicians.}  Possible starting points for mathematicians include the introductory seminar notes~\cite{Safronov}, a ``pseudo-mathematical pseudo-review''~\cite{Tachikawa-pseudo,1712.09456}, incomplete (nevertheless 200 pages long) lecture notes~\cite{Moore-FelixKlein}, a categorical version of the correspondence~\cite{1106.5698}, a review that focuses on moduli spaces of flat connections~\cite{1405.0359} and one discussing instanton counting on \ac{ALE} spaces~\cite{1507.00685}.  There are also notes on mathematical applications of the 6d $(2,0)$ \ac{SCFT} to geometric representation theory, symplectic duality, knot homology, and Hitchin systems~\cite{qchu-notes}.

\item \textbf{Generalizations of \ac{AGT}.} These include the 3d/3d correspondence reviewed in~\cite{1608.02961} and the \ac{AGT} relation between 5d $\Nsusy=1$ gauge theories and $q$-Toda correlators in~\cite{1608.02968}.
\end{itemize}

Given these numerous reviews, writing yet another set of notes is perhaps futile, but hopefully the rather different approach taken here, starting from the 6d theory, is the right one for some readers.
I apologize for omitting many directions from this review, listed in the conclusion \autoref{sec:con}, especially a broader discussion of the \ac{BPS}/\ac{CFT} correspondence and of the deep links to quantization of integrable models underlying \ac{SW} geometry, matrix models, and topological strings.

\part{Class S theories}

\section{\label{sec:6d}6d \((2,0)\) \acsfont{SCFT} of ADE type}

Superconformal algebras exist in dimensions up to~$6$, and there is by now ample evidence for the existence of 6d $\Nsusy=(2,0)$ (maximally supersymmetric) \acp{SCFT} $\Xg$, labelled by a Lie algebra~$\lie{g}$ that is simply-laced\footnote{As a reminder, simply-laced Lie algebras are $\lie{a}_{N-1}=\lie{su}(N)$, $\lie{so}(2N)=\lie{d}_N$, and the three exceptional algebras $\lie{e}_6,\lie{e}_7,\lie{e}_8$ (in each case the subscript is the rank).  This ADE classification has several beautiful avatars in theoretical physics but we will not get to explore them in this review.}\textsuperscript{,}\footnote{While different constructions of~$\Xg$ give the same condition that $\lie{g}$ is simply-laced, including some field theoretic arguments~\cite{1505.03850}, it has not been proven that $\Xg$~exhaust all 6d $\Nsusy=(2,0)$ \acp{SCFT}.  The situation is the same in 4d $\Nsusy=4$ \ac{SCFT}: there might possibly be such theories other than $\Nsusy=4$ \ac{SYM} theories.}.
Nobody knows how to actually define $\Xg$ directly in a \ac{QFT} language, for instance through a Lagrangian formulation.
It is instead obtained as a decoupling limit of certain string theory or M-theory brane setups.
\apartehere
Despite its stringy construction, the theory is expected to be a bona-fide local \ac{QFT}, for instance having a local conserved stress-tensor.\footnote{To be precise, it is a relative quantum field theory~\cite{1212.1692}.}
These constructions entail three important properties detailed
\ifjournal
\else
  to the right.
\fi

\startaparte[title={Properties of $\Xg$}]
\begin{itemize}
\item $\Xg$ has vacua whose \ac{IR} description is an abelian 6d $(2,0)$ theory of self-dual two-form gauge fields valued in the Cartan algebra of~$\lie{g}$ modulo the Weyl group.
\item $\Xg$ is a \ac{UV}-completion of 5d $\Nsusy=2$ \ac{SYM} in the sense that
  \ac{SYM} with gauge algebra~$\lie{g}$ and gauge coupling~$\gfived$ gives an \ac{IR} description of $\Xg$ compactified on a circle of radius~$\gfived^2$.
\item $\Xg$ admits codimension~$2$ half-\ac{BPS} defects labeled by nilpotent orbits in~$\lie{g}$, and codimension~$4$ half-\ac{BPS} defects labeled by representations of~$\lie{g}$.
\end{itemize}
\stopaparte
The first two properties are compatible because both 5d $\Nsusy=2$ \ac{SYM} on its Coulomb branch, and the abelian 6d theory on a circle, are described by 5d abelian vector multiplets in the Cartan of~$\lie{g}$.
The last property is compatible as well, as the defects have rather explicit descriptions when one moves along the Coulomb branch or when one places the theory on a circle.
The existence of $\Xg$ with these properties is confirmed by many consistency checks involving better-understood theories.  A major set of consistency checks is the \ac{AGT} correspondence obtained by placing these theories on the product $M_4\times C_2$ of a 4d and a 2d manifolds.

In this section we describe the symmetry algebra $\lie{osp}(8^*|4)$ (\autoref{ssec:6d-algebra}), properties of self-dual two-form gauge fields (\autoref{ssec:6d-dual}), string/M-theory constructions (\autoref{ssec:6d-string}) and extended operators (\autoref{ssec:6d-ops}) of~$\Xg$.

\subsection{\label{ssec:6d-algebra}Superconformal algebras}

Superconformal algebras in dimensions $d>2$ have been classified by Nahm~\cite{Nahm:1977tg} under certain conditions.  Their even (bosonic) part consists of the conformal algebra $\lie{so}(2,d)$ (in Lorentzian signature) and an R-symmetry algebra, and their odd (fermionic) part consists of supercharges that must transform in the spinor representation of $\lie{so}(2,d)$, and such that translations are realized as anticommutators of supercharges.

The classification is in \autoref{tab:SCA}.
In dimensions $d=3,4,6$ the conformal algebra coincides with the expected $\lie{so}(2,d)$ thanks to accidental isomorphisms\footnote{There are no such accidental isomorphisms for $d>6$, which more or less explains the lack of higher-dimensional superconformal algebras.} $\lie{so}(2,3)=\lie{sp}(4,\RR)$ and $\lie{so}(2,4)=\lie{su}(2,2)$ and $\lie{so}(2,6)=\lie{so}^*(8)$.  In each case, the spinor representation of $\lie{so}(2,d)$ is the fundamental (vector) representation of the other group.
It is known that \acp{SCFT} with more than $16$ Poincar\'e supercharges do not exist for $d\geq 4$ (and are free for $d=3$)~\cite{1612.00809}, and this leads to the bounds on~$\Nsusy$ given in the table.

\begin{table}\centering
  \caption{\label{tab:SCA}Nahm classification of superconformal algebras in Lorentzian signature.
    Here we list the superconformal algebras in each dimension, the two bosonic factors (conformal algebra and R-symmetry algebra), and the representations (of these bosonic factors) in which Poincar\'e and conformal supercharges $Q$ and~$S$ transform.}
  \smallskip
  \begin{tabular}{llllc}
    \toprule
    & Superalgebra & Conformal & R-symmetry & $Q$\&$S$ \\ \midrule
    3d $\Nsusy\leq 8$ & $\lie{osp}(\Nsusy|4)$ & $\lie{sp}(4,\RR)$ & $\lie{so}(\Nsusy)$ & $(4,\Nsusy)$ \\
    4d $\Nsusy\leq 3$ & $\lie{su}(2,2|\Nsusy)$ & $\lie{su}(2,2)$ & $\lie{su}(\Nsusy)\oplus\lie{u}(1)$ & $(4,\overline{\Nsusy})\oplus(\overline{4},\Nsusy)$\\
    4d $\Nsusy=4$ & $\lie{psu}(2,2|4)$ & $\lie{su}(2,2)$ & $\lie{su}(4)$ & $(4,\overline{4})\oplus(\overline{4},4)$\\
    5d $\Nsusy=1$ & $\lie{f}^2(4)$ & $\lie{so}(2,5)$ & $\lie{su}(2)$ & $(8,2)$ \\
    6d $\Nsusy\leq 2$ & $\lie{osp}(8^*|2\Nsusy)$ & $\lie{so}^*(8)$ & $\lie{usp}(2\Nsusy)$ & $(8,2\Nsusy)$ \\
    \bottomrule
  \end{tabular}
\end{table}

For the 6d case of interest to us, minimal spinor representation of the Lorentz algebra $\lie{so}(2,6)$ are chiral, and the superconformal algebras contain $\Nsusy=1$ or~$2$ such chiral spinors (technically, symplectic Majorana--Weyl spinors) with the same chirality.
These algebras are thus called 6d $\Nsusy=(1,0)$ and 6d $\Nsusy=(2,0)$ superconformal algebras.
There is no 6d $\Nsusy=(1,1)$ superconformal algebra.
We are interested in the largest superconformal algebra of all: the 6d $(2,0)$ algebra $\lie{osp}(8^*|4)$.

Supercharges of this algebra transform in the $(\rep{8}_s,\rep{4})$ representation\footnote{We denote irreps (irreducible representations) of a simple Lie algebra by their dimension in bold face.  When ambiguities arise there are standard decorations to distinguish them, such as overlines for conjugating the representation, or primes when there are several irreps of the same dimension and they are not related by conjugation.  A peculiar example is $\lie{so}(8)$ and other real forms thereof like $\lie{so}(p,8-p)$ as they have three dimension~$8$ irreps: the defining representation of $\lie{so}(8)$ called $8_v$, and two conjugate spinor representations $8_s$ and~$8_c$, related by the triality automorphism of $\lie{so}(8)$.} of the conformal and R-symmetry algebras $\lie{so}(6,2)\times\lie{so}(5)_{\Rsymm}$, with a reality condition.  Decomposing this into representations of the Lorentz algebra $\lie{so}(1,5)$ gives $(\rep{4},\rep{4})\oplus(\rep{4},\rep{4})$, with a symplectic reality condition.
One set $(\rep{4},\rep{4})$ consists of Poincar\'e supercharges and the other of superconformal transformations.

\subsection{\label{ssec:6d-dual}Self-dual forms}

The 6d $\Nsusy=(2,0)$ \ac{SCFT} $\Xg$ is roughly speaking a theory of self-dual two-forms gauge fields for a gauge Lie algebra~$\lie{g}$ among $\lie{a}_{N-1},\lie{d}_N,\lie{e}_6,\lie{e}_7,\lie{e}_8$, as we explain next.

\paragraph{Abelian self-dual forms.}

In even dimension $d$ there exists an interesting notion of (anti)\footnote{Self-dual and anti-self-dual cases differ by a sign, and we shall just write ``self-dual'' for simplicity.} self-dual $k$-form for $k=d/2-1$: a $k$-form $B$ with components $B_{\alpha_1\dots\alpha_k}$ (antisymmetric in $\alpha_1,\dots,\alpha_k$) such that the field strength $H=dB$ is mapped to a multiple of itself by the Hodge star, that is,
\begin{equation}\label{Hselfdual}
  H_{\alpha_0\alpha_1\dots\alpha_k}
  \coloneqq (k+1)!\, \del_{[\alpha_0} B_{\alpha_1\dots\alpha_k]}
  = \pm i^{d/2+s} \epsilon_{\alpha_0\dots\alpha_k\beta_0\dots\beta_k} \del^{[\beta_0} B^{\beta_1\dots\beta_k]} .
\end{equation}
Here indices within square brackets are antisymmetrized and the power of $i=\sqrt{-1}$ involves $s=0$ for Euclidean and $s=1$ for Lorentzian signature.
The self-duality condition regards the field strength hence is invariant under gauge transformations $B\to B+d\Lambda$ for any $k$-form $\Lambda$: explicitly this adds $k!\,\del_{[\alpha_1}\Lambda_{\alpha_2\dots\alpha_k]}$ to the component~$B_{\alpha_1\dots\alpha_k}$ of the $k$-form gauge field~$B$.

From~\eqref{Hselfdual} we see that real self-dual $k$-forms exist only if $d/2+s$ is even.
In 2d this happens in Lorentzian signature, and it corresponds to a real scalar field propagating only in one lightlike direction.  (In the Euclidean case it is a complex chiral boson depending on one holomorphic coordinate.)
In 4d with Euclidean signature, \eqref{Hselfdual}~defines self-dual gauge field configurations, also called instantons, which play a crucial role on the 4d side of the \ac{AGT} correspondence.  (In the Lorentzian case they are complex saddle-points.)
In 6d with Lorentzian signature we get a real self-dual two-form gauge field~$B_{\alpha\beta}$.

We care about 6d $(2,0)$ supersymmetry, in which case the multiplet containing $B_{\alpha\beta}$ consists of $B$, spinors~$\lambda$, and scalars~$\Phi$ that transform respectively as the singlet, the $4$-dimensional, and the $5$-dimensional representations of R-symmetry $\lie{usp}(4)=\lie{so}(5)$.

\paragraph{Compactifying on a circle.}

Let us place this 6d $(2,0)$ abelian theory of $(B,\lambda,\Phi)$ on a circle and decompose into \ac{KK} modes.
As determined in the following exercise, the five scalars~$\Phi_I$ remain scalars, the spinors~$\lambda$ as well, and the self-dual two-form gauge field~$B$ becomes a usual gauge field~$A$ in 5d.  Altogether this gives abelian 5d $\Nsusy=2$ \ac{SYM}.

We review dimensional reduction in \autoref{exe:dim-red} below.
An important aspect for the reduction from $\Xg$ to 5d is that 5d \ac{SYM} has instanton particles, namely gauge field configurations with non-trivial topological number $\int \epsilon^{\mu\nu\rho\sigma} F_{\mu\nu}F_{\rho\sigma} \rmd^4x$ on each spatial slice.  These excitations of the gauge field~$A$ play the role of the tower of \ac{KK} modes: their mass (proportional to) $1/\gfived^2$ is correctly identified with the mass $1/R$ of \ac{KK} modes.

\begin{exercise}\label{exe:dim-red}
  1. Consider a $D$-dimensional scalar field~$\varphi$, with Lagrangian $\Lag(\varphi)=\del_\alpha\varphi\del^\alpha\varphi-V(\varphi)$ (you can take $V=0$ for simplicity).
  Consider it on a $d$-dimensional Minkowski space times a $(D-d)$-dimensional torus of radius~$R$ (you can take $D-d=1$ for simplicity).
  Write a Fourier decomposition of~$\varphi$ along the circle direction and rewrite the action of~$\varphi$ as an action for these components.  In the limit $R\to 0$ notice that all Fourier modes become infinitely massive except the zero mode.

  2. Repeat the exercise for an abelian vector field $A_\alpha$ ($\alpha=0,\dots,D-1$) with Lagrangian $F_{\alpha\beta} F^{\alpha\beta}$, where $F_{\alpha\beta}=\del_\alpha A_\beta-\del_\beta A_\alpha$.  Check that the dimensionally-reduced theory has both a vector field $A_\mu$ ($\mu=0,\dots,d-1$) and $D-d$ scalar fields.  These can be gauge-invariantly understood for finite~$R$ as Wilson loops of $A_\alpha$ around coordinate circles of the torus.  How do $D$-dimensional gauge transformations act?  Deduce that the scalar fields are circle-valued.

  3. Repeat the exercise for a two-form $B_{\alpha\beta}$ reduced from 6d to 5d.  This results in a two-form $B_{\mu\nu}$ and a one-form~$A_\mu$.  By imposing the self-duality condition on $B_{\alpha\beta}$ find that $B_{\mu\nu}$ can be reconstructed (up to gauge transformations) from~$A_\mu$.
\end{exercise}

\paragraph{Nonabelian theory.}

Recall the Bianchi identity $\del_{[\mu} F_{\nu\rho]}=0$ in 4d.  It generalizes to $dH=ddB=0$.  For a self-dual form this implies the free equations of motion $d\star dB=0$, namely $\del^\mu H_{\mu\nu\dots}=0$.
How can we add interactions?
In 4d, the equation $F_{\mu\nu} = \mp \frac{1}{2} \epsilon_{\mu\nu\rho\sigma} F^{\rho\sigma}$ defining instantons makes sense even for the field strength of non-abelian gauge fields, $F=\rmd A+A\wedge A$, explicitly $F_{\mu\nu}=\del_\mu A_\nu-\del_\nu A_\mu+[A_\mu,A_\nu]$.
The non-abelian version of the Bianchi identity is $\epsilon^{\lambda\mu\nu\rho} D_\mu F_{\nu\rho}=0$.
When the gauge field configuration is self-dual this implies the standard Yang--Mills equations of motion $D_\mu F^{\mu\lambda}=0$.
In contrast, for other self-dual $k$-forms, $k\neq 1$, there is no obvious non-abelian generalization of the relation $H=dB$, hence no obvious way to introduce interactions.
Instead, we use two stringy constructions.

\subsection{\label{ssec:6d-string}Brane construction of 6d theories}

The 6d $(2,0)$ theory~$\Xg$ naturally arises as the zero string tension limit of a 6d $(2,0)$ little string theory, whose excitations are self-dual strings.
These self-dual strings were uncovered as supergravity solutions~\cite{hep-th/9306052}, then in IIB string theory at ADE orbifold singularities~\cite{hep-th/9507121}, then as D2~branes ending on NS5~branes (or M2~ending on~M5)~\cite{hep-th/9512059}.
(There also exist analogous 6d $(1,1)$ little string theories, but they have no conformal limit so we do not discuss them further.)
The 6d $(2,0)$ little string theories are labeled by a simply-laced\footnote{The ADE classification comes here from anomaly cancellation on the string worldsheet~\cite{hep-th/0405056}.} Lie algebra~$\lie{g}$, just like~$\Xg$, and we discuss two string theory constructions reviewed in~\cite{hep-th/9911147}.
\begin{itemize}
\item In the zero string coupling limit $g_s\to 0$ of a stack of $N$~coincident NS5~branes in IIA~string theory (or of M5~branes in M-theory), bulk degrees of freedom decouple, and one gets the 6d $(2,0)$ little string theory with $\lie{g}=\lie{su}(N)$.
\item General 6d $(2,0)$ little string theories arise in the zero string coupling limit $g_s\to 0$ of IIB~string theory on a $\CC^2/\Gamma$ singularity for the discrete subgroup $\Gamma\subset\SU(2)$ corresponding to~$\lie{g}$.
\end{itemize}
These approaches teach us that $\Xg$~on a circle is equivalent to 5d $\Nsusy=2$ \ac{SYM}, and how to describe~$\Xg$ upon moving on the tensor branch\footnote{The 6d $\Nsusy=(2,0)$ tensor multiplet splits into a 6d $\Nsusy=(1,0)$ tensor multiplet and a hypermultiplet.  The tensor branch and Higgs branch are vacua where scalar fields in tensor or hyper multiplets acquire a \ac{VEV} (with $(2,0)$ supersymmetry the two branches combine).  The tensor branch is sometimes called Coulomb branch because it reduces to Coulomb branches in 5d and~4d.  In 6d $\Nsusy=(1,0)$ theories one also has vector multiplets but they contain no scalars so there is no corresponding branch.}.

\paragraph{Fivebranes in IIA~or M-theory.}

We begin with the M-theory (or equivalently~IIA) construction, applicable to $\lie{g}=\lie{su}(N)$: $\XsuN$~is the world-volume theory of a stack of $N$ coincident M5~branes in M-theory, with the decoupled center of mass degrees of freedom removed.

M-theory is an $11$ dimensional theory (Lorentzian signature) with $32$ supersymmetries (one Majorana spinor).
It is related by various dualities to better-understood string theories and supergravity, for instance its low-energy limit is described by $11$-dimensional supergravity.
For our purposes, the most interesting aspect is that M-theory on a circle times a $10$-dimensional spacetime is equivalent to IIA strings on that spacetime.
The aim of this review is not to discuss the intricate web of dualities relating M-theory to IIA and other string theories, so we are quite schematic.

A standard comment on terminology: $p$~branes are $(p+1)$ dimensional objects, with $p$~space and $1$~time directions, so for instance the M5~brane is $6$-dimensional and has Lorentzian signature, as we wanted.
M-theory has two such half-\ac{BPS} objects: the M5~brane and the M2~brane.
Stacks of flat\footnote{Here we work as if spacetime were flat; the backreaction of branes on the geometry does not invalidate the conclusions.} parallel branes of the same type preserve the same half of supersymmetry (see \autoref{exe:brane-susy}), and there is no energy cost to moving the branes while keeping them flat and parallel.
While the world-volume theory of a stack of $N$ D$p$~branes has been known for a long time to be maximally supersymmetric \ac{SYM} in $p+1$ dimensions (see the review~\cite{hep-th/0307104}), the world-volume theory of stacks of branes in M-theory has proven more difficult to pin down.
\begin{itemize}
\item The world-volume theory of a stack of coincident M2~branes is now known\footnote{Depending on one's point of view, most words ``known'' in this review should be replaced by ``conjectured''.  Ultimately, since the path integral has not been properly defined in most cases of interest to physicists, almost all non-perturbative \ac{QFT} results are conjectural.  One can think about how much ``evidence'' there is for one result or another.  Results that are consistent with many others should then serve as a guide to determine if a given mathematical definition of the theories is acceptable.} to be the \ac{ABJM} Chern--Simons matter theory, an \ac{SCFT} with an explicit 3d $\Nsusy=2$ Lagrangian description, whose supersymmetry enhances to the expected 3d $\Nsusy=8$ superconformal algebra $\lie{osp}(8|4)$ preserved by the branes (see the review~\cite{0909.1580}).  The R-symmetry $\lie{so}(8)$ rotates the $11$-dimensional space around the M2~branes.  Its holographic dual is $\AdS_4\times S^7$.  That is all we will say in this review.
\item The world-volume theory of a stack of $N$~coincident M5~branes is what we call~$\XsuN$, a 6d $(2,0)$ \ac{SCFT} with no Lagrangian description.\footnote{Instead of $\lie{g}=\lie{a}_{N-1}=\lie{su}(N)$ one can realize $\lie{g}=\lie{d}_{N}=\lie{so}(2N)$ by including an O5 orbifold plane on top of the M5~branes.}  More precisely, this would give $\lie{u}(N)$, but the $\lie{u}(1)$ center of mass of the branes decouples.  The R-symmetry $\lie{so}(5)$ rotates space around the M5~branes.  The holographic dual is $\AdS_7\times S^4$, which has the expected symmetry algebra $\lie{osp}(8^*|4)$, differing only from the 3d case by some signs in the 7d and 4d parts.
\end{itemize}

\paragraph{Consequences of the M-theory construction.}

Consider now $\XsuN$ on a circle (times five-dimensional Minkowski space).
M-theory on a circle is equivalent to IIA string theory, and M5~branes wrapping the circle become D4~branes.
Thus, $\XsuN$ on a circle is equivalent to the world-volume theory of $N$ D4~branes, which is 5d $\Nsusy=2$ \ac{SYM}, as announced at the start of this \autoref{sec:6d}.
Dimensional analysis shows that the 5d gauge coupling scales as $\gfived\sim L_5^{1/2}$ in terms of the compactification circle length~$L_5$.

We move on to describing the vacua of~$\Xg$ from its M-theory construction.
Supersymmetric vacua are parametrized by the positions of the $N$~M5~branes in the $5$ transverse directions, modulo relabelling of the branes since they are indistinguishable.
The vacua are thus $(\RR^N)^5/S_N$.
At any generic vacuum, all degrees of freedom are massive (with mass proportional to the separation between the branes), except fluctuations around each individual brane, which are known to be described by one 6d abelian theory of $(B,\lambda,\Phi)$ for each brane.  The scalar fields $\Phi_I$, $I=6,\dots,10$, describe fluctuations of each of the $N$ M5~branes in the transverse directions (except the $\lie{u}(1)$ trace part).

\paragraph{IIB~strings.}

The M-theory construction gives a lot of insight on $\Xg$ for $\lie{g}=\lie{a}_{N-1}$, and can be extended to $\lie{d}_{N}$ by orbifolding, but it cannot realize the exceptional cases $\lie{e}_6,\lie{e}_7,\lie{e}_8$.  For this a dual IIB description is needed.

As understood in~\cite{hep-th/9511164}, T-duality transverse to a stack of $N$~NS5 branes in IIA~theory produces IIB~strings on an $A_{N-1}$~singularity.
This second construction of the 6d $(2,0)$ little string theory, which we will not use much, generalizes readily to all ADE cases.
Place IIB string theory on Minkowski space~$\RR^{1,5}$ times a quotient $\CC^2/\Gamma$ by a finite subgroup $\Gamma\subset\SU(2)$.
Such subgroups are classified by ADE Lie algebras~$\lie{g}$.  For instance, the $A_{N-1}$ case is $\Gamma=\ZZ_N$ acting as $(z,w)\mapsto(e^{2\pi i/N}z,e^{-2\pi i/N}w)$ on coordinates of~$\CC^2$.
The zero-coupling limit $g_s\to 0$ of this set-up yields little string theory, and the further zero-tension limit gives the 6d $(2,0)$ theory~$\Xg$.

Moving along the vacuum moduli space of the 6d theory corresponds to blowing up $\CC^2/\Gamma$ into \ac{ALE} space, namely resolving the singularity at the origin of $\CC^2/\Gamma$ into a collection of $r=\rank\lie{g}$ finite-size two-cycles.  The sizes of these two-cycles become $r$ scalar fields~$\Phi$ of the 6d theory on~$\RR^{1,5}$, in the Cartan subalgebra of~$\lie{g}$.  In fact their \ac{VEV} parametrizes the vacua of~$\Xg$.  In any vacuum, the \ac{IR} degrees of freedom are: these scalar fields~$\Phi$, the two-form~$B$ obtained by integrating the chiral four-form of IIB string theory around each of the two-cycles, and some spinors.  We end up as wanted with the 6d $(2,0)$ theory of an abelian self-dual two-form gauge field multiplet $(B,\lambda,\Phi)$ in the Cartan subalgebra of~$\lie{g}$.

In this description only $\SO(4)$ R-symmetry is manifest, and the reduction to 5d $\Nsusy=2$ \ac{SYM} is also nontrivial to see.

\subsection{\label{ssec:6d-ops}Codimension 2 and 4 defects}

We return to the M-theory construction of~$\XsuN$ and consider intersecting brane configurations with branes placed along the following directions inside~$\RR^{1,10}$.
\begin{center}
  \begin{tabular}{l*{11}{c}l}
    \toprule
    M5  & 0 & 1 & 2 & 3 & 4 & 5 & . & . & . & . & .  & $\to$ 6d $(2,0)$ theory \\
    \midrule
    M5' & 0 & 1 & 2 & 3 & . & . & 6 & 7 & . & . & .  & $\to$ codimension $2$ defect \\
    M2  & 0 & 1 & . & . & . & . & . & . & . & . & 10 & $\to$ codimension $4$ defect \\
    \bottomrule
  \end{tabular}
\end{center}
Each column is a direction in~$\RR^{1,10}$; a dot indicate that a stack of branes is localized at a given value of that coordinate, and a number indicates that the brane extends along the corresponding direction.  For instance the M5 branes are at given values of $x^6,x^7,x^8,x^9,x^{10}$ and extend in all other coordinates.
The prime on M5'~branes just helps us distinguish them from the M5~branes on which~$\Xg$ lives.
Each additional stack of branes in this table breaks half of supersymmetry (see \autoref{exe:brane-susy}).
There can be several stacks of the same kind of branes parallel to each other, in which case they don't break further supersymmetry.

The M2~branes extend only in one direction transverse to the M5~branes.  In this direction~$x^{10}$ they can either be infinite, or semi-infinite ending on one M5~brane, or finite stretching between two M5~branes.  Either way, from the point of view of~$\Xg$, stacks of M2~branes insert a half-\ac{BPS} codimension~$4$ operator, namely an operator supported on a two-dimensional slice of the 6d theory~\cite{hep-th/9711205}.

\begin{figure}\centering
  \begin{tikzpicture}
    \draw (1.7,2) -- (1,2) -- (0.5,1) -- (3.5,1) -- (4,2) -- (2.3,2);
    \draw (1.95,3.2) arc (0:-60:.6 and 2);
    \draw (2.05,3.2) arc (-180:-120:.6 and 2);
    \draw (0.7,1) -- (0.5,0.6) -- (3.5,0.6) -- (4,1.6) -- (3.8,1.6);
    \draw (1.95,-0.5) arc (0:30:.6 and 2);
    \draw (2.05,-0.5) arc (180:150:.6 and 2);
  \end{tikzpicture}
  \caption{\label{fig:trumpet-branes}Configuration of a pair of M5 branes spanning the $x^4,x^5$ directions (depicted horizontally) in the presence of an M5'~brane at a point in the $x^4,x^5$ plane.  The M5 and M5' branes merge into a complex manifold.  The $x^6,x^7$ positions (depicted vertically) diverge at one point in the $x^4,x^5$ plane.  We depicted the situation after decoupling the center of mass modes, which is why the branes diverge symmetrically.}
\end{figure}
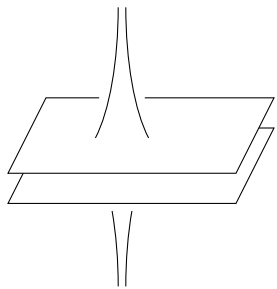

The way it is written here, it would seem the M5 and M5' branes intersect in codimension~$2$.  In truth they turn out to merge into a smooth complex manifold that asymptotes at large distances to the configuration we wrote.  For this to happen, the $x^6,x^7$ positions of the M5 branes should grow to infinity as $x^4,x^5$ get closer to the positions of M5'~branes, as depicted in \autoref{fig:trumpet-branes}.
We return to this in \autoref{sec:Lag} for concrete cases.
From the point of view of~$\Xg$, at large distance, the intersection with M5'~branes has an effective description as a four-dimensional (codimension~$2$) half-\ac{BPS} operator~\cite{hep-th/9710033}.

As we explore the \ac{AGT} correspondence in this review we learn various properties of these defects, and especially the data that describes them.  We find that:
\begin{itemize}
\item Codimension~$2$ operators are labeled by nilpotent orbits of~$\lie{g}$ \cite{0906.0359,1203.2930}.
  In the $\lie{su}(N)$ case, these amount to partitions of~$N$ specifying the way in which the $N$~M5~branes cluster into different groups as they go to infinity in the $x^6,x^7$ directions.  Additional continuous data describes the length scales in these directions.
\item Codimension~$4$ operators are labeled by representations of~$\lie{g}$.  For the $\lie{su}(N)$, recall that to each representation is associated a Young diagram, such that $\Box$ is the fundamental $N$-dimensional representation, $\Box\!\Box$ is the symmetric representation, etc.  The total number of boxes is the number of M2~branes necessary to describe the operator in M-theory.  Roughly speaking, the number of boxes in each row of the Young diagram indicates how many M2~branes can end on the same M5~brane.
\end{itemize}

\begin{exercise}\label{exe:brane-susy}
  A flat M5~brane along directions $x^0,x^1,\dots,x^5$ preserves supersymmetries with $\Gamma^{012345}\epsilon=\epsilon$ while a flat M2~brane along directions $x^0,x^1,x^{10}$ preserves supersymmetries with $\Gamma^{23456789}\epsilon=\epsilon$.  Check that the brane configurations above are such that each additional stack of branes breaks half of supersymmetry.  (Hint: check that $\Gamma^{01}$, $\Gamma^{23}$, $\Gamma^{45}$ etc.\@ commute with each other.)  What other relative orientations of the stacks of branes preserve half of the supersymmetry?
\end{exercise}

\section{\label{sec:reduce}Class S theories from 6d}

\apartehere

Our next task is to dimensionally reduce the 6d theory~$\Xg$ on a Riemann surface~$C_2$.
We explain in \autoref{ssec:reduce-twist} a partial topological twist such that the reduced theory has 4d $\Nsusy=2$ supersymmetry~\cite{0904.2715}.
The Coulomb branch and \ac{SW} curves giving the \ac{IR} physics are worked out in \autoref{ssec:reduce-Coulomb} and \autoref{ssec:reduce-SW}.
We then explain in \autoref{ssec:reduce-pieces} how the 4d theory decomposes into building blocks called tinkertoys~\cite{1008.5203}.
\startaparte[title={6d viewpoint on 4d $\Nsusy=2$ class S theories}]%
\begin{itemize}
\item Reducing (with a twist) the 6d theory~$\Xg$ on a punctured Riemann surface~$C$ gives a 4d $\Nsusy=2$ theory.
\item Its Coulomb branch is parametrized by differentials $\phi_k=u_k \rmd z^k$ of the same degrees~$k$ as the Casimir invariants of~$\lie{g}$.
\item Its \ac{SW} curve $\Sigma\subset T^*C$ is a multiple cover of~$C$.
\item Gluing punctures amounts to gauging symmetries.
\item Cutting~$C$ amounts to decoupling gauge fields.
\end{itemize}
\stopaparte

\subsection{\label{ssec:reduce-twist}Partial topological twist}

Our aim is to place the 6d $(2,0)$ theory on $\RR^4\times C_2$, where $C_2$ is an arbitrary punctured Riemann surface.
Doing this too naively would not preserve any symmetry beyond the Poincar\'e symmetry of~$\RR^4$.
We explain a procedure, the \emph{partial topological twist}, that allows 4d $\Nsusy=2$ supersymmetry to be preserved regardless of~$C_2$.

\paragraph{Generalities on topological twist.}
First we comment on the topological twist of supersymmetric theories~\cite{Witten:1988ze} in general terms from several point of views.

When placing a field theory on a curved background, the metric $g_{\mu\nu}$ acts as a source for the stress tensor~$T^{\mu\nu}$.
For a supersymmetric field theory, $T^{\mu\nu}$~typically belongs to a multiplet together with supersymmetry currents $S_\alpha^\mu$ and R-symmetry currents~$J^\mu$.  These can also be coupled to sources $\psi_\mu^\alpha$ and~$A_\mu$.
The partial topological twist consists of setting $S_\alpha^\mu=0$ and choosing $A_\mu$ equal to the spin connection derived from~$g_{\mu\nu}$.
Schematically, at linearized order around some background values of $g,\psi,A$, when these fields are changed the Lagrangian varies by
\begin{equation}
  \delta\Lag = T \delta g + J \delta A
  = T \delta g - J \del(\delta g)
  \simeq (T + \del J) \delta g .
\end{equation}
In the second step we used our choice that $A$ is related to derivatives of the metric, and in the last step we integrate by parts.

In this way the topological twist amounts to redefining the stress-tensor from $T$ to $T_{\twist}=T+\del J$ before placing the theory on a non-trivial background metric.
The twist mixes the stress-tensor~$T^{\mu\nu}$ with the R-symmetry current~$J^\mu$, but it is good to remember that it does not affect any observables of the theory in flat space, only what we call the stress-tensor.  Through the change of stress-tensor it changes how the theory is put on curved spaces.

One job of the stress-tensor is to keep track of Poincar\'e symmetries: $T^{\mu\nu}$ is the conserved current of translation symmetries, while $x^{[\mu} T^{\nu]\rho}$ is the conserved current of rotations.  Since the twist shifts $T$ by a total derivatives it is simply an improvement transformation of the translation symmetry current, and it does not change the corresponding conserved charge, the momentum operator.  In contrast, it has a non-trivial effect on what we call rotations: twisted rotation acts by a rotation plus an R-symmetry transformation, because schematically $xT \mapsto xT+x\del J=xT-J+\del(xJ)$ where $\del(xJ)$ is an improvement transformation.  Commutators between translations and the twisted rotations nevertheless coincide with those in the standard Poincar\'e algebra.

What happens to supercharges?  They typically transform as spinors under the original rotations and under R-symmetry transformations.
Under the new rotations embedded diagonally, the supercharges typically split into a scalar supercharge~$Q$ and a vector.
By virtue of $Q$~being a scalar, the stress tensor is $Q$-closed, so that placing the theory on a curved manifold using the twisted stress-tensor preserves the supersymmetry~$Q$.
The next step is typically to restrict to operators in the $Q$-cohomology.
In many cases, the twisted stress-tensor is $Q$-exact, namely $T_{\twist}^{\mu\nu} = \{Q, G^{\mu\nu}\}$ for some supersymmetry generator~$G^{\mu\nu}$, so that it vanishes in $Q$-cohomology and the correlators are described by a \ac{TQFT}.
For the twist we consider, this will not happen and there will remain non-trivial local dynamics instead.

\paragraph{Partial topological twist of 6d theories.}
The \emph{partial} topological twist we use consists of only mixing \emph{some} of the R-symmetries into \emph{some} of the rotation symmetries.
To define the specific twist we use, consider rotations $\lie{so}(1,3)\times\lie{so}(2)_{\old}$ preserving separately the two factors of a product $\RR^{1,3}\times\RR^2$, and consider the block-diagonal subalgebra $\lie{so}(2)_{\Rsymm}\times\lie{so}(3)_{\Rsymm}\subset\lie{so}(5)_{\Rsymm}$ of R-symmetry.
We define twisted rotations to be embedded diagonally into $\lie{so}(2)_{\old}\times\lie{so}(2)_{\Rsymm}$, namely we treat the following symmetries as our (twisted) Lorentz and R-symmetries:
\begin{equation}
  \lie{so}(1,3)\times\lie{so}(2)_{\twist}\times\lie{so}(3)_{\Rsymm} .
\end{equation}
This is done by changing the stress-tensor to
\begin{equation}\label{T-partial-twist}
  T_{\twist}^{\mu\nu} = T_{\old}^{\mu\nu} + \frac{1}{4} (\epsilon^{\mu\rho} \del_\rho J_{12}^\nu + \epsilon^{\nu\rho} \del_\rho J_{12}^\mu) ,
\end{equation}
where $J_{12}$ is the R-symmetry rotation generator of~$\lie{so}(2)_{\Rsymm}$ and $\epsilon^{\mu\nu}=\delta^\mu_4\delta^\nu_5-\delta^\nu_5\delta^\mu_4$ is the Levi-Civita tensor on the~$\RR^2$ factor.

\begin{exercise}
  Check that \eqref{T-partial-twist} shifts the $x^4,x^5$ rotation current $x^{[4}T^{5]\mu}$ by~$J_{12}$ up to total derivatives (an improvement term), so that the twisted rotation is a combination of rotation and R-symmetry.
\end{exercise}

Let us track supersymmetries as we twist and then compactify.
Under the $\lie{so}(1,5)$ rotations of $\RR^{1,5}$ and $\lie{so}(5)_{\Rsymm}$ R-symmetry, the Poincar\'e supersymmetries transform in the spinor representation of each, denoted $(\rep{4},\rep{4})$, with a symplectic reality condition that we hide for simplicity.
Each 6d Weyl spinor, namely each representation $\rep{4}$ of $\lie{so}(1,5)$ decomposes into a pair of 4d Weyl spinors of opposite chirality $(\rep{2},\rep{1})\oplus(\rep{1},\rep{2})$ under $\lie{so}(1,3)$, and these spinors have opposite charges $1/2$ and $-1/2$ under $\lie{so}(2)_{\old}$.
Each spinor $\rep{4}$ of $\lie{so}(5)$ decomposes into two $\rep{2}$ of $\lie{so}(3)_{\Rsymm}$ with $\lie{so}(2)_{\Rsymm}$ charges $\pm 1/2$.
Altogether we denote this as follows, with subscripts denoting charges under the two $\lie{so}(2)$ algebras:
\begin{equation}
  \begin{aligned}
    (\rep{4},\rep{4})
    & = \Bigl((\rep{2},\rep{1})_{\frac{1}{2}}\oplus (\rep{1},\rep{2})_{-\frac{1}{2}}\Bigr)\otimes \Bigl(\rep{2}_{\frac{1}{2}}\oplus \rep{2}_{-\frac{1}{2}}\Bigr) \\
    & = (\rep{2},\rep{1},\rep{2})_{\frac{1}{2},\frac{1}{2}}
    \oplus (\rep{2},\rep{1},\rep{2})_{\frac{1}{2},-\frac{1}{2}}
    \oplus (\rep{1},\rep{2},\rep{2})_{-\frac{1}{2},\frac{1}{2}}
    \oplus (\rep{1},\rep{2},\rep{2})_{-\frac{1}{2},-\frac{1}{2}} .
  \end{aligned}
\end{equation}
By construction the charge under $\lie{so}(2)_{\twist}$ is the sum of those under $\lie{so}(2)_{\old}$ and $\lie{so}(2)_{\Rsymm}$.
Thus, under the $\lie{so}(1,3)\times\lie{so}(3)_{\Rsymm}\times\lie{so}(2)_{\twist}$ symmetry of $\RR^{1,5}$ that we are concentrating on, Poincar\'e supercharges transform as
\begin{equation}
  (\rep{2},\rep{1};\rep{2})_1 \oplus (\rep{2},\rep{1};\rep{2})_0
  \oplus (\rep{1},\rep{2};\rep{2})_0 \oplus (\rep{1},\rep{2};\rep{2})_{-1} .
\end{equation}
We denote them respectively as
\begin{equation}
  Q_z^{\alpha A}, Q^{\alpha A} , \bar{Q}^{\dot{\alpha}A} , \bar{Q}_{\bar{z}}^{\dot{\alpha}A} ,
\end{equation}
where $\alpha,\dot{\alpha},A$, ranging from $1$ to~$2$, are indices for spinors of $\lie{so}(1,3)$ of the two chiralities and spinors of $\lie{so}(3)_{\Rsymm}$, respectively,
while $z$~is a complex coordinate on the $\RR^2$~factor that keeps track of $\lie{so}(2)_{\twist}$ charges $\pm 1$ of the first and last supercharges $Q_z,\bar{Q}_{\bar{z}}$.

The middle two supercharges $Q,\bar{Q}$ are scalars under $\lie{so}(2)_{\twist}$ rotations, so that deforming the metric on~$\RR^2$ to any curved metric preserves these supercharges.
Altogether, upon compactifying on $\RR^{1,3}\times C$ with the partial topological twist we obtain a system that preserves $\lie{iso}(1,3)$ Poincar\'e symmetry, supercharges $Q^{\alpha A}$ and $\bar{Q}^{\dot{\alpha}A}$, and the $\lie{so}(3)_{\Rsymm}=\lie{su}(2)$ R-symmetry.
Together these form the 4d $\Nsusy=2$ Poincar\'e supersymmetry algebra.

In the limit where $C$~has zero size, we thus obtain a 4d $\Nsusy=2$ theory, generically.\footnote{The system at finite area of~$C$ has a certain moduli space of vacua, and in the scaling limit where the area is sent to zero one must specify around which vacuum to expand.  If $C$ has ``enough'' handles or punctures, then its Higgs branch has a maximally symmetric point around which it is natural to expand, and the 4d $\Nsusy=2$ limit is well-defined.  If~$C$ is a sphere with ``too few punctures'' or is a torus without punctures, there is no maximally symmetric point and the situation is more subtle, as explained in~\cite{1110.2657}.}
Twisting~\eqref{T-partial-twist} does not preserve the tracelessness of~$T$, so even though the original 6d rotation symmetry extends to conformal symmetry, this is not the case of the twisted rotation symmetry.
In the zero area limit, 4d conformal symmetry can be restored and we get an \ac{SCFT} unless data at punctures of~$C$ carry an intrinsic mass scale.

\subsection{\label{ssec:reduce-Coulomb}Coulomb branch}

The Coulomb branch of a 4d $\Nsusy=2$ theory is described by giving a \ac{VEV} to Coulomb branch operators, namely (gauge-invariant) operators of the 4d theory that are annihilated by all antichiral Poincar\'e supercharges~$\bar{Q}^{\dot{\alpha}A}$.  Let us identify these operators starting from the 6d theory~$\Xg$, following roughly~\cite[section~3]{0907.3987}.

Importantly, the resulting Coulomb branch~$\cB$ obtained in~\eqref{Coulomb-branch} \emph{only depends on the complex structure of~$C$}, not on its metric.  This lets us deform the Riemann surface in various ways to understand the resulting 4d theory, and it underlies the appearance of 2d \ac{CFT} objects on~$C$ in the \ac{AGT} correspondence.

General supersymmetry considerations allow the vacuum moduli space of 4d $\Nsusy=2$ theories~\cite{hep-th/9603042} to be a union of mixed branches $\mathcal{C}_\alpha\times\mathcal{H}_\alpha$, which include the pure Coulomb and pure Higgs branches as special cases.
The special K\"ahler manifolds~$\mathcal{C}_\alpha$ are parametrized by Coulomb branch operators and the hyper-K\"ahler manifolds~$\mathcal{H}_\alpha$ are parametrized by Higgs branch operators.
Higgs branch chiral ring relations for class~S theories were explored in~\cite{0810.4541,1305.5250,1410.6868,1411.3252}.
Determining all branches as done in~\cite{1206.4700,1404.7521} for class~S theories is in general difficult, so we will concentrate solely on the Coulomb branch (for which $\mathcal{H}_\alpha$~is a point).

\paragraph{Coulomb branch operators.}

The vacua of~$\Xg$ are parametrized by the \ac{VEV} of scalar fields~$\Phi_I$, $I=6,\dots,10$, in the Cartan subalgebra of~$\lie{g}$ (modulo the Weyl group).
The low-energy theory in a given vacuum is described by fluctuations of these fields as well as spinors~$\lambda$ and a self-dual two-form~$B$.
Under the $\bigl(\lie{so}(1,3)\times\lie{so}(3)_{\Rsymm}\bigr)\times\lie{so}(2)_{\old}\times\lie{so}(2)_{\Rsymm}$ symmetry algebra of interest to us just before the twist, these fields transform as
\begin{equation}
  \begin{aligned}
    & B \mathrlap{{}\in (\rep{3},\rep{1},\rep{1})_{0,0}\oplus(\rep{1},\rep{3},\rep{1})_{0,0}
      \oplus(\rep{2},\rep{2},\rep{1})_{\pm 1,0}\oplus(\rep{1},\rep{1},\rep{1})_{0,0},} \\
    & \lambda \in(\rep{2},\rep{1},\rep{2})_{\frac{1}{2},\pm\frac{1}{2}}\oplus (\rep{1},\rep{2},\rep{2})_{-\frac{1}{2},\pm\frac{1}{2}} , \qquad\qquad &
    \Phi_z \coloneqq\Phi_6+i\Phi_7 & \in(\rep{1},\rep{1},\rep{1})_{0,1} , \\
    & \Phi_8,\Phi_9,\Phi_{10} \in(\rep{1},\rep{1},\rep{3})_{0,0} , &
    \Phi_{\bar{z}} \coloneqq\Phi_6-i\Phi_7 & \in (\rep{1},\rep{1},\rep{1})_{0,-1} .
  \end{aligned}
\end{equation}
On the other hand the supercharges $\bar{Q}^{\dot{\alpha}A}$ transform as $(\rep{1},\rep{2},\rep{2})_{-1/2,1/2}$.  We deduce that
\begin{equation}\label{barQ-Phiz}
  \bar{Q}^{\dot{\alpha}A}\Phi_z = 0
\end{equation}
because no component of~$\lambda$ has the appropriate $\lie{so}(2)_{\Rsymm}$ charge $3/2$.

Really, we should be working with the corresponding gauge-invariant operators, such as traces $\Tr(\Phi_z^j)$ in classical cases $\lie{su}(N)$ and $\lie{so}(2N)$.
These are the Casimirs of~$\lie{g}$, polynomials $P_k(\Phi_z)$ of various degrees $d_k$ for $k=1,\dots,\rank\lie{g}$.
Concretely, for classical gauge groups these gauge-invariant operators annihilated by $\bar{Q}^{\dot{\alpha}A}$ are
\begin{equation}
  \begin{aligned}
    & \Tr(\Phi_z^j), \quad j=2,3,4,\dots,N & & \text{for }\lie{su}(N) , \\
    & \Tr(\Phi_z^j), \quad\ j=2,4,6,\dots,2N-2 , \text{ and} \quad \Pfaff(\Phi_z) & & \text{for }\lie{so}(2N) .
  \end{aligned}
\end{equation}
(We recall that the Pfaffian is a square root of the determinant.)
For reference, the degrees of Casimirs of $\lie{su}(N)$ are $2,3,\dots,N$;
of $\lie{so}(2N)$ are $2,4,6,\dots,2N-2$ and~$N$;
of $\lie{e}_6$ are $2,5,6,8,9,12$;
of $\lie{e}_7$ are $2,6,8,10,12,14,18$;
of $\lie{e}_8$ are $2,8,12,14,18,20,24,30$.

It is often convenient to replace $\Phi_z$ by $\Phi_z \rmd z$ to soak up the $z$~index and obtain a tensor.  Then we work with the order~$d_k$ differentials $P_k(\Phi_z)\rmd z^{d_k}$ on the holomorphic curve (aka Riemann surface)~$C$.
A somewhat different basis is more practical: for instance for $\lie{su}(N)$ one expands
\begin{equation}\label{detXPhi}
  \det(X-\Phi_z\rmd z) = X^N - \sum_{j=2}^N \cO_j\,X^{N-j} .
\end{equation}

\begin{exercise}
  Check that $\cO_2 = \Tr(\Phi_z^2/2)\rmd z^2$, $\cO_3 = \Tr(\Phi_z^3/3)\rmd z^3$, and perhaps check that $\cO_4 = \Tr(\Phi_z^4/4)\rmd z^4-\cO_2^2/2$.  Why is there no $\cO_1$?
\end{exercise}

\paragraph{Coulomb branch.}

What \ac{VEV} can we give $\cO_j$ to define a vacuum?  Denote it by\footnote{The notation is slightly ill-defined in the case of $\lie{so}(4K)$ because there are then two Casimirs of the same degree $2K$, leading to two order~$2K$ differentials: $\phi_{2K}$ defined from traces of powers of~$\Phi_z$, and $\tilde\phi_{2K}=\vev{\Pfaff(\Phi_z)}\,\rmd z^{2K}$.}
\begin{equation}
  \phi_j \coloneqq \vev{\cO_j} .
\end{equation}
It should be constant along $\RR^{1,3}$ to avoid breaking Poincar\'e symmetry.
Next we use the anticommutator $\{\bar{Q}^{\dot{\alpha}A},\bar{Q}_{\bar{z}}^{\dot{\beta}B}\}\sim\epsilon^{\dot{\alpha}\dot{\beta}}\epsilon^{AB}\del_{\bar{z}}$ to deduce that $\phi_j$ must depend holomorphically on~$z$:
\begin{equation}
  \epsilon^{\dot{\alpha}\dot{\beta}}\epsilon^{AB} \del_{\bar{z}} \phi_j
  \sim \vev*{\bar{Q}^{\dot{\alpha}A}\bigl(\bar{Q}_{\bar{z}}^{\dot{\beta}B}\cO_j\bigr)}
  + \vev*{\bar{Q}_{\bar{z}}^{\dot{\beta}B}\bigl(\bar{Q}^{\dot{\alpha}A}\cO_j\bigr)} = 0 .
\end{equation}
The first term vanishes because the twisted compactification preserves the supercharge~$\bar{Q}$.  The second term vanishes by construction of~$\cO_j$.

The Coulomb branch of the 4d $\Nsusy=2$ theory is thus parametrized by degree~$d_k$ differentials $\phi_{d_k}$ on~$C$ for $k=1,\dots,\rank\lie{g}$.
When there are no punctures,
\begin{equation}\label{Coulomb-branch}
  \cB = \bigoplus_{k=1}^r H^0(C, K^{\otimes d_k}) , \qquad
  \phi_j \in H^0(C, K^{\otimes j}) ,
\end{equation}
where $K$ is the canonical bundle on the curve~$C$ and $H^0(C,\cL)$ is the vector space of sections of the line bundle~$\cL$ on~$C$.
When~$C$ has punctures, the sections~$\phi_{d_k}$ have prescribed behaviours at each puncture.
Starting in \autoref{sec:Lag} we explain, for concrete choices of~$C$ giving usual 4d $\Nsusy=2$ gauge theories, how to relate the parametrization~\eqref{Coulomb-branch} to the usual description in terms of scalars in 4d $\Nsusy=2$ vector multiplets.

\paragraph{Hitchin system.}

We would like to say intuitively that the 4d Coulomb branch is parametrized by the ``\ac{VEV}'' of the adjoint-valued holomorphic one-form $\Phi_z\rmd z$, a putative element of~$H^0(C,K\otimes\lie{g})$, modulo gauge transformations.  Of course, \acp{VEV} of non-gauge-invariant operators don't make sense (or are automatically zero, depending on your point of view) so talking about them is an abuse of language.
Nevertheless in our case there is a construction of the so-called Hitchin field (or Higgs field), a holomorphic one-form $\varphi=\varphi_z\rmd z$ with component $\varphi_z\in\lie{g}$, whose Casimirs give $\Tr(\Phi_z^j)\rmd z^j$ in the $\lie{su}(N)$ case and likewise in other cases.
For convenience we occasionally use $\varphi$ rather than the gauge-invariants~$\phi_j$ in some explanations.

The story is a bit longer: one compactifies the 4d theory further on~$S^1$.  Coulomb branch vacua of the 3d theory are given by solutions $(A,\varphi)$ of the Hitchin system on~$C$,
\begin{equation}\label{Hitchin}
  F+[\varphi,\bar\varphi] = 0 , \quad
  \bar\del_A\varphi = 0 , \quad
  \del_A\bar\varphi = 0 ,
\end{equation}
modulo $G$~gauge transformations.
The resulting Coulomb branch~$\cM$ (the Hitchin moduli space) admits a projection onto the Coulomb branch~$\cB$ of the 4d theory by mapping $(A,\varphi)$ to Casimirs of~$\varphi$.
The Hitchin equations~\eqref{Hitchin} are equivalent to flatness of the $G_{\CC}$~connection $A+\varphi_z \rmd z+\bar\varphi_{\bar z} d\bar z$ together with a gauge-fixing condition, so that $\cM$~can also be described as the moduli space of complex $G_{\CC}$ flat connections on~$C$ modulo $G_{\CC}$~gauge transformaions.

\paragraph{Comment on the \ac{IR} behaviour.}\label{IRbehaviour}

The low-energy limit of the 6d theory in a generic vacuum is given by an abelian 6d $(2,0)$ theory valued in the vacuum moduli space.
Likewise, in a Coulomb branch vacuum described by a given choice of differentials~$\phi_j$ in~\eqref{Coulomb-branch}, the effective description of the 4d $\Nsusy=2$ theory includes massless scalar fields describing fluctuations of the~$\phi_j$.
Together with similar dimensional reductions of $B_{\mu\nu}$ and $\lambda$, these scalar fields form 4d $\Nsusy=2$ abelian vector multiplets.

How many?
The scalar fields have the Coulomb branch~$\cB$ as their target, so we should expect an infrared description as a 4d $\Nsusy=2$ gauge theory with gauge group $\Lie{U}(1)^{\dim_{\CC}\cB}$.
At particular points on the Coulomb branch there are additional massless particles charged under this gauge group.
The Coulomb branch typically features points that are even more singular, where the low-energy dynamics are not abelian.

\subsection{\label{ssec:reduce-SW}Seiberg--Witten curve}

\paragraph{\acl{SW} curve.}
In the $\lie{su}(N)=\lie{a}_{N-1}$ case we can repackage the data of~$\phi_k$ in a geometric way in terms of the \ac{SW} curve~$\Sigma$ and \ac{SW} differential~$\lambda$ defined next.

Consider the canonical line bundle $T^*C\twoheadrightarrow C$, whose fiber at a point in~$C$ consists of one-forms at that point.
In a local coordinate $z$ on~$C$ the total space $T^*C$ admits coordinates $(z,x)$ where $x\in\CC$ describes a one-form $x\rmd z$.
There is a natural injection $C\hookrightarrow T^*C$, the ``zero section'', that maps $z\in C$ to $(z,x=0)$.
We define the (complex) curve $\Sigma\subset T^*C$ as the locus $(z,x)$ such that
\begin{equation}\label{SWcurve}
  \vev{\det(x - \Phi_z)} = \underbrace{\det(x-\varphi_z)}_{\text{see \eqref{Hitchin}}} = x^N - \sum_{j=2}^N u_j(z) x^{N-j} = 0
\end{equation}
where we used the construction~\eqref{detXPhi} of $\cO_j$ and wrote $\phi_j=\vev{\cO_j}=u_j(z)\rmd z^j$ for each exponent~$j$ of~$\lie{g}$.
Note that \eqref{SWcurve} is consistent with transformation properties of $x$ and of the $u_j$ since each term is (the sole component of) a holomorphic $N$-form.
At generic points $z\in C$ this equation~\eqref{SWcurve} has $N$~solutions, which locally gives an $N$~sheeted cover of~$C$.  Generically, at certain isolated points on~$C$ two sheets intersect with a branch point of order~$2$.
We have constructed in this way an $N$-sheeted ramified cover~$\Sigma$ of~$C$.

As we will see in concrete examples, $\Sigma$~turns out to be the \ac{SW} curve of the 4d $\Nsusy=2$ theory in the given Coulomb branch vacuum, and the \ac{SW} differential is the holomorphic one-form~$\lambda$ defined as $\lambda=x\rmd z$ in coordinates $(z,x)$ of $T^*C$.
The fact that our $(\Sigma,\lambda)$ matches the usual one is easier to see for concrete theories later on, but we can give some intuition.
Besides indirectly giving the prepotential for the low-energy $\Lie{U}(1)^{\dim_{\CC}\cB}$ vector multiplets, one of the jobs of the \ac{SW} curves is to calculate the central charge of particles (which puts a \ac{BPS} lower bound on their mass) in terms of their electric, magnetic, and flavour charges: it should be obtained by integrating $\lambda$ along closed contours in~$\Sigma$.
Let us confirm this from the M-theory perspective in the A-type case.

\paragraph{M-theory perspective on \ac{SW} curve.}
We recall that $\XsuN$ is the world-volume theory of $N$~M5~branes (with the decoupled center of mass modes removed).  The R-symmetry is then realized geometrically as transverse rotations.  The topological twist corresponds to combining the 2d rotations with 2d transverse rotations, and one finds that the full geometrical set-up corresponding to $\XsuN$ partially twisted on $\RR^{1,3}\times C$ is to consider M-theory on $\RR^{1,3}\times T^*C\times\RR^3$ and to place M5~branes along $\RR^{1,3}\times C$, the zero section.\footnote{More generally, $T^*C$ can be replaced by a four-dimensional hyper-K\"ahler manifold and $C$ by a holomorphic cycle inside~$Q$.}

Moving onto the Coulomb branch corresponds to shifting the M5~branes away from each other along the fibers of~$T^*C$.  Since the branes are indistinguishable they generically reconnect into an $N$-sheeted ramified cover $\Sigma\subset T^*C$ of~$C$.  Supersymmetry requires it to be holomorphic and we thus reproduce the above classification of Coulomb branch vacua.
We emphasize that the \ac{UV} curve $C$ characterizes the theory, while the \ac{IR} curve  (or \ac{SW} curve) $\Sigma$ depends on (and characterizes) the given Coulomb branch vacuum.

Excitations of the brane system include massless fluctuations along the Coulomb branch of course, but also very interesting massive excitations coming from M2~branes ending on the M5~branes.  Consider a two-dimensional surface $D\subset T^*C$ whose boundary lies in the \ac{SW} curve, $\del D\subset\Sigma$, and let us place an M2~brane along $D\times\RR$ where $\RR$ is the time direction in 4d Minkowski space.
From the 4d point of view this describes a particle sitting still as time passes.
Its mass~$m$ is simply the area of~$D$,
\begin{equation}
  m = \int_D \abs{\rmd z\rmd x}
  \geq \abs[\Big]{\int_D \rmd z\rmd x} = \abs[\Big]{\int_D \rmd(x\rmd z)} = \abs[\Big]{\int_{\del D} \lambda} .
\end{equation}
This reproduces the \ac{BPS} lower bound expected from the \ac{SW} curve and differential $(\Sigma,\lambda)$.
In fact, realizing the \ac{SW} curve~$\Sigma$ as a fibration over~$C$ gives slightly finer control of the \ac{BPS} spectrum than just knowing~$\Sigma$ (and~$\lambda$).
Indeed, some closed curves on~$\Sigma$ are not the boundary of any two-dimensional $D\subset T^*C$, so that the M-theory setup ``knows'' that no \ac{BPS} state with these charges exist, while the data of $(\Sigma,\lambda)$ only would not know it.

These M-theory considerations suggest that we found the right notion of \ac{SW} curve and differential for class~S theories.  But we have yet to explain any concrete description of the 4d theories, rather than only their \ac{IR} behaviour on the Coulomb branch.  We turn to this next.

\subsection{\label{ssec:reduce-pieces}Tubes and tinkertoys}

So far we only discussed the low-energy effective description of $\TgCD$ on its Coulomb branch.
We now study how the class~S theory can be described without moving along its Coulomb branch.
Our guide to find such a description is that it should reproduce the aforementioned \ac{IR} physics (it also reproduces some protected observables), and that different descriptions we find should be (exactly) dual to each other.
Recall that the partial twist ensures that 4d physics we are interested in only depends on the complex structure of the Riemann surface $C$ on which we compactify.
We can thus pick any metric compatible with this complex structure.

\paragraph{Gluing.}
Consider two punctures $p_1,p_2\in\fullC$ of the same (or of different) punctured Riemann surface $C=\fullC\setminus\{p_i\}$ and consider disks around~$p_1$ and~$p_2$.
As far as the complex structure is concerned, these punctured disks are the same as semi-infinite cylinders thanks to the exponential map (expressed here in coordinates centered at~$p_i$)
\begin{equation}\label{expo-map}
  \begin{aligned}
    \exp\colon (-\infty,\rho_i]\times(\RR/(2\pi\ZZ))
    \xrightarrow{\sim} \bigl\{z\bigm| |z|\leq e^{\rho_i}\bigr\}\setminus\{0\} . \\
    \begin{tikzpicture}
      \node at (-.2,0){$p_i$};
      \draw (3,-.2) -- (0,-.2);
      \draw (0,.2) -- (3,.2);
      \draw (3,0) circle [x radius=.1, y radius=.2];
      \node at (3,-.2) [below] {$\rho_i$};
      \draw (5,0) circle[radius=.6];
      \filldraw (5,0) circle[radius=.05] node[right]{$p_i$};
    \end{tikzpicture}
    \qquad\qquad
  \end{aligned}
\end{equation}
We can glue two such semi-infinite cylinders by cutting their infinite end off at some finite distance and identifying the cutoff points on the left side of the following diagram:
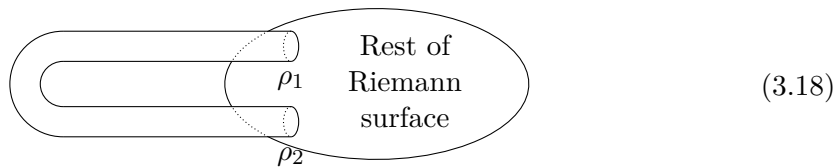
\begin{equation}\label{gluing-cylinder}
  \mathtikz{
    \draw (3,-.2) -- (0,-.2);
    \draw (0,.2) -- (3,.2) arc (90:-90:.1 and .2);
    \draw[densely dotted] (3,.2) arc (90:270:.1 and .2);
    \node at (3,-.2) [below] {$\rho_1$};
    \begin{scope}[shift={(0,-1)}]
      \draw (3,-.2) -- (0,-.2);
      \draw (0,.2) -- (3,.2) arc (90:-90:.1 and .2);
      \draw[densely dotted] (3,.2) arc (90:270:.1 and .2);
      \node at (3,-.2) [below] {$\rho_2$};
    \end{scope}
    \draw (0,-.2) arc (90:270:.3);
    \draw (0,.2) arc (90:270:.7);
    \draw[densely dotted] (2.7,.2) arc (135.573:162.5423:2 and 1);
    \draw (2.7,.2) arc (135.573:-135.573:2 and 1);
    \draw[densely dotted] (2.7,-1.2) arc (-135.573:-162.5423:2 and 1);
    \draw (2.22,-.2) arc (162.5423:197.4576:2 and 1);
    \node at (4.5,-.5) {\begin{tabular}{c}Rest of\\Riemann\\surface\end{tabular}};
  }
\end{equation}
where the ``rest'' can remain connected or be disconnected upon removing the tube.
In terms of complex coordinates $w$ and $z$ around $p_1$ and $p_2$ respectively (with $p_1$ at $w=0$ and $p_2$ at $z=0$), the identification is
\begin{equation}
  zw=q
\end{equation}
for some parameter~$q$.
The modulus $\abs{q}$ gives the aspect ratio (length over circumference) $(-\log|q|)/2\pi$ of the tube, while the phase of~$q$ indicates how the cylinders are rotated before gluing.
The coordinates $w,z$ are only locally defined so $\abs{q}$ cannot be too big: the tube can be arbitrarily long/thin but not too short/thick, as the description otherwise breaks down.

\begin{exercise}[On punctured spheres]\label{exe:punct-sph}
  Choose a coordinate~$w$ on the complex projective plane $\CP^1$ (the two-sphere), where $w\in\CC\cup\{\infty\}$.

  1. For $n\geq 3$ arbitrary distinct points $w_j\in\CC\cup\{\infty\}$, $j=1,\dots,n$,
  define a new coordinate $z(w)\coloneqq\frac{(w-w_1)(w_2-w_3)}{(w-w_3)(w_2-w_1)}$.
  Check that $w\mapsto z$ is bijective on~$\CP^1$ so that the definition gives a good coordinate on~$\CP^1$.
  Check that $w_1,w_2,w_3$ are mapped to $0,1,\infty$.
  The coordinate $z(w_j)$ for $j>3$ is called cross-ratio of $w_1,w_2,w_3,w_j$.

  2. In the four-punctured sphere, how does the cross-ratio~$q$ change when $w_1,w_2,w_3,w_4$ are permuted?

  3. Construct the four-punctured sphere $\CP^1\setminus\{0,q,1,\infty\}$ by gluing two three-punctured spheres $\CP^1\setminus\{0,1,\infty\}$.  (Hint: let $x,y$ be coordinates on the two three-punctured spheres; identify $qx=y$ for some region $1<|x|<1/|q|$.)
  Generalize to the $n$-punctured sphere.
  
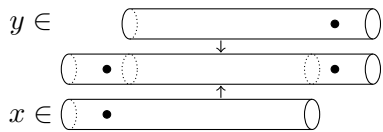
\begin{center}
    \begin{tikzpicture}
      \begin{scope}[shift={(0,.6)}]
        \draw (4,-.2) -- (0.8,-.2) arc (-90:-270:.1 and .2) -- (4,.2);
        \draw[densely dotted] (0.8,-.2) arc (-90:90:.1 and .2);
        \draw (4,0) circle [x radius=.1, y radius=.2];
        \filldraw[radius=.05] (3.5,0) circle;
        \node at (-.5,0) {$y\in$};
      \end{scope}
      \begin{scope}[shift={(0,-.6)}]
        \draw (3.2,-.2) -- (0,-.2) arc (-90:-270:.1 and .2) -- (3.2,.2);
        \draw[densely dotted] (0,-.2) arc (-90:90:.1 and .2);
        \draw (3.2,0) circle [x radius=.1, y radius=.2];
        \filldraw[radius=.05] (.5,0) circle;
        \node at (-.5,0) {$x\in$};
      \end{scope}
      \draw[->] (2,.38) -- (2,.22);
      \draw[->] (2,-.38) -- (2,-.22);
      \draw (4,-.2) -- (0,-.2) arc (-90:-270:.1 and .2) -- (4,.2);
      \draw[densely dotted] (0,-.2) arc (-90:90:.1 and .2);
      \draw (4,0) circle [x radius=.1, y radius=.2];
      \filldraw[radius=.05] (.5,0) circle;
      \filldraw[radius=.05] (3.5,0) circle;
      \draw[densely dotted, x radius=.1, y radius=.2] (0.8,0) circle;
      \draw[densely dotted, x radius=.1, y radius=.2] (3.2,0) circle;
    \end{tikzpicture}
  \end{center}
\end{exercise}

\paragraph{Vector multiplets.}
Despite how it is drawn in~\eqref{gluing-cylinder}, the cylinder connecting the two punctures is flat and of constant circumference $2\pi L_5$ (for some metric).
Let us denote the directions along and around the cylinder as $x^4,x^5$.
We know that the 6d theory $\Xg$ reduced on a circle gives 5d $\Nsusy=2$ \ac{SYM} with gauge algebra~$\lie{g}$ and coupling $\gfived^2\simeq L_5$.
We should thus expect that part of the system obtained by reducing $\Xg$ on the glued surface~\eqref{gluing-cylinder} is 5d $\Nsusy=2$ \ac{SYM} on an interval of length $L_4\sim (-\log|q|)L_5$.
In the limit where $C$ shrinks to a point, the 5d term $\Tr(F^2)$ does not depend much on the $x^4$ direction, thus the 4d Lagrangian ought to have a term
\begin{equation}
  \frac{1}{\gfived^2} \int_I \Tr(F^2) = \frac{1}{\gfourd^2} \Tr(F^2) , \qquad
  \frac{1}{\gfourd^2} = \frac{L_4}{\gfived^2} \simeq \frac{L_4}{L_5} = -\log|q|.
\end{equation}

What about the phase of~$q$, which implements a translation along the circle direction, namely a rotation around the cylinder?
The instanton current of a 5d gauge field is defined as
\begin{equation}
  J^{\inst}_\mu = \epsilon_{\mu\nu\rho\sigma\tau} \Tr(F^{\nu\rho} F^{\sigma\tau}) .
\end{equation}
As we mentioned earlier, in the reduction from $\Xg$ to 5d $\Nsusy=2$ \ac{SYM} the \ac{KK} (Kaluza--Klein) modes correspond to instanton particles of the 5d theory, namely the \ac{KK} momentum in~$x^5$ is equal to the charge under~$J^{\inst}$.
Thinking of~$x^4$ as Euclidean time, the translation operator~$P_5$ is given as an integral of the ``time'' component~$J_4^{\inst}$ over the 4d ``spatial'' directions $x^0,\dots,x^3$.
Twisting the cylinder by an angle $\theta=\Im\log|q|$ thus contributes a term $\theta \Tr(F\wedge F)$ to the 4d Lagrangian when we eventually reduce $C$~to a point.

Altogether we expect that a long cylinder as in~\eqref{gluing-cylinder} should yield a 4d $\Nsusy=2$ vector multiplet with complexified gauge coupling~$\tau$ given by $\log q$:
\begin{equation}
  \tau=\frac{\theta}{2\pi} + \frac{4\pi i}{g^2} , \qquad
  q \sim e^{2\pi i\tau} .
\end{equation}
The relation is made more precise later in concrete geometries.

\paragraph{Pants decomposition and S-duality.}

Vector multiplets must gauge flavour symmetries of some matter sector, and our next task is to understand where that matter comes from.
For this, the key is to send gauge couplings to zero, because in this limit the vector multiplet decouples and leaves behind the matter sector with its flavour symmetries.

\begin{exercise}
  As a toy model, consider a scalar field~$\phi$ transforming in some representation of a group~$G$, and gauge the symmetry~$G$ using a gauge field~$A$.
  We denote by $D=\rmd+A$ the covariant derivative and $F=\rmd A+A\wedge A$, and ignore numerical factors.
  By introducing a field $\tilde{A}=g^{-1}A$ with canonically normalized kinetic term, show how
  \begin{equation}
    \Lag = \frac{1}{g^2} \Tr(F^2) + |D\phi|^2 \xrightarrow{g\to 0}
    \Tr(\rmd\tilde{A})^2 + |\rmd\phi|^2 .
  \end{equation}
  Note that in the limit the flavour symmetry~$G$ of~$\phi$ is not gauged any longer.
  The original gauge theory can be then restored (up to the free gauge field~$\tilde{A}$) by gauging this flavour symmetry with a new gauge field.
  Check the same decoupling happens for spinors ($\bar\psi\gamma^\mu D_\mu\psi$).
\end{exercise}

Any punctured Riemann surface~$C$ with genus~$g$ and $n$~punctures, except for $(g,n)$ among $(0,0)$, $(0,1)$, $(0,2)$, $(1,0)$, can be decomposed into three-punctured spheres (pairs of pants) glued as described above.  Such a decomposition is called a pants decomposition.
For each pants decomposition of~$C$ there is a corresponding cusp in the moduli space $\cM_{g,n}$ of Riemann surfaces with genus~$g$ and $n$~punctures.
At this cusp, $C$~is described by three-punctured spheres joined by infinitely thin tubes.
Each such tube yields an infinitely weakly coupled vector multiplet in the 4d theory, so that in this limit we can expect 6d fields ``localized'' on each pair of pants to decouple from each other since the 4d vector multiplets joining them become free:
\begin{equation}
  \mathtikz{
    \begin{scope}[x radius=1, y radius=.5]
      \draw (0,0) circle;
      \draw (4.5,0) circle;
      \draw (8.5,0) circle;
    \end{scope}
    \begin{scope}[radius=0.05]
      \filldraw (-.42,.28) circle;
      \filldraw (-.42,-.28) circle;
      \filldraw (4.92,.28) circle;
      \filldraw (8.08,.28) circle;
      \filldraw (9.2,0) circle;
    \end{scope}
    \filldraw[white] (.7,.05) rectangle (3.8,-.05);
    \draw (.7,0) circle [x radius=0.03, y radius=0.05];
    \draw (3.8,0) circle [x radius=0.03, y radius=0.05];
    \draw (.7,.05) -- (3.8,.05);
    \draw (.7,-.05) -- (3.8,-.05) node [midway,above] {free vector};
    \filldraw[white] (4.92,-.33) rectangle (8.08,-.23);
    \draw (4.92,-.28) circle [x radius=0.03, y radius=0.05];
    \draw (8.08,-.28) circle [x radius=0.03, y radius=0.05];
    \draw (4.92,-.33) -- (8.08,-.33) node [midway,above] {free vector};
    \draw (4.92,-.23) -- (8.08,-.23);
  }
\end{equation}
As in the toy model, the symmetries gauged by the vector multiplet are restored as flavour symmetries in the zero coupling limit.

The picture that emerges is as follows.  The building blocks of $\TgCD$ are class~S theories called tinkertoys associated to three-punctured spheres.  These (4d $\Nsusy=2$) tinkertoys have flavour symmetries associated to each puncture, which we study carefully later.  For each tube, consider the flavour symmetry groups $F_1$ and~$F_2$ associated to the two punctures that it connects, and gauge a suitable diagonal subgroup $F\subset F_1\times F_2$ using a 4d $\Nsusy=2$ vector multiplet.  This yields $\TgCD$.
This description of $\TgCD$ for each pants decomposition of~$C$ can be written schematically as
\begin{equation}\label{TgCD-pants-tubes}
  \TgCD = \biggl(\prod_{\text{pants}} \Theory(\lie{g},\text{sphere}\setminus\text{3pt}) \biggr) \biggm/ \biggl( \prod_{\text{tubes}} \text{gauge group} \biggr) .
\end{equation}
A large part of the work in understanding the \ac{AGT} correspondence is to classify tinkertoys obtained from three punctured spheres with different types of punctures.
Their flavour symmetry can be rather intricate, which is why we cannot make the gauge groups more explicit in~\eqref{TgCD-pants-tubes} in such generality.

When all punctures are so-called full tame punctures (explained later), all building blocks are the same tinkertoy~$T_{\lie{g}}$.  This theory is an isolated\footnote{An \ac{SCFT} with a certain amount of supersymmetry is \emph{isolated} if it does not have any exactly marginal deformation with the same supersymmetry (such as gauge couplings in 4d).} \ac{SCFT} with (at least) $\lie{g}^3$ flavour symmetry associated to its three punctures.  For $\lie{g}=\lie{su}(2)$ it consists of four free hypermultiplets, while for other~$\lie{g}$ it has no 4d $\Nsusy=2$ Lagrangian description.

Of course, a given Riemann surface has many inequivalent decompositions into pairs of pants.  Each one leads to a description of $\TgCD$ as a weakly coupled gauge theory in one corner of parameter space.  At strong coupling (short tubes) another description may be weakly coupled hence more useful.  In concrete cases this reproduces known 4d $\Nsusy=2$ S-dualities.
Here is an exercise to get an intuition about pants decompositions.

\begin{exercise}[Combinatorics of pants decompositions]
  1. Given a surface $C_{g,n}$ with genus~$g$ and $n$~punctures, check that all pants decompositions use the same number of three-punctured spheres.

  2. Draw the three topologically different\footnote{Two pants decompositions are the same in this sense if the closed curves cutting the surface into pieces with three boundaries can be deformed into each other without (self)\llap{-}intersection or crossing punctures.} pants decompositions of a four-punctured sphere (assuming punctures are distinguishable).  How many pants decompositions does an $n$-punctured sphere have?  Does a once-punctured torus have a finite number of pants decompositions?

  3. Return to point~3 of \autoref{exe:punct-sph} and construct the sphere with $n=4$ (or~$5$) punctures by gluing three-punctured spheres in all possible ways.

  4. We don't need to degenerate the Riemann surface completely down to pairs of pants: as soon as~$C$ involves one long tube the theory $\TgCD$ can be written in terms of a weakly coupled vector multiplet gauging flavour symmetries of a ``smaller'' class~S theory.  What Riemann surface (genus, punctures) underlies the latter theory?  There are two cases depending on whether the surface disconnects.
\end{exercise}

\section{\label{sec:Lag}Lagrangians for class S theories}

After discussing the tame punctures that arise when pinching tubes, we argue in \autoref{ssec:Lag-trif} that $\Xsuii$ on a sphere with three tame punctures yields $4$~free hypermultiplets, with a flavour symmetry $\SU(2)^3$ made manifest.
In \autoref{ssec:Lag-SQCD} we glue two such building blocks to learn how $\Xsuii$ on a sphere with four tame punctures reproduces known aspects of 4d $\Nsusy=2$ \ac{SQCD} with gauge group $\SU(2)$ and $N_f=4$ flavours.
This is the conventional starting point of \ac{AGT} reviews: one usually studies S-duality of $\SU(2)$ \ac{SQCD}~\cite{hep-th/9408099} and of quivers gauge theories~\cite{hep-th/9910125}, before explaining the unifying 6d point of
view~\cite{0904.2715}. We extend the discussion in \autoref{ssec:Lag-SU2} to generalized $\SU(2)$ quivers arising
\apartehere
from $\Xsuii$ on arbitrary punctured Riemann surfaces.

In \autoref{ssec:Lag-linear} we realize as class~S theories some $\SU(N)$ linear quiver gauge theories including $\SU(N)$ \ac{SQCD} with $N_f=2N$ flavours.
This teaches us that there are several types of tame punctures hence several types of codimension~$4$ operators in the 6d theory.
All theories we consider in this section are such that gauge couplings have vanishing one-loop beta function, and this implies that the couplings have vanishing beta function at all orders~\cite{1005.3546}.

\startaparte[title={Description of punctures and some tinkertoys}]
\begin{itemize}
\item The $\lie{su}(2)$ tinkertoy consist of a trifundamental half-hypermultiplet of $\SU(2)^3$.
\item The $\lie{su}(N)$ tinkertoy with two full and one simple puncture is an $\SU(N)^2$ bifundamental hypermultiplet.
\item For $\lie{su}(N)$, at a full tame puncture~$z_0$ the differentials obey $\phi_k=\bigl(\frac{\sigma_k}{(z-z_0)^k}+O(\frac{1}{(z-z_0)^{k-1}})\bigr)\rmd z^k$ for $2\leq k\leq N$ where $\sigma_k$ has mass dimension~$k$.
  Other tame punctures arise by restricting the pole coefficients down to orders $<k-1$.
\item Many linear quivers have both class~S and IIA descriptions; Coulomb branches and \ac{SW} curves work out.
\end{itemize}
\stopaparte

\subsection{\label{ssec:Lag-trif}Trifundamental tinkertoy}

We discuss tame punctures; for $\lie{su}(2)$ there is only one type.
We then consider $\Xsuii$ on a sphere $\CP^1$ with three tame punctures at $0,1,\infty$ and we argue that the resulting tinkertoy $T_2=T_{\lie{su}(2)}$, which is the main building block of $\lie{su}(2)$ class~S theories, is a collection of four free hypermultiplets.
There is no first principle derivation of that fact, but we will see many checks of it, especially correct postdictions of Coulomb branch and \ac{SW} curves of many gauge theories, as well as consistency with string theory dualities.

\paragraph{Tame punctures.}
We describe punctures in terms of their effect on the Hitchin field $\varphi(z)$ parametrizing the Coulomb branch, or gauge-invariantly in terms of the higher-order differentials~$\phi_{d_k}$, $k=1,\dots,\rank\lie{g}$.

Punctures can arise from pinching a thin tube.
In a complex coordinate $w\in\RR\times S^1$ describing this tube, the $\phi_{d_k}$ often tend to constants (times $dw^{d_k}$) inside the thin tube.
Cutting the cylinder (the opposite of what we did in~\eqref{gluing-cylinder}) and applying the exponential map~\eqref{expo-map} $z=e^w$, we generically expect
\begin{equation}\label{tame-phidk}
  \phi_{d_k} \simeq \frac{\rmd z^{d_k}}{z^{d_k}} + \dots
\end{equation}
with some coefficients, in a local coordinate~$z$ in which the puncture is at $z=0$.

This motivates us to work with \emph{tame punctures}, namely points where $\varphi(z)\rmd z$ has a first order pole with a prescribed residue, of course up to gauge conjugation: the prescribed residue translates generically to a prescribed leading coefficient in~\eqref{tame-phidk} ---a \emph{full} tame puncture.
We study other punctures later in \autoref{sec:gen}: tame punctures in which~$\phi_{d_k}$ have lower-order poles instead of~\eqref{tame-phidk}, and wild punctures defined as having higher-order poles.

\paragraph{Massive and massless tame punctures for $\lie{su}(2)$.}

For now we focus on $\lie{su}(2)$: there is then a single type of tame puncture.

This case has a single Casimir, the quadratic differential $\phi_2=\tfrac{1}{2}\Tr(\varphi^2)\rmd z^2$.
We impose the residue of the Hitchin field~$\varphi$ up to conjugation (which we denote $\sim$):
for non-zero $m\in\CC$,
\begin{equation}\label{tame-massive-SU2}
  \varphi(z) \sim \biggl(\frac{\diag(m,-m)}{z-z_i} + O(1)\biggr)\rmd z
  \implies \phi_2(z) = \biggl(\frac{m^2}{(z-z_i)^2} + O\biggl(\frac{1}{z-z_i}\biggr)\biggr)\rmd z^2 .
\end{equation}
We call $m\neq 0$ the \emph{mass parameter} of the puncture for the following reason.
The sheets of~$\Sigma$ defined in~\eqref{SWcurve} behave as $x_{\pm}(z)=\pm m/(z-z_i)+O(1)$, and integrating the \ac{SW} differential~$\lambda$ around~$z_i$ on one of the two sheets picks up the residue $\pm m$.  This means $m$~appears as a contribution to the central charge hence to masses of \ac{BPS} particles.

Naively, taking the $m\to 0$ limit in the $\varphi(z)$ asymptotics changes~$z_i$ into a regular point.
In the $\phi_2$ equation however, the puncture remains as a first order pole.
This is explained from the $\varphi(z)$ point of view by noting that it is only defined up to conjugation.
Conjugating the diagonal matrix $\diag(m,-m)$ before taking the $m\to 0$ limit can yield a non-zero value,
\begin{equation}
  \begin{pmatrix} m & 0 \\ 0 & -m \end{pmatrix}
  \sim \begin{pmatrix} m & 1 \\ 0 & -m \end{pmatrix}
  \xrightarrow{m\to 0} \begin{pmatrix} 0 & 1 \\ 0 & 0 \end{pmatrix} .
\end{equation}
Indeed, we find a consistent \emph{massless} tame puncture
\begin{equation}\label{tame-massless-SU2}
  \varphi(z) \sim \Biggl( \begin{pmatrix} 0 & 1 \\ 0 & 0 \end{pmatrix} \frac{1}{z-z_i} + O(1)\Biggr) \rmd z
  \implies \phi_2(z) = O\biggl(\frac{1}{z-z_i}\biggr)\rmd z^2
\end{equation}
where the pole has a free coefficient.  Interestingly, the sheets of~$\Sigma$ defined by $x^2\rmd z^2=\phi_2$ admit a branch point at such a massless puncture.

\begin{exercise}
  1. For any $\alpha\in\CC$ find an invertible matrix $g\in\Lie{SL}(2,\CC)$ such that
  $g^{-1}\diag(m,-m)g=\bigl(\begin{smallmatrix} m & \alpha \\ 0 & -m \end{smallmatrix}\bigr)$.

  2. Check that all $\bigl(\begin{smallmatrix} 0 & \alpha \\ 0 & 0 \end{smallmatrix}\bigr)$, $\alpha\neq 0$, are conjugate to each other.

  3. How does the coefficient of $1/(z-z_i)$ arise in \eqref{tame-massive-SU2} and \eqref{tame-massless-SU2} from components of the $(z-z_i)^0$ term in the expansion of~$\varphi$?
\end{exercise}

We interpret~\eqref{tame-massive-SU2} as follows:
the massless puncture~\eqref{tame-massless-SU2} carries $\SU(2)$ flavour symmetry, and turning on a constant scalar $\phi_{\text{background}}=m$ in a background vector multiplet coupled to that symmetry changes the puncture to the massive one~\eqref{tame-massive-SU2}.

\paragraph{Three-punctured sphere and symmetries.}

Now consider $T_2$, the result of placing $\Xsuii$ on a sphere $\CP^1$ with three tame punctures.
What do we know for sure about~$T_2$?

First, it should have at least $\SU(2)^3$ flavour symmetry, one $\SU(2)$ per puncture.
We learn in the next exercise that $4$ free hypermultiplets indeed have an $\USp(8)$ flavour symmetry, which contains $\SU(2)\times\Spin(4)=\SU(2)^3$.
The $\USp(8)$ flavour symmetry is in this context an emergent symmetry that is only present in the limit where $C$~shrinks to a point; it is not a symmetry of the 6d setup.

\begin{exercise}\label{exe:free-symm}
  1. Check that $k$~free scalar fields carry $\Lie{O}(k)$ flavour symmetry.
  Check that $p$ free hypermultiplets contain $4p$~free scalars hence have $\Lie{O}(4p)$ symmetries (and more from spinors).
  Out of these, check a $\USp(2p)$ subgroup commutes with $\SU(2)_{\Rsymm}$ R-symmetry.  (Hint: as an intermediate step, the $\Lie{U}(2p)$ subgroup commutes with $J_3\in\SU(2)_{\Rsymm}$.)

  2. Now gauge an $\SU(2)=\USp(2)$ flavour symmetry embedded diagonally into $\USp(2)^p\subset\USp(2p)$.  The gauged $\lie{su}(2)$ times $\lie{su}(2)_{\Rsymm}$ combine into $\lie{so}(4)$.  Check that the $4p$ scalars organize as $p$ copies of the fundamental representation of $\lie{so}(4)$.  Deduce that the remaining flavour symmetry is $SO(p)$.  It is in fact $\Spin(p)$ because of the action on spinors in the hypermultiplets.
\end{exercise}

\paragraph{The trifundamental half-hypermultiplet.}

All three $\SU(2)$ symmetries of the four hypermultiplets can be made manifest, at the cost of hiding $\Nsusy=2$ supersymmetry.
Split each hypermultiplet into a pair of $\Nsusy=1$ chiral multiplets $(q,\tilde{q})$.
The four hypermultiplets split into eight $\Nsusy=1$ chiral multiplets $q_{aiu}$ where $a,i,u$ (ranging from $1$ to~$2$) are indices for the three independent~$\SU(2)$.
To reconstruct the hypermultiplets as $(q,\tilde{q})$ simply introduce the notation
\begin{equation}\label{trifund-half-hyper}
  \tilde{q}^{aiu} = \epsilon^{ab}\epsilon^{ij}\epsilon^{uv}q_{bjv} .
\end{equation}
The hypermultiplets are thus in a trifundamental representation of $\SU(2)^3$ with a reality property~\eqref{trifund-half-hyper} that halves the number of components.
This set of matter fields is called a half-hypermultiplet.

If background vector multiplet scalars (i.e., masses $m_1,m_2,m_3$) are turned on for the three $\SU(2)^3$, then the underlying $8$~chiral multiplets have complex masses $\pm m_1 \pm m_2 \pm m_3$ for all choices of signs.
In particular one of the $4$~hypermultiplets becomes massless when $m_2=\pm m_1\pm m_3$.  This is important later.

\paragraph{\acl{SW} curve of \(T\sb 2\).}

We denote by $m_1,m_2,m_3$ the mass parameters of punctures at $0,1,\infty$ in the sense of \eqref{tame-massive-SU2} or~\eqref{tame-massless-SU2}.\footnote{We don't know at this stage that they are the same parameters as in the last paragraph about the trifundamental half-hypermultiplet.}
The Coulomb branch (if any) of the 4d theory is parametrized by holomorphic quadratic differential $\phi_2(z)$ that have second order poles~\eqref{tame-massive-SU2} or first-order in the massless case~\eqref{tame-massless-SU2} at each of the punctures, and no other pole.
The puncture at infinity translates to a condition as $z\to\infty$:
\begin{equation}\label{tame-T2-infinity}
  \phi_2(z) = \biggl(\frac{m_3^2}{z^2} + O\biggl(\frac{1}{z^3}\biggr)\biggr)\rmd z^2 .
\end{equation}

We recall Liouville's theorem regarding entire functions (holomorphic functions on~$\CC$ with no pole): if an entire function~$f$ is bounded as $\abs{f(z)}<Kz^p$ for some constant~$K$ and exponent~$p$ then $f$ is a polynomial of degree at most~$p$.

\begin{exercise}\label{exe:curve-T2}
  Find a quadratic differential $\phi_2(z)=u_2(z)\rmd z^2$ that has the prescribed second order poles at $0,1,\infty$ and no other singularity and show it is unique.
  (Hint: write it as $u_2(z)=f(z)/(z^2(z-1)^2)$, change variables to $w=1/z$ to polynomially bound~$f$ at infinity and use Liouville's theorem to bound the degree of~$f$, then compare with the prescribed asymptotics to fix coefficients.)
\end{exercise}

The Coulomb branch is thus a single point, which is consistent with the lack of vector multiplet in our description of $T_2$ as free hypermultiplets.  Explicitly,
\begin{equation}
  \phi_2(z) = u_2(z)\rmd z^2, \qquad u_2(z)=\frac{-m_1^2}{z^2(z-1)}+\frac{m_2^2}{z(z-1)^2} + \frac{m_3^2}{z(z-1)} .
\end{equation}

Let us find the \ac{IR} description of $T_2$ at this unique Coulomb branch vacuum.
As we commented on \autopageref{IRbehaviour}, the low-energy theory is generically a 4d $\Nsusy=2$ abelian gauge theory with the vector multiplet scalars living in the Coulomb branch~$\cB$.
Here there is no Coulomb branch hence no gauge fields in the \ac{IR}.
There may be hypermultiplets: for this we have to study the \ac{SW} curve~$\Sigma$ defined by $x^2=u_2$ and the \ac{SW} differential $\lambda=x\rmd z$.  The integral of $\lambda$ over closed cycles tells us about masses of \ac{BPS} states.

The curve $\Sigma$ is a ramified double cover of~$\CP^1$.  How many branch points does it have?
Branch points are where the two sheets $x=\pm\sqrt{u_2}$ rejoin, namely where $u_2=0$.  This happens at the (generically) two roots of the quadratic polynomial
\begin{equation}\label{u2quadpol}
  z^2(z-1)^2u_2(z) = -(z-1)m_1^2+zm_2^2+z(z-1)m_3^2 .
\end{equation}
Altogether, $\Sigma$~wraps the sphere twice, with a single branch cut.
It is thus topologically a sphere.
The three punctures at $0,1,\infty\in\CP^1$ become six point on~$\Sigma$ where the \ac{SW} differential~$\lambda$ blows up:
\begin{equation}
  \mathtikz{
    \def\tmp#1{
      \draw (0,0) circle [x radius=1.5, y radius=.5];
      \filldraw[radius=.05] (-.8,-.2) circle;
      \filldraw[radius=.05] (-.8,.2) circle;
      \filldraw[radius=.05] (1,0) circle;
      \filldraw[densely dotted,radius=.03] (.6,.1) coordinate (#1A) circle -- (.25,.13) coordinate (#1B) circle;
    }
    \node at (-2,0) {$\Sigma\colon$};
    \begin{scope}[shift={(0,.6)}]
      \tmp{SWt}
    \end{scope}
    \begin{scope}[shift={(0,-.6)}]
      \tmp{SWb}
    \end{scope}
    \fill[color=gray,opacity=.5] (SWtA.south)--(SWbA.north)--(SWbB.north)--(SWtB.south)--cycle;
    \draw[->] (1.7,.5) -- (2.3,.2);
    \draw[->] (1.7,-.5) -- (2.3,-.2);
    \begin{scope}[shift={(4,0)}]
      \tmp{C}
    \end{scope}
    \node at (3.4,-.2) {\scriptsize\strut $0$};
    \node at (3.4,+.2) {\scriptsize\strut $1$};
    \node at (5,-.2) {\scriptsize\strut $\infty$};
    \node at (5.8,0) {$\colon\!C$};
  }
\end{equation}

\paragraph{BPS spectrum.}
Contour integrals of~$\lambda$ give integer\footnote{We ignore the factor of $2\pi i$ in the residue theorem.} linear combinations of residues of $\lambda=x\rmd z=\pm\sqrt{u_2}\,\rmd z$ at the poles $z=0,1,\infty$.  By construction these residues are $\pm m_1,\pm m_2,\pm m_3$, so we find that masses (or rather central charges) of \ac{BPS} states take the form $Z=f_1m_1+f_2m_2+f_3m_3$ for $f_1,f_2,f_3\in\ZZ$.
On the other hand, the trifundamental half-hypermultiplet only has \ac{BPS} states with integer linear combinations of $\pm m_1\pm m_2\pm m_3$: this imposes the further restriction that $f_1=f_2=f_3\mod{2}$.
Does the tinkertoy~$T_2$ also have that restriction?

In \autoref{ssec:reduce-SW} we learned that M-theory instructs us to only integrate $\lambda$ over contours $\gamma$ in $\Sigma\subset T^*C$ that can be written as the boundary $\gamma=\del D$ of some two-dimensional surface $D\subset T^*C$.

\begin{exercise}
  1. First choose $D$ to be a small circle around one of the punctures, times the interval connecting the two sheets of~$\Sigma$.  Its boundary is a pair of circles picking up twice the same residue~$m_i$ (from different sheets).  Deduce that $2m_1,2m_2,2m_3$ and all their integer linear combinations are in the spectrum.

  2. Next, choose $D$ such that $\del D$ is a contour from one branch point to the other (on one sheet) and back via the other sheet.  Note that the contour $\del D$ can be deformed to a contour staying on one sheet and surrounding the cut.  Deduce that the integral of~$\lambda$ is one of the combinations $\pm m_1\pm m_2\pm m_3$ (three poles are on each side of the contour) and conclude that the \ac{BPS} spectrum of $T_2$ contains all $Z=f_1m_1+f_2m_2+f_3m_3$ with $f_1=f_2=f_3\mod{2}$.

  3. (Mathematical.) For any $D\subset T^*C$ with boundary $\del D\subset\Sigma$, consider the projection $\pi\colon T^*C\to C$ and deduce that $\pi(\del D)=\del(\pi(D))$ cannot surround a pole.  Deduce that the \ac{BPS} spectrum of $T_2$ is exactly that of the trifundamental half-hypermultiplet.
\end{exercise}

Generically, all of these \ac{BPS} particles are massive so that the low-energy theory is empty.
An interesting case is the limit $m_2\to\pm(m_1\pm m_3)$ where one of the four hypermultiplets in the trifundamental half-hypermultiplet becomes massless.  Then the \ac{SW} curve degenerates because the two branch points collide: indeed, $u_2$~has a double root~\eqref{u2quadpol}
\begin{equation}\label{u2quadpolfac}
  z^2(z-1)^2u_2(z) = (m_1\pm zm_3)^2 .
\end{equation}
The contour we considered in point~2 of the above exercise shrinks to zero size while $\lambda$ itself remains finite, so the integral is indeed zero, consistent with the vanishing mass.
We will run this kind of easy consistency checks for the more complicated theories.

\subsection{\label{ssec:Lag-SQCD}4d \(\Nsusy=2\) \(\SU(2)\) \(N\sb f=4\) \acsfont{SQCD}}

After so many generalities let us study the Coulomb branch, \ac{SW} curve and S-dualities of our first interesting concrete theory: $\Xsuii$~on a sphere with four tame punctures.

\paragraph{Identifying the 4d theory (spoilers in the title above).}
We place the four punctures at $z_1=0$, $z_2=q$, $z_3=1$, $z_4=\infty$ on the two-sphere~$\CP^1$.
The three degeneration limits $q\to 0,1,\infty$ of the four-punctured sphere correspond to all ways of clustering the punctures pairwise.
Since the three limits are identical up to permuting the punctures we concentrate on $q\to 0$.
In this limit, we expect on general grounds that the 4d theory consists of two tinkertoys $T_{\lie{su}(2)}$ and one $\SU(2)$ vector multiplet gauging an $\SU(2)$ flavour symmetry of each tinkertoy.
After this gauging each tinkertoy should still carry at least $\lie{su}(2)\times\lie{su}(2)$ flavour symmetries associated to its two remaining punctures.
We can depict this as a (generalized) quiver making all symmetries explicit:
\begin{equation}\label{SQCD-generalized-quiver}
  \Theory\bigl(\lie{su}(2),\CP^1\setminus\{0,q,1,\infty\}\bigr)
  = \mathtikz{
    \node(A)[flavor-group] at (-1.5,.5) {$\SU(2)$};
    \node(B)[flavor-group] at (-1.5,-.5) {$\SU(2)$};
    \node(C)[color-group] at (0,0) {$\SU(2)$};
    \node(D)[flavor-group] at (1.5,.5) {$\SU(2)$};
    \node(E)[flavor-group] at (1.5,-.5) {$\SU(2)$};
    \draw(-1,0)--(A);
    \draw(-1,0)--(B);
    \draw(-1,0)--(C);
    \draw(1,0)--(C);
    \draw(1,0)--(D);
    \draw(1,0)--(E);
  }
\end{equation}
Here the round node denotes a gauge group while square nodes denote flavour symmetries.
Each junction $\mathtikz{\draw(0,0)--(.2,0);\draw(-.15,.1)--(0,0)--(-.15,-.1);}$ represents our favorite tinkertoy $T_2$, the trifundamental half-hypermultiplet, i.e., four hypermultiplets transforming as two doublet representations of the $\SU(2)$ gauge group.

We thus get two flavours from the left junction and two flavours from the right junction, hence the theory is $\SU(2)$ \ac{SQCD} with $N_f=4$ flavours.
While each tinkertoy in~\eqref{SQCD-generalized-quiver} has $\Spin(4)$ flavour symmetry after gauging $\SU(2)$, the full theory has $N_f=4$ doublets of $\SU(2)$ on an equal footing hence has flavour symmetry $\Spin(8)$, larger than~$\Spin(4)^2$.
This symmetry of the 4d theory only emerges in the limit where $C$ shrinks to a point.

\paragraph{Coulomb branch.}

We denote by $m_1,m_2,m_3,m_4$ the mass parameters at each of these punctures in the sense of \eqref{tame-massive-SU2} or~\eqref{tame-massless-SU2}.
Coulomb branch vacua of the 4d theory are parametrized by holomorphic quadratic differential $\phi_2(z)$ that have second order poles~\eqref{tame-massive-SU2} or first-order in the massless case~\eqref{tame-massless-SU2} at each of the punctures, and no other pole.
The puncture at infinity translates to a condition as $z\to\infty$:
\begin{equation}\label{tame-SU2-infinity}
  \phi_2(z) = \biggl(\frac{m_4^2}{z^2} + O\biggl(\frac{1}{z^3}\biggr)\biggr)\rmd z^2 .
\end{equation}
We parametrize the possible $\phi_2(z)$ in the next exercise, starting with the massless case $m_1=m_2=m_3=m_4=0$ for which $\phi_2$ has first-order poles.

\begin{exercise}\label{exe:curve-SU24}
  1. Find all quadratic differentials $\phi_2(z)=u_2(z)\rmd z^2$ that have first order poles at $0,q,1,\infty$ and no other.  (Hint: after writing $u_2(z)=f(z)/(z(z-q)(z-1))$, change variables to $w=1/z$ to deduce a polynomial bound on $f(z)$, then use Liouville's theorem mentioned above.)

  2. Find one quadratic differential $\phi_2$ that has leading behaviour $m_i^2/(z-z_i)^2$ for $i=1,2,3$ and $m_4^2/z^2$ at infinity as per~\eqref{tame-SU2-infinity}.  Combining with the massless case deduce all such quadratic differentials.
\end{exercise}

We find a one-dimensional Coulomb branch $\cB=\CC$ with vacua\footnote{The variable~$u$ parametrizing the Coulomb branch can be freely redefined, hence you may have gotten a slightly different expression in \autoref{exe:curve-SU24}.}
\begin{equation}\label{four-punctured-phi2}
  \phi_2 = u_2 \rmd z^2, \qquad
  u_2(z) = \frac{\frac{q}{z}m_1^2 + \frac{q(q-1)}{z-q}m_2^2 + \frac{z-q}{z-1}m_3^2 + z m_4^2 - u}{z(z-q)(z-1)}
\end{equation}
labeled by $u\in\cB=\CC$.
A zero-th order check that we did not go astray is that we got the correct dimension (namely~$1$) for the Coulomb branch of $\SU(2)$ \ac{SQCD} with $N_f=4$ flavours.

\paragraph{Degeneration limit.}
As $q\to 0$, the surface degenerates, and we should obtain in a suitable sense two disconnected three-punctured spheres.
For $\abs{q},\abs{z}\ll 1$ at fixed masses, \eqref{four-punctured-phi2} behaves as
\begin{equation}\label{phi2-limit-3pt}
  \phi_2(z) \simeq \frac{\frac{-q}{z}m_1^2 + \frac{q}{z-q}m_2^2 + u}{z(z-q)} \rmd z^2 ,
\end{equation}
which is precisely the quadratic differential on a sphere with three tame punctures of masses squared $m_1^2$, $m_2^2$, and~$u$.
Likewise, for $\abs{q}\ll\abs{z},1$ \eqref{four-punctured-phi2} behaves as the quadratic differential on a three-punctured sphere with masses squared $u$, $m_3^2$ and~$m_4^2$.
This is consistent with how we introduced tame punctures in \autoref{ssec:Lag-trif}.

Since masses are background values of vector multiplet scalars, we learn from~\eqref{phi2-limit-3pt} the identification
\begin{equation}
  u = \frac{1}{2} \vev*{\Tr\phi^2}
\end{equation}
in the weakly-coupled limit $\abs{q}\ll 1$, where $\phi$ is the (dynamical) vector multiplet scalar corresponding to the $\SU(2)$ gauge group.
In other words $u$~is the usual parametrization of the Coulomb branch of \ac{SQCD}.

\paragraph{\acl{SW} curve.}
We now return to general~$q$.
The \ac{SW} curve and differential are defined by $\Sigma=\{(x,z)\in T^*\CP^1\mid x^2 = u_2(z)\}$ and $\lambda=x\rmd z$.

The curve $\Sigma$ is a ramified double cover of~$\CP^1$.  How many branch points does it have?
Branch points are where the two sheets $x=\pm\sqrt{u_2}$ rejoin, namely where $u_2=0$.  This happens at the (generically) four roots of the polynomial $z^2(z-q)^2(z-1)^2u_2(z)$, which is quartic.
Altogether, $\Sigma$~wraps the sphere twice, with four branch points joined by two branch cuts.
It is thus topologically a torus.
In addition to these branch cuts we have four punctures at $0,q,1,\infty\in\CP^1$, hence eight point on~$\Sigma$ where the \ac{SW} differential~$\lambda$ blows up:
\begin{equation}
  \mathtikz{
    \def\tmp#1{
      \draw (0,0) circle [x radius=1.5, y radius=.5];
      \filldraw[radius=.05] (-1,-.2) circle;
      \filldraw[radius=.05] (-1,.2) circle;
      \filldraw[radius=.05] (1,.2) circle;
      \filldraw[radius=.05] (1,-.2) circle;
      \filldraw[densely dotted,radius=.03] (.6,.1) coordinate (#1A) circle -- (.25,.13) coordinate (#1B) circle;
      \filldraw[densely dotted,radius=.03] (-.6,.1) coordinate (#1C) circle -- (-.25,.13) coordinate (#1D) circle;
    }
    \node at (-2,0) {$\Sigma\colon$};
    \begin{scope}[shift={(0,.6)}]
      \tmp{SWt}
    \end{scope}
    \begin{scope}[shift={(0,-.6)}]
      \tmp{SWb}
    \end{scope}
    \fill[color=gray,opacity=.5] (SWtA.south)--(SWbA.north)--(SWbB.north)--(SWtB.south)--cycle;
    \fill[color=gray,opacity=.5] (SWtC.south)--(SWbC.north)--(SWbD.north)--(SWtD.south)--cycle;
    \draw[->] (1.7,.5) -- (2.3,.2);
    \draw[->] (1.7,-.5) -- (2.3,-.2);
    \begin{scope}[shift={(4,0)}]
      \tmp{C}
    \end{scope}
    \node at (3.2,-.2) {\scriptsize\strut $0$};
    \node at (3.2,.2) {\scriptsize\strut $q$};
    \node at (4.85,.2) {\scriptsize\strut $1$};
    \node at (4.75,-.2) {\scriptsize\strut $\infty$};
    \node at (5.8,0) {$\colon\!C$};
  }
\end{equation}

\begin{exercise}
  1. By changing coordinates as $x=\tilde{x}/z+m_2/(z-q)+m_3/(z-1)$, rewrite the curve $x^2=u_2$ in a form that only has simple poles at $z=0,q,1$.
  Show that $\tilde{\lambda}\coloneqq\tilde{x}\rmd z/z$ differs from the \ac{SW} differential $\lambda=x\rmd z$ by a $u$-independent term whose contour integrals (residues) are linear combinations of masses.
  Recall the \ac{BPS} mass formula $\oint\lambda=na+ma_D+f_im_i$ and check what changing $\lambda$ to $\tilde{\lambda}$ amounts to a redefinition of flavour charges.
  Up to simple changes of coordinates perhaps\footnote{I have not checked yet what form Martone uses in his notes.} match with more conventional expressions of the \ac{SW} curve and differential of $\SU(2)$ $N_f=4$ \ac{SQCD} given in~\cite{2006.14038}.
  The match confirms that we correctly identified the tinkertoy $T_{\lie{su}(2)}$.
\end{exercise}

\paragraph{Singularities on the Coulomb branch.}

As we described on \autopageref{IRbehaviour}, the low-energy theory is generically a 4d $\Nsusy=2$ abelian gauge theory with the vector multiplet scalars living in the Coulomb branch~$\cB$.
For generic values of~$u$ and of masses, we thus get a $\Lie{U}(1)$ vector multiplet, but at special values of the parameters some branch points may collide, which leads to interesting low-energy behaviours.
We already saw that near~\eqref{u2quadpolfac} in our study of the tinkertoy: we found particular values of the masses where a pair of branch points of the \ac{SW} curve collide.  This collision made a certain contour shrink to zero size, hence lead to a massless \ac{BPS} particle which remains present in the \ac{IR} theory.
For \ac{SQCD} such collisions of branch points enrich the \ac{IR} theory by adding one or more massless hypermultiplets charged under the low-energy $\Lie{U}(1)$.

\begin{exercise}[On the discriminant]
  The discriminant of a degree~$d$ polynomial $P(z)=p_d\prod_{a=1}^d (z-z_a)$ is $\Delta_P=p_d^{2d-2}\prod_{a<b} (z_a-z_b)^2$.
  It vanishes by construction exactly when $P(z)$ has double roots.
  It is known that $\Delta_P$ can be expressed as a polynomial of degree $2d-2$ in the coefficients $p_j$ of $P(z)=\sum_{j=0}^d p_jz^j$.
  Check this for quadratic polynomials.
\end{exercise}

Our question is to find when $P(z)=z^2(z-q)^2(z-1)^2u_2(z)$, which is a quartic polynomial given explicitly in~\eqref{four-punctured-phi2}, has double roots (hence when two branch points collide).
The discriminant $\Delta_P$ is then of degree~$6$ in the coefficients of~$P$.
Since $P$ depends linearly on~$u$ we find that $\Delta_P$ is of degree~$6$ in~$u$ (the leading coefficient turns out to be nonzero).
We should thus expect $6$ singularities on the Coulomb branch.

Four of these can be seen concretely in the weak coupling limit (with fixed masses).
Then $\phi_2$ is roughly given by the quadratic differential on each pair of pants, connected by a long tube where $\phi_2$ is suitably constant, see~\eqref{phi2-limit-3pt}.  Each three-punctured sphere has two zeros of $\phi_2$, hence one branch cut of the \ac{SW} curve.
Consider the pair of pants with masses squared $m_1^2,m_2^2,u$, for definiteness.
Its branch cut shrinks to zero size whenever any combination $\pm m_1\pm m_2\pm\sqrt{u}$ of the mass parameters vanishes.
We thus find four of the six singular points of the Coulomb branch:
\begin{equation}
  u = (m_1\pm m_2)^2 + O(q) \quad \text{and} \quad u = (m_3\pm m_4)^2 + O(q) .
\end{equation}
The remaining two points are not so easy to determine from the explicit quadratic differential~\eqref{four-punctured-phi2} of the class~S theory, partly because they correspond to the collision of branch points that sit in different pair of pants in our decomposition above.
A tedious series expansion shows that at\footnote{It would be nice to understand the formulae better from our 6d construction.}
\begin{equation}\label{uplanesingularity-2}
  u = \pm 2 \bigl(q (m_2^2-m_1^2) (m_3^2-m_4^2)\bigr)^{1/2} + O(q)
\end{equation}
two branch points collide at $z=\mp 2 \bigl(q (m_2^2-m_1^2)/(m_3^2-m_4^2)\bigr)^{1/2}+O(q)$, with the sign being correlated to that of~$u$.

From the point of view of \ac{SQCD} with $N_f=4$ flavours, what happens is as follows.
The four doublet hypermultiplets have mass parameters $m_1+m_2,m_1-m_2,m_3+m_4,m_3-m_4$, so when the ``\ac{VEV}'' of the vector multiplet~$\phi$ matches one of these we get a massless hypermultiplet; its charge is $+1$ or~$-1$ under the low-energy $\Lie{U}(1)$ because that is how the diagonal $\Lie{U}(1)\subset\SU(2)$ acts on a doublet.
At low energies $\abs{u}\ll\abs{m_i},\abs{m_i\pm m_j}$, all hypermultiplets are massive and can be integrated out, leaving behind pure $\SU(2)$ \ac{SYM}, whose Coulomb branch is known to have two singular points at $u=\pm 2\Lambda$, the monopole and dyon points.
Incidentally we learn that the dynamically generated scale is at $1/2$ times the value~\eqref{uplanesingularity-2}.
The main takeaway for our purposes is that the 6d perspective reproduces all the expected physics of \ac{SQCD}.

By tuning more than just~$u$ we can get more than two branch points to collide, hence more than one set of fields to become massless.  Such limits can lead in the \ac{IR} to non-trivial \ac{SCFT} including the \ac{AD} theory, which we return to in \autoref{sec:gen}.  The limits are also interesting on the 2d side.

\paragraph{Coupling constants.}

It may be puzzling how $q=e^{2\pi i\tau}$, which ranges over $\CC\setminus\{0,1\}$, reproduces the complexified gauge coupling $\tauLag=\theta/(2\pi) + 4\pi i/g^2$ that appears in a Lagrangian description of $\SU(2)$ \ac{SQCD} with $N_f=4$, especially the fact that $\Im\tauLag>0$.
To relate them, we consider the massless limit where all $m_i\to 0$, such that the theory is an \ac{SCFT} and the couplings do not run.
An equivalent limit is the large~$\abs{u}$ region of the Coulomb branch, specifically $\abs{u}\gg\abs{m_i}^2$.
From this point of view the running is stopped very quickly at the large energy scale~$\abs{u}$.
Thus, the \ac{IR} gauge coupling $\tauIR$ captured by the \ac{SW} curve obeys $\tauIR=2\tauLag$, with a factor due to how $\Lie{U}(1)$ embeds into $\SU(2)$.

The \ac{SW} curve~$\Sigma$ is a torus double-covering the sphere, with four branch points found at zeros of~$u_2$.
At most one of the first four terms in $u_2$ given in~\eqref{four-punctured-phi2} can be large at once, so to compensate for the large value of~$u$ one must have $z$ close to one of the punctures $0,q,1,\infty$: the four branch points are thus at
\begin{equation}
  z = O\biggl(\frac{m_1^2}{u}\biggr) , \quad z = q + O\biggl(\frac{m_2^2}{u}\biggr) , \quad
  z = 1 + O\biggl(\frac{m_3^2}{u}\biggr) , \quad z = O\biggl(\frac{u}{m_4^2}\biggr) .
\end{equation}
The modular parameter $\tauIR$ of the torus~$\Sigma$ gives the complexified gauge coupling, which automatically obeys $\Im\tauIR>0$.
In terms of the modular lambda function~$\lambda$, one has~\cite{hep-th/0702187}
\begin{equation}\label{couplings-SQCD}
  \begin{aligned}
    q & = \lambda(\tauIR) = 16 e^{\pi i\tauIR} - 128 e^{2\pi i\tauIR} + \dots , \qquad \\
    \tau & = \frac{1}{2} \tauIR + \frac{\log 16}{2\pi i} - \frac{8}{2\pi i} e^{\pi i\tauIR} + \dots ,
  \end{aligned}
\end{equation}
where the expansion holds for small gauge coupling~$g$.
Both $\tau$ and $\tauLag=\frac{1}{2}\tauIR$ are perfectly good definitions of gauge coupling, which amount to two different renormalization scheme, differing by a constant shift and by instanton corrections.
The freedom to redefine couplings appears again in~\eqref{couplings-redef}.

\paragraph{S-duality.}
The four-punctured sphere $\CP^1\setminus\{0,q,1,\infty\}$ has three pants decompositions hence three Lagrangian descriptions.  The descriptions are identical except for permutations of masses $m_1,m_2,m_3,m_4$ and changing $q\to 1/q$ or $q\to 1-q$.  This is S-duality of \ac{SQCD}~\cite{hep-th/9408099}.  In the notation of~\eqref{SQCD-generalized-quiver},
\begin{equation}\label{elementary-SU2-duality}
  \mathtikz{
    \node(A)[flavor-group] at (-1.5,-.5) {$\SU(2)_1$};
    \node(B)[flavor-group] at (-1.5,.5) {$\SU(2)_2$};
    \node(C)[color-group] at (0,0) {$\SU(2)$};
    \node at (0,-.5) {$q$};
    \node(D)[flavor-group] at (1.5,.5) {$\SU(2)_3$};
    \node(E)[flavor-group] at (1.5,-.5) {$\SU(2)_4$};
    \draw(A)--(-1,0)--(B);
    \draw(-1,0)--(C)--(1,0);
    \draw(D)--(1,0)--(E);
  }
  =
  \mathtikz{
    \node(A)[flavor-group] at (-1.5,-.5) {$\SU(2)_1$};
    \node(B)[flavor-group] at (-1.5,.6) {$\SU(2)_2$};
    \node(C)[color-group] at (0,0) {$\SU(2)$};
    \node at (0,-.5) {$1/q$};
    \node(D)[flavor-group] at (1.5,.6) {$\SU(2)_3$};
    \node(E)[flavor-group] at (1.5,-.5) {$\SU(2)_4$};
    \draw(A)--(-1,0) to [bend left=20] (D);
    \draw(-1,0)--(C)--(1,0);
    \draw(E)--(1,0) to [bend right=20] (B);
  }
  =
  \mathtikz{
    \node(A)[flavor-group] at (-1.5,-.6) {$\SU(2)_1$};
    \node(B)[flavor-group] at (-1.5,.6) {$\SU(2)_2$};
    \node(C)[color-group] at (0,0) {$\SU(2)$};
    \node at (1,0) {\strut $1{-}q$};
    \node(D)[flavor-group] at (1.5,.6) {$\SU(2)_3$};
    \node(E)[flavor-group] at (1.5,-.6) {$\SU(2)_4$};
    \draw(A)--(0,-.4)--(E);
    \draw(0,-.4)--(C)--(0,.4);
    \draw(B)--(0,.4)--(D);
  }
\end{equation}
While these equalities are manifest in the 6d perspective they hide deep non-perturbative physics, as they are equalities between \acp{QFT} involving completely different elementary gauge fields and matter fields (the gauge field $A_\mu$ in some description is unrelated to $A_\mu$ in another).

\subsection{\label{ssec:Lag-SU2}Generalized \(\SU(2)\) quivers}

We have given all the ingredients to determine $\lie{su}(2)$ class~S theories arising from $\Xsuii$ on an arbitrary punctured Riemann surface~$C$ with tame punctures.\footnote{We exclude the sphere with no puncture, one puncture (a plane), or two punctures (a cylinder), and the torus without punctures, as they are pathological.}
This subsection will thus consist essentially of exercises.

\paragraph{Five-punctured sphere.}
We consider here $C=\CP^1\setminus\{0,z_1,z_2,1,\infty\}$; note that we shifted indices of punctures~$z_i$ a bit compared to our earlier conventions.  For any decomposition into three-punctured spheres the Lagrangian has the form
\begin{equation}\label{SU2SU2-quiver}
  \Theory\bigl(\lie{su}(2),\CP^1\setminus\{0,z_1,z_2,1,\infty\}\bigr)
  = \mathtikz{
    \node(A)[flavor-group] at (-1.5,.5) {$\SU(2)$};
    \node(B)[flavor-group] at (-1.5,-.5) {$\SU(2)$};
    \node(C)[color-group] at (0,0) {$\SU(2)$};
    \node(F)[flavor-group] at (1,.6) {$\SU(2)$};
    \node(C2)[color-group] at (2,0) {$\SU(2)$};
    \node(D)[flavor-group] at (3.5,.5) {$\SU(2)$};
    \node(E)[flavor-group] at (3.5,-.5) {$\SU(2)$};
    \draw(A)--(-1,0)--(B);
    \draw(1,0)--(F);
    \draw(-1,0)--(C)--(C2)--(3,0);
    \draw(D)--(3,0)--(E);
  }
\end{equation}
Contrarily to spheres with fewer punctures, the $\SU(2)^5$ flavour symmetry manifest from~6d does not enhance in the 4d theory (as far as I know).
S-dualities of this theory were studied in~\cite{hep-th/9910125} before class~S theories and their S-dualities were uncovered in~\cite{0904.2715}.

\begin{exercise}
  Write~$C$ as the gluing of three pairs of pants with gluing parameters $z_1/z_2$ and $z_2$ following \autoref{exe:punct-sph}.
\end{exercise}

\begin{exercise}
  For each pants decomposition of~$C$ check that the Lagrangian description is~\eqref{SU2SU2-quiver}, with gauge group $\SU(2)^2$ and twelve hypermultiplets.
  In what representations of the $\SU(2)^2$ gauge group do they transform?
  What flavour symmetries do these representations carry?
\end{exercise}

\begin{exercise}
  1. Each $\SU(2)$ gauge group is coupled to four doublet hypermultiplets.
  When the other gauge group is weakly coupled the theory is thus \ac{SQCD} coupled to further matter by a weakly coupled gauge field.
  ``Apply'' S-duality to this \ac{SQCD} theory and check that the resulting description is the description one would have written for some pair of pants of the five-punctured sphere.

  2. Check that elementary S-dualities~\eqref{elementary-SU2-duality} applied to different gauge nodes do not commute so that the S-duality group(oid) of the $\SU(2)^2$ gauge theory is not the product of S-duality groups of two \ac{SQCD} theories.
\end{exercise}

\paragraph{Punctured sphere.}
Next we consider $\CP^1\setminus\{z_0,\dots,z_{n-1}\}$ with $n$ punctures, with $z_0=0$, $z_{n-2}=1$, $z_{n-1}=\infty$.  Denote by $m_0,m_1,\dots,m_{n-1}$ the mass parameters of the punctures.

\begin{exercise}\label{exe:curve-sph}
  1. By using Liouville's theorem as in \autoref{exe:curve-T2} and \autoref{exe:curve-SU24}, find all quadratic differentials $\phi_2(z)$ that have the prescribed second order poles at punctures.  Deduce that the Coulomb branch is $\cB=\CC^{n-3}$.

  2. Write a $\SU(2)^{n-3}$ linear quiver description of the theory that is weakly coupled in the regime $\abs{z_1}\ll\abs{z_2}\ll\dots\ll\abs{z_{n-3}}\ll 1$.

  2. Expand $\phi_2(z)$ in this regime for $z$ in an annulus $\abs{z_{i-1}}\ll\abs{z}\ll\abs{z_{i+1}}$ ($i=1,\dots,n-2$).  Check that $\phi_2$ reduces to the differential of~$T_2$ on each of these pair of pants building blocks.  Check that $\cB=\CC^{n-3}$ can be parametrized by the parameters $u_i$, $i=1,\dots,n-3$ of punctures in these pants.  Identify $u_i=\frac{1}{2}\Tr\phi_i^2$ where $\phi_i$ is the vector multiplet scalar of the $i$-th vector multiplet.

  3. Check that starting at $n=6$ pants decompositions can be topologically distinct beyond just the permutation of punctures.
\end{exercise}

\paragraph{Punctured torus.}  We repeat a similar exercise for genus $g=1$.  One could also study theories associated to higher-genus curves, but the relevant mathematics are out of scope of this review.

\begin{exercise}
  1. The once-punctured torus is obtained by gluing two punctures of the same pair of pants.
  Write the theory as an $\SU(2)$ gauge theory and note that there is a decoupled gauge singlet in addition to the adjoint hypermultiplet.\footnote{It is not immediately clear to me how such gauge singlets work out when considering different Lagrangian descriptions of $n$-punctured tori or of higher genus surfaces.  Indeed, some channels include adjoint hypermultiplets, hence gauge singlets, while for others the singlets are not manifest.}

  2. Write the theory associated to an $n$-punctured torus as a circular $\SU(2)^n$ quiver with a bifundamental hypermultiplet for each pair of neighboring groups.  The weak gauge coupling regime corresponds to a long torus with well-separated punctures.

  3. If you know enough about elliptic functions determine all quadratic differentials with prescribed second order poles at the punctures.  Expand them in the weak gauge coupling limit as in \autoref{exe:curve-sph}.
\end{exercise}

\subsection{\label{ssec:Lag-linear}Linear quiver \(\lie{su}(N)\) theories}

We move on briefly to $\lie{su}(N)$ class~S theories, specifically a particular subclass that is ad-hoc from the 6d perspective but leads to linear quiver gauge theories in~4d, as can be understood using brane constructions.
These will be useful for our discussion of instantons.

\paragraph{Conformal \acs{SQCD}.}
Let us try and realize \(\SU(N)\) \ac{SQCD} with $N_f=2N$ flavours (the number of flavours needed for a vanishing beta function) as a class~S theory.
Its flavour symmetry is $\lie{u}(N_f)=\lie{u}(2N)$ (enhanced to $\lie{so}(8)$ when $N=2$).
In analogy to the $N=2$ case we expect the gauge group to correspond to a tube joining two three-punctured spheres, so we split the $2N$ flavours as two groups of~$N$, where each group should come from one of the two three-punctured sphere.
The flavour symmetry of each group is $\lie{u}(N)=\lie{u}(1)\times\lie{su}(N)$, so that this split makes $\lie{su}(N)^2\times\lie{u}(1)^2$ flavour symmetry manifest.
In analogy to the $N=2$ case we associate each of the four factors to one puncture and write an analogue of~\eqref{SQCD-generalized-quiver}:\footnote{As in various other places in this review there are inaccuracies about the global structure of groups.  Corrections welcome.}
\begin{equation}\label{SUN-SQCD-quiver}
  \Theory\bigl(\lie{su}(N),\CP^1\setminus\text{4pt},\text{suitable data}\bigr)
  = \mathtikz{
    \node(A)[flavor-group] at (-1.5,-.5) {$\SU(N)$};
    \node(B)[flavor-group] at (-1.5,.5) {$\Lie{U}(1)$};
    \node(C)[color-group] at (0,0) {$\SU(N)$};
    \node(D)[flavor-group] at (1.5,.5) {$\Lie{U}(1)$};
    \node(E)[flavor-group] at (1.5,-.5) {$\SU(N)$};
    \draw(A)--(-1,0)--(B);
    \draw(-1,0)--(C)--(1,0);
    \draw(D)--(1,0)--(E);
  }
\end{equation}
In the $N=2$ case the $\lie{u}(N)=\lie{u}(2)$ symmetry enhances to $\lie{so}(4)$, namely the $\lie{u}(1)$ factor enhances to $\lie{su}(2)$.
For $N>2$ in contrast we have to deal with the presence of different kinds of punctures.
We delay the full story to \autoref{ssec:gen-rank}.
For now we shall be content with using two types of tame punctures: full punctures that carry $\lie{su}(N)$ flavour symmetry and simple punctures that carry $\lie{u}(1)$.

From the 6d point of view, the $\lie{u}(2N)$ flavour symmetry of conformal $\lie{su}(N)$ \ac{SQCD} is an accidental \ac{IR} symmetry, as it is not a symmetry of the 6d $\Nsusy=(2,0)$ setup.

\paragraph{Free hypermultiplets.}
The left and right sides of the quiver~\eqref{SUN-SQCD-quiver} consist of $N^2$~hypermultiplets that each have $\lie{u}(1)\times\lie{su}(N)^2$ flavour symmetry (actually more before gauging), of which one $\lie{su}(N)$ factor is gauged.
This collection of $N^2$~free hypermultiplets is the tinkertoy associated to a sphere with two full and one simple puncture.

\paragraph{Punctures.}
Following the general ideas from the $\lie{su}(2)$ case the full punctures are described as a boundary condition like~\eqref{tame-massive-SU2}:
\begin{equation}\label{full-tame-massive-SUN}
  \varphi(z) \sim \biggl(\frac{m_i}{z-z_i} + O(1)\biggr)\rmd z
  \implies \phi_k(z) = \biggl(\frac{(-1)^{k+1}\sigma_k(m_i)}{(z-z_i)^k} + O\biggl(\frac{1}{(z-z_i)^{k-1}}\biggr)\biggr)\rmd z^k ,
\end{equation}
for $m_i\in\lie{su}(N)$, where the symmetric polynomials $\sigma_k(m_i)$ are defined by expanding $\det(X-m_i)=X^N+\sum_{k\geq 2}X^{N-k}(-1)^k\sigma_k(m_i)$.  The condition on $\varphi(z)$ should be understood modulo conjugation, hence only the conjugacy class of~$m_i$ is important.

We return in \autoref{ssec:gen-rank} to a description of conjugacy classes in $\lie{su}(N)_{\CC}=\lie{sl}(N,\CC)$.
For now, it suffices to mention simple punctures, whose mass parameters take the form $m_i=\diag((N-1)\mu,-\mu,\dots,-\mu)$.
As a result, the massless limit where $m_i$ becomes nilpotent (see~\eqref{tame-massless-SU2} for the $N=2$ case) is different for full and simple punctures:
\begin{equation}\label{tame-massless-SUN}
  \begin{aligned}
    \phi_k(z) & = O\biggl(\frac{1}{(z-z_i)^{k-1}}\biggr)\rmd z^k \qquad (\text{massless full puncture}) , \\
    \phi_k(z) & = O\biggl(\frac{1}{(z-z_i)}\biggr)\rmd z^k \qquad (\text{massless simple puncture}) .
  \end{aligned}
\end{equation}

\begin{exercise}\label{exo:SUN-linear}
  1. Massless case.  Find the most general degree $k\geq 2$ differential on the three-punctured sphere with a simple pole at $z=1$ and poles of order $k-1$ at $z=0,\infty$.  Deduce the \ac{SW} curve of the class~S theory corresponding to a sphere with one simple and two full punctures and deduce the theory has no Coulomb branch, consistent with free hypermultiplets.

  2. Massive case.  For $N=3$, evaluate $\det(x-\varphi_z)$ near a simple puncture $\varphi(z) \sim \bigl((z-1)^{-1} m + p\bigr)\rmd z$ with $m=\diag(2\mu,-\mu,-\mu)$.  Observe that the $(z-1)^{-k}$ terms in $u_d(z)$, for $k,d=2,3$, are expressed in terms of the $(z-1)^{-1}$ terms and of~$\mu$.
  Deduce that there is again no Coulomb branch.
  
  3. Massless case.  Write the most general degree $k\geq 2$ differential on the $n$-punctured sphere with order $k-1$ poles (full punctures) at $0,\infty$ and simple poles (simple punctures) at $q_1,\dots,q_{n-3},1$.  Write the \ac{SW} curve and check that the Coulomb branch has dimension $(n-3)(N-1)$, consistent with the quiver~\eqref{SUN-linear} below.
\end{exercise}

\paragraph{Linear quiver gauge theory.}

Starting with collections of $N^2$ free hypermultiplets, identifying pairs of $\lie{su}(N)$ symmetries, and gauging them using vector multiplets, we find
\begin{equation}\label{SUN-linear}
  \begin{aligned}
    & \Theory\bigl(\lie{su}(N),\CP^1\setminus\{0,\underline{z_1},\dots,\underline{z_{n-2}},\infty\}\bigr) \\
    & \quad = \mathtikz{
      \node(A)[flavor-group] at (-1.5,-.5) {$\SU(N)$};
      \node(B)[flavor-group] at (-1.5,.5) {$\Lie{U}(1)$};
      \draw(A)--(-1,0)--(B);
      \node(C)[color-group] at (0,0) {$\SU(N)$};
      \draw(-1,0)--(C);
      \node(D)[flavor-group] at (1,.6) {$\Lie{U}(1)$};
      \draw(1,0)--(D);
      \node(E)[color-group] at (2,0) {$\SU(N)$};
      \draw(C)--(E);
      \node(F)[flavor-group] at (3,.6) {$\Lie{U}(1)$};
      \draw(3,0)--(F);
      \node(Cdots) at (3.6,0) {${\cdots}$};
      \draw(E)--(Cdots);
      \node(G)[flavor-group] at (4.2,.6) {$\Lie{U}(1)$};
      \draw(4.2,0)--(G);
      \node(H)[color-group] at (5.2,0) {$\SU(N)$};
      \draw(Cdots)--(H);
      \draw(H)--(6.2,0);
      \node(I)[flavor-group] at (6.7,.5) {$\Lie{U}(1)$};
      \node(J)[flavor-group] at (6.7,-.5) {$\SU(N)$};
      \draw(I)--(6.2,0)--(J);
    }
  \end{aligned}
\end{equation}
where we have underlined the simple punctures (so that only $0$ and $\infty$ are full punctures).
This linear quiver gauge theory description corresponds to a specific degeneration limit of the Riemann surface.
We emphasize that other limits would involve more elaborate tinkertoys, which do not typically have any Lagrangian description unless every pair of pants involves a simple puncture.
As found in \autoref{exo:SUN-linear}, the \ac{SW} curve in the massless case reads
\begin{equation}\label{SW-quiver-massless}
  x^N = \sum_{k=2}^N \frac{P_{n-4}^{(k)}(z)}{(z-z_1)\cdots(z-z_{n-2})z^{k-1}}x^{N-k}
\end{equation}
where $P_{n-4}^{(k)}$ denote polynomials of degree $n-4$.

\begin{exercise}
  Consider the degeneration limit where each $z_{i+1}/z_i$ is kept fixed (and $|z_{i+1}/z_i|>1$) except $z_{j+1}/z_j\to+\infty$ for some~$j$.  Check that the part of the curve with finite $x/z_j$ takes the form~\eqref{SW-quiver-massless} with $n$ replaced by $j+2$ and additional mass terms.  In particular the new puncture is a full puncture.
\end{exercise}

\paragraph{Brane construction.}

We know that the 6d $(2,0)$ theory $\XsuN$ is the world-volume theory of a stack of $N$ coincident M5~branes (minus the center of mass).
The Riemann surface in~\eqref{SUN-linear} can be taken to be a cylinder, with some punctures.
Then the brane setup can be described by $N$~M5~branes wrapping the cylinder, with the insertion of transverse M5'~branes at $n-2$ points on the cylinder.

Now, M-theory on a circle is IIA string theory, M5~branes wrapping the circle become D4~branes, while M5'~branes at points on the circle become NS5~branes.
This gives a well-known brane set-up~\cite{hep-th/9703166} with $N$ D4~brane segments stretching between each pair of neighboring NS5~branes:
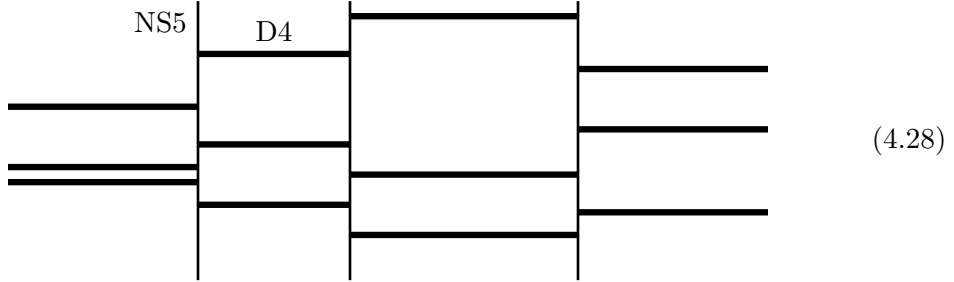
\begin{equation}
  \mathtikz[brane/.style={line width=1pt}, d4/.style={line width=2.4pt}]{
    \draw[d4] (-1.5,1.3) -- (1,1.3);
    \draw[d4] (-1.5,1.5) -- (1,1.5);
    \draw[d4] (-1.5,2.3) -- (1,2.3);
    \draw[brane] (1,0) -- (1,3.7) node [below left] {NS5};
    \draw[d4] (1,1) -- (3,1);
    \draw[d4] (1,1.8) -- (3,1.8);
    \draw[d4] (1,3) -- (3,3) node [midway, above] {D4};
    \draw[brane] (3,0) -- (3,3.7);
    \draw[d4] (3,.6) -- (6,.6);
    \draw[d4] (3,1.4) -- (6,1.4);
    \draw[d4] (3,3.5) -- (6,3.5);
    \draw[brane] (6,0) -- (6,3.7);
    \draw[d4] (6,0.9) -- (8.5,0.9);
    \draw[d4] (6,2.0) -- (8.5,2.0);
    \draw[d4] (6,2.8) -- (8.5,2.8);
  }
\end{equation}
The world-volume description of this brane diagram is known to be the linear quiver gauge theory~\eqref{SUN-linear}.  Mass parameters of the two $\SU(N)$ flavour symmetries are positions (vertically in the figure) of the semi-infinite D4 branes on either end.  Mass parameters of all $\Lie{U}(1)$ flavour symmetries are distances between centers of masses of each collection of $N$~D4~branes.
The remaining vertical positions are dynamical and appear on the gauge theory side as Coulomb branch parameters.
The \ac{SW} curve and differential of the linear quiver can be extracted from this construction~\cite{hep-th/9703166} and coincides with what is found from the 6d perspective.

It is very easy in the brane diagram to accomodate gauge group ranks that are not all the same.  We outline in \autoref{ssec:gen-rank} how to realize such theories in class~S.

\part{\acsfont{AGT} correspondence}

\section{\label{sec:loc}Localization for 4d quivers}

Up to this point we have been working with 4d $\Nsusy=2$ class~S theories in Minkowski space.
We now turn\footnote{We shall ignore possible difficulties with Wick rotation.} to Euclidean signature.
Our aim in this section and the next is to explain both sides of the \ac{AGT} relation \eqref{AGT-ZS4} for the case $\lie{g}=\lie{su}(2)$ with tame punctures:
\begin{equation}\label{AGT-bis}
  Z_{S^4_b}\bigl(\Theory(\lie{su}(2),C,m)\bigr)
  = \vev*{\Vhat_{\alpha_1}(z_1)\dots \Vhat_{\alpha_n}(z_n)}_{\fullC}^{\Liouville} .
\end{equation}
We postpone to \autoref{sec:AGT} the description of the right-hand side, a 2d Liouville \ac{CFT} correlator on the Riemann surface~$C$.
For now we concentrate on the sphere (and squashed sphere) partition function of $\Theory(\lie{su}(2),C,m)$.
We explain how 4d $\Nsusy=2$ theories are placed on this curved background geometry in \autoref{ssec:loc-S4}, by coupling with
\apartehere
supergravity~\cite{1105.0689,1205.1115,1209.5408} (see also~\cite{1209.4043,1211.1367} for other early explorations).
In \autoref{ssec:loc-loc} we reduce the infinite-dimensional path integral to a finite-dimensional one (a matrix model) in the Lagrangian case using supersymmetric localization.
The resulting expression is built from Nekrasov instanton partition functions, which we explore in \autoref{ssec:loc-inst}.
Supersymmetric localization implies that some factorization properties remain true even for non-Lagrangian theories, see \autoref{ssec:loc-cut}.

\startaparte[title=Localization and factorization]
\begin{itemize}
\item 4d $\Nsusy=2$ theories can be put supersymmetrically on $S^4_b=\{y_5^2+b^2(y_1^2+y_2^2)+b^{-2}(y_3^2+y_4^2)=r^2\}\subset\RR^5$.
\item For any gauge theory, supersymmetric localization reduces the $S^4_b$~partition function exactly to an integral $Z_{S^4_b}=\int \rmd a\,Z_{\cl}(a,q\qbar) \,Z_{\oneloop}(a) \,Z_{\inst}(a,q) \,Z_{\inst}(a,\qbar)$
over the gauge group's Cartan algebra.
\item The classical contribution is $Z_{\cl}=|q|^{r^2|a|^2}$.
\item The one-loop contribution $Z_{\oneloop}$ only depends on matter and is known for any Lagrangian theory.
\item The instanton contribution $Z_{\inst}$ is holomorphic in~$q$ and known for many theories such as linear quivers.
\end{itemize}
\stopaparte

Localization on the round sphere was done in~\cite{0712.2824} and extended to the squashed sphere in~\cite{1206.6359} based on some analogous developments on 3d squashed spheres~\cite{1012.3512,1102.4716}.
The supergravity background of~\cite{1206.6359} was generalized to complex~$b$ in~\cite{1310.5939}.
The partition function admits alternate ``Higgs branch localization'' expressions~\cite{1506.04390,1508.07329,1711.06150,1807.11900} which we will not need.
See~\cite{1608.02957} for a review of curved-space localization, and \cite{1608.02962}~more specifically for 4d $\Nsusy=2$ theories.

\subsection{\label{ssec:loc-S4}Theories on an ellipsoid}

\paragraph{An easy case: the round sphere.}
Placing an \ac{SCFT} on a round sphere $S^4$ is in principle\footnote{Identifying operators in flat space and on the sphere is subtle, as there can be some mixing involving curvature tensors.  This was understood for 4d $\Nsusy=2$ theories in~\cite{1602.05971}.} straightforward: just apply a conformal transformation to the flat space theory since the sphere is conformally flat.
The 4d $\Nsusy=2$ superconformal algebra on~$S^4$ is then the same as on~$\RR^4$, namely $\lie{su}^*(4|2)$, whose bosonic part is the conformal algebra $\lie{su}^*(4)=\lie{so}(5,1)$ times the R-symmetry algebras $\lie{u}(1)$ and $\lie{su}^*(2)=\lie{su}(2)$.

The class~S theories we study (for tame punctures) are mass deformations of \acp{SCFT}.
They can thus be placed on $S^4$ by conformally mapping the \ac{SCFT} to the sphere, then turning on masses as background values for vector multiplet scalars coupled to the various flavour symmetries.
Mass terms turn out to break half of supersymmetry, break the conformal algebra to the rotation algebra $\lie{so}(5)=\lie{usp}(4)$, and the R-symmetry to $\lie{so}(2)=\lie{so}^*(2)$.
Altogether one can work out that massive theories preserve the supersymmetry subalgebra
\begin{equation}
  \lie{osp}^*(2|4) \subset \lie{su}^*(2|4) .
\end{equation}
Note that this differs quite a bit from the Poincar\'e algebra preserved by massive theories on~$\RR^4$: for instance spatial isometries of $\RR^4$ are $\lie{iso}(4)=\RR^4\rtimes\lie{so}(4)$ while here we have the $\lie{usp}(4)=\lie{so}(5)$ rotation algebra.

The \ac{AGT} correspondence involves an ellipsoid (often called squashed sphere)~$S^4_b$ and not only~$S^4$.
The squashed sphere is not conformally flat, and defining theories on this background requires technology that we now explain.

\paragraph{Conformal Killing vectors.}
We are interested in \acp{QFT} on rigid curved spaces (no dynamical gravity).
Placing a Poincar\'e-invariant \ac{QFT} on a curved space is done by coupling the theory to gravity and freezing the value of the metric\footnote{To be precise, if the theory has spinors one must additionally give a spin structure rather than only the metric (for instance giving a vielbein is enough).}.
Invariance under changes of coordinates leads to the existence of a conserved stress tensor ($\nabla_\nu T^{\mu\nu}=0$).
As the next exercise shows, the resulting curved-space theory preserves some space-time (Poincar\'e) symmetries provided the metric admits \emph{Killing vectors} $Y_\mu$, defined by the Killing vector equation $\nabla_\mu Y_\nu+\nabla_\nu Y_\mu=0$.
More generally if the flat-space \ac{QFT} is conformal, isometry and conformal symmetries of the \ac{CFT} are given by \emph{conformal Killing vectors}
\begin{equation}\label{conformal-Killing-vector}
  \nabla_\mu Y_\nu + \nabla_\nu Y_\mu = \frac{2}{d} g_{\mu\nu} \nabla_\rho Y^\rho .
\end{equation}

\begin{exercise}
  1. A Poincar\'e-invariant local \ac{QFT} has a conserved stress-tensor~$T^{\mu\nu}$ that is symmetric.
  Check that the current $Y_\mu T^{\mu\nu}$ is conserved if $Y$ is a Killing vector.

  2. If the flat-space \ac{QFT} is conformally invariant, $T^{\mu\nu}$ is traceless as well.
  Check that $Y_\mu T^{\mu\nu}$ is then conserved provided~$Y$ is a conformal Killing vector.
  Hint: explain the factor $2/d$ in~\eqref{conformal-Killing-vector} by taking the trace of the equation.
\end{exercise}

\paragraph{Conformal Killing spinors.}

Consider now a supersymmetric theory.
This means that there are conserved supersymmetry currents $G_\alpha^\mu$ and $\tilde{G}^{\dot{\alpha}\mu}$, where $\mu$ is a vector index of the $\SO(4)$ rotation group, and $\alpha=1,2$ and $\dot{\alpha}=1,2$ are spinor indices with both chiralities.
Leaving the spinor index of $G^\mu$ implicit, the conservation equation reads
\begin{equation}\label{conserved-Gmualpha}
  D_\mu G^\mu \coloneqq \nabla_\mu G^\mu + \frac{1}{4} \omega_\mu{}^{ab} \gamma_a\gamma_b G^\mu = 0
\end{equation}
where $\nabla$ is the Levi--Civita connection of the given metric, $\omega$~its spin connection, $a,b$ are vielbein indices, and $\gamma$~are Dirac matrices.

To get a usual conserved translation current from the conserved stress-tensor in flat space one contracts $T^{\mu\nu}$ with a constant translation vector~$a_\mu$.
Likewise here we have usual currents $\xi G^\mu=\xi^\alpha G_\alpha^\mu$ and $\tilde{\xi}\tilde{G}^\mu=\tilde{\xi}_{\dot{\alpha}} G^{\dot{\alpha}\mu}$ for constant\footnote{Here we use a common abuse of language: talking about constant spinors requires a choice of vielbein, for which we choose the standard Cartesian one on flat space.} spinors~$\xi$.
In curved space we can check that $\xi G^\mu$ is conserved provided $\xi$ is a Killing spinor:
\begin{equation}\label{Killing-spinor}
  D_\mu \xi \coloneqq \biggl(\del_\mu + \frac{1}{4} \omega_\mu^{ab}\gamma_a\gamma_b\biggr) \xi = 0 .
\end{equation}
(Note that we put $\del$ instead of $\nabla$ because $\xi$ has no vector index.)

Just as a theory is conformally invariant if $x_\mu T^{\mu\nu}$ is conserved, a theory is superconformally invariant if $x^\nu\gamma_\nu G^\mu$ is conserved in the same sense as~\eqref{conserved-Gmualpha}.
When put on curved space, the theory now has super(conformal) symmetries if the spacetime admits a conformal Killing spinor
\begin{equation}\label{conformal-Killing-spinor}
  D_\mu \xi = \frac{1}{d} \gamma_\mu \gamma^\nu D_\nu \xi .
\end{equation}

\begin{exercise}
  Check that \eqref{conformal-Killing-spinor} indeed leads to a conserved current $\xi G$ if the theory is superconformal.
\end{exercise}

\paragraph{Generalized Killing spinors.}

We defined the (partial) topological twist in \autoref{ssec:reduce-twist} as a mixing of some rotations and R-symmetries, which allowed us to compactify the 6d $(2,0)$ theory on a Riemann surface~$C$.
This idea is refined and generalized as follows whenever the flat-space theory has an R-symmetry current~$J^\mu$.
When placing the \ac{QFT} on curved space we can turn on a background gauge field~$V_\mu$ coupled to~$J^\mu$ in addition to the metric~$g_{\mu\nu}$ that is coupled to~$T^{\mu\nu}$ (we typically don't turn on fermionic backgrounds coupled to supersymmetry currents).

In such a background, the Killing spinor equation~\eqref{Killing-spinor} generalizes by including the R-symmetry gauge field in the covariant derivative:
\begin{equation}\label{generalized-Killing-spinor}
  D_\mu \xi \coloneqq \biggl(\del_\mu + \frac{1}{4}\omega_\mu^{ab}\gamma_a\gamma_b + V_\mu\biggr) \xi = 0 .
\end{equation}
The conformal Killing spinor equation generalizes in the same way to \eqref{conformal-Killing-spinor} with the new~$D_\mu$.
Here $V_\mu\xi$ should be suitably weighted by the R-charge under the given R-symmetry, as is standard for covariant derivatives in the presence of a gauge field.
The (partial) topological twist consists of choosing $V_\mu$ to cancel some component of $\omega_\mu^{ab}$ so that the corresponding component of~$\xi$ can simply be chosen to be constant (along~$C$).
More general choices for the background gauge field~$V$ can make it possible to preserve some supersymmetries even if the curved manifold of interest does not have any (conformal) Killing spinors.

\paragraph{Squashed sphere.}

The supergravity background found in~\cite{1206.6359} to place 4d $\Nsusy=2$ theories on~$S^4_b$ is rather complicated and gives non-zero values to most bosonic fields in the supergravity multiplet.  We point to the review~\cite{1608.02962} for actual expressions.
For our purposes we only need two aspects.
The metric is the one induced from that of Euclidean~$\RR^5$ in the embedding
\begin{equation}\label{squashed-sphere}
  S^4_b \coloneqq \bigl\{y_5^2+b^2(y_1^2+y_2^2)+b^{-2}(y_3^2+y_4^2)=r^2\}\subset\RR^5 .
\end{equation}
Parts of 4d $\Nsusy=2$ supersymmetry remain: $\Lie{U}(1)^2$ rotations $M_{12}$ and~$M_{34}$ in the $y_1,y_2$ and $y_3,y_4$ planes, and a supercharge $Q$ (and its conjugate) such that
\begin{equation}
  Q^2 = \frac{b}{r} (M_{12}-\tfrac{1}{2}J_3^{\Rsymm}) + \frac{1}{br} (M_{34}-\tfrac{1}{2}J_3^{\Rsymm})
\end{equation}
where $J_3^{\Rsymm}$ is the Cartan generator of $\lie{su}(2)_{\Rsymm}$.

\subsection{\label{ssec:loc-loc}Supersymmetric localization}

The idea of supersymmetric localization is several decades old when applied to scalar supercharges of topologically twisted field theories.
It received a new life since Pestun's calculation in 2007~\cite{0712.2824} of the sphere partition function of 4d $\Nsusy=2$ theories, and of Wilson loop expectation values.
In the following decade the technique was sucessfully applied to many dimensions (from 1d to 7d and even continuous dimensions) and geometries (such as spheres $S^d$, products $S^{d-1}\times S^1$, hemispheres and other spaces with boundaries), as summarized in the 2016 review volume~\cite{1608.02952}.
Besides the applications to understanding supersymmetric theories and black hole state counting, the resulting expressions are often complicated special functions with an interest of their own.
We introduce here the technique and present in \autoref{ssec:loc-cut} a variant that explains various factorization properties.

\paragraph{Set-up for supersymmetric localization.}

Our goal is to compute a path integral
\begin{equation}
  \vev{\cO} = \int[\rmD\phi] e^{-S} \cO
\end{equation}
that is invariant under some supercharge~$Q$ (we denote collectively all the fields as~$\phi$).
This means that the action and path integration measure are $Q$-invariant ($QS=0$ and $\int[\rmD\phi]Q(\text{anything})=0$) and that the observable also is ($Q\cO=0$).
Roughly speaking,
the integrand is invariant along orbits of~$Q$ in the space of field configurations, so the
integral on each non-trivial orbit is the Grassmann integral of a constant hence vanishes.
\apartehere
Intuitively, this ought to reduce the integral to $Q$-invariant field configurations.

Supersymmetric localization is based on the path integral so we need the supersymmetry~$Q$ to be realized \emph{off-shell}, not only on-shell.  For highly supersymmetric theories (such as 4d $\Nsusy=2$), realizing all supersymmetries on-shell requires infinitely many auxiliary fields, but only a finite number are typically needed to realize a single supercharge off-shell.

The key idea of supersymmetric localization is to deform the action~$S$ in a way that does not affect~$\vev{\cO}$ yet suppresses contributions from most configurations, thus reducing the path integral down to a smaller space of configurations.
Typically one arranges to make this smaller space finite-dimensional so as to get a well-defined finite-dimensional integral, but it can also be interesting to get a lower-dimensional field theory.

Concretely, one needs some functional of the fields with three properties: it is $Q$-exact (namely of the form $QV$), $Q$-closed (namely $V$~is invariant under the bosonic symmetry~$Q^2$), and has nonnegative bosonic part on the path integration contour we wish to consider.
A typical choice is roughly speaking a sum over all fermions of the theory (collectively denoted as $\psi$) of the form
\begin{equation}\label{Vsusyloc}
  V = \sum_{\text{all spinors }\psi} \psi \overline{Q\psi}
  \quad \implies \quad (QV)_{\bosonic} = \sum_\psi |Q\psi|^2
\end{equation}
for a suitable definition of the conjugate $\overline{Q\psi}$ which ensures positivity of $(Q\psi)\overline{Q\psi}$ along the path integral contour.

\paragraph{Saddle-point calculation.}

Once such a term is chosen, we deform the action by $tQV$ for $t\in[0,+\infty)$ and notice that the observable is unaffected since
\begin{equation}\label{dtOt}
  \begin{aligned}
    \vev{\cO}_t & = \int [\rmD\phi] e^{-S-tQV} \cO \\
    \implies \del_t \vev{\cO}_t & = - \! \int [\rmD\phi] e^{-S-tQV} \cO\,QV = - \! \int [\rmD\phi] Q\bigl(e^{-S-tQV}\cO\,V\bigr) = 0 .
  \end{aligned}
\end{equation}
Here we used that $Q\cO=0=Q(S+tQV)$ to write the integrand as $Q$ of something.

The observable is $t$-independent, so we can take the limit $t\to\infty$, in which limit the saddle-point approximation becomes exact.
In addition, any saddle with $(QV)_{\bosonic}>0$ is infinitely suppressed by~$e^{-tQV}$.
Since we assumed $QV\geq 0$, in this limit we are left with an integral over field configurations with $QV=0$ and the Gaussian integral of quadratic fluctuations around it:
\begin{equation}\label{localization}
  \vev{\cO} = \vev{\cO}_{t=0} = \lim_{t\to\infty} \vev{\cO}_t
  = \int_{QV=0} [\rmD\phi] e^{-S[\phi]} Z_{\oneloop}[\phi] \cO[\phi] .
\end{equation}
Here we wrote schematically $\int_{QV=0}$, but this may also involve discrete sums if the space of zeros of $QV$ is disconnected.
Here $Z_{\oneloop}[\phi]$ is the result of a Gaussian integral of $\exp(-tQV)$ around a field configuration~$\phi$ that is a zero of~$QV$.

Calculating the one-loop determinant~$Z_{\oneloop}$ is often the most technical step.
It is determined by the quadratic terms $(QV)_2$ in~$QV$, and more precisely is a ratio of determinants of the fermionic and bosonic parts of~$(QV)_2$.
In sufficiently simple geometries one can explicitly diagonalize these operators by listing all the modes and regularize the infinite products over modes using e.g.\ zeta-function regularization.
The calculation involves huge cancellations, due to how the supercharge~$Q$ pairs up most bosonic and fermionic modes together.
Roughly, only modes that are annihilated by~$Q$ contribute, and this observation helps extend the cases where one-loop determinants can be evaluated in a pedestrian manner.
Finally, this latter calculation can be much simplified by using powerful index theorems.
We refer to the review volume~\cite{1608.02952} for details.

\begin{exercise}
  1. Remind yourself that for a symmetric $n\times n$ matrix~$\Delta$ we have the Gaussian integral $\int_{\RR^n} \rmd^nx \exp(-x^T\Delta x)=\sqrt{\det(\pi/\Delta)}$.

  2. Consider next $n$ pairs of Grassmann variables $\theta_1,\bar{\theta}_1,\dots,\theta_n,\bar{\theta}_n$ and an $n\times n$ matrix~$D$, and evaluate the Berezin integral $\int \rmd^n\theta \rmd^n\bar{\theta}\exp(-\bar{\theta}.D.\theta) = \pm \det D$.

  In concrete \ac{QFT} applications, the bosonic operator $\Delta$~is roughly the Laplacian and the fermionic operator $D$~is roughly the Dirac operator, so that $\Delta\sim D^2$ and their one-loop determinants mostly cancel.
\end{exercise}

\startaparte[title=Recipe for supersymmetric localization]
\begin{itemize}
\item Choose a supergravity background on~$M$ with at least one generalized conformal Killing spinor~$\xi$, so that the theory on~$M$ has at least one supersymmetry~$Q$.  Realize it off-shell.
\item Choose a fermionic functional~$V$ that is $Q^2$-invariant and has $(QV)_{\bosonic}\geq 0$ on the path integral contour.
\item Find zeros of $(QV)_{\bosonic}$, which will be the resulting integration locus, often finite-dimensional.
\item Expand $(QV)_{\bosonic}$ to quadratic order around these zeros and compute the Gaussian integral (one-loop determinants).
\end{itemize}
\stopaparte

\paragraph{Saddle-points on ellipsoid.}

We now apply localization to 4d $\Nsusy=2$ theories on the squashed sphere~$S^4_b$.
We take the standard deformation term~\eqref{Vsusyloc} where the sum ranges over quarks~$\psi$ (hypermultiplet spinors) and gauginos~$\lambda$ (vector multiplet spinors).
The resulting $QV$ is pretty similar to the 4d $\Nsusy=2$ action of these multiplets, and we only mention what is needed to determine the space $(QV)_{\bosonic}=0$.

Supersymmetric localization relies on the existence of a supercharge~$Q$ that is an \emph{off-shell} symmetry.  This requires the addition of some auxiliary fields~$K$ to the 4d $\Nsusy=2$ theory.
For now we focus on the sphere~\cite{0712.2824}, restoring the squashing only in the final expressions~\cite{1206.6359}.

For the hypermultiplet we have
\begin{equation}
  (QV_{\hyper})_{\bosonic} = |Dq|^2+|D\tilde{q}|^2+\dots+ \frac{R}{6}|q|^2+\frac{R}{6}|\tilde{q}|^2+|K_q|^2 .
\end{equation}
Here, $R$ is the Ricci scalar, which is positive: this term arises upon conformally mapping $|Dq|^2$ from flat space to the sphere.  The ``$\dots$'' are also a sum of squares, so the zero locus has the whole hypermultiplet set to zero:
\begin{equation}\label{loc-locus-hyper}
  q = \tilde{q} = K_q = 0 ,
\end{equation}
and fermions as well since they are Grassmann variables.

For the vector multiplet we have similar terms with $q,\tilde{q}$ replaced by the vector multiplet scalar~$\phi$, but we also have terms like $|F_{\mu\nu}|^2$ and terms due to the supergravity background.  Eventually (the bosonic part of) the deformation term can be massaged to a sum of squares of the form
\begin{equation}\label{QVvecbos}
  \begin{aligned}
    (QV)_{\vect,\bosonic}
    & = \frac{r-x^0}{2r} \bigl(F^-_{\mu\nu} + w^-_{\mu\nu} \Re\phi\bigr)^2
    + \frac{r+x^0}{2r} \bigl(F^+_{\mu\nu} + w^+_{\mu\nu} \Re\phi\bigr)^2 \\
    & + |D\phi|^2 + [\phi,\phi^\dagger]^2 + |K_{\phi,i}+w_i\Im\phi|^2 .
  \end{aligned}
\end{equation}
Here $F^-$ and $F^+$ are the (anti)\llap{-}self-dual parts of the gauge field strength, $K_{\phi,i}$, $i=1,2,3$ are auxiliary fields (a triplet of $\lie{su}(2)_{\Rsymm}$), and $w^{\pm}_{\mu\nu}$ and~$w_i$ are determined by the supergravity background.

Let us find zeros of~\eqref{QVvecbos}.  Each term must vanish, so in particular $\phi$ is covariantly constant ($D\phi=0$).  Away from the poles $x^0=\pm r$, we have $F^{\pm}_{\mu\nu}=-w^{\pm}_{\mu\nu}\Re\phi$ so $F_{\mu\nu} = -w_{\mu\nu}\Re\phi$ where $w=w^++w^-$.
Then the Bianchi identities imply
\begin{equation}\label{Bianchi-why-phi-imaginary}
  0 = D^\mu F_{\mu\nu} = D^\mu(-w_{\mu\nu}\Re\phi)
  = - (\del^\mu w_{\mu\nu}) \Re\phi - w_{\mu\nu} \Re(D^\mu\phi)
\end{equation}
and the last term vanish since $D^\mu\phi=0$.
In the specific supergravity background we have here, $\del^\mu w_{\mu\nu}\neq 0$, so we learn that $\Re\phi=0$, hence $F_{\mu\nu}=0$.
Thus, in a suitable gauge the gauge field vanishes and
\begin{equation}\label{loc-locus-vec}
  A_{\mu} = 0 , \quad \phi = a , \quad K_{\phi,i}=-w_ia \qquad \text{away from poles,}
\end{equation}
for a constant~$a$.

At the poles $x^0=\pm r$, on the other hand, we only have one of the two equations $F^{\pm}_{\mu\nu}=-w^{\pm}_{\mu\nu}\Re\phi$, while the other part of $F_{\mu\nu}$ is unconstrained.
This suggests to include point-like instanton configurations at the poles:
\begin{equation}\label{loc-locus-inst}
  \text{instantons (} F^+=0 \text{) at } x^0=r ; \quad
  \text{anti-instantons (} F^-=0 \text{) at } x^0=-r .
\end{equation}
Let us concentrate on the North pole $x^0=r$.
Instanton configurations are insensitive to matter, so that the instanton moduli space $\Mcal_{\inst}$ is a product, over simple gauge group factors, of a moduli space of instantons for each gauge group.
This, in turn, splits as a union of infinitely many connected components, labeled by the instanton number $k=\#\int\Tr(F\wedge F)\in\ZZ_{\geq 0}$ (for some calculable constant~$\#$), with one instanton number per gauge group.
Altogether, denoting the gauge group by $\prod_I G_I$,
\begin{equation}
  \Mcal_{\inst} = \prod_I \biggl( \bigsqcup_{k_I\geq 0} \Mcal_{G_I,k_I} \biggr).
\end{equation}

\paragraph{Reality of Coulomb branch parameter and masses.}
The condition $\Re a=0$ simply means that $a\in\lie{g}$ rather than the complexification thereof.
We gauge-fix it so that it belongs to the Cartan subalgebra~$\lie{h}$ modulo the Weyl group.
The reality condition is then that
\begin{equation}\label{mass-reality}
  \vev{w,a}\in i\RR , \qquad \text{for every weight~$w$.}
\end{equation}
For instance, $a$~is anti-Hermitian for $\lie{g}=\lie{su}(N)$.

This reality condition arose here from studying saddle-points of the deformation term, but the same condition also derives from requiring the configuration to be $Q$-invariant.
As explained previously, hypermultiplet masses~$m$ simply amount to constant background values for vector multiplet scalars.  Thus, the reality condition~\eqref{mass-reality} is also required for masses in order for them to preserve the supercharge $Q$~used by supersymmetric localization.
In other words, curved-space supersymmetry on $S^4_b$ requires masses to be imaginary.

\paragraph{Result for round and squashed sphere.}
Saddle-point configurations defined by \eqref{loc-locus-hyper}, \eqref{loc-locus-vec}, \eqref{loc-locus-inst} are thus characterized by a choice, for each gauge group, of an imaginary Coulomb branch parameter~$a$ and point-like (anti)\llap{-}instanton configurations at the poles.
For any such saddle-point we compute the classical action
\begin{equation}
  S_{\cl} = - \Re(2\pi i\tau) \Tr(ra)^2 + 2\pi i \tau n - 2 \pi i \overline{\tau} \overline{n} , \qquad \tau=\frac{\theta}{2\pi}+\frac{4\pi i}{g^2} ,
\end{equation}
where the radius~$r$ of the (squashed) sphere~\eqref{squashed-sphere} makes the first term dimensionless.

We refer to~\cite{1608.02962} for the computation of one-loop determinants (Gaussian integral) capturing the effect of quadratic fluctuations around the saddle-points.
For the zero (anti)\llap{-}instanton saddle, one gets
\begin{equation}
  Z_{\oneloop} = Z_{\oneloop}^{\vect} Z_{\oneloop}^{\hyper}
  \quad \text{for } k=\kbar=0 .
\end{equation}
Here, the vector multiplet one-loop determinant is a product over roots~$\alpha$ of all gauge group factors (non-zero weights of the adjoint representation),\footnote{To be precise, we have included here in $Z_{\oneloop}^{\vect}$ the Vandermonde determinant $\prod_{\alpha\in\Delta}(\vev{\alpha,ra})$ that arises when converting from an integral over the whole gauge algebra~$\lie{g}$ to its Cartan subalgebra.}
\begin{equation}
  Z_{\oneloop}^{\vect} = \prod_{\alpha\in\Delta} \Upsilon_b(\vev{\alpha,ra}) ,
\end{equation}
where $\Upsilon_b$ is a special function defined in \autoref{sec:special}.  We emphasize that its argument here is purely imaginary.
The hypermultiplet one-loop determinant is a product over weights~$w$ (with multiplicity) of the representation in which the hypermultiplet transforms,
\begin{equation}\label{oneloop-hyper}
  Z_{\oneloop}^{\hyper} = \prod_{w\in R} \frac{1}{\Upsilon_b\bigl(\frac{b+1/b}{2}+\vev{w,ra}\bigr)} .
\end{equation}
As explained previously, hypermultiplet masses are simply background values for vector multiplet scalars corresponding to flavour symmetries, so adding a mass~$m$ in~\eqref{oneloop-hyper} simply changes
\begin{equation}
  \Upsilon_b\biggl(\frac{b+1/b}{2}+\vev{w,ra}\biggr) \to \Upsilon_b\biggl(\frac{b+1/b}{2}+\vev{w,ra}+rm\biggr) .
\end{equation}
Importantly, hypermultiplets in the representations $R$ and $\overline{R}$ are equivalent (with mass $m\to-m$) and one checks that the symmetry $\Upsilon_b(b+1/b-x)=\Upsilon_b(x)$ ensures that the one-loop determinant~\eqref{oneloop-hyper} computed with both presentations is the same.
For a half-hypermultiplet in a pseudoreal representation $R\simeq\overline{R}$ one should keep only one factor for each pair of conjugate weights; thanks to the same symmetry of~$\Upsilon_b$ it does not matter which weight one selects in each pair.
As expected all factors are invariant under the $b\to 1/b$ symmetry of~$S^4_b$ thanks to $\Upsilon_b=\Upsilon_{1/b}$.

One-loop determinants can be further understood as products of contributions from both hemispheres, essentially by decomposing each $\Upsilon_b$ function as $\Upsilon_b(x)=1/(\Gamma_b(x)\allowbreak\Gamma_b(b+1/b-x))$.
(Anti)\llap{-}instantons at each pole only affect one-loop contributions from the corresponding hemisphere, which leads to a factorization property of the form
\begin{equation}\label{Zoneloop-factorizes}
  Z_{\oneloop}(a,k,\xi,\kbar,\xi) = Z_{\oneloop}(a) Z_{\oneloop,\inst}(a,k,\xi) Z_{\oneloop,\inst}(a,\kbar,\xibar)
\end{equation}
where $\xi\in\Mcal_{G,k}$ and $\xibar\in\Mcal_{G,\kbar}$ parametrize the instanton configurations and $Z_{\oneloop}(a)$ is the ratio of $\Upsilon_b$ written above.

Altogether, collecting all (anti)\llap{-}instanton contributions together, including the classical contributions expressed in terms of $q=\exp(2\pi i\tau)$, the partition function reads
\begin{equation}\label{ZS4b-final-form}
  Z_{S^4_b} = \int \rmd a\, Z_{\cl}(a,q\qbar) Z_{\oneloop}(a) Z_{\inst}(a,q) Z_{\inst}(a,\qbar) .
\end{equation}
Here, the integral ranges over the Cartan algebras of all gauge groups, $Z_{\oneloop}$~is given above, and the exponent in $Z_{\cl}=\exp(-S_{\cl})=|q|^{r^2|a|^2}$ involves the Killing form $|a|^2=-\Tr(a^2)$ of the Lie algebra and is positive since $a$~is imaginary.
We collected into~$Z_{\inst}$ the shift of~$S_{\cl}$ due to instantons, and~$Z_{\oneloop,\inst}$, suitably integrated over the instanton moduli space,
\begin{equation}
  Z_{\inst}(a,q) = \sum_{\text{all }k_I\geq 0} \biggl(\prod_I q_I^{k_I}\biggr) \int_{\prod_I\Mcal_{G_I,k_I}} Z_{\oneloop,\inst}(a,k,\xi) \, d\xi .
\end{equation}
There remains to compute this instanton partition function.

\subsection{\label{ssec:loc-inst}Instanton partition functions}

\paragraph{Omega background.}

The point-like configurations in~\eqref{loc-locus-inst} are only sensitive to the leading expansion of the supergravity background around the poles.
This supergravity background has a flat metric and a non-trivial graviphoton, and coincides with the Omega background $\RR^4_{\epsilon_1,\epsilon_2}$ discovered by Nekrasov~\cite{hep-th/0206161}, with parameters $\epsilon_1=b/r$, $\epsilon_2=1/(rb)$.
It was thus naturally conjectured in~\cite{0712.2824,1206.6359} (and later works) that in the expression~\eqref{ZS4b-final-form} of $Z_{S^4_b}$, the function $Z_{\inst}(a,q)=1+O(q^1)$ to be included is the partition function of the 4d $\Nsusy=2$ theory on the Omega background $\RR^4_{\epsilon_1,\epsilon_2}$, called Nekrasov's instanton partition function~\cite{hep-th/0206161,hep-th/0306238}.
We refer to reviews~\cite{1412.7121,Song:2012kgc} for a more detailed introduction to $Z_{\inst}$.

The Omega background tends to~$\RR^4$ as $\epsilon_1,\epsilon_2\to 0$, and can be understood as a regulator for \ac{IR} divergences due to non-compactness of~$\RR^4$.
In fact, as explained in~\cite{hep-th/0206161,hep-th/0306238} and the appendix of~\cite{hep-th/0401184}, the partition function gives the low-energy prepotential of the gauge theory:\footnote{The prepotential is the Lagrangian density in $\Nsusy=2$ superspace, and it encodes fully the low-energy dynamics of $\Lie{U}(1)$ gauge fields at a generic point on the Coulomb branch.  We point to reviews such as~\cite{1312.2684} for more discussion of this crucial function.}
\begin{equation}\label{prepotential}
  F(a,q) = F_{\pert}(a,q) + \lim_{\epsilon_1,\epsilon_2\to 0} \Bigl(\epsilon_1\epsilon_2\log Z_{\inst}(a,q;\epsilon_1,\epsilon_2)\Bigr) ,
\end{equation}
where $F_{\pert}$ results from a one-loop computation and can be extracted from~$Z_{\oneloop}$.
In this way, the instanton partition function gives access to the low-energy dynamics of the 4d $\Nsusy=2$ theory at a point~$a$ along the Coulomb branch.
The whole \ac{SW} curve can then be rigorously derived, as done in~\cite{hep-th/0306238,1211.2240,1910.10104}.
The derivation relies on a link with the theory of random partitions.

The Omega background can also be obtained as the $\beta\to 0$ limit of a 5d background $S^1_\beta\times_{\epsilon_1,\epsilon_2}\RR^4$ defined as the quotient of $\RR\times\CC\times\CC$ under the identification $(x,z_1,z_2)\sim(x+\beta,e^{i\beta\epsilon_1}z_1,e^{i\beta\epsilon_2}z_2)$.
Many 4d $\Nsusy=2$ Lagrangian theory can be lifted to a 5d $\Nsusy=1$ theory, in which case the instanton partition function~$Z_{\inst}$ has a 5d analogue defined as the partition function on $S^1_\beta\times_{\epsilon_1,\epsilon_2}\RR^4$.
It is also worth mentioning developments in how to construct the Omega background in string theory and M-theory~\cite{1002.0888,1204.4192,1304.3488,1309.7350,1409.1219}.

\paragraph{Computation methods.}

The instanton partition function is an integral over the moduli space of instantons with an integrand depending on matter.
As mentioned above, this moduli space decomposes as a product over simple gauge group factors, and the moduli space for each gauge group factor has one connected component for each instanton number $k\geq 0$, with dimensions growing with~$k$.
There are several methods to compute the instanton partition function~\cite{Song:2012kgc}.\footnote{I thank Jaewon Song for correspondence on some methods.}
\begin{itemize}
\item \textbf{\acs{ADHM} construction.}  For classical gauge groups $\Lie{U}$ (rather than $\SU$), $\USp$, $\SO$ the $n$-instanton moduli space can be realized as a symplectic quotient through the \ac{ADHM} construction~\cite{Atiyah:1978ri}, while no such constructions are available in general for the exceptional groups $\Lie{E}_6,\Lie{E}_7,\Lie{E}_8,\Lie{F}_4,\Lie{G}_2$, nor for~$\SU$.
  Physically, the \ac{ADHM} construction expresses the instanton moduli space as the moduli space of a supersymmetric matrix model, describing D$(-1)$ brane instantons in a background of D3~branes and O3~planes that realize 4d $\Nsusy=2$ vector multiplets~\cite{hep-th/9410052,hep-th/9511030,hep-th/9512077}.

  The matrix model is known for hypermultiplets in suitable representation (e.g.\ bifundamental).
  Localization then reduces $Z_{\inst}$ from an integral over the whole moduli space down to a discrete sum of contributions from collections of point-like instantons respecting certain $\Lie{U}(1)$ symmetries.
  We give the resulting formula for $\Lie{U}(N)$ gauge theories in~\eqref{Zinst-quiver-formula}, below.  This \ac{LMNS} formula was obtained in~\cite{hep-th/9712241,hep-th/9711108,hep-th/9801061}, derived in~\cite{hep-th/0206161}, and extended to other classical groups in~\cite{hep-th/0404125,hep-th/0404225,hep-th/0404125,math/0609841}, to quivers in~\cite{1211.2240}, and to supergroup theories in~\cite{1905.01513}.
  These methods apply to 5d $\Nsusy=1$ lifts of these theories: the $S^1_\beta\times_{\epsilon_1,\epsilon_2}\RR^4$ instanton partition function matches the supersymmetric index of a quantum mechanics analogue of the matrix model (see e.g.~\cite{1305.5684,1601.06841}).

\item \textbf{Pure 4d $\Nsusy=2$ \acs{SYM}.}
  For exceptional gauge groups there is no \ac{ADHM} construction.
  The one-instanton moduli space is ($\CC^2$~times) the orbit under~$G$ of the highest weight vector of its adjoint representation.  This allows a group-theoretic calculation of one-instanton partition functions for arbitrary gauge groups in pure \ac{SYM}~\cite{1005.3026,1111.5624,1509.01294}.  This was extended to two instantons in~\cite{1205.4741}.

  Instead of realizing the $k$-instanton moduli space as a Higgs branch as in the \ac{ADHM} construction, it can be realized as the Coulomb branch of a 3d $\Nsusy=4$ \ac{SCFT}.
  The Coulomb branch Hilbert series (a specialization of the superconformal index) of this 3d theory then gives the $k$-instanton partition function of pure \ac{SYM} on $S^1_\beta\times_{\epsilon_1,\epsilon_2}\RR^4$ with an arbitrary gauge group~\cite{1408.6835}.
  The instanton partition function for pure $\Lie{E}_n$ \ac{SYM} theory can be determined from the Hall--Littlewood index of the $\Lie{E}_n$ Minahan--Nemeschansky theory, calculated using the \ac{TQFT} realization of this index discussed in \autoref{ssec:dim-index}.

\item \textbf{Mass-deformed 4d $\Nsusy=4$ \ac{SYM}.}
  For 4d $\Nsusy=2^*$ \acs{SYM}, which interpolates between $\Nsusy=4$ and $\Nsusy=2$ \ac{SYM} as the mass is varied, the prepotential (or its Omega-deformation) is expected to obey a modular anomaly equation that describes how it transforms under S-duality~\cite{1302.0686,1307.6648}.
  This led to a formula for the prepotential in terms of Eisenstein series for arbitrary gauge groups~\cite{1406.7255,1507.07709,1507.08027,1602.00273}.
  This technique extends somewhat, for instance to conformal \ac{SQCD}~\cite{1507.07476,1601.01827}.

\item \textbf{Recursion relations.}
  Comparing the instanton partition function to the partition function on the blow-up of~$\CC^2$ yields recursion relations (see~\cite{1205.4722,1411.4222,1908.11276} and references therein).
  For hypermultiplets (but not half-hypermultiplets) in a large class of representations of both classical and exceptional gauge groups, one can solve these recursion relations and deduce $Z_{\inst}$ from the perturbative (one-loop) partition function.
\end{itemize}
Some matter representations escape all the available methods: most notably, general Lagrangians constructed from $\SU(2)$ vector multiplets and trifundamental half-hypermultiplets, which are the Lagrangian descriptions of $\SU(2)$ class~S theories.
Among these theories, $Z_{\inst}$~is known whenever the theory can be written with gauge groups $\SU(2)$ and $\SO(4)=(\SU(2)\times\SU(2))/\ZZ_2$ and bifundamental matter (of two $\SU(2)$ groups or of an $\SU(2)$ and an $\SO(4)$ group); see~\cite{1012.4468,1107.0973}.
A possible way forward relies on the topological vertex formalism~\cite{1906.06351}.

Nekrasov partition functions have many generalizations, such as on \ac{ALE} space~\cite{1303.5765,1309.0812,1312.5554,1405.6992,1406.3008,1508.06813,1804.00771} and more general spaces~\cite{1306.0432,1502.07876,1606.07148}, spiked instantons~\cite{1611.03478},
and a proposed generalization to 4d $\Nsusy=1$ theories~\cite{1812.11188}.

\paragraph{Explicit formula for linear quiver gauge theories.}
We focus now on the linear quiver gauge theory
\begin{equation}\label{bis-SU-quiver}
  \quiver{
    \node (A) [color-group] at (0,0) {$\SU(N_1)$};
    \node (B) [color-group] at (1.8,0) {$\SU(N_2)$};
    \node (C) at (3.2,0) {${\cdots}$};
    \node (D) [color-group] at (4.6,0) {$\SU(N_p)$};
    \node (Am) [flavor-group] at (0,.7) {$M_1$};
    \node (Bm) [flavor-group] at (1.8,.7) {$M_2$};
    \node (Dm) [flavor-group] at (4.6,.7) {$M_p$};
    \draw (Am)--(A)--(B)--(C)--(D)--(Dm);
    \draw (B)--(Bm);
  }
\end{equation}
which is a slight generalization (allowing $N_i$~to be distinct) of those in \autoref{ssec:Lag-linear}.
This theory has gauge group $\prod_{i=1}^p \SU(N_i)$, one hypermultiplet in each bifundamental representation $N_i\otimes\overline{N_{i+1}}$, and $M_i$~hypermultiplets transforming in the fundamental representation~$N_i$ of each group.
If the theory is conformal or asymptotically free, namely the beta-function coefficient $b_j=2N_j-N_{j-1}-N_{j+1}-M_j$ is non-negative, then it can be realized in class~S using quiver tails, see~\eqref{quiver-tail}.

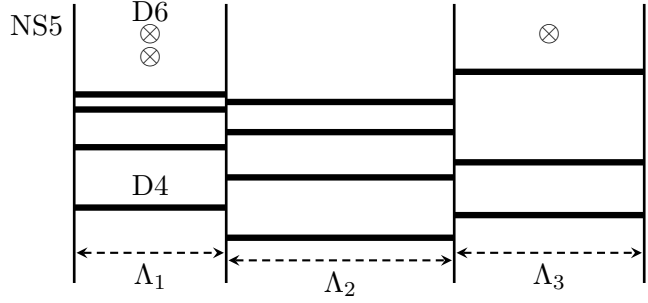
\begin{figure}
  \centering
  \begin{tikzpicture}[brane/.style={line width=1pt}, d4/.style={line width=2.4pt}]
    \draw[brane] (1,0) -- (1,3.7) node [below left] {NS5};
    \draw[d4] (1,1) -- (3,1) node [midway, above] {D4};
    \draw[d4] (1,1.8) -- (3,1.8);
    \draw[d4] (1,2.3) -- (3,2.3);
    \draw[d4] (1,2.5) -- (3,2.5);
    \node at (2,3.6) {D6};
    \node at (2,3.3) {$\otimes$};
    \node at (2,3) {$\otimes$};
    \draw[densely dashed,{stealth}-{stealth},thick] (1.03,.4) -- (2.97,.4) node [midway, below] {$\Lambda_1$};
    \draw[brane] (3,0) -- (3,3.7);
    \draw[d4] (3,.6) -- (6,.6);
    \draw[d4] (3,1.4) -- (6,1.4);
    \draw[d4] (3,2.0) -- (6,2.0);
    \draw[d4] (3,2.4) -- (6,2.4);
    \draw[densely dashed,{stealth}-{stealth},thick] (3.03,.3) -- (5.97,.3) node [midway, below] {$\Lambda_2$};
    \draw[brane] (6,0) -- (6,3.7);
    \draw[d4] (6,0.9) -- (8.5,0.9);
    \draw[d4] (6,1.6) -- (8.5,1.6);
    \draw[d4] (6,2.8) -- (8.5,2.8);
    \node at (7.25,3.3) {$\otimes$};
    \draw[densely dashed,{stealth}-{stealth},thick] (6.03,.4) -- (8.47,.4) node [midway, below] {$\Lambda_3$};
    \draw[brane] (8.5,0) -- (8.5,3.7);
  \end{tikzpicture}
  \caption{\label{fig:a_quiver}Brane construction of a $\Lie{U}(4) \times \Lie{U}(4) \times \Lie{U}(3)$ gauge theory.  The leftmost $\Lie{U}(4)$ factor has two fundamental hypermultiplets (inserted by transverse D6~branes depicted as~$\otimes$) and the rightmost $\Lie{U}(3)$ factor has one fundamental hypermultiplet.  In this example each gauge group is asymptotically free.}
\end{figure}

Extending the gauge groups from $\SU(N_i)$ to $\Lie{U}(N_i)$ allows the theory to be realized in IIA~string theory, as the world-volume theory of groups of $N_i$~parallel D4~branes stretched between consecutive NS5~branes, together with $M_i$~transverse D6~branes that give rise to fundamental hypermultiplets, as depicted in \autoref{fig:a_quiver}.
Instantons are described as D0~branes also stretching between NS5~branes, and $Z_{\inst}$ is the partition function of their world-volume theory, summed over the number of D0~branes.
Eventually, one must remove spurious factors caused by the additional $\Lie{U}(1)$ gauge groups.

We denote the Coulomb branch parameter $a^i=\{ a^i_1,\dots,a^i_{N_i} \}$ for $1\leq i\leq p$, and
by $k_i$~the number of instantons for $\Lie{U}(N_i)$, namely the number of D0~branes in the $i$-th interval between NS5~branes.
Then
\begin{equation}\label{Zinst-expansion}
  Z_{\inst}\Bigl[\prod\Lie{U}(N_i)\Bigr]
  = \sum_{k_1\geq 0,\dots,k_p\geq 0} z_1^{k_1}\cdots z_p^{k_p} Z_{\inst}^{(k)} ,
\end{equation}
where the counting parameters~$z_j$ encode the dynamical scale~$\Lambda_j$ or gauge coupling~$\tau_j$ of the $j$-th gauge group.\footnote{If $b_j=2N_j-N_{j-1}-N_{j+1}-M_j$ is positive then $z_j=\Lambda_j^{b_j}$ while if $b_j=0$ then $z_j$ is essentially $e^{2\pi i\tau_j}$.  To be precise there is some renormalization scheme ambiguity in what we mean by~$\tau_j$, as we discuss momentarily on \autopageref{par:renorm-schemes}.  The logic is rather to extract from $Z_{\inst}$ the prepotential as a function of~$z_j$, deduce how \ac{IR} couplings relate to~$z_j$, and finally invert the map if we want to express $z_j$~in terms of physically meaningful quantities.}
The $k$-instanton contribution to $Z_{\inst}$ is the matrix model contour integral
\begin{equation}\label{Zinst-quiver-formula}
  Z_{\inst}^{(k)} = \!\int\! \rmd\phi
  \frac{\prod\limits_{i,F,I} (m^i_F-\phi^i_I)
    \prod\limits_{i=1}^{p-1} \biggl[\prod\limits_{I,J} S(\phi^{i+1}_J-\phi^i_I)
    \prod\limits_{A,J} (\phi^{i+1}_J-a^i_A+\epsilon_1+\epsilon_2)
    \prod\limits_{I,B} (a^{i+1}_B-\phi^i_I)\biggr]}
  {\prod\limits_i \biggl[ (\frac{\epsilon_1\epsilon_2}{\epsilon_1+\epsilon_2})^{k_i}
    \prod\limits_{I\neq J} S(\phi^i_I-\phi^i_J)
    \prod\limits_{A,I} (\phi^i_I-a^i_A+\epsilon_1+\epsilon_2)(a^i_A-\phi^i_I)\biggr]}
\end{equation}
where $S(\phi)=-(\phi+\epsilon_1)(\phi+\epsilon_2)/[\phi(-\phi-\epsilon_1-\epsilon_2)]$, indices run over the ranges that are natural given where they appear ($1\leq i\leq p$, $1\leq F\leq M_i$, $1\leq I\leq k_i$, $1\leq J\leq k_{i+1}$, $1\leq A\leq N_i$, $1\leq B\leq N_{i+1}$), and $\rmd\phi$ denotes a product of all $\rmd\phi^i_I$.
The first factor in the numerator captures the effect of fundamental hypermultiplets, the rest of the numerator comes from bifundamental hypermultiplets, and the denominator from vector multiplets.

The integrand can be understood from the IIA string theory construction depicted in \autoref{fig:a_quiver}.
The vector multiplet contribution (denominator) arises from strings stretching between different D0 branes in a given interval ($1/S$ factors) and strings stretching between D0 and D4 branes (the remaining factors).
The fundamental hypermultiplet contribution arises from strings connecting D0 and D6 branes.
Strings joining D0~or D4~branes on two sides of an NS5~brane yield the bifundamental hypermultiplet contribution, in which one omits the D4-D4 interactions because they are already taken into account in~$Z_{\oneloop}$.

The same formula can be obtained from first principles using the mathematically rigorous \ac{ADHM} construction.
The contour in~\eqref{Zinst-quiver-formula} is such that it enclose poles at
\begin{equation}\label{polesenclosed}
  \{\phi^i_I\mid 1\leq I\leq k_i\} = \{a^i_A+(r-1)\epsilon_1+(s-1)\epsilon_2\mid(r,s)\in\lambda^i_A\}
\end{equation}
for each collection of Young diagrams $\lambda^i_A$, $1\leq i\leq p$, $1\leq A\leq N_i$ with a total number of boxes equal to the instanton number, $\sum_{A=1}^{N_i} \lvert\lambda^i_A \rvert=k_i$.
The set of poles~\eqref{polesenclosed} arises from a prescription called the \ac{JK} residue prescription, which follows from equivariant localization~\cite{math/0609841}.
This prescription was also obtained from localization of the aforementioned \ac{ADHM} supersymmetric matrix model~\cite{1406.6793,1406.7853,1407.2567,1502.04188}.

\begin{exercise}
  Specialize these formulas to $\Lie{U}(2)$ \ac{SQCD} with $N_f=4$ flavours ($p=1$, $N_1=2$, $M_1=4$).
  Note that the four mass parameters $m^1_B$, $B=1,\dots,4$ can all be shifted by shifting the Coulomb branch parameters $a^1_A$, $A=1,2$.
  This is a shadow of the fact that in the $\Lie{U}(2)$ theory we have only $\SU(N_f)$ flavour symmetry, not $\SO(2N_f)$ like for $\SU(2)$ \ac{SQCD}.
  Compute the one-instanton contribution in~$Z_{\inst}$.
  Start computing the two-instanton contribution to get an idea of the complexity: write the integrand and list the poles.
\end{exercise}

\paragraph{\(\Lie{U}(1)\) factor.}

The brane realization of the theory required the gauging of $\Lie{U}(1)$ symmetries to complete $\SU(N_i)$ into $\Lie{U}(N_i)$ gauge groups.
In this process, $\Lie{U}(1)$ mass parameters become dynamical vector multiplet scalars.
Thankfully, $Z_{\inst}$~is calculated for constant values of these scalars, so that we can simply set these Coulomb branch parameters to the $\Lie{U}(1)$ mass parameters we started with.

The major problematic effect of the gauging is to introduce spurious instanton contributions from the $\Lie{U}(1)$ gauge factor.
After spending a week puzzled about a mismatch with 2d \ac{CFT} conformal blocks, the authors of~\cite{0906.3219} proposed to divide out this spurious $\Lie{U}(1)$ instanton contribution, schematically\footnote{References welcome: it would be good to lay down the procedure nicely on the instanton partition function side, and I likely missed the correct reference.}
\begin{equation}\label{using-U1-factor}
  Z_{\inst}\Bigl[\prod\SU(N_i)\Bigr]
  = Z_{\inst}\Bigl[\prod\Lie{U}(N_i)\Bigr] \Bigm/ \bigl(\Lie{U}(1)\text{ factor}\bigr) .
\end{equation}
This $\Lie{U}(1)$ factor is more readily singled out on the \ac{CFT} side.
Instanton partition functions for linear quivers of $\SU(N)$ gauge groups are expected to match conformal blocks of the $W_N$~algebra (Virasoro for $N=2$), while $\Lie{U}(N)$ gauge groups correspond to conformal blocks of $W_N$~times the Heisenberg chiral algebra.
The difference (or rather ratio) of these two situations is thus given by a conformal block of the Heisenberg chiral algebra, namely a free field correlation function of chiral vertex operators.
This point of view was understood in~\cite{1012.1312} for $N=2$ and later extended in~\cite{1109.4042}.
For the linear quiver considered here, the factor is schematically
\begin{equation}\label{U1-factor}
  \bigl(\Lie{U}(1)\text{ factor}\bigr)
  \simeq \prod_{1\leq i\leq j\leq p} (1 - z_i\cdots z_j)^{c_{ij}}
\end{equation}
for some exponents~$c_{ij}$ that are quadratic in the mass parameters.

\paragraph{\label{par:renorm-schemes}Renormalization schemes: an example.}

Given the issue of extracting the $\Lie{U}(1)$ factor, and additional restrictions on which theories can be treated by brane constructions, people have sought other methods to determine~$Z_{\inst}$.
We present one method~\cite{1012.4468} that exemplifies a subtlety in comparing different results.

Brane setups with orientifold planes gives access to $Z_{\inst}$ for linear quivers with alternating $\USp$ and $\Spin$ groups.
We return to such quivers and their class~S construction at the end of \autoref{ssec:gen-rank}.
One of the simplest examples is in fact a different description of $\SU(2)$ \ac{SQCD} with $N_f=4$ flavors:
\begin{equation}\label{SQCD-as-Spin4}
  \quiver{
    \node(a)[flavor-group] at (0,0) {\scriptsize $\Spin(4)$};
    \node(b)[color-group] at (1.5,0) {\scriptsize $\USp(2)$};
    \node(c)[flavor-group] at (3,0) {\scriptsize $\Spin(4)$};
    \draw(a)--(b);
    \draw(b)--(c);
  }
  =
  \quiver{
    \node(At)[flavor-group] at (0,.4) {\scriptsize $\SU(2)$};
    \node(Ab)[flavor-group] at (0,-.4) {\scriptsize $\SU(2)$};
    \node(B)[color-group] at (1,0) {\scriptsize $\SU(2)$};
    \node(Ct)[flavor-group] at (2,.4) {\scriptsize $\SU(2)$};
    \node(Cb)[flavor-group] at (2,-.4) {\scriptsize $\SU(2)$};
    \draw(At)--(.3,0)--(Ab);
    \draw(.3,0)--(B)--(1.7,0);
    \draw(Ct)--(1.7,0)--(Cb);
  } .
\end{equation}
More generally, linear quivers of $\USp(2)=\SU(2)$ and $\Spin(4)=\SU(2)\times\SU(2)$ gauge groups are equivalent to generalized $\SU(2)$ quivers.
In simple enough cases such as~\eqref{SQCD-as-Spin4} the instanton partition function can also be computed through the $\Lie{U}(2)$ \ac{ADHM} construction and removing $\Lie{U}(1)$ factors~\eqref{using-U1-factor}.
Surprisingly, the two methods yield different series in powers of the exponentiated gauge coupling $\qLag=e^{2\pi i\tauLag}$, which would naively suggest an inconsistency!

The same issue showed up a long time ago when comparing \ac{SW} solutions coming from different constructions of 4d $\Nsusy=2$ theories.
It was resolved~\cite{hep-th/9911255} by noting a renormalization scheme ambiguity in the definition of the couplings.
We have already encountered this ambiguity in~\eqref{couplings-SQCD} when comparing the gauge coupling $\tauLag$ appearing in the Lagrangian of $\SU(2)$ \ac{SQCD} to the position $q=e^{2\pi i\tau}$ arising in the class~S construction of this same theory and found a relation $\tau=\tauLag+\cdots$ with a constant shift and an infinite tower of instanton corrections.\footnote{In the \ac{AGT} context, an early reference pointing out these instanton corrections is~\cite{0909.3338}, see also~\cite{1008.5240} for an F-theory derivation.}

Let us concentrate on (mass-deformed) \acp{SCFT} such as $\SU(2)$ $N_f=4$ \ac{SQCD}, whose gauge couplings do not run, and let us simplify expressions by assuming there is a single gauge coupling.
In two renormalization schemes that agree at leading order, the gauge coupling constants $\tau,\tilde\tau$ are related as
\begin{equation}\label{couplings-redef}
  \tilde\tau = \tau + \sum_{n\geq 0} c_n e^{2\pi i n\tau} ,
\end{equation}
where the $\tau$-independent coefficients~$c_n$ are dimensionless.
In concrete cases these coefficients are independent of Coulomb branch and mass parameters; assuming that the schemes agree at leading order throughout the parameter space, a possible argument is that the coefficients $c_n$~should be bounded functions of the Coulomb branch and mass parameters, hence must be constants by the Liouville theorem on analytic functions.

The $\Lie{U}(2)$ and $\USp(2)$--$\Spin(4)$ approaches to $Z_{\inst}$ are related in precisely this way, as understood in~\cite{1012.4468}.
Consider $\SU(2)$ \ac{SQCD} with $N_f=4$ massless hypermultiplets.
The quiver descriptions~\eqref{SQCD-as-Spin4} correspond to seeing \ac{SQCD} either as a reduction of $\mathcal{X}(\lie{so}(4))=\Xsuii\otimes\Xsuii$ with $\ZZ_2$-automorphism twist defects, or a reduction of~$\Xsuii$:
\begin{equation}
  \Theory\Bigl[\lie{so}(4),
    \quiver{
      \draw[densely dotted] (0,-.2) arc (-90:90:.1 and .2);
      \draw (3,-.2)--(0,-.2) arc (270:90:.1 and .2)--(3,.2);
      \filldraw (1,0) circle (.04) -- (2,0) circle (.04);
      \draw[densely dotted] (1.5,0) circle (.1 and .2);
      \draw (3,0) circle (.1 and .2);
    }
  \Bigr]
  =
  \Theory\left[\lie{su}(2),
    \quiver{
      \foreach\s in {+,-} {
        \draw[densely dotted] (0,\s.4-.2) arc (-90:90:.1 and .2);
        \draw (0,\s.4-.2) arc (270:90:.1 and .2);
        \draw (2,\s.4) circle (.1 and .2);
        \draw (0,\s.6)
        .. controls +(.5,0) and +(-.5,0) .. (1,\s.2)
        .. controls +(.5,0) and +(-.5,0) .. (2,\s.6);
      }
      \draw (0,-.2) arc (-90:90:.4 and .2);
      \draw (2,-.2) arc (-90:-270:.4 and .2);
      \draw[densely dotted] (1,0) circle (.1 and .2);
    }
  \right] .
\end{equation}
The double-cover of the $\lie{so}(4)$ curve (cylinder with a branch cut) is the $\lie{su}(2)$ curve (four-punctured sphere).
The relation between the modular parameter $q_{\lie{so}(4)}$ of the cut cylinder and the \ac{IR} coupling is fixed through the same considerations as for the modular parameter $q_{\lie{su}(2)}$ of the four-punctured sphere~\eqref{couplings-SQCD}.
This allows to relate them to each other,
\begin{equation}
  \begin{aligned}
    q_{\lie{so}(4)} & = \sqrt{16\lambda(2\tauIR)} = 16 e^{\pi i\tauIR} + O(e^{3\pi i\tauIR}) , \\
    q_{\lie{su}(2)} & = \lambda(\tauIR) = q_{\lie{so}(4)} \Bigl( 1 + \frac{q_{\lie{so}(4)}}{4} \Bigr)^{-2} = 16 e^{\pi i\tauIR} - 128 e^{2\pi i\tauIR} + O(e^{3\pi i\tauIR}) .
  \end{aligned}
\end{equation}
When comparing Nekrasov instanton partition functions, the non-trivial map between the counting parameters $q_{\lie{su}(2)}$ and $q_{\lie{so}(4)}$ must be taken into account.

\subsection{\label{ssec:loc-cut}Cutting by localization}

\paragraph{Supersymmetric localization without a Lagrangian.}

We have presented supersymmetric localization so far for Lagrangian theories, as the technique relies on a path-integral formulation.
In the absence of a path-integral formulation we cannot deform the action $S\to S+tQV$ as before, but we can still deform an expectation value $\vev{\cO}$ to $\vev{\cO e^{-tQV}}$.
In the same way as in~\eqref{dtOt} this expectation value is $t$-independent provided the supercharge~$Q$ is not spontaneously broken in the state of interest, and $Q\cO=0$.
In the $t\to+\infty$ limit the observable $\vev{\cO}$ should reduce in some sense to ``zeros of $(QV)_{\bosonic}$'', whenever such a notion can be defined.

In the following,\footnote{The ideas mostly appear in the literature, but may not have been collected previously in this way.} using a deformation term for vector multiplets only, we decompose the partition function of a general class~S theory into simple building blocks for any pants decomposition of~$C$.
On an orthogonal note, using a deformation term restricted to $S^3_b\subset S^4_b$, we factorize the partition function as the integral of a product of (anti)\llap{-}holomorphic functions of~$q$ as in~\eqref{ZS4b-final-form}, for any theory whose 3d restriction is Lagrangian.

\paragraph{Cutting \(C\) by localizing vector multiplets.}

Higher-rank class~S theories are typically non-Lagrangian, obtained by gauging common flavour symmetries of certain isolated \acp{SCFT} called tinkertoys (most notably~$T_N$).
Gauging symmetries, however, is performed using standard vector multiplets that are described by a standard path integral, with a gauge coupling~$\tau$ just as in the Lagrangian case.
This raises the prospect of applying supersymmetric localization to these vector multiplets.

Let us make more explicit what we mean by gauging a flavour symmetry group~$G$ of some theory~$\Theory$ (a product of tinkertoys).
The $G$~symmetry current multiplet in~$\Theory$ is first coupled to a non-dynamical vector multiplet $(A,\lambda,\phi)$.
The partition function $Z_{\Theory}[A,\lambda,\phi]$ depends on this background vector multiplet, whose components may be given any values (constant or not).
Gauging consists of performing the path integral over the vector multiplet fields, so that the partition function of the gauged theory is
\begin{equation}
  Z_{\Theory,\gauged} = \int [\rmD A\,\rmD\lambda\,\rmD\phi]\,e^{-S_{\SYM}[A,\lambda,\phi]} Z_{\Theory}[A,\lambda,\phi] ,
\end{equation}
where the \ac{SYM} action includes the gauge coupling and theta term, and the path integral measure implicitly accounts for gauge fixing.
Similar expressions are available for any other observable of the gauged theory, in terms of observables of the non-gauged theory.

The idea then is to add the deformation term~\eqref{Vsusyloc} for the vector multiplet, without adding any deformation terms for the tinkertoys.
Supersymmetric localization restricts the vector multiplet path integral as
\begin{equation}
  \begin{aligned}
    Z_{\Theory,\gauged}
    & = \lim_{t\to+\infty} \int [\rmD A\,\rmD\lambda\,\rmD\phi]\,e^{-S_{\SYM}-tQV_\vect} Z_{\Theory}[A,\lambda,\phi]
    \\
    & = \int_{QV_\vect=0} [\rmD A\,\rmD\phi] e^{-S_{\SYM}[A,\phi]} Z_{\oneloop}^{\vect}[A,\phi] Z_{\Theory}[A,\phi] .
  \end{aligned}
\end{equation}
The locus $QV_\vect=0$ cannot in principle have the fermion~$\lambda$ turned on, which is why $\lambda$~disappears in the last line.
The locus is described by~\eqref{loc-locus-vec} away from the poles and~\eqref{loc-locus-inst} at poles.
Namely, zeros of the vector multiplet deformation term are characterized by a constant imaginary value $\phi=a\in\lie{g}$ and by (anti)\llap{-}instantons at the poles.

In our applications $\Theory$~splits into decoupled sectors $\Theory=\prod_L\Theory_L$.
Decomposing $G=\prod_I G_I$ into simple factors and denoting by $\lie{h}_I$ their Cartan algebra, we find
\begin{equation}\label{Ztheory-gauged}
  \begin{aligned}
    & Z_{\Theory,\gauged}
    = \prod_I \Biggl[ \sum_{k_I,\kbar_I\geq 0} \int_{\lie{h}_I\times\Mcal_{G_I,k_I}\times\Mcal_{G_I,\kbar_I}} \rmd a_I \rmd\xi_I \rmd\xibar_I \Biggr]
    \biggl\{ \\
    & \qquad \prod_I \biggl[ |q_I|^{r^2|a_I|^2} q_I^{k_I} \qbar_I^{\kbar_I} Z_{\oneloop}^{\vect}[G_I;a_I,k_I,\xi_I,\kbar_I,\xibar_I] \biggr] \, \prod_L Z_{\Theory_L}[a,k,\xi,\kbar,\xibar] \biggr\} .
  \end{aligned}
\end{equation}
For class~S theories, this matches precisely how a correlator on~$C$ would be calculated by inserting, on each tube of a pants decomposition of~$C$, a complete set of states labeled by $a_I,k_I,\xi_I,\kbar_I,\xibar_I$.
Indeed, we recognize the sum over states $\sum_{k,\kbar}\int_{a,\xi,\xibar}$ for each tube~$I$, the inverse norm of that state, and the remaining three-point functions $Z_{\Theory_L}$ for each three-punctured sphere~$L$.
We know that~$Z_{\Theory_L}$ is only sensitive to $a_I,k_I,\xi_I,\kbar_I,\xibar_I$ for groups under which $\Theory_L$ is charged; in class~S these correspond to the tubes that connect to the given tinkertoy.

\paragraph{Holomorphic factorization by localizing on \(S^3_b\).}

We now turn to a method to cut and glue partition functions using supersymmetric localization.  This was most deeply explored in \cite{1807.04274,1807.04278} where many previously observed factorisation properties were derived (see also~\cite{1708.04631} for an earlier application of the idea).
Our aim here is to explain the holomorphic dependence in~$q$ by cutting the sphere in halves along the equator $S^3_b\subset S^4_b$.  This is nicely complementary to how we have cut the Riemann surface~$C$ above.

As a warm-up we revise the Lagrangian case.
The keys for factorization in that case are that
\begin{itemize}
\item the $q$ and~$\qbar$ dependence arises from instantons at the poles;
\item these instantons do not interact through~$Z_{\Theory_L}$.
\end{itemize}
To be precise, $\Theory$~consists of hypermultiplets and supersymmetric localization with the hypermultiplet deformation term sets all fields of~$\Theory$ to zero, and quadratic fluctuations yield the product of a function of $a,k,\xi$ and a function of $a,\kbar,\xibar$, as for the vector multiplet~\eqref{Zoneloop-factorizes}.
Thus, the $q,k,\xi$ and $\qbar,\kbar,\xibar$ dependence completely decouple from each other in~\eqref{Ztheory-gauged}.
Collecting together the sum-integral over $k,\xi$ into an instanton partition function, we retrieve the expected factorization as an integral over~$a$ of a holomorphic times an antiholomorphic function of~$q$, as stated in~\eqref{ZS4b-final-form}.

We now rephrase this in terms of localization.
The $S^3_b$~restriction of the 4d fields are 3d $\Nsusy=2$ vector and hypermultiplets, and the full 4d theory is alternatively described as two (identical) 4d theories on the hemispheres of~$S^4_b$, coupled together through their boundary condition: for instance the partition function reads
\begin{equation}
  Z_{S^4_b} = \int [\rmD\Phi_{\text{3d}}] e^{-S_{\text{3d}}[\Phi_{\text{3d}}]} Z_{HS^4_b,\text{North}}[\Phi_{\text{3d}},q,\qbar] Z_{HS^4_b,\text{South}}[\Phi_{\text{3d}},q,\qbar] ,
\end{equation}
where $\Phi_{\text{3d}}$ denotes collectively all fields of the 3d $\Nsusy=2$ theory on the equator and $Z_{HS\dots}$ denote the path integrals over fields on each of the two hemispheres.
This expression can be seen as the expectation value of (very complicated) observables $Z_{HS\dots}$ of the 3d theory.

The equator theory can then be localized by adding to it the usual deformation term that is used for localization on~$S^3_b$.
The 3d hypermultiplets get localized to zero as in 4d~\eqref{loc-locus-hyper}.
The 3d vector multiplets get localized to constant values $\phi=a\in\lie{g}$.
As for any other observable, the $Z_{HS\dots}$ factors are evaluated on the localization locus so that
\begin{equation}
  Z_{S^4_b} = \int \rmd a\, Z_{\cl,\text{3d}}[a] Z_{\oneloop,\text{3d}}[a] Z_{HS^4_b,\text{North}}[a,q] Z_{HS^4_b,\text{South}}[a,\qbar] .
\end{equation}
The hemisphere partition functions are evaluated here with Dirichlet boundary conditions with a constant boundary value~$a$ for the vector multiplet scalars.
The absence of gauge-coupling dependence in 3d is standard: \ac{YM} terms in 3d are $Q$-exact.  More generally, \ac{YM} and theta terms are $Q$-exact everywhere away from the North/South poles where their sum/difference is not $Q$-exact in a smooth way.  This explains why the hemisphere contributions only depend on~$q$ or~$\qbar$.

This idea of cutting the theory along the equator generalizes beyond Lagrangian theories.
Even though $Z_{\Theory_L}$ may not have a path integral description, general axioms of \ac{QFT} still apply and one can insert a complete set of states along the hypersurface $S^3_b\subset S^4_b$.
The partition function is then seen as the overlap of two states prepared by the path integral over hemispheres,
\begin{equation}
  Z_{S^4_b} = \sum_{\ket{\Phi}\in\Hcal[S^3_b]} \bigl\langle HS^4_b\text{ North}\bigm|\Phi\bigr\rangle
  \bigl\langle\Phi\bigm| HS^4_b\text{ South}\bigr\rangle .
\end{equation}
Then, $Q$-invariance of the set-up implies that the sum restricts to $Q$-invariant states,
\begin{equation}
  Z_{S^4_b} = \sum_{\ket{a}\in\Hcal[S^3_b],Q\ket{a}=0} \bigl\langle HS^4_b\text{ North}\bigm|a\bigr\rangle
  \bigl\langle a\bigm| HS^4_b\text{ South}\bigr\rangle ,
\end{equation}
and $Q$-exactness of the \acs{YM}${}\pm{}$theta terms on each hemisphere shows that the two factors are (anti)\llap{-}holomorphic functions of~$q$ as wanted.

We observe that the Coulomb branch integral over $a\in\lie{h}$ is in general replaced by a sum/integral over $Q$-invariant states in the $S^3_b$~Hilbert space of the 4d theory.  This is typically a larger space, especially in the presence of non-Lagrangian tinkertoys.

\section{\label{sec:AGT}\acsfont{AGT} for \(\SU(2)\) quivers}

The \ac{AGT} relation~\eqref{AGT-bis} relates observables of two different dimensional reductions of the 6d $(2,0)$ theory $\Xsuii$.
We have explained in \autoref{sec:Lag} how reducing $\Xsuii$
on a Riemann surface $C=\fullC\setminus\{z_1,\dots,z_n\}$ yields a 4d $\Nsusy=2$ $\lie{su}(2)$ Sicilian\footnote{Thus named because of a resemblance between trinions and the triskelion featuring prominently on the flag of Sicily.} quiver gauge theory that depends on~$C$, with gauge couplings related to the complex structure of~$C$.
Likewise, we expect that reducing $\Xsuii$ on $S^4_b$ should yield a 2d theory with a coupling constant~$b$, and the codimension~$2$ defects of $\Xsuii$ inserted at punctures $z_i$ of~$C$ should become local operators in~2d.
Since this 2d dimensional reduction is somewhat technical, we postpone it to \autoref{ssec:AGT-CJ}, explaining there briefly why one should expect 6d observables to be computable both on the 4d and 2d sides.

Before that we determine the relevant 2d theory in a more historically accurate way in \autoref{ssec:AGT-dict}: its correlators should reproduce the $S^4_b$~partition functions of class~S theories.
These partition functions only depend on the complex structure of~$C$, hence only on the conformal class of the metric on~$C$.  This means that the 2d theory we seek should be a \ac{CFT}, and it turns out to be Liouville \ac{CFT}.
As a result, we begin by reviewing Liouville \ac{CFT} and 2d \ac{CFT} basics in \autoref{ssec:AGT-Liou}.
We summarize aspects of the correspondence in \autoref{tab:corr}.

\ifjournal\else\medskip\fi

\begin{apartetable}{tab:corr}{Basics of \ac{AGT}. In Liouville \acs{CFT} conventions, change $Q$ to~$Q/2$.}
  \begin{tabular}{cll}
    & Gauge theory & Toda/Liouville \ac{CFT} \\
    \midrule
    & Partition function on~$S^4_b$ & Toda correlator on~$C$ \\
    \midrule
    \ \clap{\smash{\rotatebox[x=4.1em,y=\baselineskip]{90}{\begin{tabular}{@{}l@{}}Parameters\end{tabular}}}}\ \
    & Squashing parameter $b=\sqrt{\epsilon_1/\epsilon_2}$ & Toda coupling constant~$b$ \\
    & Gauge couplings in~4d & Complex structure of~$C$ \\
    & Dual gauge theory descriptions & Pants decompositions of~$C$ \\
    & Full tame puncture of mass~$m$ & Vertex operator $\Vhat_{Q+rm}$ \\
    \midrule
    \ \clap{\smash{\rotatebox[x=5.65em,y=\baselineskip]{90}{\begin{tabular}{@{}l@{}}Building\,blocks\end{tabular}}}}\ \
    & Coulomb branch parameter~$a$ & Internal momentum $\alpha=Q+ra$ \\
    & Gauge one-loop determinant & Inverse two-point function \\
    & Matter one-loop determinant & Three-point structure constant \\
    & Classical action contribution & Conformal block leading term \\
    & Instanton partition function & Conformal block power series
  \end{tabular}
\end{apartetable}

\subsection{\label{ssec:AGT-Liou}2d \acsfont{CFT} and Liouville \acsfont{CFT}}

We refer to reviews such as~\cite{1406.4290,1609.09523} (and references therein) for an introduction to 2d \ac{CFT} and in particular to Liouville \ac{CFT}.
The \acf{YRISW} course~\cite{2006.13892} also provides a brief introduction to 2d \ac{CFT} and their symmetry algebras (chiral algebras), which curiously also show up independently as a protected subsector of 4d $\Nsusy=2$ theories.

\paragraph{Virasoro algebra.}

In contrast to other dimensions, the conformal symmetry algebra in 2d is the product $\Vir\times\overline{\Vir}$ of two infinite-dimensional algebras.
The Virasoro algebra $\Vir$ is spanned by $L_n$, $n\in\ZZ$ and a central element~$C$, subject to
\begin{equation}\label{Vir-comm}
  [L_m,L_n] = (m-n) L_{m+n} + \frac{1}{12} \delta_{m+n=0} (m^3 - m) C .
\end{equation}
This algebra, and the other copy spanned by $\overline{L}_n$, $n\in\ZZ$ and $\overline{C}$, acts on the Hilbert space of the theory on a circle.
For any given 2d \ac{CFT} the elements $C,\overline{C}$ of the conformal algebra act as multiplications by constants called central charges and denoted $c,\overline{c}$ or $c_L,c_R$.

In a \ac{CFT}, radial quantization identifies such states to their infinite-radial-past limit, which is a local operator~$\Ocal$ at the center of radial quantization.
The state corresponding to $\Ocal$ under this state-operator correspondence is denoted by $\ket{\Ocal}$.
Under this correspondence, the action of $\Vir\times\overline{\Vir}$ on states translates to an action on local operators by commutator.

\begin{exercise}
  Using \eqref{Vir-comm}, check $[L_m,[L_n,L_p]]+[L_n,[L_p,L_m]]+[L_p,[L_m,L_n]]=0$.
  This is the Jacobi identity, essential for consistency of the Lie algebra~$\Vir$.
  Check that the factor $m^3-m$ is fixed by the Jacobi identity up to mixing $L_n$ with~$C$ and scaling~$C$.

  Check that $\tilde{C}=k C$ and $\tilde{L}_n=L_{kn}/k+\delta_{n=0}(k-1/k)C/24$ obey the Virasoro commutation relations for any $k\in\ZZ\setminus\{0\}$, so that the Virasoro algebra contains infinitely many Virasoro subalgebras.
  Conversely, the Virasoro algebra can be embedded as the integer modes of a larger Virasoro algebra with fractional modes, useful in symmetric product orbifolds~\cite{1911.08485}.
\end{exercise}

\paragraph{Conformal dimensions.}

Dilations and rotations around the center of radial quantizations are generated by $L_0\pm\overline{L}_0$, hence the dimension $\Delta=h_{\Ocal}+\overline{h}_{\Ocal}$ and spin $h_{\Ocal}-\overline{h}_{\Ocal}\in\frac{1}{2}\ZZ$ of a local operator are given by the action of $L_0$ and $\overline{L}_0$:
\begin{equation}
  [L_0,\Ocal] = h_{\Ocal} \Ocal , \qquad
  [\overline{L}_0,\Ocal] = \overline{h}_{\Ocal} \Ocal .
\end{equation}
Despite the notation, the conformal dimensions $h_{\Ocal}$ and $\overline{h}_{\Ocal}$ are independent numbers.
In a unitary \ac{CFT} they are both real and non-negative.
We find it more convenient to mostly work with states, in which case $\Delta_{\Ocal}$ and $h_{\Ocal}-\overline{h}_{\Ocal}$ are the energy and momentum of the state.

Interestingly, the commutator $[L_0,L_n]=-nL_n$ implies that if $\ket{\Ocal}$ has conformal dimensions $(h_{\Ocal},\overline{h}_{\Ocal})$ then $L_n\ket{\Ocal}$ has conformal dimensions $(h_{\Ocal}-n,\overline{h}_{\Ocal})$.
For this reason, $L_n$, $n\geq 1$ are called lowering operators, and $L_n$, $n\leq -1$ are called raising operators.
The same applies to~$\overline{L}_n$.

\paragraph{Primary operators.}
The Hilbert space of a given theory organizes into conformal families, namely representations of $\Vir\times\overline{\Vir}$.
In a unitary \ac{CFT}, states have a non-negative energy so each conformal family has a state $\ket{V}$ of minimal dimension.
Such a state is annihilated by all lowering operators $L_n$, $\overline{L}_n$, $n\geq 1$ and is called a \emph{primary state} (the corresponding operator~$V$ is called a \emph{primary operator}):
\begin{equation}
  L_n\ket{V} = \overline{L}_n\ket{V} = 0 , \quad n\geq 1 .
\end{equation}
From this state of conformal dimensions $(h,\overline{h})$ one can construct a tower of states of higher dimensions by acting with $L_{-n}$ and~$\overline{L}_{-n}$, $n\geq 1$.
Using the Virasoro commutator~\eqref{Vir-comm} these raising operators can be ordered, and the set of descendants is spanned by the following states
\begin{equation}\label{LLketV}
  \LL_{-Y} \overline{\LL}_{-\overline{Y}} \ket{V}
  \coloneqq L_{-m_1}\dots L_{-m_k} \overline{L}_{-n_1}\dots \overline{L}_{-n_l} \ket{V} ,
\end{equation}
where $Y,\overline{Y}$ are two Young diagrams, $m_1\geq m_2\geq \dots\geq m_k\geq 1$ are the successive lengths of rows of~$Y$, and likewise $n_1\geq\dots\geq 1$ the rows of~$\overline{Y}$.
The conformal dimensions of~\eqref{LLketV} are $\bigl(h + m_1 + \dots + m_k , \overline{h} + n_1 + \dots + n_l\bigr)$.
These states are called \emph{descendants} of $\ket{V}$.
In fact, the whole representation of $\Vir\times\overline{\Vir}$ is spanned (as a vector space) by~\eqref{LLketV}, and generically these states are linearly independent.

\begin{exercise}
  1. For $\ket{V}$ a primary state, and for $m,n\in\ZZ$, rewrite $L_m L_n \ket{V}$ as a linear combination of terms~\eqref{LLketV} with properly sorted indices.

  2. Check that for any $Y,\overline{Y}$, acting with any $L_n$ or $\overline{L}_n$ on $\LL_{-Y} \overline{\LL}_{-\overline{Y}} \ket{V}$ yields a linear combination of such descendants.
\end{exercise}

\paragraph{Two and three-point functions.}

Conformal symmetry constrains correlators of local operators.
Denoting by $z_{ij}=z_i-z_j$, the two-\ and three-point functions of primary operators $V_i$ of conformal dimensions $(h_i,\overline{h}_i)$ take the form
\begin{equation}\label{2pt3pt}
  \begin{aligned}
    \vev{V_1(z_1,\overline{z}_1)\, V_2(z_2,\overline{z}_2)}
    & = g_{12}\,\delta_{h_1=h_2}\delta_{\overline{h}_1=\overline{h}_2} \, z_{12}^{-2h_1}\overline{z}_{12}^{-2\overline{h}_1} , \\
    \vev{V_1(z_1,\overline{z}_1)\, V_2(z_2,\overline{z}_2) \, V_3(z_3,\overline{z}_3)}
    & = C_{123} \, z_{23}^{h_1-h_2-h_3} \, z_{31}^{h_2-h_3-h_1} \, z_{12}^{h_3-h_1-h_2} \\
    & \qquad\qquad \times
    \overline{z}_{23}^{\overline{h}_1-\overline{h}_2-\overline{h}_3} \,
    \overline{z}_{31}^{\overline{h}_2-\overline{h}_3-\overline{h}_1} \,
    \overline{z}_{12}^{\overline{h}_3-\overline{h}_1-\overline{h}_2} ,
  \end{aligned}
\end{equation}
where $g_{12},C_{123}$ are constants depending on the primary operators involved.
Descendant operators can be written as Virasoro generators $L_n,\overline{L}_n$ acting on primary operators, and these generators act on~\eqref{2pt3pt} as various differential operators.

If the \ac{CFT} has a single primary operator of each conformal dimension $(h,\overline{h})$, which is the case for Liouville \ac{CFT}, then $g_{12}$ can be absorbed in a normalization of that operator.
Despite this possibility, the standard normalization of primary operators~$V_\alpha$ has non-trivial~$g_{12}$, and our \ac{AGT}-friendly normalization~$\Vhat_{\alpha}$ given below also does.
Once this normalization is chosen, the theory is characterized by its spectrum of primary operators (their conformal dimensions) and by three-point functions $C_{123}$ as explained next.

\paragraph{Correlators and conformal blocks.}
The $n\geq 4$ point functions of primary operators are then fixed by conformal invariance~\cite{Belavin:1984vu}.
For concreteness, consider a $4$-point function, and insert a complete set of states, which we separate according to representations of $\Vir\times\overline{\Vir}$:
\begin{equation}\label{four-point}
  \vev{V_1V_2V_3V_4}
  = \sum_{\substack{V,V'\text{ primaries}\\\mathclap{Y,\overline{Y},Y',\overline{Y}'\text{ Young diagrams}}}}
  \vev{V_1V_2\LL_{-Y}\overline{\LL}_{-\overline{Y}}|V}\;
  g^{-1}\bigl(Y,\overline{Y},V;Y',\overline{Y}',V'\bigr)\;
  \vev{V'|\LL_{Y'}\overline{\LL}_{\overline{Y}'}V_3V_4} ,
\end{equation}
where $g^{-1}(Y,\overline{Y},V;Y',\overline{Y}',V'\bigr)$ denotes components of the (matrix) inverse of the ``matrix'' with components $\vev{V'|\LL_{Y'}\overline{\LL}_{\overline{Y}'}\LL_{-Y}\overline{\LL}_{-\overline{Y}}|V}$.
Translating all Virasoro generators to differential operators acting on the position dependence in~\eqref{2pt3pt} gives
\begin{equation}
  \vev{V_1V_2V_3V_4}
  = \sum_{V_5,V_6\text{ primaries}}
  C_{125} g_{56}^{-1} C_{634} \Fcal(h_1,\dots,h_5;z_1,\dots,z_4)
  \Fcal(\overline{h}_1,\dots,\overline{h}_5;\overline{z}_1,\dots,\overline{z}_4) .
\end{equation}
Up to unimportant factors, the (locally) holomorphic factor~$\Fcal$ is called a \emph{conformal block}.
It is entirely determined by dimensions $h_1,\dots,h_4$ of the \emph{external operators}, and the dimension $h_5=h_6$ of the \emph{internal operator} inserted as part of the complete set of states.

We could have inserted a complete set of states with a different choice of which pair of operators $V_i$ lies on the two sides of the inserted states.  More generally, the possible ways to insert complete set of states to reduce a sphere $n$-point function (down to the constants $g$, $C$, and conformal blocks) correspond to the ways of decomposing the $n$-punctured sphere into three-punctured spheres: a complete set of states is inserted along each closed loop cutting the sphere into pieces.
In all cases, conformal blocks are purely representation-theoretic objects; they depend on dimensions of the $n$ external operators and of $n-3$ internal operators inserted along cuts.

\paragraph{Liouville theory.}

Liouville theory describes a single scalar field subject to the action
\begin{equation}
  S[\phi]=\frac{1}{4\pi}\int \rmd^2z\,\sqrt{g}\,(\del_\nu\phi\del^\nu\phi+QR\phi+4\pi\mu\,e^{2b\phi})
\end{equation}
where $R$ is the Ricci scalar.
Provided $Q=b+1/b$ this theory is conformal, with holomorphic stress-tensor $T=(\del\phi)^2+Q\del^2\phi$
and central charges $c=\overline{c}=1+6Q^2\geq 25$.

While it looks like the cosmological constant~$\mu$ is a coupling constant, it turns out to only appears in trivial ways in correlators: instead there is interesting dependence on $b>0$, with $b\to 0$ being the semiclassical limit.
The Liouville \ac{CFT} admits a (non-manifest) duality $b\to 1/b$ while keeping $\lambda=\bigl(\frac{\pi\Gamma(b^2)}{\Gamma(1-b^2)}\mu\bigr)^{1/b}$ fixed.

One can check that $V_\alpha={:}e^{2\alpha\varphi}{:}$ are conformal primary operators of left/right-moving dimension $h(\alpha)=\alpha(Q-\alpha)=Q^2/4+P^2$, for $\alpha=(b+1/b)/2+iP$, $P\in\RR$.
The invariance $h(Q-\alpha)=h(\alpha)$ suggests the identification $V_\alpha=R(\alpha)V_{Q-\alpha}$.
The \emph{reflection coefficient} can be determined (using conformal bootstrap) to be
\begin{equation}\label{Liou-R}
  R(\alpha) = - \lambda^{Q-2\alpha}
  \frac{\Gamma(b(2\alpha-Q))\Gamma(\frac{1}{b}(2\alpha-Q))}
  {\Gamma(b(Q-2\alpha))\Gamma(\frac{1}{b}(Q-2\alpha))} .
\end{equation}
The two-point function is then
\begin{equation}\label{Liou-2pt}
  \vev{V_{\alpha_1}V_{\alpha_2}} = \delta_{\alpha_1+\alpha_2=Q} + R(\alpha_1) \delta_{\alpha_1=\alpha_2} .
\end{equation}
The three-point function is known to be given by the \ac{DOZZ} formula \cite{hep-th/9403141,hep-th/9506136}
\begin{equation}\label{Liou-3pt}
  \begin{aligned}
    & C_{\alpha_1\alpha_2\alpha_3}
    = \vev{ V_{\alpha_1} V_{\alpha_2} V_{\alpha_3} }
    \\
    & = \frac{(b^{2/b-2b}\lambda)^{Q-\alpha_1-\alpha_2-\alpha_3}\Upsilon_b'(0)
      \Upsilon_b(2\alpha_1)\Upsilon_b(2\alpha_2)\Upsilon_b(2\alpha_3)}
    {\Upsilon_b(\alpha_1+\alpha_2+\alpha_3-Q)\Upsilon_b(\alpha_1+\alpha_2-\alpha_3)
      \Upsilon_b(\alpha_2+\alpha_3-\alpha_1)\Upsilon_b(\alpha_3+\alpha_1-\alpha_2)} .
  \end{aligned}
\end{equation}
Four-point functions for instance read
\begin{equation}
  \vev{ V_{\alpha_1}(0) V_{\alpha_2}(q) V_{\alpha_3}(1) V_{\alpha_4}(\infty) }
  = \frac{1}{2} \int_{Q/2+i\RR} \rmd\alpha_{\text{s}} \,
  C_{\alpha_1\alpha_2\alpha_{\text{s}}} C_{(Q-\alpha_{\text{s}})\alpha_3\alpha_4}
  \Bigl| q^{h_{\text{s}}-h_1-h_2} (1 + O(q))\Bigr|^2
\end{equation}
where the factor of $1/2$ cancels the double-counting from the identification $\alpha_{\text{s}}\sim Q-\alpha_{\text{s}}$, and $1+O(q)$ denotes an infinite series in positive integer power of~$q$, the normalized conformal block.

\begin{exercise}
  1. Check that the \ac{DOZZ} formula~\eqref{Liou-3pt} respects the expected $b\to 1/b$ duality, and the symmetries $\alpha_i\to Q-\alpha_i$ for any of the~$i$, up to the appropriate reflection coefficient.

  2. Using properties of $\Upsilon_b$ listed in~\autoref{sec:special}, show that at fixed generic $\alpha_1,\alpha_2$, the $\alpha_3\to 0$ limit of $C_{\alpha_1\alpha_2\alpha_3}$ vanishes.  Show that for $\alpha_1=\alpha_2$ the limit is infinite, while $(\alpha_3/2) C_{\alpha\alpha\alpha_3}\to g_{\alpha\alpha}=R(\alpha)$.
\end{exercise}

\subsection{\label{ssec:AGT-dict}Finding the \acsfont{AGT} dictionary}

We expect a relation of the form
\begin{equation}
  Z_{S^4_b}\bigl(\Theory(\lie{su}(2),C,m)\bigr)
  = \vev*{\Vhat_{\alpha_1}(z_1)\dots \Vhat_{\alpha_n}(z_n)}_{\fullC}
\end{equation}
for any number~$n$ of puncture, where $\Vhat_{\alpha_i}(z_i)$ are the reductions of codimension~$2$ operators of the 6d theory down to points.
In this section we use known $S^4_b$ partition function to determine that the relevant 2d \ac{CFT} is Liouville \ac{CFT} described above, and that $\Vhat_\alpha$ are suitable rescalings of vertex operators~$V_{\alpha}$.

\paragraph{Three-point functions and normalization.}

A 2d \ac{CFT} is characterized by its spectrum (left and right conformal dimensions of primary operators) and \ac{OPE} structure constants (equivalently, three-point functions of conformal primary operators).
When constructing class~S theories from $\Xsuii$, the data associated to a puncture is a mass parameter $m\in i\RR/\ZZ_2$.
We thus want local operators $V$ with a continuous parameter.
For consistency with earlier notation we denote this (dimensionless) parameter as $\alpha=Q/2+rm$, where $Q=b+1/b$.

Determining the conformal dimension of $\Vhat_\alpha$ will have to wait; let us begin with three-point functions.
We know that the theory associated to a three-punctured sphere is a trifundamental half-hypermultiplet.
Its partition function is a hypermultiplet one-loop determinant~\eqref{oneloop-hyper}, so that the three-point function is
\begin{equation}\label{three-point-from-4d}
  \begin{aligned}
    & \Chat_{\alpha_1\alpha_2\alpha_3} \coloneqq \vev{\Vhat_{\alpha_1}\Vhat_{\alpha_2}\Vhat_{\alpha_3}}
    = \prod_{\pm\pm} \frac{1}{\Upsilon_b\bigl(\alpha_1\pm(\alpha_2-Q/2)\pm(\alpha_3-Q/2)\bigr)}\\
    & = \frac{1}{\Upsilon_b(\alpha_2+\alpha_3-\alpha_1)\Upsilon_b(\alpha_3+\alpha_1-\alpha_2)
      \Upsilon_b(\alpha_1+\alpha_2-\alpha_3)\Upsilon_b(\alpha_1+\alpha_2+\alpha_3-Q)} ,
  \end{aligned}
\end{equation}
in which we used the invariance $\Upsilon_b(x)=\Upsilon_b(Q-x)$.
This matches precisely the denominator of the \ac{DOZZ} formula~\eqref{Liou-3pt}, and the numerator can be absorbed (except for an $\alpha$-independent factor) by the normalization
\begin{equation}
  \Vhat_\alpha = \frac{(b^{2/b-2b}\lambda)^{\alpha-Q/2}}{\Upsilon_b(2\alpha)} V_\alpha .
\end{equation}
With this normalization one can check that $\Vhat_\alpha=\Vhat_{Q-\alpha}$ and that the two-point function reads
\begin{equation}
  \widehat{g}_{\alpha\alpha'}
  = \vev{\Vhat_{\alpha}\Vhat_{\alpha'}}
  = \frac{\delta_{\alpha+\alpha'=Q} + \delta_{\alpha=\alpha'}}{\Upsilon_b(Q-2\alpha) \Upsilon_b(2\alpha-Q)} .
\end{equation}

\paragraph{Four-point functions and dimensions.}
To determine the conformal dimension of $\Vhat_{\alpha}$ we consider a four-punctured sphere and cut it in a channel suitable for the $q\to 0$ limit, where $q$ is the cross-ratio of the four punctures.
The gauge theory corresponding to a four-punctured sphere is $\lie{su}(2)$ $N_f=4$ \ac{SQCD}, and its partition function, computed using supersymmetric localization, takes the form~\eqref{ZS4b-final-form}
\begin{equation}
  Z_{S^4_b} = \int_{i\RR/\ZZ_2} \rmd a \, |q|^{r^2|a|^2} Z_{\oneloop}(a) Z_{\inst}(a,q) \overline{Z_{\inst}}(a,\qbar) .
\end{equation}
In the $q\to 0$ limit, $Z_{\inst}\to 1$.
This expression should be compared to the decomposition of a four-point function in 2d \ac{CFT},
\begin{equation}
  \begin{aligned}
    & \vev{\Vhat_{\alpha_1}(0)\Vhat_{\alpha_2}(q)\Vhat_{\alpha_3}(1)\Vhat_{\alpha_4}(\infty)}
    \\
    & \quad = \int_{Q/2+i\RR/\ZZ_2} \rmd\alpha\, q^{h(\alpha)} \qbar^{\overline{h}(\alpha)} \frac{\Chat_{\alpha_1\alpha_2\alpha} \Chat_{\alpha\alpha_3\alpha_4}}{\ghat_{\alpha\alpha}} \Fcal(\alpha_i,\alpha;q) \Fcal(\alpha_i,\alpha;\qbar)
  \end{aligned}
\end{equation}
in which $\Fcal$ are conformal blocks that depend (anti)\llap{-}holomorphically on the cross-ratio~$q$, and tend to~$1$ as $q\to 0$.

We have already identified the three-point functions $C$ to hypermultiplet one-loop determinants.
In turn, the inverse two-point function $\ghat_{\alpha\alpha}^{-1}$ is equal to the vector multiplet one-loop determinant.
It is thus natural to expect the conformal blocks to match instanton partition functions, and to identify the powers of~$q$, namely $h(\alpha)=\overline{h}(\alpha)=Q^2/4+P^2$ and $r^2|a|^2$, up to a harmless shift by $Q^2/4$.

\paragraph{Conformal blocks and proofs of \ac{AGT}.}

The key remaining piece to check the \ac{AGT} dictionary is to verify that conformal blocks do indeed match instanton partition functions,\footnote{Interestingly, the case of $\Nsusy=4$ \ac{SYM} caused some early confusion in the literature, clarified in~\cite{1004.1222}: this theory has $Z_{\inst}=1$, and it corresponds to a torus with a non-trivial vertex operator insertion $\alpha=Q/2$.}  as tested at low orders (in powers of~$q$) in~\cite{0906.3219,0907.2189,0908.2064,0908.2569,0912.2535,1001.1407,1007.0601,1111.1899,1312.5732}.
There have been many approaches to this (see for instance \cite[section 5.3]{1412.7145} for a short review).

One set of approaches relies on exhibiting an action of the Virasoro algebra (and many generalizations) on the instanton moduli space.
See~\cite{math/0401409,math/0409441} for an early example, and generalizations in \cite{1004.2814,1012.1312,1106.4088,1111.2803,1202.2756,1211.1287,1301.1977,1306.1523,1309.4775,1404.5304,1405.6992,1406.2381,1407.8341,1509.00075,1509.01000,1510.05482,1512.02492,1512.05388,1512.08016,1703.10990,1907.13005,1911.02963}.
In particular, one can construct~\cite{1012.1312,1107.4784,1109.4042,1110.1101,1111.2803} (see also \cite{0907.3946,0908.2190,1003.5752}) an orthonormal basis of conformal descendants of $\ket{\Vhat_\alpha}$ such that inserting these states in a four-point function as in~\eqref{four-point} yields term by term the expression of Nekrasov instanton partition functions as sums over $\Lie{U}(1)$-invariant point-like instanton configurations.
The Virasoro algebra and W-algebras also appear in a 6d context in~\cite{1112.0260,1205.6820,1403.6454,1404.1079}.
See also our discussion of more elaborate symmetry algebras on \autopageref{par:symmetries} in \autoref{sec:con}.

Recursion relations are studied in~\cite{1004.1841,1010.0528,1207.5658,1306.1523,1411.4222}.
One difficulty is for instance the presence of spurious poles in terms of the instanton expansion, which disappear when summing all contributions~\cite{1605.00077}.
The large~$c$ limit is investigated in \cite{0909.3531,1307.8174}.
A free-field approach based on Dotsenko--Fateev representations of \ac{CFT} correlators is given in~\cite{1003.5752,1007.4100,1011.5629,1012.3137,1102.0343,1105.0948,1307.2576,1512.06701}.
A string-theory derivation of the \ac{AGT} dictionary (from a 5d generalization) is given in~\cite{1309.1687,1403.3657} and reviewed in~\cite{1412.7132}.
A rather different approach is based on characterizing both conformal blocks and instanton partition functions as solutions to Riemann--Hilbert problems~\cite{1302.3778}.

\subsection{\label{ssec:AGT-CJ}Liouville from 6d}

We have argued that the relevant 2d theory for the \ac{AGT} correspondence is Liouville \ac{CFT}, and numerous checks of the \ac{AGT} correspondence validate this.
Could we see it directly from 6d?

\paragraph{The approach.}

Deformations of the metric of~$C$ that preserve its conformal class (or equivalently complex structure) are $Q$-exact with respect to the supercharge $Q$ that we used for supersymmetric localization~\cite{1605.03997}.
Thus, such deformations do not affect the partition function of the 6d theory, which can be computed in the limits where $C$ is infinitely smaller or larger than~$S^4_b$.
Importantly, this argument holds also in the presence of any $Q$-closed observables such as the loops, surfaces, or walls that we consider in \autoref{sec:ops}.
Remembering that the 6d theory is conformally invariant, these limits are equivalent to dimensionally reducing on either one of the factors.
We should thus expect to obtain Liouville \ac{CFT} (or its higher-rank generalization, Toda \ac{CFT} discussed further in \autoref{ssec:gen-Toda}) by dimensionally reducing the 6d theory~$\Xsuii$ (or~$\XsuN$) along~$S^4_b$.

In~\cite{1305.2891,1605.03997}, C\'ordova and Jafferis have performed this reduction in three steps: $\Xg$~reduced on~$S^1$ yields 5d $\Nsusy=2$ \ac{SYM}; $\Xg$~reduced on~$S^3_b$ (or quotients thereof\footnote{The reduction of~$\Xg$ on $S^1\times S^2$ also yields complex Chern--Simons, at a different level~\cite{1305.0291,1305.2429}.}) yields 3d complex Chern--Simons theory; $\Xg$~reduced on~$S^4_b$ yields 2d complex Toda \ac{CFT}.
They conjectured that this complexified version of Toda \ac{CFT} is dual to ordinary Toda \ac{CFT}.
The derivation was extended in~\cite{1708.07840} to include orbifold surface operators (see also~\cite{1811.03649} for another approach).

\paragraph{Reduction to complex Chern--Simons theory.}

The reduction to 3d is relevant for the 3d/3d analogue of the \ac{AGT} correspondence that we will describe in \autoref{ssec:dim-3d}.
We place the 6d theory on $S^3_b\times C_3$, where the squashed sphere is described for instance by its isometric embedding into~$\RR^4$ as $S^3_b=\{b^2(y_1^2+y_2^2)+b^{-2}(y_3^2+y_4^2)=r^2\}\subset\RR^4$.
Preserving supersymmetry requires a partial topological twist, which amounts to including suitable background values for supergravity fields, determined in~\cite{1305.2886}.
The approach in~\cite{1305.2891} was to work with a different squashing of the sphere~$S^3$ that preserves $\Lie{U}(1)\times\SU(2)$ isometries instead of $\Lie{U}(1)\times\Lie{U}(1)$.  We will gloss over this, as the backgrounds differ by suitably $Q$-exact terms that do not affect partition functions eventually.

The Hopf fibration of~$S^3$, namely an $S^1$ fibration over~$S^2$, is compatible with the squashing.
Thus, $\Xg$~can be reduced first on the $S^1$~fibers, obtaining 5d $\Nsusy=2$ \ac{SYM} theory.
Thankfully, the non-abelian 5d~theory has a Lagrangian description, hence can be further dimensionally reduced explicitly, in contrast to~$\Xg$, which has no known Lagrangian description.

A further reduction on the $S^2$~base of the Hopf fibration gives the following light fields, all valued in the adjoint representation of~$\lie{g}$.
\begin{itemize}
\item One 3d gauge field~$A$ arising from components of the 5d gauge field along~$C_3$.
  It has a 3d Chern--Simons term at level $k=1$, which arises because the 5d graviphoton has one unit of flux through~$S^2$ in the supergravity background.  This in turn stems from the Hopf fibration; when reducing on $S^2\times S^1$ instead, $k=0$.
\item Zero modes of the five vector multiplet scalars of 5d $\Nsusy=2$ \ac{SYM}.  Because the twist identifies an $\lie{so}(3)\subset\lie{so}(5)$ subgroup of R-symmetry with 3d rotations, these zero modes combine into a one-form~$X$ and a pair of scalars~$Y_i$.
\item Four fermions~$\lambda$ with a two-derivative Lagrangian $-\lambda(\nabla^A)^2\lambda+[X,\lambda]^2$.  In terms of $\Delta=(\nabla^A)^2+(\ad_X)^2$, the quadratic path integral over~$\lambda$ yields a factor of $(\det\Delta)^2$, while the pair of scalars~$Y_i$ yields $1/\det\Delta$ since their Lagrangian is~$-Y\Delta Y$.
\end{itemize}
Altogether, $Y$ and~$\lambda$ give a factor of $\det\Delta$, which matches the Faddeev--Popov determinant for gauge fixing $(\nabla^A)_\mu X^\mu=0$ the ``imaginary'' gauge transformation $(A_\mu,X_\mu)\mapsto (A_\mu-[X_\mu,g], X_\mu+\nabla^A_\mu g)$ for a local gauge parameter~$g$.

The final 3d theory has a pair of one-forms, hence a complex one-form $\cA=A+iX\in\lie{g}_{\CC}$ with action
\begin{equation}
  S = \frac{q}{8\pi} \int_{C_3} \Tr\biggl(\cA\wedge \rmd\cA + \frac{2}{3} \cA\wedge\cA\wedge\cA\biggr)
  + \frac{\tilde q}{8\pi} \int_{C_3} \Tr\biggl(\cAbar\wedge \rmd\cAbar + \frac{2}{3} \cAbar\wedge\cAbar\wedge\cAbar\biggr)
\end{equation}
subject to $G_{\CC}$~gauge invariance stemming from the standard and the imaginary gauge invariances.
The action for~$X$ arises from numerous supergravity fields necessary to ensure supersymmetry.
The coupling constants $q=k+is$ and $\tilde q=k-is$ encode the geometry as
\begin{equation}\label{qkis}
  k = 1 , \qquad s = \frac{1-b^2}{1+b^2} .
\end{equation}
More generally the theory makes sense for $k\in\ZZ$ and $s\in\RR\cup i\RR$.
The $\SU(2)\times\Lie{U}(1)$ preserving squashing of $S^3$ used in~\cite{1305.2891} has a parameter $\ell\in(0,+\infty)$ and $s=\sqrt{1-\ell^2}$ can also take imaginary values.
Other values of~$k$ arise from changing $S^3_b$ to $S^2\times S^1$ for $k=0$, or to the (squashed) Lens space $L(k,1)_b=S^3_b/\ZZ_k$.
The 3d/3d correspondence is discussed in \autoref{ssec:dim-3d}.

\paragraph{Reduction to complex Toda theory.}

The idea in~\cite{1605.03997} is to treat~$S^4_b$ as a squashed three-sphere~$S^3_b$ fibered over an interval.
In the notation of~\eqref{squashed-sphere} the interval is parametrized by $y_5\in[-r,r]$ and the~$S^3_b$ has squared radius $r^2-y_5^2$, namely it degenerates to a point at both ends.
The product metric~$g$ on $S^4_b\times C$ is mapped by a Weyl transformation to $S^3_b\times C_3$, where $C_3$ is a warped product of $C$~with an interval:
\begin{equation}
  \begin{aligned}
    g & = dy_5^2 + (r^2-y_5^2) \,g_{S^3_b} + g_C , & \qquad
    \frac{1}{r^2-y_5^2} \, g & = g_{S^3_b} + \frac{dy_5^2+g_C}{r^2-y_5^2} .
  \end{aligned}
\end{equation}
The resulting metric is singular at $y_5=\pm r$, which leads to boundary conditions for the theory on $C_3=[-r,r]\times_w C$.
The edge modes coming from each extremity are then understood to be described by chiral complex Toda theory.
Combining these two chiral theories gives complex Toda \ac{CFT} on~$C$.
This theory describes a complex boson~$\Phi$ in the complexification of the Cartan subalgebra $\lie{h}\subset\lie{g}$, with an exponential potential,
\begin{equation}
  S_{\lie{g}_{\CC}\text{ Toda}} = \frac{q}{8\pi} \int_C \biggl( \vev{\del\Phi,\delbar\Phi} + \sum_{j=1}^r e^{\vev{e_j,\Phi}}\biggr)\,\rmd^2z
  + \frac{\tilde q}{8\pi} \int_C \biggl( \vev{\del\Phibar,\delbar\Phibar} + \sum_{j=1}^r e^{\vev{e_j,\Phibar}}\biggr)\,\rmd^2z ,
\end{equation}
where $e_j$~are the simple roots of~$\lie{g}$, $r=\rank\lie{g}$, and the Killing form and the pairing of $\lie{h}^*$ and~$\lie{h}$ are both denoted $\vev{\ ,\ }$.
The coupling constants $q=k+is$, $\tilde q=k-is$ encode the geometry as in~\eqref{qkis}.

\paragraph{Relation to ordinary Toda theory.}

The conjecture is then that complex Toda \ac{CFT} is related to an earlier proposal of~\cite{1106.1172} (based on~\cite{1105.5800} in the $k=2$ case) for the \ac{AGT} correspondence on $S^4_b/\ZZ_k$.
For $\lie{g}=\lie{su}(N)$, the 2d \ac{CFT} proposed in~\cite{1106.1172} consists of two decoupled theories: an $\hat{\lie{su}}(k)_N/\lie{u}(1)^{k-1}$ coset, and real parafermionic Toda \ac{CFT} with parameters $N$, $k$, and $b=\sqrt{\tilde{q}/q}$ (coinciding with the squashing parameter).
The latter theory describes parafermions and real bosons $\psi,\varphi\in\lie{h}$, where the parafermions are described by another coset model $\hat{\lie{su}}(N)_k/\hat{\lie{u}}(1)^{N-1}$ and are coupled through dimension $1-1/k$ operators $\psi_j\bar\psi_j$ to the real bosons,
\begin{equation}
  S_{\text{para-Toda}} = S\biggl(\frac{\hat{\lie{su}}(N)_k}{\hat{\lie{u}}(1)^{N-1}}\biggr) + \int_C \biggl( \vev{\del\varphi,\delbar\varphi} + \sum_{j=1}^{N-1} \psi_j\bar\psi_j e^{(b/\sqrt{k})\vev{e_j,\varphi}}\biggr)\,\rmd^2z .
\end{equation}
For $k=1$ both the decoupled coset and the parafermions trivialize and we are left with ordinary Toda \ac{CFT} with coupling~$b$, as stated by the standard \ac{AGT} correspondence.

The conjectured duality between complex Toda \ac{CFT} and coset plus para-Toda \ac{CFT} has only been checked explicitly for the simplest case of $\lie{g}=\lie{su}(2)$ with $k=1$ in~\cite{1409.0857}, and in~\cite{1111.2803,1210.1856} for the case $N=k=2$ where it essentially boils down to bosonization.
See \autopageref{par:local-orbifold} for a further discussion of the orbifold case.

\part{Extensions of \acsfont{AGT}}

\section{\label{sec:gen}General class~S theories}

In this section we enter the realm of non-Lagrangian theories: while all class~S theories arising from $\Xsuii$ with tame punctures can be realized by coupling vector and hypermultiplets, we now extend the story in two ways.

The 6d $(2,0)$ theory~$\Xg$ is labeled by an arbitrary simply-laced simple Lie algebra~$\lie{g}$,
so it is no wonder that the \ac{AGT} correspondence~\cite{0906.3219} extends beyond $\lie{su}(2)$ to $\lie{su}(N)$~\cite{0907.2189} and
general gauge algebras~\cite{1012.4468,1107.0973,1111.5624,1404.3737}, with Liouville \ac{CFT} generalizing to the Toda \ac{CFT}.
After reviewing this \ac{CFT} in \autoref{ssec:gen-Toda} with an eye towards its connections to gauge
theory, we describe an example of \ac{AGT} correspondence and important considerations
\apartehere
about punctures in \autoref{ssec:gen-rank}.

Second, in \autoref{ssec:gen-irreg}, we consider interesting limits where two punctures collide
while the parameters describing the defects are appropriately scaled.  The resulting wild
punctures allow to realize asymptotically\hskip 0pt\relax\nobreak-\hskip 0pt\relax free gauge theories (such as $\SU(2)$ \ac{SQCD} with
$N_f<4$), and \ac{AD} theories~\cite{hep-th/9505062} as part of class~S\@.

\startaparte[title={}]
\begin{itemize}
\item Most higher-rank class~S theories are non-Lagrangian.
\item Partial Higgsing gives a hierarchy of tame punctures.  Some are described by quiver tails of $\SU$ groups.
\item Collisions of tame punctures give wild punctures.  This often results in \ac{AD} theories.
\item Tame punctures map to Toda semi-degenerate primaries.
\item Wild punctures map to Toda \ac{CFT} irregular operators.
\end{itemize}
\stopaparte

\subsection{\label{ssec:gen-Toda}Toda \acsfont{CFT}}

\paragraph{Lagrangian and symmetries.}

We cannot do justice to the fifty year history of Toda theory, starting from the Toda lattice \cite{toda1967vibration,toda1967wave} in 1967, which consists of particles with nearest-neighbor exponential interactions (see \cite{Olshanetsky:1981dk} for an early review).
Its \ac{QFT} version was defined in 1982 in~\cite{Mansfield:1982sq} and found to be conformal.
Given a simply-laced\footnote{Toda \ac{CFT} is defined for an arbitrary simple Lie algebra, but we only present the simply-laced case because only that case is relevant for the \ac{AGT} correspondence.  Correlators in non-simply-laced Toda \ac{CFT} presumably correspond to correlators of suitable outer automorphism twist operators in the gauge theory on~$S^4_b$.} Lie algebra~$\lie{g}$ and its Cartan Lie algebra~$\lie{h}$, Toda \ac{CFT} describes a real scalar field $\phi\in\lie{h}$ subject to the Lagrangian density
\begin{equation}\label{action}
  \cL = \frac{1}{8\pi} \bigl( \hat{g}^{ab} \vev{\del_a\phi,\del_b\phi} + 2 \vev{Q,\varphi} \hat{R} \bigr) + \mu \sum_{i=1}^{\rank\lie{g}} e^{b\vev{e_i,\phi}} .
\end{equation}
Here, $e_i$~are the simple roots of~$\lie{g}$ and $\vev{\ ,\ }$ denotes both the pairing of $\lie{h}^*$ and~$\lie{h}$ and the Killing form.
The background charge vector~$Q$ that multiplies the scalar curvature~$\hat{R}$ of the background metric~$\hat{g}$ is set to $Q = (b + \frac{1}{b}) \rho$ where $\rho$ is the Weyl vector, namely the half-sum of positive roots, equivalently the sum of all fundamental weights~$\varpi_j$.
Vertex operators $V_\alpha=e^{\vev{\alpha,\phi}}$ (we suppress normal ordering in this notation) are labeled by $\alpha\in\lie{h}^*_{\CC}$ and have holomorphic conformal dimension\footnote{The standard conventions for $Q$~in Liouville and Toda \ac{CFT} differ by a factor of~$2$, so that typical Liouville momenta take the form $\alpha=Q/2+iP$ while typical Toda momenta are $\alpha=Q+iP$.}
\begin{equation}\label{h-alpha}
  h(\alpha) = \frac{1}{2} \vev{\alpha,2Q-\alpha} .
\end{equation}
In particular $h(be_i)=1$, which ensures that the exponential potential terms are exactly marginal.
Their coupling~$\mu$ is redundant and amounts to shifting~$\phi$ by a multiple of~$\rho$.
The coupling $b>0$ however plays an essential role: for instance the central charge $c = \rank\lie{g} + 12\vev{Q,Q}$ depends on it.
For simply-laced~$\lie{g}$ the theory is (expected to be) dual under $b\mapsto 1/b$, while keeping $\lambda=[\pi\Gamma(b^2)\mu/\Gamma(1-b^2)]^{1/b}$ fixed.
This is quite satisfactory for the \ac{AGT} correspondence since $S^4_b$ and $S^4_{1/b}$ are isometric.

Beyond the infinite-dimensional Virasoro symmetries of 2d \ac{CFT}, Toda \ac{CFT} has (anti)\llap{-}holomorphic $W_{\lie{g}}$~symmetries.
This chiral algebra was uncovered in~\cite{Zamolodchikov:1985wn} for $\lie{g}=\lie{su}(3)$, and more broadly in \cite{Fateev:1987vh,Fateev:1987zh,Lukyanov:1987xg} as a symmetry of minimal models.  See \cite{Lukyanov-1990tf} for an early review.\footnote{The contemporary \cite{Fateev:1985mm} by the same authors is supposedly relevant, but not available to me.}
It can be realized by quantum Drinfeld--Sokolov reduction of an affine Lie algebra \cite{Bershadsky:1989mf,Feigin:1990pn,hep-th/9210010,hep-th/9302006}.  (See \cite{hep-th/9204093,hep-th/9309095} for applications to W-strings.)
In the $\lie{g}=\lie{su}(N)$ case it can be realized as a truncation of a more general chiral algebra~$W_\infty$ generated by infinitely many conserved currents, as reviewed in~\cite{1411.7697}.
As a chiral algebra, $W_{\lie{g}}$~is generated by $\rank\lie{g}$ conserved currents $W^{(k)}(z)$ whose spins~$k$ are the degrees of Casimir invariants of~$\lie{g}$.  The quadratic Casimir invariant yields the holomorphic stress-tensor $T(z)=W^{(2)}(z)$ which generates the Virasoro subalgebra of~$W_{\lie{g}}$.

\paragraph{Primary operators and normalization.}

The vertex operators $V_\alpha=e^{\vev{\alpha,\phi}}$ have definite quantum numbers $w^{(k)}(\alpha)$ under zero-modes of all~$W^{(k)}$, namely the \ac{OPE} starts as
\begin{equation}\label{WkValphaOPE}
  W^{(k)}(z) V_\alpha(0) = \frac{w^{(k)}(\alpha)}{z^k} V_\alpha(0) + \dotsm
\end{equation}
with for instance $w^{(2)}(\alpha)=h(\alpha)$ given in~\eqref{h-alpha}.
The conserved current $W^{(k)}(z)$ translates on the gauge theory side to the degree~$k$ differential~$\phi_k$ that shows up in the construction~\eqref{intro-SW} of the \ac{SW} curve.
The momenta $\Im\alpha$ are $r$~times the (diagonalized) mass parameter $m\in\lie{g}$ at a given tame puncture.
In the classical limit $r\to\infty$ the quantum numbers $w^{(k)}(\alpha)$ simplify to Casimir invariants of~$\lie{g}$ and the \ac{OPE}~\eqref{WkValphaOPE} becomes the singularities of~$\phi_k$ near tame punctures~\cite{0909.4031}.

The quantum numbers $w^{(k)}(\alpha)$ are invariant under the Weyl group action $\alpha\mapsto Q+w(\alpha-Q)$ for any Weyl group element $w\colon\lie{h}^*\to\lie{h}^*$.
Thus, $V_\alpha$ and $V_{Q+w(\alpha-Q)}$ have the same quantum numbers; in Toda \ac{CFT} they are the same operator up to a normalization called reflection amplitude and determined in \cite{hep-th/9907072,hep-th/0002213,hep-th/0103014}.
For generic~$\alpha$, the expressions can be recast as the statement that~\cite{0907.2189}\footnote{The reflection amplitude in \cite{hep-th/9907072,hep-th/0002213,hep-th/0103014} and in later work \cite{0709.3806} (by one of the same authors) seems to differ by a sign.  I take the latter sign to be correct as it seems to agree with Liouville \ac{CFT}.}
\begin{equation}\label{Vhat-def}
  \Vhat_\alpha = \frac{\lambda^{\vev{\rho,\alpha-Q}}}{\prod_{e>0}\Upsilon_b(\vev{Q-\alpha,e})} V_\alpha(z)
\end{equation}
is invariant under Weyl reflections of~$\alpha$, where the product ranges over all positive roots~$e$ and we recall $\lambda=[\pi\Gamma(b^2)\mu/\Gamma(1-b^2)]^{1/b}$.
While often convenient, the normalization~\eqref{Vhat-def} does not make sense for values of~$\alpha$ where the denominator blows up.

The operator spectrum of Toda \ac{CFT} consists of vertex operators $V_{Q+a}$~with $a\in\lie{h}$ (purely imaginary in our conventions), modulo the Weyl group.
Each of them is the highest-weight of a Verma module of the $W_{\lie{g}}$~algebra, with no null states.
As for the Virasoro algebra, there are some values of momenta (away from this line) for which the vertex operators have null descendants.
The precise condition\footnote{\label{foot:Toda-degeneration}This is obtained using screening charges, see e.g.~\cite{0911.4787}.  Reading modern references, I think that the null descendants start at level $n_1n_2$, but I have not found the suitable references in old Toda \ac{CFT} literature.  Help welcome.} is that $V_\alpha$ has null descendants if $\vev{\alpha-Q,e}=-n_1b-n_2/b$ for any root~$e$ and positive integers $n_1,n_2>0$.

\paragraph{Correlators in Toda \ac{CFT}.}

The $W_{\lie{g}}$ symmetry severely constrains two and three-point functions in Toda \acs{CFT}.
A two-point function of primary operators $V_{\alpha_1}$ and $V_{\alpha_2}$ can only be non-vanishing if their quantum numbers obey $w^{(k)}(\alpha_1)=(-1)^kw^{(k)}(\alpha_2)$, hence $\alpha_1=2Q-\alpha_2$ modulo the Weyl group.
Taking into account our preferred normalization~\eqref{Vhat-def} one has
\begin{equation}\label{Toda-2pt}
  \vev{\Vhat_\alpha(z,\bar{z})\Vhat_{\alpha'}(0)}
  = |z|^{-4h(\alpha)} \frac{\sum_{w\in\text{Weyl}}\delta_{Q-\alpha'=w(\alpha-Q)}}{\prod_{\text{roots }e}\Upsilon_b(\vev{Q-\alpha,e})} ,
\end{equation}
where the sum of delta functions simply ensures Weyl invariance and working with the unnormalized~$V_\alpha$ (which is necessary to treat partially degenerate momenta) would simply introduce some reflection amplitudes in this sum.
The Shapovalov matrix of two-point functions of $W_{\lie{g}}$-descendants follows in the standard way by commuting the W-algebra modes~$W^{(k)}_n$.
Our normalization choice is pleasant because, as in the $\lie{su}(2)$ case, the inverse two-point function $\prod_e\Upsilon_b(\vev{Q-\alpha,e})$ matches the one-loop determinant of a vector multiplet for the gauge algebra~$\lie{g}$.

The three-point functions of primaries are encoded in coefficients $\Chat_{123}=\Chat(\alpha_1,\alpha_2,\alpha_3)$
\begin{equation}
  \vev{\Vhat_{\alpha_1}(z_1,\bar{z}_1)\Vhat_{\alpha_2}(z_2,\bar{z}_2)\Vhat_{\alpha_3}(z_3,\bar{z}_3)}
  = \frac{\Chat(\alpha_1,\alpha_2,\alpha_3)} 
  {|z_1-z_2|^{2h_{12}}|z_1-z_3|^{2h_{13}}|z_2-z_3|^{2h_{23}}},
\end{equation}
where $h_{ij}=h(\alpha_i)+h(\alpha_j)-h(\alpha_k)$.
The general three-point function is not known, and in addition three-point functions of most $W_{\lie{g}}$-descendants cannot be expressed in terms of $\Chat_{123}$, unlike the standard case of Virasoro descendants.
Only certain special cases~\cite{hep-th/0505120,0709.3806,0810.3020} discussed below have been determined.

\paragraph{Under-determined conformal blocks.}

This has a knock-on effect on higher-point Toda \acs{CFT} correlators, as the conformal blocks describing how $W_{\lie{g}}$~descendants contribute are not fixed by symmetry.  Consider for instance
\begin{equation}\label{Vhat-fourpoint-split}
  \vev{\Vhat_{\alpha_1}\Vhat_{\alpha_2}\Vhat_{\alpha_3}\Vhat_{\alpha_4}}
  = \int_{a\in\lie{h}/\text{Weyl}} \rmd a\, \sum_Y \vev{\Vhat_{\alpha_1}\Vhat_{\alpha_2}\Vhat^{[Y]}_{Q-a}} \frac{1}{\vev{\Vhat_{Q-a}\Vhat_{Q+a}}} \vev{\Vhat^{[Y]}_{Q+a}\Vhat_{\alpha_3}\Vhat_{\alpha_4}} ,
\end{equation}
where we suppressed the spatial dependence,
the integral ranges over primary operators~$\Vhat_\alpha$ in the spectrum,
and the sum ranges over their descendants, orthogonalized and normalized to have the same norm $\vev{\Vhat_{Q-a}\Vhat_{Q+a}}$ as primaries.
For generic~$\alpha_i$, the only three-point functions $\vev{\Vhat_{\alpha_1}\Vhat_{\alpha_2}\Vhat^{[Y]}_{Q-a}}$ that are determined by primary three-point functions are those where $\Vhat^{[Y]}_{Q-a}$ is in fact a Virasoro descendant of~$\Vhat_{Q-a}$.

The class~S theory coming from a three-punctured sphere with full tame punctures is the non-Lagrangian tinkertoy~$T_{\lie{g}}$, and supersymmetric localization has nothing to say on its sphere partition function.
A very powerful roundabout way is to consider the 5d lift and work out the limit of $S_b^4\times S^1$ partition function when the circle radius~$\beta$ shrinks \cite{1310.3841,1409.6313,1412.3395,1506.04183}.
In principle this provides conjectural expressions for $\Chat_{123}$ and all descendant three-point functions~\cite{1712.10225}, but it is not clear that the $\beta\to 0$ limits exist, and it is not clear how to relate parameters of the topological vertex formalism to bases of descendants, as explained in detail in~\cite{1906.06351}.
Higher-point correlators correspond to theories obtained by gauging together copies of~$T_{\lie{g}}$.
Since the gauge group is simply one factor~$G$ per tube, the instanton moduli space is known, and the instanton partition function is some integral over this space.  The integrand, however, depends on the matter theories~$T_{\lie{g}}$, whose reaction to instantons is not known.
This is exactly analogous to how the sum over descendants is known but the requisite three-point functions do not derive from~$\Chat_{123}$.

\subsection{\label{ssec:gen-rank}Higher rank \acsfont{AGT} correspondence}

The main building block of class~S theories is the tinkertoy~$T_{\lie{g}}$ for three full tame punctures, reviewed in~\cite{1504.01481}.  This tinkertoy is non-Lagrangian\footnote{Interestingly, its 3d $\Nsusy=4$ dimensional reduction is mirror to a Lagrangian theory described by a star-shaped quiver~\cite{1007.0992}.} for $\lie{g}\neq\lie{su}(2)$.
As a result, another type of tame punctures (called simple punctures) is needed for the simplest examples of higher-rank \ac{AGT} correspondence, such as the matching of an $\SU(N)$ \ac{SQCD} partition function with an $\lie{su}(N)$ Toda \ac{CFT} four-point function.

Besides full and simple tame punctures, there are other tame punctures (and of course a host of wild punctures) studied in \cite{0904.2715,0905.4074,0911.4787,1006.3486,1007.0601,1012.1352,1404.3737}
A large program to classify tinkertoys has been carried out by Chacaltana and Distler and collaborators in~\cite{1008.5203,1106.5410,1203.2930,1212.3952,1309.2299,1403.4604,1412.8129,1501.00357,1601.02077,1711.04727} (a warning though, their use of ``irregular'' is non-standard in this context).
The punctures that can arise in a limit where one of the tubes in the Riemann surface becomes pinched were studied in~\cite{1110.2657,1702.00939}.

The various tame punctures correspond to Toda \ac{CFT} vertex operators~$V_\alpha$ that are partially degenerate, as we explain for $\lie{g}=\lie{su}(N)$.
We also describe how punctures can be ``partially closed'' by tuning their parameters, which on the gauge theory side corresponds to a partial Higgsing.
Finally, we outline how to include non-simply-laced gauge groups.

\paragraph{Wyllard relation: \(\SU(N)\) linear quiver and a \(\lie{su}(N)\) Toda \ac{CFT} correlator.}

For simplicity we now focus on the $\lie{g}=\lie{su}(N)$ case, which is understood best.
The chiral algebra is denoted variously $W_{\lie{su}(N)}=W\!A_{N-1}=W_N$.

In \autoref{ssec:Lag-linear} we considered a linear quiver gauge theory whose hypermultiplets transform in bifundamental representations of $n-1$ successive $\SU(N)$ groups, of which the middle $n-3$ are gauged.  Besides the two $\SU(N)$ flavour symmetries at the ends of the quiver, each of the $n-2$ hypermultiplets has a $\Lie{U}(1)$ flavour symmetry.
As stated in~\eqref{SUN-linear}, this theory is realized by reducing $\XsuN$ on a sphere with $n$ punctures corresponding to these $n$ flavour symmetry factors.
The $\SU(N)$ flavour symmetries correspond to full tame punctures, at which each differential~$\phi_k$ has a pole of order $k-1$ in the massless case, or $k$~when $\SU(N)$ masses are turned on.
The $\Lie{U}(1)$ flavour symmetries correspond to simple tame punctures where each~$\phi_k$ has a simple pole in the massless case (the massive case is more complicated).

The \ac{AGT} correspondence proposed in~\cite{0907.2189} takes the form
\begin{equation}\label{AGTW-quiver}
  Z_{S^4_b}\left(
    \mathtikz[every node/.style={font=\scriptsize}]{
      \node(A)[flavor-group] at (-1,-.5) {$\SU(N)$};
      \node(B)[flavor-group] at (-1,.5) {$\Lie{U}(1)$};
      \draw(A)--(-.7,0)--(B);
      \node(C)[color-group] at (0,0) {$\SU(N)$};
      \draw(-.7,0)--(C);
      \node(D)[flavor-group] at (.7,.6) {$\Lie{U}(1)$};
      \draw(.7,0)--(D);
      \node(Cdots) at (1.2,0) {${\cdots}$};
      \draw(C)--(Cdots);
      \node(G)[flavor-group] at (1.7,.6) {$\Lie{U}(1)$};
      \draw(1.7,0)--(G);
      \node(H)[color-group] at (2.4,0) {$\SU(N)$};
      \draw(Cdots)--(H);
      \draw(H)--(3.1,0);
      \node(I)[flavor-group] at (3.4,.5) {$\Lie{U}(1)$};
      \node(J)[flavor-group] at (3.4,-.5) {$\SU(N)$};
      \draw(I)--(3.1,0)--(J);
    }
  \right)
  = \vev*{\Vhat_{\alpha_1}\Vhat_{\mu_2}\Vhat_{\mu_3}\cdots\Vhat_{\mu_{n-1}}\Vhat_{\alpha_n}}^{\Toda(\lie{su}(N))}
\end{equation}
where $\alpha_j$ ($j=1,n$) encode the imaginary $\SU(N)$ mass parameters of the two full punctures, $m_j=(m_{j1},\dots,m_{jN})$ with $\sum_p m_{jp}=0$,
while $\mu_j$ ($j=2,\ldots,n-1$) encodes the $j$-th $\Lie{U}(1)$ mass parameter $m_j\in i\RR$ as
\begin{equation}\label{Wyllard-momentum-dictionary}
  \begin{aligned}
    \alpha_j & = Q + r m_j ,  & & j = 1, n , \\
    \mu_j & = \biggl(\frac{N}{2}\Bigl(b+\frac{1}{b}\Bigr) + r m_j\biggr) \varpi_1 , & & j=2,\cdots n-1 .
  \end{aligned}
\end{equation}
For instance in the case $n=4$ this identifies a Toda \ac{CFT} four-point function to the partition function of $\SU(N)$ \ac{SQCD} with $N_f=2N$ flavours.

To understand these momenta, we discuss the corresponding vertex operators~$V_\alpha$ and their $W_N$~descendants.
We have already encountered momenta in $Q+\lie{h}$ which describe normalizable states that occur in the spectrum of Toda \ac{CFT}.
Momenta proportional to the first weight~$\varpi_1$ are called semi-degenerate momenta: as uncovered starting in \cite{hep-th/9206075,hep-th/9309146} they have null $W_N$~descendants at level~$1$.
Consider a three-point function $\vev{V_{\alpha}V_{\varkappa\varpi_1}V_{\alpha'}}$ with a semi-degenerate vertex operator.
Thanks to null vectors, the action of arbitrary $W_N$~generators can be converted to Virasoro generators, hence to differential operators acting on the known coordinate-dependence.
This means that three-point functions of descendants of $V_{\alpha}$, $V_{\varkappa\varpi_1}$, and~$V_{\alpha'}$ are uniquely fixed as a multiple of the three-point function of primaries.

\paragraph{Checking the Wyllard relation.}

As in the $\lie{su}(2)$ case, the matching~\eqref{AGTW-quiver} is most directly checked in the S-duality frame corresponding to the s-channel decomposition of the sphere correlator.
Each three-punctured sphere piece has one simple puncture and two full punctures (and corresponds on the gauge theory side to a bifundamental hypermultiplet).
Expanding the correlator in this channel, rewriting descendant three-point functions in terms of the primaries, and collecting the descendant's contributions into a conformal block, we have
\begin{equation}\label{Toda-correlator-split}
  \begin{aligned}
    & \vev{\Vhat_{\alpha_1}\Vhat_{\mu_2}\cdots\Vhat_{\mu_{n-1}}\Vhat_{\alpha_n}} \\
    & \quad = \int_{\substack{\ a_j\in\lie{h}/\text{Weyl}\\2\leq j\leq n-2}} \rmd a\,
    \frac{\vev{\Vhat_{\alpha_1}\Vhat_{\mu_2}\Vhat_{Q+a_2}} \vev{\Vhat_{Q-a_2}\Vhat_{\mu_3}\Vhat_{Q+a_3}} \cdots \vev{\Vhat_{Q-a_{n-2}}\Vhat_{\mu_{n-1}}\Vhat_{\alpha_n}}} {\vev{\Vhat_{Q+a_2}\Vhat_{Q-a_2}}\cdots\vev{\Vhat_{Q+a_{n-2}}\Vhat_{Q-a_{n-2}}}}
    \Fcal(z) \Fcal(\zbar) ,
  \end{aligned}
\end{equation}
for some conformal blocks $\Fcal(z)$ that are (in principle) calculable as series in powers of complex structure parameters of~$C$, and that depend on all external and internal momenta.
The very fact that conformal blocks are calculable (for these momenta) matches nicely with the fact that the 4d theory is Lagrangian hence its partition function is calculable by supersymmetric localization.

One key part of the matching is that the $n-2$ three-point functions in~\eqref{Toda-correlator-split} should match with the one-loop determinants of the $n-2$ bifundamental hypermultiplets in the quiver.
Thankfully, the three-point functions $\vev{V_{\alpha_1}V_{\varkappa\varpi_1}V_{\alpha}}$ of two non-degenerate and one semi-degenerate vertex operators were worked out in \cite{hep-th/0505120,0709.3806,0810.3020} by inserting a fully degenerate vertex operator $V_{-b\varpi_1}$ into the correlator and solving a differential equation that results.\footnote{This bootstrap technique was introduced by Teschner to solve Liouville \ac{CFT}.  See also \cite{1407.1852,1602.03870,1610.07993,1711.04361} for further explorations in the Toda \ac{CFT} context with more general fully degenerate vertex operators.  See \cite{1504.07556,1606.02535} for other related correlators.}
Consider the normalization~\eqref{Vhat-def} of non-degenerate operators and an ad-hoc normalization
\begin{equation}\label{Vhat-def-semi}
  \Vhat_{\varkappa\varpi_1} = \frac{\lambda^{\vev{\varkappa\varpi_1,\rho}}}{(\Upsilon_b(b))^{N-1}\Upsilon_b(\varkappa)} V_{\varkappa\varpi_1}
\end{equation}
which is Weyl-invariant but is an abuse of notation since $\Vhat$ does not relate to~$V$ in the same way as in~\eqref{Vhat-def}.
Then the three-point function is~\cite{0709.3806}
\begin{equation}
  \Chat(Q+a_1,\varkappa\varpi_1,Q+a_2)
  = \frac{1}{\prod_{i,j=1}^N \Upsilon_b(\varkappa/N - \vev{a_1,h_i} - \vev{a_2,h_j})}
\end{equation}
where $h_i$ are the weights of the fundamental representation of~$\lie{su}(N)$.
The product of Upsilon functions correctly coincides with the one-loop determinants of $N^2$~hypermultiplets on~$S^4_b$ with suitable masses.
The inverse two-point functions match with vector multiplet one-loop determinants, see below~\eqref{Toda-2pt}.
The classical action also reproduces correctly the leading $z$-dependence of the Toda \ac{CFT} correlator.
Finally, one can tediously match conformal blocks with instanton partition functions order by order, to confirm~\eqref{AGTW-quiver}.

The Toda correlator in~\eqref{AGTW-quiver} can be decomposed in principle in many other channels.
For instance taking the \ac{OPE} of the semi-degenerate vertex operators $\Vhat_{\mu_1}$ and $\Vhat_{\mu_2}$ in the four-point function ($n=4$) yields a t-channel decomposition.
In the $N=3$ case, the internal momenta produced by this fusion are non-degenerate, so that the decomposition involves general three-point functions.
The corresponding S-duality frame of $\SU(3)$ \ac{SQCD} consists of the $T_{\lie{su}(3)}$~tinkertoy (the $\lie{e}_6$ Minahan--Nemeschansky \ac{SCFT}) coupled to some hypermultiplets by a gauge group that turns out to be $\SU(2)$.
For general~$N$, such fusions lead to numerous types of punctures intermediate between the full and the simple puncture.

\paragraph{Partial Higgsing.}

Consider the linear quiver in~\eqref{AGTW-quiver}.
Its Higgs branch consists of supersymmetric vacua where hypermultiplet scalars get a \ac{VEV}.
Upon moving to any given point on the Higgs branch, the hypermultiplet \ac{VEV} may break gauge symmetry to a smaller group, thus reducing the quiver to a smaller one.

We denote scalars in the bifundamental hypermultiplets as $(Q_j,\tilde{Q}_j)$ for $j=2,\dots,n-1$.
An important class of vacua are obtained by imposing a nilpotent \ac{VEV} to the moment map (see e.g.\ the appendix of~\cite{1409.1908})
\begin{equation}
  \mu_1 = \tilde{Q}_2 Q_2 - \frac{1}{N} \Tr(\tilde{Q}_2Q_2)
\end{equation}
of the leftmost $\SU(N)$ flavour symmetry group.
Nilpotent matrices in $\lie{su}(N)_{\CC}=\lie{sl}(N,\CC)$ are classified up to conjugation by a partition of~$N$, or equivalently a Young diagram~$Y$ with $N$~boxes.
Denoting by $n_k$ the number of columns of length~$k$ in~$Y$, the nilpotent \ac{VEV} we consider takes the block-diagonal form
\begin{equation}\label{vevmu1}
  \vev{\mu_1} = J_1^{\oplus n_1}\oplus J_2^{\oplus n_2}\oplus\cdots\oplus J_{\ell}^{\oplus n_{\ell}}
\end{equation}
with $n_k$ Jordan blocks $J_k$ of size $k\times k$ (for instance $J_1=(0)$).
The \ac{VEV} $\vev{\mu_1}$ cannot be imposed in isolation, as the $F$-term relations lead to non-vanishing values for the hypermultiplets $(Q_j,\widetilde{Q}_j)$ for $j=2,\cdots,\ell$.

The Higgs mechanism thus breaks multiple gauge symmetries.
Specifically, it reduces the first few groups of the quiver to a \emph{quiver tail}
\begin{equation}\label{quiver-tail}
  \quiver{
    \node(cdots) at (0,0) {${\cdots}$};
    \node(Nl) [color-group] at (1,0) {$N_{\ell}$};
    \node(Nl1) [color-group] at (2.5,0) {$N_{\ell-1}$};
    \node(cdots2) at (3.5,0) {${\cdots}$};
    \node(N1) [color-group] at (4.5,0) {$N_1$};
    \node(Ml) [flavor-group] at (1,1) {$n_{\ell}$};
    \node(Ml1) [flavor-group] at (2.5,1) {$n_{\ell-1}$};
    \node(M1) [flavor-group] at (4.5,1) {$n_1$};
    \draw(Nl)--(Ml);
    \draw(Nl1)--(Ml1);
    \draw(N1)--(M1);
    \draw(Nl)--(Nl1)--(cdots2)--(N1);
  }
\end{equation}
where round nodes are $\SU(N_j)$ gauge groups, rectangles count fundamental hypermultiplets, and $N_j$~is determined by $N_0=0$ and $N_j-N_{j-1}=n_j+n_{j+1}+\cdots+n_{\ell}$ (which is the length of the $j$-th row).  In other words, $N_j$~counts all boxes in rows $1,\cdots,j$ and in particular $N_{\ell}=N$.
The flavour symmetry is
\begin{equation}\label{puncture-flavour-symm}
  \biggl[\prod_{k=1}^{\ell} U(n_k)\biggr] / U(1)_{\text{diag}} .
\end{equation}
The puncture has $n_1+\cdots+n_{\ell}-1=N_1-1$ mass parameters.
The nilpotent matrix~\eqref{vevmu1} can also be understood as the image of the raising operator under an embedding $\rho\colon\SU(2)\to\SU(N)$.  In that equivalent description, the flavour symmetry~\eqref{puncture-flavour-symm} arises as the commutant of~$\rho$.

\paragraph{Quiver tails and punctures.}

Starting from a quiver tail~\eqref{quiver-tail} and its \ac{SW} solution obtained from M-theory~\cite{hep-th/9703166}, Gaiotto understood in~\cite{0904.2715} the relevant patterns of pole orders for the differentials~$\phi_k$.
This helped determine that tame punctures are labeled by partitions of~$N$.
Linear quivers with arbitrary quiver tails are realized in class~S as the reduction of~$\XsuN$ on a sphere with arbitrarily many simple tame punctures and with two (general) tame punctures.

In fact, the partial Higgsing procedure we described replaces the full tame puncture (that we started with) by precisely the tame puncture labeled by~$Y$.
Just as full tame punctures carry $\SU(N)$ flavour symmetry, the puncture carries flavour symmetry~\eqref{puncture-flavour-symm}.
The dictionary between quiver tails, punctures, and the order of poles, is nicely written in~\cite{0905.4074}.

Punctures can be closed entirely.  This is most easily seen for the simple punctures in~\eqref{AGTW-quiver}: on the gauge theory side, two neighboring $\SU(N)$ groups are reduced to their diagonal subgroup, and we are left with a shorter linear quiver.

\paragraph{Partially degenerate vertex operators.}

Partial Higgsing translates on the Toda \ac{CFT} side to changing a non-degenerate vertex operator to a partially degenerate one.

According to \autoref{foot:Toda-degeneration}, level~$1$ null $W_N$~descendants of a vertex operator~$V_\alpha$ are characterized by roots~$e$ for which $\vev{\alpha-Q,e}=-b-1/b$.
Up to a Weyl group transformation of $\alpha-Q$, we choose these roots be simple roots~$e_j$ only.
The condition then reduces to $\vev{\alpha,e_j}=0$, namely $\vev{\alpha,h_j}=\vev{\alpha,h_{j+1}}$ in terms of the weights~$h_i$ of the defining representation of~$\lie{su}(N)$.
The components $\vev{\alpha,h_i}$ of~$\alpha$ organize as
\begin{equation}\label{alpha-partial-degenerate}
  \alpha = \bigl(\underbrace{\alpha_{(1)},\cdots,\alpha_{(1)}}_{l_1},\cdots,\underbrace{\alpha_{(r)},\cdots,\alpha_{(r)}}_{l_r}\bigr)
\end{equation}
where $l_k$, $1\leq k\leq r$ denote the number of equal components, and we can reorder the components such that $l_1\geq l_2\geq\cdots\geq l_r\geq 0$ defines lengths of columns of some Young diagram~$Y$.
This is the same Young diagram classification as for the punctures.
For instance the momentum has $r-1$ parameters (because of tracelessness), which is the length of the first row of~$Y$ (minus one) thus matches the counting in~\eqref{puncture-flavour-symm}.

The precise proposal in~\cite{0911.4787} is that a tame puncture labeled by~$Y$ corresponds to a vertex operator with momentum
\begin{equation}
  \alpha = \Bigl(b+\frac{1}{b}\Bigr) \rho_Y + m_Y ,
\end{equation}
with $m_Y$ the mass parameters for the flavour symmetry~\eqref{puncture-flavour-symm} and $\rho_Y$ the projection of the Weyl vector~$\rho$ onto the subspace with the multiplicities~\eqref{alpha-partial-degenerate}.
For the full puncture case $Y=1^N$ we have $\rho_Y=\rho$, while for the simple puncture $Y=(N-1)+1$ one finds $\rho_Y=(N/2)\varpi_1$.  In both cases the proposal reproduces~\eqref{Wyllard-momentum-dictionary}.

\paragraph{Punctures: nilpotent orbits, Nahm, Hitchin, and Toda.}

For the A-type case $\lie{g}=\lie{su}(N)$ we have seen that tame punctures are labeled by partitions of~$N$.
A broader perspective is that tame codimension~$2$ defects of~$\XsuN$ are labeled by such a partition.
The D-type case and $\USp$--$\SO$ quiver tails are considered in~\cite{0905.4074}.

For general 6d $(2,0)$ theories~$\Xg$, there are (at least) three sets of data that equivalently characterize the defect.\footnote{Help welcome to complete the references.}
\begin{itemize}
\item \textbf{Nahm data.} A nilpotent orbit $\Ocal_{\text{N}}\subset\lie{g}_{\CC}$ that describes a Nahm pole boundary condition for \ac{SYM} with gauge group~$G$.  This arises by considering $\Xg$ on $\RR^{2,1}\times\text{cigar}\times S^1$ with the defect at the tip of the cigar.  Reducing first on the circle direction of the cigar gives 5d $\Nsusy=2$ \ac{SYM} with gauge group~$G$, with a Nahm pole boundary condition.

\item \textbf{Hitchin data.} A nilpotent orbit $\Ocal_{\text{H}}\subset\lie{g}_{\CC}$ with additional discrete data.
  Within the same $\RR^{2,1}\times\text{cigar}\times S^1$ geometric setup, reducing first on~$S^1$ gives a codimension~$2$ defect in 5d, then reducing the cigar to a half-line gives the S-dual of the Nahm pole boundary condition.
  This data was studied early on in~\cite{0911.1990}.

\item \textbf{Toda data.} A partially degenerate primary operator of $\lie{g}$~Toda \ac{CFT} specified by its null $W_{\lie{g}}$~descendants.
\end{itemize}
Nahm and Hitchin data were related in~\cite{1203.2930}.
The relation to Toda data was understood in~\cite{1404.3737} for the case where $e$ is principal nilpotent inside some Levi subalgebra of~$\lie{g}$, see also~\cite{1502.06311,1810.10652}.

The same classification holds for codimension~$2$ operators in the 6d $(2,0)$ little string theory with Lie algebra~$\lie{g}$ (which implies the classification for~$\Xg$).
The approach in~\cite{1506.04183,1608.07279,1711.11065} is to realize little strings as IIB~strings on a $\CC^2/\Gamma$ singularity, where the defect consists of D5~branes wrapping certain \emph{non-compact} two-cycles in $\CC^2/\Gamma$.

\paragraph{Non-simply-laced gauge groups.}

\begin{figure}\centering
  \begin{tikzpicture}[semithick,tight/.style={inner sep=0ex}]
    \begin{scope}
      \node(a)[flavor-group] at (0,0) {\scriptsize $\Spin$};
      \node(b)[color-group] at (1.2,0) {\scriptsize $\USp$};
      \node(c)[color-group] at (2.4,0) {\scriptsize $\Spin$};
      \node(d)[color-group] at (3.6,0) {\scriptsize $\USp$};
      \node(e)[color-group] at (4.8,0) {\scriptsize $\Spin$};
      \node(f)[flavor-group] at (6,0) {\scriptsize $\USp$};
      \draw(a)--(b);
      \draw(b)--(c);
      \draw(c)--(d);
      \draw(d)--(e);
      \draw(e)--(f);
      \begin{scope}[yshift=-2cm]
        \draw[densely dotted] (0,-.2) arc (-90:90:.1 and .2);
        \draw (6,-.2)--(0,-.2) arc (270:90:.1 and .2)--(6,.2);
        \filldraw (.7,0) circle (.04) -- (1.7,0) circle (.04);
        \filldraw (3.1,0) circle (.04) -- (4.1,0) circle (.04);
        \filldraw (5.5,0) circle (.04) -- (5.9,0);
        \foreach\x in {1.2,2.4,3.6,4.8} { \draw[densely dotted] (\x,0) circle (.1 and .2); }
        \draw (6,0) circle (.1 and .2);
      \end{scope}
    \end{scope}
    \begin{scope}[xshift=7.5cm]
      \node(At)[flavor-group,tight] at (0,.4) {\scriptsize $\SU(2)$};
      \node(Ab)[flavor-group,tight] at (0,-.4) {\scriptsize $\SU(2)$};
      \node(B)[color-group,tight] at (1,0) {\scriptsize $\SU(2)$};
      \node(Ct)[color-group,tight] at (2,.4) {\scriptsize $\SU(2)$};
      \node(Cb)[color-group,tight] at (2,-.4) {\scriptsize $\SU(2)$};
      \node(D)[color-group,tight] at (3,0) {\scriptsize $\SU(2)$};
      \node(Et)[color-group,tight] at (4,.4) {\scriptsize $\SU(2)$};
      \node(Eb)[color-group,tight] at (4,-.4) {\scriptsize $\SU(2)$};
      \node(F)[flavor-group,tight] at (5,0) {\scriptsize $\SU(2)$};
      \draw(At)--(.3,0)--(Ab);
      \draw(.3,0)--(B)--(1.7,0);
      \draw(Ct)--(1.7,0)--(Cb);
      \draw(Ct)--(2.3,0)--(Cb);
      \draw(2.3,0)--(D)--(3.7,0);
      \draw(Et)--(3.7,0)--(Eb);
      \draw(Et)--(4.3,0)--(Eb);
      \draw(4.3,0)--(F);
      \begin{scope}[yshift=-2cm]
        \foreach\s in {+,-} {
          \draw[densely dotted] (0,\s.4-.2) arc (-90:90:.1 and .2);
          \draw (0,\s.4-.2) arc (270:90:.1 and .2);
          \draw[densely dotted] (2,\s.4) circle (.1 and .2);
          \draw[densely dotted] (4,\s.4) circle (.1 and .2);
          \draw (0,\s.6)
          .. controls +(.5,0) and +(-.5,0) .. (1,\s.2)
          .. controls +(.5,0) and +(-.5,0) .. (2,\s.6)
          .. controls +(.5,0) and +(-.5,0) .. (3,\s.2)
          .. controls +(.5,0) and +(-.5,0) .. (4,\s.6)
          .. controls +(.5,0) and +(-.5,0) .. (5,\s.2);
        }
        \draw (0,-.2) arc (-90:90:.4 and .2);
        \draw (2,0) circle (.4 and .2);
        \draw (4,0) circle (.4 and .2);
        \draw[densely dotted] (1,0) circle (.1 and .2);
        \draw[densely dotted] (3,0) circle (.1 and .2);
        \draw (5,0) circle (.1 and .2);
      \end{scope}
    \end{scope}
  \end{tikzpicture}
  \caption{\label{fig:Spin4}Left: linear quiver with alternating $\USp(2N-2)$ and~$\Spin(2N)$ groups and its $\lie{so}(2N)$ class~S curve with branch cuts.
    Right: for $N=2$, same theory as a $\SU(2)$ generalized quiver, and its $\lie{su}(2)$ class~S curve, which is a double-cover of the $\lie{so}(4)$ curve on the left.}
\end{figure}
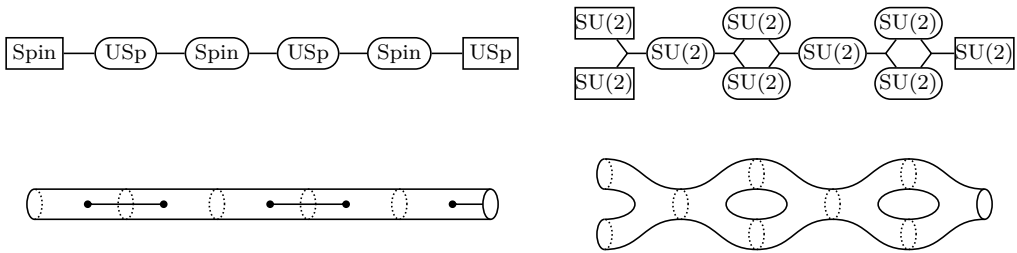

Reducing $\Xg$ on a circle yields 5d $\Nsusy=2$ \ac{SYM} with gauge algebra~$\lie{g}$, which is simply-laced.
Non-simply-laced gauge groups are achieved by twisting, namely changing the periodicity of fields,\footnote{This notion of twist is unrelated to the partial topological twist used to preserve supersymmetry when reducing $\Xg$ on~$C$.} by an outer automorphism of~$\lie{g}$.
Likewise, class~S includes 4d $\Nsusy=2$ theories with arbitrary gauge groups: these are obtained by including (outer automorphism) twist lines that may end on punctures.
This direction was explored in \cite{1012.4468,1107.0973,1111.5624,1404.3737} (see also~\cite{1311.2945}).
See \autoref{fig:Spin4} for an example with $\lie{g}=\lie{so}(2N)$ and its reformulation as a $\lie{su}(2)$ class~S theory for $N=2$ since $\lie{so}(4)=\lie{su}(2)^2$.

\subsection{\label{ssec:gen-irreg}Wild punctures and \acsfont{AD} theories}

Our investigations so far only involved so-called ``tame'' codimension~$2$ defects of the 6d $(2,0)$ theory.  They admit a broad generalization to ``wild'' defects, introduced by Witten~\cite{0710.0631} in the context of surface operators in 4d $\Nsusy=4$ \ac{SYM}.  These defects impose a stronger blow-up near their support for the ``fields'' of the 6d theory.

\paragraph{Wild punctures from collisions.}

We recall the massive tame puncture~\eqref{tame-massive-SU2}
\begin{equation}
  \varphi(z) \sim \biggl(\frac{\diag(m,-m)}{z-z_i} + O(1)\biggr)\rmd z
  \implies \phi_2(z) = \biggl(\frac{m^2}{(z-z_i)^2} + O\biggl(\frac{1}{z-z_i}\biggr)\biggr)\rmd z^2
\end{equation}
and its massless version~\eqref{tame-massless-SU2}.
Colliding $l$ such simple poles of~$\varphi$, while scaling appropriately the mass parameters, leads to a pole of order~$l$, hence generically to $\phi_2\sim \rmd z^2/(z-z_i)^{2l}$.
Just as their tame counterparts, the resulting \emph{wild punctures} of level~$l$ can be partially closed by imposing some relations between eigenvalues in the series expansion of~$\Phi_z$, so that the pole of~$\phi_2$ has an order lower than~$2l$.

The collision limits can have two main effects on the 4d gauge theory:
decoupling some hypermultiplets by making them massive while keeping the dynamical scale~$\Lambda$ fixed, or
tuning the theory to an \ac{AD} point on the Coulomb branch~\cite{hep-th/9505062,hep-th/9511154,hep-th/9608047,hep-th/9610076} at which point the theory becomes a strongly-coupled isolated \ac{SCFT}.

\begin{figure}
  \centering
  \begin{tikzpicture}
    \draw (0,-.2) -- (1.4,-.2) node [midway,above] {$\scriptstyle T_2$} -- (1.7,-.2);
    \draw (0,-.1) circle (.05 and .1) node [left] {\scriptsize tame};
    \draw (0,0) arc (-90:0:.6 and .6) arc (180:90:.6 and .6);
    \draw (1.2,1.2) -- (2.2,1.2) node [midway,below=-3.5pt]{$\scriptstyle X_{p_1}$};
    \draw (1.7,0) -- (1.4,0) arc (270:180:.6 and .6) arc (180:90:.4 and .4) to [bend right=30] (2.2,1);
    \draw (1.2,1.1) circle (.05 and .1) node [above] {$\scriptstyle \lie{su}(2)$};
    \draw (2.2,1.1) circle (.05 and .1) node [right] {$\scriptstyle p_1>2$};
    \draw (1.7,-.1) circle (.05 and .1) node [below] {$\scriptstyle \lie{su}(2)$};
    \draw (1.7,-.2) -- (3.7,-.2) node [midway,above] {$\scriptstyle T_2$};
    \draw (1.7,0) -- (2,0) arc (-90:0:.6 and .6);
    \draw (2.7,.6) circle (.1 and .05) node [right] {\scriptsize tame};
    \draw (3.7,0) -- (3.4,0) arc (270:180:.6 and .6);
    \draw (3.7,-.1) circle (.05 and .1) node [below] {$\scriptstyle \lie{su}(2)$};
    \draw (3.7,-.2) -- (4.7,-.2) node [midway,above=-3.5pt]{$\scriptstyle X_{p_2}$};
    \draw (3.7,0) to [bend left=30] (4.7,0);
    \draw (4.7,-.1) circle (.05 and .1) node [right] {$\scriptstyle p_2>2$};
  \end{tikzpicture}
  \caption{\label{fig:T2-Xp-Yp}Decomposition of a typical $\lie{su}(2)$ class~S theory (associated to a sphere with two tame and two wild punctures) into $T_2$ and $X_p$ building blocks coupled together by gauging diagonal $\lie{su}(2)$ flavour symmetries.}
\end{figure}
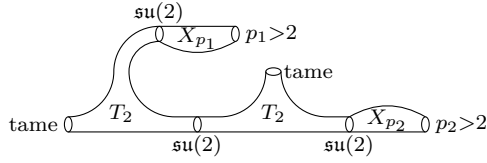

For the case $\lie{g}=\lie{su}(2)$ that we consider for now, wild punctures are labeled by the order of the pole of~$\phi_2$ (which is $2$~for a tame puncture), and of course by coefficients of the expansion at these poles.
By cutting the Riemann surface along circles as in the tame case, $\lie{su}(2)$ class~S theories can be constructed by gauging $\SU(2)$ flavour symmetries of the trifundamental half-hypermultiplet~$T_2$ (corresponding to a sphere with three tame punctures) and of theories~$X_p$ corresponding to a sphere with a tame puncture and a wild puncture at which $\phi_2$~has a pole of order $p>2$.  See \autoref{fig:T2-Xp-Yp}.  Spheres with a single wild puncture cannot be cut into these building blocks and lead to other interesting theories~$Y_p$.
This exhausts $\lie{su}(2)$ class~S\@.

\paragraph{Examples of theories with wild punctures.}
Just for this explanation we denote by $(p_1\,p_2\dots p_k)$ the class~S theory obtained for a sphere with~$k$ punctures at which $\phi_2$ has poles of order $p_1,\dots,p_k$, respectively.
Let us exemplify both effects above starting from $\SU(2)$ $N_f=4$ \ac{SQCD}, realized as $(2\,2\,2\,2)$ in this notation, namely by taking~$C$ to be a sphere with four tame punctures.
We first decouple hypermultiplets.
\begin{itemize}
\item $\SU(2)$ $N_f=3$ \ac{SQCD} arises from $(2\,2\,4)$, a sphere with two tame punctures and one wild puncture of order~$4$, obtained as a collision of two tame punctures.
\item $\SU(2)$ $N_f=2$ \ac{SQCD} appears in two ways in class~S\@.  First, as $(4\,4)$ obtained from $(2\,2\,4)$ by colliding the two tame punctures.  Alternatively, as $(2\,2\,3)$: one can decouple the hypermultiplet by tuning a mass parameter of the wild puncture in the $(2\,2\,4)$ description of the $N_f=3$ theory, and this reduces the pole of $\phi_2$ at the wild puncture from order~$4$ to order~$3$.
  A consistency check is that the two constructions lead to equivalent \ac{SW} geometry.
\item $\SU(2)$ $N_f=1$ \ac{SQCD} then appears as $(4\,3)$.
\item Pure $\SU(2)$ \ac{SYM} appears as $(3\,3)$ with two minimally wild punctures.
\end{itemize}

There are further collision limits, which turn out to realize \ac{AD} theories.
By colliding the two wild punctures in the $(4\,3)$ realization of $\SU(2)$ $N_f=1$ \ac{SQCD} we get a single wild puncture of rather high order~$(7)$: this is the $Y_7$ theory mentioned above.
The \ac{AD} point (most singular point) of the Coulomb branch of $\SU(2)$ $N_f=2$ is obtained by colliding punctures $(4\,4)\to(8)$ or $(2\,2\,3)\to(2\,5)$, both punctured curves~$C$ turning out to give the same 4d \ac{SCFT} (namely $Y_8\simeq X_5$).
For $\SU(2)$ $N_f=3$ we find the collision $(2\,2\,4)\to(2\,6)$, which is~$X_6$.
Of course, these limits all translate to tuning parameters on the gauge theory side and were thus found a long time ago~\cite{hep-th/9505062,hep-th/9511154}, but the class~S realization embeds them in a broader setting.

\begin{exercise}
  Recall the \ac{SW} curve~\eqref{four-punctured-phi2} of the $\lie{su}(2)$ class~S theory for a four-punctured sphere: $x^2=u_2(z)$ with
  \begin{equation}
    u_2(z) = \frac{\frac{q}{z}m_1^2 + \frac{q(q-1)}{z-q}m_2^2 + \frac{z-q}{z-1}m_3^2 + z m_4^2 - u}{z(z-q)(z-1)} .
  \end{equation}
  This theory has a description as $\SU(2)$ \ac{SQCD} with gauge coupling $\tau=(\log q)/(2\pi i)$ and $N_f=4$ flavours of masses $m_1\pm m_2$ and $m_3\pm m_4$.

  1. Decouple one hypermultiplet: take $m_1+m_2\to\infty$, keeping $m_1-m_2$ and $m_3\pm m_4$ and $\Lambda=q(m_1+m_2)$ fixed.  You should get $u_2=P(z)/(z^4(z-1)^2)$ for some quartic polynomial~$P$.

  2. Decouple a second hypermultiplet in two ways.
  First, take $m_1-m_2\to\infty$, keeping $\Lambda'^2=\Lambda(m_1-m_2)$ and other masses fixed.
  Second, instead, take $m_3+m_4\to\infty$, keeping $\tilde{z}=z(m_3+m_4)$ and $\tilde{x}=x/(m_3+m_4)$ and $\Lambda'^2=\Lambda(m_3+m_4)$ and other masses fixed.
  Map one \ac{SW} curve to the other and check the difference of \ac{SW} differentials~$\lambda$ is inessential (residues are masses, no Coulomb branch dependence).

  3. Decouple a third and a fourth hypermultiplet and rescale $z\to z\Lambda^2$ to get the well-known curve of pure $\SU(2)$ \ac{SYM}: $z^2x^2 = u + \Lambda^2(z+1/z)$ with $\lambda=x\rmd z$.
\end{exercise}

\paragraph{\ac{CFT} side.}

On the 2d \ac{CFT} side, the limits that produce wild punctures correspond to colliding primary vertex operators~$\Vhat_\alpha$.
The collision of $n\geq 2$ vertex operators yields irregular operators denoted~$I_{n-1}$, which depend on the original $n$~momenta (suitably rescaled).
Their Ward identities with the stress tensor involve poles of the same order~$2n$ as the pole of~$\phi_2$ on the gauge theory side that arises in the same collision.
This matching is consistent with the fact that $\phi_2$ can be understood as the semiclassical limit of~$T$:
\begin{equation}
  T(z) I_{n-1}(0) = \sum_{k=n-1}^{2n-2} \frac{\Lambda_k}{z^{k+2}} I_{n-1}(0) + \dotsm
  \underset{r\to\infty}{\simeq} r^2 \phi_2(z) I_{n-1}(0) + \dots .
\end{equation}
By the state-operator correspondence, these operators give coherent states of the Virasoro algebra~\cite{0908.0307} (alternatively called Whittaker vectors or Gaiotto states) and generalizations thereof called irregular states (sometimes \ac{BMT} states)~\cite{1112.1691,1203.1052}.

\paragraph{Wild punctures and \ac{AD} theories.}

The decoupling of hypermultiplets starting from $\SU(2)$ $N_f=4$ \ac{SQCD} or $\Nsusy=2^*$ \ac{SYM}, and its effect in the \ac{AGT} corespondence, were studied in \cite{0908.0307,0909.2052,0911.0363,0911.4797,1003.1049,1008.1861,1209.6009,1407.0305,1812.03387}.
Higher-level irregular states (\ac{BMT} states) of the Virasoro algebra were investigated in~\cite{1112.1691,1112.4453,1203.1052,1207.4480,1301.0721,1312.5535,1411.4453}, and generalizations to W-algebras in \cite{0912.4789,1111.5624,1206.2844,1301.0721,1301.5342}.
On the \ac{CFT} side, collisions of primary operators and direct definitions of irregular states were made by Rim and collaborators~\cite{1207.4480,1210.7925,1312.5535,1405.3141,1411.4453,1504.07910,1506.02421,1506.03561,1510.09060,1608.05027,1612.00348}, and others~\cite{1402.2946,1407.1852,1407.8423,1505.02398,1601.07756,1604.08741,1608.04921,1611.08971,1708.03162,1710.01051,1905.03795}.

\paragraph{Further directions.}

Some \ac{AD} theories were found at particular points on the Coulomb branch of Lagrangian gauge theories, and as appearing in S-dual descriptions in~\cite{hep-th/9505062,hep-th/9511154,hep-th/9608047,hep-th/9610076,0711.0054,0810.4541}.
Class~S constructions of a variety of \ac{AD} theories and related topics are in~\cite{0905.4074,1009.0339,1011.4568,1110.2657,1203.6357,1210.2886,1303.3149,1411.6026,1804.09143,1912.09348}
and in Xie's work with collaborators~\cite{0907.1651,1005.1350,Xie-2011egw,1204.2270,1207.6112,1301.0210,1509.00847,1602.03529,1606.06306,1701.01123,1711.06684,1712.03244,1805.08839}.
It is not expected that class~S theories exhaust all possible 4d $\Nsusy=2$ theories (see e.g.~\cite{1404.3736}).
Besides a classification of Lagrangian field theories~\cite{1108.2315,1309.5160}, and a program to classify theories according to their Coulomb branch geometry \cite{hep-th/0504070,hep-th/0510226,1505.04814,1601.00011,1602.02764,1609.04404,1801.01122,1801.06554,1804.03152},
other constructions of 4d $\Nsusy=2$ theories have been explored \cite{1006.3435,1103.5832,1203.6734,1204.2270,1210.2886,1301.0210,1507.01799,1509.00847,1510.01324,1604.07843,1606.06306,1611.08602,1612.08065,1704.05110,1707.08981,1712.00464,1906.03912}.

An interesting tool to check the \ac{AGT} correspondence even in the absence of Lagrangian descriptions of the class~S theories is to compute central charges and anomalies.
The central charge of Toda \ac{CFT} (and generalizations) was matched with a reduction to 2d of the anomaly $8$-form of the 6d $(2,0)$ theory~$\Xg$ in \cite{0904.4466,0909.4031,0909.4776,1106.1172}.
Reducing instead on~$C$ gives the $a$ and~$c$ conformal anomalies of the 4d class~S theories \cite{1203.2930,1310.5033,Balasubramanian-2014dia,1504.01481,1803.00136,1806.06066,1812.04016}.

\section{\label{sec:ops}Operators of various dimensions}

Wilson~\cite{Wilson:1974sk} and 't~Hooft~\cite{tHooft:1977nqb} loop operators, and dyonic loops combining them~\cite{hep-th/0501015}, play an important role in studying phases of 4d gauge theories.  Surface operators are less studied yet very rich; for instance the moduli space of some surface operators is the \ac{UV} curve of the class~S theory.  Finally, domain walls describe interfaces between two 4d theories, or boundary phenomena.
A large source of such operators in the \ac{AGT} correspondence are the half-\ac{BPS} codimension~$2$ and codimension~$4$ defects of the 6d $(2,0)$ theory~$\Xg$.
Another source is to orbifold the 6d setup, with an orbifold group that must
\apartehere
respect orientation since $\Xg$~is a chiral theory.
These defects can be inserted into the \ac{AGT} correspondence with various orientations relative to the product spacetime $M_4\times C$.
We have already covered at length the case where a codimension~$2$ defect inserted at a point in~$C$ wraps the whole 4d spacetime: indeed these are simply the punctures and twist operators described throughout this review, especially in \autoref{sec:gen}.
\startaparte[title={Main examples of gauge theory and \ac{CFT} operators}]
\begin{itemize}
\item Codimension~$2$ case: coupling to a tinkertoy matches with inserting a vertex operator; symmetry-breaking walls match with Verlinde loops; \acl{GW}~surface operators change the Toda~\ac{CFT}.
\item Codimension~$4$ case: vortex string surface operators match with degenerate vertex operators; Wilson--'t~Hooft loops match with degenerate Verlinde loops; orbifold singularities at poles of~$S^4_b$ change the Toda~\ac{CFT}.
\end{itemize}
\stopaparte
\ifjournal\else\par\fi
We refer to \autoref{tab:ops} in the introduction for a full list of possibilities that have been investigated, and references.
Here, we organize our discussion by increasing dimension on the 4d side, starting with a discussion of point-like operators in \autoref{ssec:local}, then line and loop operators in \autoref{ssec:line} (see the review~\cite{1412.7126}), 2d operators in \autoref{ssec:surf} (see the review~\cite{1412.7127}), and 3d walls and interfaces in \autoref{ssec:wall}.

\subsection{\label{ssec:local}Local operators in 4d}

Codimension~$4$ operators of~$\Xg$ are labeled by representations of~$\lie{g}$ and the effect of wrapping such an operator over all of~$C$ has not been fully understood.
In fact, for most of the geometries we describe in the following, the main ideas have been understood in certain theories such as for $\lie{g}=\lie{su}(2)$, but not in full generality.

\paragraph{\label{par:local-Coulomb}Coulomb branch operators.}

The order~$k$ holomorphic differentials $\phi_k(z)=u_k(z)\rmd z^k$ that define the \ac{SW} curve can be calculated from the classical limit of Toda \ac{CFT} correlators.
In this limit, where the radius of~$S^4_b$ is large, or $\epsilon_1,\epsilon_2\to 0$, or equivalently 2d \ac{CFT} conformal dimensions are large, the quadratic differential $u_2$~is given by the insertion of the energy-momentum tensor, and more generally $u_k$~by the spin~$k$ current~$W_k$:\footnote{The numerical coefficient depends on conventions.}
\begin{equation}\label{SW-from-classical}
  u_k(z) \propto \frac{\vev{W_k(z)\Vhat_{\mu_1}\dots \Vhat_{\mu_n}}}{\vev{\Vhat_{\mu_1}\dots \Vhat_{\mu_n}}} \qquad \text{as }\epsilon_1,\epsilon_2\to 0 .
\end{equation}
For $\lie{su}(2)$ see the original \ac{AGT} paper~\cite{0906.3219} or the more explicit~\cite{1502.05581} for instance.

Consider now a pants decomposition of~$C$, and the corresponding description of $\TgCD$ as a collection of tinkertoys and vector multiplets gauging some symmetries.
Let $\varphi$ be the scalar in one of the vector multiplets (corresponding to a tube), and consider gauge-invariants such as $\Tr\varphi^l$ in the A-type case, and more generally all Casimirs of~$\lie{g}$.
Classically, they appear as coefficients of the differentials~$\phi_k$ and can thus be retrieved as certain weighted integrals of~$\phi_k$.
Going back to general $\epsilon_1,\epsilon_2$, the \emph{operator} $\Tr\varphi^l$ on the 4d side corresponds to a suitable weighted integral of currents~$\tilde{W}_l$~\cite{0909.4031}.\footnote{The spin~$l$ current $\tilde{W}_l$ is polynomial in the~$W_k$ to account for the difference between Casimirs $\Tr\varphi^l$ and coefficients in a characteristic polynomial $\det(x-\varphi)$.}
For instance, inserting $\Tr\varphi^2$ takes a derivative of $Z_{S^4_b}$ with respect to gauge couplings~\cite{hep-th/0302191,hep-th/0403057,hep-th/0510173}, namely to the shape of~$C$, which indeed translates to an integrated insertion of the holomorphic stress-tensor $T=\tilde{W}_2$.

Correlation functions on $S^4_b$ with (products of) $\Tr\varphi^j$ inserted at one pole and $\Tr\bar\varphi^k$ at the other can be computed by supersymmetric localization, although the operators complicate instanton counting.
By a conformal transformation the round case $b=1$ leads to results on flat space correlators with exactly one antichiral operator~\cite{1602.05971,1607.07878,1610.07612,1701.02315,1810.00840,1901.09693}, which provide detailed checks of various field theory ideas such as resurgence~\cite{1602.05971,1603.06207}, large-charge expansions~\cite{1604.07416,1710.07336,1803.00580,1804.01535,1809.06280,1810.10483,1908.10306,2001.06645}, and more~\cite{1911.05827}.  These specific correlators have not been pursued on the 2d \ac{CFT} side of the correspondence.

\paragraph{\label{par:local-orbifold}Orbifold \texorpdfstring{$\CC^2/\ZZ_M$}{C\string^2/Z\string_M}.}

Next we consider another operation whose effect is to make one point singular inside~$\RR^4$, or both poles of~$S^4_b$: orbifolding by a group $\ZZ_M$ acting as $(\exp(2\pi i/M),\exp(-2\pi i/M))$ on~$\CC^2$.
Supersymmetric localization still works: one must simply restrict all modes to $\ZZ_M$-invariant ones, and instanton counting to $\ZZ_M$-invariant instanton counting.
This appears to correspond to the coset \ac{CFT}
\begin{equation}\label{orbifold-coset}
  \frac{\widehat{\lie{su}}(N)_k\times\widehat{\lie{su}}(N)_M}{\widehat{\lie{su}}(N)_{k+M}}\times \frac{\widehat{\lie{su}}(M)_N}{\widehat{\lie{u}}(1)^{M-1}} , \quad k = -N-\frac{Mb^2}{1+b^2} ,
\end{equation}
with a fractional level~$k$,
which for $M=1$ reduces nontrivially to the usual Toda \ac{CFT}.
The case $N=M=2$, essentially super-Liouville \ac{CFT}, is studied in~\cite{1105.5800,1106.2505,1106.4001,1107.4609,1110.2176,1111.2803,1205.0784,1210.7454,1306.4227,1706.07474}, see also~\cite{1111.7095} with a surface operator.
Instanton counting on $\CC^2/\ZZ_2$ and on its blow up, where one has instantons at two fixed point of an $\Lie{U}(1)$ isometry, are related~\cite{1209.1486,1303.5765,1406.3008,1410.2742}.
This leads to a decomposition of super-Liouville \ac{CFT} into a Liouville and time-like Liouville pieces~\cite{1210.1856,1312.4520,Hadasz:2014hza,1510.01773}.
The general $N,M$ extension is partially worked out in~\cite{1106.1172,1109.4264,1110.5628,1208.0790,1306.3938} and the coset~\eqref{orbifold-coset} studied further in~\cite{1212.6600,1511.08265,1912.04407}.
Another perspective is to realize $\RR^4/\ZZ_M$ by dimensional reduction of $\RR^4\times S^1$, which corresponds to taking $q$ to a root of unity~\cite{1105.6091,1211.2788,1308.2068,1409.3465,1705.03628} in the $q$-deformed \ac{AGT} correspondence of \autoref{ssec:dim-5d}.

\subsection{\label{ssec:line}Line operators}

We now place a codimension~$4$ operator of the 6d theory along $L\times\gamma$, where $L$~is one of the circles $\{y_3=y_4=0,y_5=\text{const}\}$ or $\{y_1=y_2=0,y_5=\text{const}\}$ in~$S^4_b$ allowed by supersymmetry, while $\gamma$~is a closed loop in~$C$ with no self-intersection.
Upon dimensional reduction, the operator inserts loop operators in the \ac{AGT} correspondence: a loop operator labeled by~$\gamma$ and placed on $L\subset S^4_b$ in the partition function, and a loop operator on~$\gamma$ in the Toda \ac{CFT} correlator.
This setup is studied for $\lie{su}(2)$ in~\cite{0907.2593,0909.0945,0909.1105,0911.1922,0912.5535,1003.1112}, for more general~$\lie{g}$ in~\cite{1003.1151,1008.4139,1108.0242,1110.3354,1206.6896,1304.2390,1312.5001,1504.00121}, for networks of such operators~\cite{1312.5001,1404.0332,1504.00121,1603.02939,1701.04090}, and for the algebra of line operators~\cite{1505.05898,1712.10225,Coman-Lohi:2018mgj}, see also~\cite{Petkova:2013yoa,1602.07476}.
Line and loop operators play an important role in characterizing phases of 4d gauge theories~\cite{Wilson:1974sk,tHooft:1977nqb,hep-th/0501015}, and refining the global structure of the gauge group~\cite{1305.0318}; for class~S see~\cite{0907.2593,1304.2390,1307.4381,1309.0697,1312.3371,1312.5001,1504.00121,1507.04743,1603.03044,1604.08222,1908.08027,1911.13264}.
Exact expectation values of loop operators in 4d $\Nsusy=2$ gauge theories are calculated using supersymmetric localization in~\cite{0712.2824,1105.2568,1111.4221} and reviewed in~\cite{1412.7126}, with important subtleties being clarified later in~\cite{1801.01986,1806.00024,1810.07191,Brennan:2019vqz,1903.00376,1905.11305}.
Other considerations about line defect observables in 4d theories include~\cite{0909.4272,1006.0146,1202.6393,1309.1213,1311.2058,1711.10799,1907.04345}.

\paragraph{\label{par:line-dyon}Wilson loop operators.}

Since $\gamma\subset C$ has no self-intersection we can cut $C$ along it and get a (possibly disconnected) surface~$C'$ with two additional punctures (with some data, say $D_1,D_2$).  As discussed near~\eqref{intro-split}, the corresponding class~S theory $\TgCD$ is obtained from the theory $\Theory\bigl(\lie{g},C',\{D,D_1,D_2\}\bigr)$ corresponding to~$C'$ by gauging a diagonal subgroup of the flavour symmetries $F_1,F_2$ associated to~$D_1,D_2$ as in~\eqref{intro-split}.
In this way each non-self-intersecting loop~$\gamma$ is associated to a gauge group $G_\gamma=(F_1\times F_2)_{\text{diag}}$ in some description of $\TgCD$.

In the limit of weak coupling for~$G_\gamma$, the loop~$\gamma$ wraps a thin tube.
The reduction of~$\Xg$ on this tube gives 5d $\Nsusy=2$ \ac{SYM} on an interval, and the codimension~$4$ defect gives a line operator, specifically a Wilson loop, which depends additionally on a choice of representation~$R$ of~$\lie{g}$.
In fact this is the clearest way to see that codimension~$4$ operators of~$\Xg$ should depend on such a representation of~$\lie{g}$.
Further reduction to 4d gives a half-\ac{BPS} Wilson loop measuring the holonomy of the corresponding gauge field~$A_\gamma$ along~$L$, plus some contribution from scalar superpartners to ensure supersymmetry:
\begin{equation}\label{Wilson-loop}
  W_{\gamma,R} = \Tr_R \Bigl( \Pexp \int_L \bigl(A_\gamma + \text{scalars}_\gamma\bigr) \Bigr) .
\end{equation}
The gauge group~$G_\gamma$ and the Wilson loop only depend on the homotopy class of~$\gamma$.

On the 2d \ac{CFT} side the corresponding object is a certain 1d~topological defect along~$\gamma$ called a degenerate Verlinde loop operator.
Verlinde loops are constructed as monodromies of a vertex operator~$V_\omega$.
The specific choice corresponding to $W_{\gamma,R}$ is to take a momentum $\omega=-b^{\pm 1}\Omega_R$, where $\pm$ depends on the choice of circle~$L$, while $\Omega_R$ is the highest weight\footnote{Weights are in~$\lie{h}^*$, which we identify with the Cartan subalgebra $\lie{h}\subset\lie{g}$ using the Killing form.} of the representation~$R$.
For this choice of~$\omega$, the vertex operator~$V_\omega$ is degenerate in the sense that it is annihilated by various combinations of W-algebra generators.
Incidentally, the most general degenerate momentum $\omega=-b\Omega-b^{-1}\Omega'$ corresponds to inserting Wilson loops along both allowed circles.

Concrete checks of the correspondence are straightforward.
The Wilson loop is compatible with supersymmetric localization~\cite{0712.2824} and inserts a simple $a$-dependent factor in~\eqref{ZS4-localized}.
The Verlinde loop~$\cL_\gamma$ acts diagonally on a complete set of states inserted along~$\gamma$ hence appears in the correlator~\eqref{Liouville-into-blocks} simply as a function of the internal momentum~$\alpha$ (related to~$a$).
They match:
\begin{equation}\label{WL-diag}
  \begin{aligned}
    \vev{W_{\gamma,R}}_{S^4_b}^{\TgCD}
    & = \int \rmd a\, \Tr_R(e^{a_\gamma}) \, Z_{\cl}(a,q,\qbar)\,Z_{\oneloop}(a) Z_{\inst}(a,q) Z_{\inst}(a,\qbar) \\
    & = \int \rmd\alpha\,f(\alpha)\,C(\alpha) \mathcal{F}(\alpha,q) \mathcal{F}(\alpha,\qbar) \\
    & = \vev*{\Vhat_{\mu_1}\dots \Vhat_{\mu_n}\cL_\gamma}_{\fullC}^{\Toda(\lie{g})} .
  \end{aligned}
\end{equation}

\paragraph{Dyonic loop operators.}
Now consider a pants decomposition of~$C$ that does not include~$\gamma$ among its cuts.
The 2d \ac{CFT} side is still given by a Verlinde loop along~$\gamma$, but its expression in the given basis of conformal blocks is no longer diagonal: it is
\begin{equation}\label{loop-2}
  \vev*{\Vhat_{\mu_1}\dots \Vhat_{\mu_n}\cL_\gamma}_{\fullC}^{\Toda(\lie{g})}
  = \int \rmd\alpha\,C(\alpha) \mathcal{F}(\alpha,q)\, \sum_h \cL_\gamma(\alpha,\alpha+h) \mathcal{F}(\alpha+h,\qbar)
\end{equation}
where $h$ ranges over a finite collection\footnote{The fusion of a degenerate operator~$V_\omega$ with another vertex operator has a finite number of terms.} of momenta related to the weights of~$R$.

The corresponding loop operator in 4d is described in this S-duality frame as a 't~Hooft or dyonic loop instead of a Wilson loop.
Rather than being defined by an insertion in the path integral like the Wilson loop~\eqref{Wilson-loop}, a 't~Hooft loop on~$L$ is defined by imposing a singular boundary condition on the gauge field that prescribes a non-zero monopole charge $\frac{1}{2\pi}\int F$ on a two-sphere~$S^2$ surrounding~$L$.
Dyonic loops involve additionally a Wilson loop insertion along the same circle~$L$.
The path integral ranges over such singular field configurations instead of the usual smooth ones, and supersymmetric localization still applies~\cite{1105.2568} and reproduces~\eqref{loop-2}, provided one correctly accounts for monopole bubbling~\cite{1801.01986,1806.00024,1810.07191,Brennan:2019vqz,1903.00376,1905.11305}.

Interestingly, Verlinde loops must be generalized to Verlinde networks (involving fusion of degenerate vertex operators) to reproduce half-\ac{BPS} dyonic loops with the most general electric and magnetic charges~\cite{1008.4139,1312.5001,1504.00121,1507.06318}.

\paragraph{Algebra of line operators.}

In light of~\eqref{loop-2}, dyonic loops or Verlinde networks can be understood as difference operators acting on functions of internal momenta~$\alpha$, or equivalently acting on functions on the Coulomb branch~$\cB$.
Inserting dyonic loops at different latitudes~$y^5$ yields a product of difference operators, which is non-commutative because the loops cannot be reordered while preserving supersymmetry.
The \ac{OPE} of loop operators provides skein relations that express these products as linear combinations of dyonic loops and defines an algebra $\fA_{\epsilon_1,\epsilon_2}$ of loop operators (we recall $\epsilon_1=b/r$ and $\epsilon_2=1/(rb)$).
Teschner~\cite{1005.2846,1302.3778} emphasized early on the importance of this algebra in the \ac{AGT} correspondence; see the reviews \cite{1412.7140,Coman-Lohi:2018mgj}.  The earlier work~\cite{0710.2097} concentrated on 4d $\Nsusy=4$ \ac{SYM}.

Each pants decomposition of~$C$ gives a representation of $\fA_{\epsilon_1,\epsilon_2}$ as difference operators, in which certain loops are diagonalized~\eqref{WL-diag} while others remain nondiagonal~\eqref{loop-2}.
The modular kernels, which relate conformal blocks in different pants decompositions, simply map between eigenbases of various loop operators, and they can be worked out from $\fA_{\epsilon_1,\epsilon_2}$.
The eigenbases themselves (instanton partition functions) are then solutions of a Riemann--Hilbert type problem that can be solved.
Their $\epsilon_1,\epsilon_2\to 0$ limit is then the low-energy prepotential~\eqref{prepotential}.
All of this rich content hidden in~$\fA_{\epsilon_1,\epsilon_2}$ led the authors of~\cite{1302.3778} to call it the ``non-perturbative skeleton'' of $\TgCD$.

In the \ac{NS} limit $\epsilon_2\to 0$ of the algebra~$\fA_{\epsilon_1,\epsilon_2}$, the difference operators become differential operators on~$\cB$, which further become coordinates on a torus fibration over~$\cB$, in the classical limit $\epsilon_1,\epsilon_2\to 0$.\footnote{Similar to how momentum $p\sim i\del_x$ in quantum mechanics is a coordinate in classical mechanics.}
The geometry simplifies in the \ac{NS} limit, where $S^4_b$~degenerates to $\RR^2\times S^2_b$.
Near the loops along the equator of~$S^2_b$ the geometry is $\RR^3\times_b S^1$ with a twisted periodicity around~$S^1$, and the twist is removed in the classical limit.
Precisely this geometry was considered in~\cite{0807.4723,0907.3987,1006.0146,1111.4221,1301.3065,1307.7134,1308.6829} (on ``framed \ac{BPS} states'').
On untwisted $\RR^3\times S^1$, vacuum expectation values of the loop operators define coordinates on the Coulomb branch~$\cM$ of the class~S theory on $\RR^3\times S^1$, which is the aforementioned torus fibration over~$\cB$.
The twisted periodicity of $\RR^3\times_b S^1$ quantizes this algebra of coordinates into an algebra of differential operators, the \ac{NS} limit of $\fA_{\epsilon_1,\epsilon_2}$.

As described near~\eqref{Hitchin} the $\RR^3\times S^1$ Coulomb branch~$\cM$ can be seen alternatively as the Hitchin moduli space on~$C$, or as the moduli space of flat $G_{\CC}$ connections on~$C$ (where the Lie algebra of~$G_{\CC}$ is the complexification of~$\lie{g}$).
The last point of view fits nicely with the 3d/3d correspondence discussed in \autoref{ssec:dim-3d}, which involves $G_{\CC}$ Chern--Simons theory.
Further works on the Hitchin system, opers, and Darboux coordinates on~$\cM$ include \cite{0911.1990,1103.3919,1106.4550,1112.1691,1203.4573,1207.6112,1301.0192,1304.2390,1310.7943,1403.6137,Nekrasov:2014yra,1505.05898,1512.02617,1712.10225,1806.08270,1810.00928,1903.08172,1904.10491} (some are reviewed in~\cite{1412.7120});
a different technique is based on spectral networks, which abelianize flat connections on~$C$ \cite{1204.4824,1205.2512,1209.0866,1312.2979,1409.2561,1603.05258,1611.00150,1611.06177,1704.04204,1710.04438,1710.08449,1711.04038,1906.04271}; see also~\cite{1206.6896,1703.04786,1703.06449,1902.08586}.

\subsection{\label{ssec:surf}Surface operators}

Surface operators are reviewed in~\cite{1412.7127} (for early references, see \cite{hep-th/0612073,0804.1561}).
Surface operators compatible with supersymmetric localization on~$S^4_b$ can be inserted along two squashed spheres intersecting at the poles or some two-tori, expressed in Cartesian coordinates of an~$\RR^5$ as follows:
\begin{equation}
  \begin{aligned}
    S^4_b & \coloneqq \bigl\{y_5^2 + b^2(y_1^2+y_2^2)+b^{-2}(y_3^2+y_4^2)=r^2\bigr\} , \\
    S^2_b & \coloneqq \bigl\{y_1=y_2=0,y_5^2 + b^{-2}(y_3^2+y_4^2)=r^2\bigr\} , \\
    S^2_{1/b} & \coloneqq \bigl\{y_5^2 + b^2(y_1^2+y_2^2)=r^2,y_3=y_4=0\bigr\} , \\
    T^2_{\theta,\varphi} & \coloneqq \bigl\{y_5=r\cos\theta, y_1^2+y_2^2=(rb^{-1}\sin\theta\cos\varphi)^2, y_3^2+y_4^2=(rb\sin\theta\sin\varphi)^2\bigr\} .
  \end{aligned}
\end{equation}
The latter case has not been studied so we concentrate here on the spheres, on which the surface operators preserve a 2d $\Nsusy=(2,2)$ subalgebra of 4d $\Nsusy=2$.
Such operators arise either from a codimension~$4$ operator of~$\Xg$ at a point $z\in C$ or from a codimension~$2$ operator wrapping all of~$C$.
(See also \cite{1507.06318}~on foams of surface operators, \cite{1404.2929,1610.03663}~on duality defects, \cite{1911.02185}~for a holographic approach.)
The space of couplings of the first type of surface operators is exactly the \ac{UV} curve~$C$, thus it gives a definition of~$C$ directly from the 4d $\Nsusy=2$ theory.

\paragraph{\label{par:surf-vort}Vortex string operators.}

We begin with surface operators arising from a codimension~$4$ operator of the 6d theory placed at a point $z\in C$ and one of the two possible spheres $S^2_b$ (sign ``$+$'' below) or $S^2_{1/b}$ (sign ``$-$'' below).
This class of operators is sometimes called M2-brane surface operators because it arises from the addition of M2 branes in the M-theory construction of class~S theories.  They can also arise via Higgsing a larger 4d theory if the Higgsed field has a non-trivial space-dependent profile~\cite{1207.3577,1711.03561,1808.00725}.
Their \ac{AGT} interpretation is explored in \cite{0909.0945,0911.1316,1004.2025,1006.4505,1007.2524,1008.0574,1011.4491,1102.0184,1107.2787,1111.7095,1307.6612,1403.3657,1407.1852,1602.02772,1608.02849,1610.03501,1710.06970,1711.11582}, in part based on their exact partition functions, studied in \cite{1206.2356,1206.2606,1307.2578,1308.2217,1407.4587,1412.0530,1412.2781,1412.3872,1610.03501,1612.04839,Lamy-Poirier-2016yjn,1702.02833,1702.03330,1809.10485,1811.04901}.  The same operators are important in the 5d version and $S^1\times S^3_b$ version of the correspondence \cite{1303.2626,1303.4460,1312.1294,1401.3379,1412.3872,1612.04839,1801.00960,1807.11900,1811.11772}; see also \cite{1103.2598,1204.4824,1209.0866,1503.04806,1601.02633,1709.04926,1712.08140,1804.09694,1811.12375,1812.00923,1812.08745,1905.02724} for other considerations on this class of surface operators.

As we have learned from studying loops, codimension~$4$ operators carry a choice of representation~$R$ of~$\lie{g}$.
On the 2d \ac{CFT} side we thus want a point operator labeled by~$R$: the natural guess is a degenerate vertex operator $V_\omega$ with $\omega=-b^{\pm 1}\Omega_R$, the sign $\pm$ being determined by which squashed two-sphere we use on the 4d side.
This suggests an equality
\begin{equation}\label{vortex-AGT}
  \vev{\text{surface operator}}_{S^4_b}^{\TgCD} = \vev*{\Vhat_{\mu_1}\dots \Vhat_{\mu_n}V_{-b^{\pm 1}\Omega_R}}_{\fullC}^{\Toda(\lie{g})} .
\end{equation}
The right-hand side can be written as an analytic continuation of an $(n+1)$-point correlator $\vev{\Vhat_{\mu_1}\dots \Vhat_{\mu_{n+1}}}$ of non-degenerate vertex operators.
The analytic continuation in the corresponding class~S theory~$T'$ was first understood in~\cite{1207.3577} in Lagrangian cases\footnote{It would be good to clarify the situation for the most general class~S theories.}: it amounts to considering a supersymmetric ``vortex string'' configuration in~$T'$ in which certain hypermultiplet scalars acquire space-dependent \acp{VEV} concentrated in codimension~$2$.  In the low-energy limit, the non-zero scalars Higgs some gauge symmetries of~$T'$ down to those of~$T$, and the configuration is effectively described by a surface operator in the theory~$T$.

\paragraph{Description as a 4d-2d coupled system.}

Besides this vortex string construction of surface operators obtained from codimension~$4$ operators of~$\Xg$, these surface operators can be described by coupling to the 4d theory a 2d $\Nsusy=(2,2)$ gauge theory living on the defect.
In this context the left-hand side of~\eqref{vortex-AGT} is the partition function of the 4d-2d coupled system on squashed spheres.
A simple example is that of $\SU(2)$ \ac{SQCD} with $N_f=4$ and a defect labeled by the $K$-th symmetric representation.  The 2d theory then consists of chiral multiplets in doublet representations of 4d flavour and gauge groups, and in fundamental and antifundamental representations of a 2d $\Lie{U}(K)$ gauge group:
\begin{equation}\label{simplest-vortex-AGT}
  Z_{S^2_b\subset S^4_b}\left[
    \quiver{
      \node (Nf) at (0,0) [flavor-group] {$2$};
      \node (Nf') at (1,0) [color-group] {$2$};
      \node (Nf'') at (2,0) [flavor-group] {$2$};
      \node (Nn) at (.5,-.7) [color-group] {$K$};
      \draw (Nf) -- (Nf');
      \draw (Nf') -- (Nf'');
      \draw[->-=.55] (Nf) -- (Nn);
      \draw[->-=.55] (Nn) -- (Nf');
      \node (4d) at (-.5,0) {4d};
      \node (2d) at (-.5,-.7) {2d};
      \draw [color=gray, dashed, rounded corners]
      (-.75,-.3) rectangle (2.3,.3);
      \draw [color=gray, dashed, rounded corners]
      (-.75,-.3) rectangle (2.3,-1);
    }
  \right]
  = \vev*{\Vhat_{\mu_1}\Vhat_{\mu_2}\Vhat_{\mu_3}\Vhat_{\mu_4}V_{-Kb/2}}_{S^2}^{\Liouville} .
\end{equation}
The position $z$ of $V_{-Kb/2}$ matches the \ac{FI} parameter of the 2d $\Lie{U}(K)$ gauge group.
Such a 2d description of the most general~$R$ in $\lie{su}(N)$ Lagrangian theories is proposed in~\cite{1407.1852} and checked by comparing limits $z\to z_i$ in gauge theory to the known \ac{OPE} of $V_{-b^{\pm 1}\Omega_R}$ and~$\Vhat_{\mu_i}$.
More general degenerate insertions $V_{-b\Omega-\Omega'/b}$ translate to intersecting defects with extra 0d fields living at the poles where the defects intersect~\cite{1610.03501}.
An important difficulty in checking equalities like~\eqref{simplest-vortex-AGT} is to compute contributions $Z_{\inst,\vort}$ from the poles of~$S^4_b$, which involve both instantons of the 4d theory and vortices of the 2d theory~\cite{1702.03330}.
Incidentally, in an $\epsilon_1,\epsilon_2\to 0$ limit this 4d-2d analogue of Nekrasov's partition function gives both the 4d theory's effective prepotential~$F$ and the 2d theory's effective twisted superpotential~$\cW$, obtained earlier in~\cite{0911.1316,1307.2578}:
\begin{equation}
  \log Z_{\inst,\vort} = \frac{F}{\epsilon_1\epsilon_2} + \frac{\cW}{\epsilon_1} + O(1) .
\end{equation}

\paragraph{\label{par:surf-mono}Gukov--Witten operators: monodromy defects and orbifolds.}

Next we discuss surface operators called M5-brane surface operators, or codimension~$2$ operators, or orbifold surface defects, studied in~\cite{hep-th/0612073,0710.0631,1005.4469,1008.1412,1011.0289,1011.0292,1012.1355,1102.0076,1105.0357,1203.1427,1205.3091,1209.2992,1301.0940,1305.0266,1404.3737,1405.6992,1408.4132,Balasubramanian-2014dia,1412.3872,1412.6081,1509.07516,1512.01084,1608.05350,Mori-2016lbn,1612.02008,RajanJohn-2017zic,1810.10652,1811.04901,1907.02771} and in~\cite{1608.07279,1711.11065} from the little string theory viewpoint.
We have already encountered codimension~$2$ defects of~$\Xg$, since they are the origin of tame punctures that impose certain boundary conditions on the differentials~$\phi_k$.
Wrapping these codimension~$2$ operators on~$C$ thus gives surface operators that impose certain singular boundary conditions on the 4d fields.
Specifically, this yields $\Nsusy=2$ versions of \ac{GW} surface operators~\cite{hep-th/0612073}, which impose that 4d gauge fields~$A$ behave as $A\sim\alpha d\theta$ as $r\to 0$ with a prescribed $\alpha\in\lie{t}$ in the Cartan algebra of~$\lie{g}$, where $(r,\theta)$ are polar coordinates transverse to the defect.

If $A$ is an $\SU(N)$ gauge field, say, let us denote eigenvalues of $\alpha=\diag(\alpha_1,\dots)$ as $\alpha_i$ with multiplicities~$N_i$, $i=1,\dots,M$ so that $\sum_i N_i\alpha_i=0$ and $\sum_i N_i=N$.
Then the 4d gauge group breaks to $\bigl(\prod_i \Lie{U}(N_i)\bigr)/\Lie{U}(1)$ at the defect.
The instanton moduli space with such a monodromy defect is equivalent as a complex manifold
to the moduli space of instantons on an orbifold $\CC\times(\CC/\ZZ_M)$~\cite{biquard1996fibres}.
Here, $\ZZ_M$~embeds into both rotations of~$\CC$ with charge~$+1$ and the gauge group $\Lie{SU}(N)$ with charges $i$ with multiplicity~$N_i$, thus reproduces the expected symmetry breaking.
The Nekrasov partition function~$Z_{\inst}$ is obtained from the usual one by restricting to $\ZZ_M$-invariant instantons.
It matches conformal blocks of the affine $\Lie{SL}(N)$ algebra (for the full defect that has all $N_i=1$)~\cite{1005.4469,1008.1412}, or its \ac{DS} reductions~\cite{1011.0289} more generally.

The \ac{GW} defects can also be described by coupling suitable 2d $\Nsusy=(2,2)$ gauge theories to the 4d theory.  For pure 4d $\Nsusy=2$ \ac{SYM},
\begin{equation}\label{simplest-GW-AGT}
  Z_{S^2_b\subset S^4_b}\left[
    \quiver{
      \node (Nf) at (0,0) [color-group] {$N$};
      \node (Kn1) at (1.4,0) [color-group,inner sep=.2pt] {$K_{n-1}$};
      \node (dots) at (2.7,0) {${\cdots}$};
      \node (K1) at (4,0) [color-group,inner sep=.2pt] {$K_1$};
      \draw[->-=.55] (K1) -- (dots);
      \draw[->-=.55] (dots) -- (Kn1);
      \draw[->-=.55] (Kn1) -- (Nf);
      \node (4d) at (-.55,0) {4d};
      \draw [color=gray, dashed, rounded corners]
      (0.5,.3) rectangle (-.8,-.3);
      \node (2d) at (4.6,0) {2d};
      \draw [color=gray, dashed, rounded corners]
      (0.5,.3) rectangle (4.85,-.3);
    }
  \right]
  = \vev*{\Vhat_{\mu_1}\Vhat_{\mu_2}\Vhat_{\mu_3}\Vhat_{\mu_4}}_{S^2}^{\text{generalization of Toda \ac{CFT}}}
\end{equation}
where $K_i=N_1+\dots+N_i$.
While the symmetry algebra is understood, the relevant (non-chiral) \acp{CFT} generalizing Toda \ac{CFT} is not.

Interestingly, conformal blocks of the affine $\Lie{SL}(2)$ algebra are related to conformal blocks of the Virasoro algebra with additional degenerate vertex operators, as pointed out early on in an \ac{AGT} setting in~\cite{0912.1930}.
This, and its $N>2$ analogues, leads to some identifications between the two types of surface operators up to a suitable integral kernel~\cite{1506.07508,1806.08270}.

One should be careful in reading some early 2010's literature on surface operators in the \ac{AGT} context, as the two types of surface operators are hard to distinguish in the $\lie{su}(2)$ case.
The ``codimension~$2$'' orbifold $\CC\times(\CC/\ZZ_M)$ considered here should also not be confused with the ``codimension~$4$'' orbifold $\CC^2/\ZZ_M$ discussed on \autopageref{par:local-orbifold}, for which $\ZZ_M$~rotates both factors.
Just as for the $\CC^2/\ZZ_M$ orbifold, the $\CC\times(\CC/\ZZ_M)$ orbifold ought to arise as a limit of a 5d gauge theory on $\CC^2\times_q S^1$ for a suitable root of unity limit of $q,t$.

\subsection{\label{ssec:wall}Domain walls}

The \ac{AGT} correspondence also allows for half-\ac{BPS} 3d operators that separate the 4d spacetime into two parts or give it a boundary.
A good starting point is~\cite{1003.1112}.

\paragraph{\label{par:wall-symm}Symmetry-breaking wall.}

A first construction~\cite{1003.1112} is to place a (tame) codimension~$2$ defect of~$\Xg$ on the equator of~$S^4_b$, times a closed loop $\gamma\subset C$.
The gauge theory description is understood in terms of the gauge group~$G_\gamma$ defined on \autopageref{par:line-dyon}.
The boundary conditions for the 4d vector multiplet are similar to those near a \ac{GW} surface defect, and they define a symmetry-breaking wall where gauge symmetries reduce to a subgroup $H\subset G_\gamma$.
On the \ac{CFT} side the situation is similar to loop operators, and one gets the Verlinde loop on~$\gamma$ constructed by inserting the vertex operator~$V_\alpha$ associated to the defect and moving it around~$\gamma$.

The continuous parameters of~$\alpha$ (of the codimension~$2$ defect) are \ac{FI} parameters of the unbroken gauge symmetry~$H$ on the wall.
General vertex operators can also include a discrete part, and correspondingly tame codimension~$2$ defects that are not full can be dressed with additional codimension~$4$ defects living on their world-volume.  In the present construction this leads to Wilson loop operators in representations $R_1$ and~$R_2$ of~$H$ on the two circles of \autoref{ssec:line}.
These loops are stuck on the domain wall unless~$R_1$ (resp.~$R_2$) are representations of~$G$ itself.

\paragraph{Janus wall.}

Our second construction does not involve codimension~$2$ or~$4$ operators.
Instead, we place $\Xg$ on $S^4_b\times C$ with the complex structure of~$C$ varying with the latitude of~$S^4_b$~\cite{1003.1112}.
This preserves half of the supersymmetry and in the limit where the variation happens sharply at the equator (or a parallel) we get a so-called Janus domain wall~\cite{hep-th/0304129,hep-th/0407073,hep-th/0506265,hep-th/0603013,0705.0022,0804.2907} in the 4d theory.
This is a half-\ac{BPS} interface between class~S theories with different gauge couplings.
The partition function with this interface has the usual factorized form~\eqref{ZS4-localized} with holomorphic and antiholomorphic contributions from the poles, but the gauge couplings used in each factor are not complex conjugates.
Correspondingly, the \ac{CFT} correlator~\eqref{Liouville-into-blocks} changes to using different complex structures for the holomorphic and antiholomorphic factors.

\paragraph{\label{par:wall-dual}S-duality wall.}

Tuning gauge couplings we can get theories that are S-dual.  By switching to the same S-duality frame on both sides we get a 3d operator called the S-duality wall~\cite{0807.3720} that has the same theory (and same gauge couplings) on both sides.
Inserting an S-duality wall in a 4d theory amounts on the 2d side to performing a modular transformation (fusion, braiding, S-move) on holomorphic (or antiholomorphic) conformal blocks.
This is related to special cases~\cite{1103.5748,1302.0015,1304.6721,1305.0937} of the 3d/3d correspondence we discuss in \autoref{ssec:dim-3d}.
The S-duality wall has an interplay with loop operators: it translates in a suitable sense from Wilson loops on one side of the wall to 't~Hooft loops on the other side.

For instance, the S-duality wall of 4d $\Nsusy=4$ \ac{SYM} is equivalent to coupling a 3d $\Nsusy=4$ theory $T[G]$ to \ac{SYM} on both sides of the wall~\cite{0807.3720}, and the S-move kernel is expected~\cite{1009.0340} to match the $S^3_b$~partition function of~$T[G]$.  For $\lie{g}=\lie{su}(N)$, the wall theory is a 3d $\Nsusy=4$ linear quiver,
\begin{equation}
  Z_{S^3_b}\Bigl[
    \quiver{
      \node (Nf) at (0,0) [flavor-group] {$N$};
      \node (N1) at (1.2,0) [color-group,inner sep=.2pt] {$N{-}1$};
      \node (dots) at (2.4,0) {${\cdots}$};
      \node (1) at (3.5,0) [color-group,inner sep=.2pt] {$1$};
      \draw (1) -- (dots);
      \draw (dots) -- (N1);
      \draw (N1) -- (Nf);
    }
  \Bigr]
  = \text{($W_N$ algebra S-move kernel).}
\end{equation}

The known braiding kernel~\cite{hep-th/9911110,math/0007097} for Virasoro four-point conformal blocks led to a description~\cite{1202.4698} of the S-duality wall of 4d $\Nsusy=2$ $\SU(2)$ \ac{SQCD} with $N_f=4$, as 3d $\Nsusy=2$ $\SU(2)$ \ac{SQCD} with $N_f=6$, suitably coupled to the 4d theories on both sides.

Bootstrapping the braiding kernel of $W_N$~four-point conformal blocks led to a description~\cite{1512.09128} for the S-duality wall of $\SU(N)$ \ac{SQCD} as $\Lie{U}(N-1)$ \ac{SQCD} (with a 3d monopole superpotential understood in~\cite{1703.08460}), coupled by cubic superpotentials to the 4d theories on both sides of the wall,
\begin{equation}
  Z \!\left[
    \quiver{
      \node (1)   [color-group] {$\scriptstyle\Lie{U}(N-1)$};
      \node (R)   [flavor-group, right=1em of 1] {$2N$};
      \node (A)   [color-group, above=3ex of 1] {$N$};
      \node (B)   [color-group, below=3ex of 1] {$N$};
      \draw[->-=.55] (R) -- (1);
      \draw[->-=.7] (1) -- (A);
      \draw[->-=.7] (1) -- (B);
      \draw (A) -- (R) node [pos=.2, above right=-4pt] {4d};
      \draw (B) -- (R) node [pos=.2, below right=-4pt] {4d};
    }
  \right]
  =
  \text{($W_N$ algebra braiding kernel)}.
\end{equation}
Just like the $T[G]$ theories, this 3d $\Nsusy=2$ theory (in isolation, after decoupling the 4d fields) admits numerous dualities~\cite{1909.02832}.
Duality walls of 5d theories were studied in~\cite{1503.05159,1506.03871}.

\paragraph{\label{par:wall-BCFT}Boundary \acs{CFT}.}

We mentioned orbifolds earlier.  Instead of orbifolding the 4d space one can orbifold by a $\ZZ_2$ symmetry that acts as a reflection with respect to the equator of~$S^4_b$ and a reflection on the Riemann surface (so as to preserve chirality of~$\Xg$).  This leads to an \ac{AGT} correspondence for Riemann surfaces with boundaries and for non-orientable surfaces~\cite{1708.04631,1710.06283}.
On the gauge theory side, one obtains simultaneously some gauge fields defined on the squashed hemisphere~$HS^4_b$ and others on the squashed projective space~$\RP^4_b$.
The hemisphere partition function was evaluated in~\cite{1611.04804} while the projective space partition function is obtained in~\cite{1708.04631} by the gluing technique explored further in~\cite{1807.04274,1807.04278}.

The inclusion of boundary operators has not been fully elucidated.
It would also be interesting to go beyond the $\lie{g}=\lie{su}(N)$ case treated so far, for instance understanding how the double-cover construction discussed below~\eqref{couplings-redef} interacts with the $\ZZ_2$~quotient that produces the Riemann surface boundaries.

\ifjournal\else
\apartehere
\leavevmode\vspace{-\baselineskip}
\fi

\section{\label{sec:dim}Other geometries}

Class~S theories $\TgCD$ are obtained by compactifying the 6d $(2,0)$ theory of type~$\lie{g}$ on $M_4\times C$ with $C$ a Riemann surface with punctures which extra data~$D$.  So far we have extensively discussed the case $M_4=S^4_b$ and its building block $M_4=\RR^4_{\epsilon_1,\epsilon_2}$, for which the partition function is equal to a 2d \ac{CFT} correlator or conformal block, respectively.

We first discuss 5d lifts of these 4d observables to (deformations of) $\RR^4\times S^1$, $S^4\times S^1$, and~$S^5$, which are connected to $q$-deformations\footnote{The parameter~$q$ appearing in the 5d lift is unrelated to the gauge couplings parameters describing the complex structure of~$C$.  We will actually not need a notation for this gauge coupling any longer.} (\autoref{ssec:dim-5d}).
We then change the geometry, first relating the supersymmetric index, which is the partition function on $M_4=S^3\times S^1$, to (a generalization of) a $q$-Yang--Mills \ac{TQFT} correlator (\autoref{ssec:dim-index}), then compactifying instead on products $M_3\times C_3$ and $M_2\times C_4$ in which the ``internal'' manifold~$C$ is a hyperbolic three-manifold (\autoref{ssec:dim-3d}) or a four-manifold (\autoref{ssec:dim-other}).
In this last subsection we also mention generalizations with less supersymmetry and a few geometric setups that have been less fruitful.

\startaparte[after skip=-20pt,title={\ac{AGT} correspondence in some other geometries}]
\begin{itemize}
\item For 5d $\Nsusy=1$ lifts of class~S theories, $Z_{\inst}$ and $Z_{S^5}$ match $q$-deformed Toda conformal blocks and correlators.  A lift to 6d matches $(q,t)$-deformed Toda theory.
\item The $S^3\times S^1$ partition function (superconformal index) of $\TgCD$ matches a \ac{TQFT} correlator on~$C$.
\item Twisted reductions of $\Xg$ on~$C_3$ are 3d $\Nsusy=2$ theories whose $Z_{S^2\times S^1}$, $Z_{S^3}$ or lens space partition functions match complex $\lie{g}_{\CC}$ Chern--Simons theory on~$C_3$ at level~$0$, $1$ or~$k$.
\item The 6d $(1,0)$ \ac{SCFT} of M5~branes probing a $\ZZ_k$ singularity defines class~S$_k$ 4d $\Nsusy=1$ theories similar to class~S.
\end{itemize}
\stopaparte

\subsection{\label{ssec:dim-5d}Lift to 5d and \(q\)-Toda}

Here we briefly survey how lifting the 4d $\Nsusy=2$ theories to 5d $\Nsusy=1$ amounts to a $q$-deformation of the 2d theories.  For a review, see~\cite{1608.02968}.\footnote{I thank Fabrizio Nieri for answering my questions thoroughly.}

\paragraph{Instanton partition functions.}
The 4d $\Nsusy=2$ Omega background used to define Nekrasov's instanton partition function~\cite{hep-th/0206161,hep-th/0306238} is conveniently expressed in terms of a 5d $\Nsusy=1$ lift: placing the theory on $\RR^4\times S^1$ with twisted boundary conditions around~$S^1$ such that $\RR^4$ rotates by $q$ and~$t$ in two two-planes, and with an additional twist by an R-symmetry to preserve some supersymmetry.  More precisely, this definition for $|q|=|t|=1$ can be extended to complex $q,t$ by turning on additional supergravity fields.
The 5d lift deforms all factors in $Z_{\inst}$ from rational functions to trigonometric functions of masses and Coulomb branch parameters.  It is natural to ask how the Toda \ac{CFT} side of the \ac{AGT} correspondence can be deformed to accomodate for this.

One finds that the 5d (also called K-theoretic) $Z_{\inst}$ is a chiral block for a $q$-deformed W-algebra~\cite{0910.4431,1004.5122,1005.0216,1308.2068,1403.7016,1409.6313,1412.3395,1906.06351}: the relevant deformations of the Virasoro algebra and of W-algebras were constructed long ago~\cite{q-alg/9505025,q-alg/9507034,q-alg/9508009,q-alg/9508011} (see \cite{1512.08533} for a modern construction).\footnote{As in 4d, working with $\Lie{U}(N)$ rather than $\SU(N)$ gauge groups makes $Z_{\inst}$ more tractable; correspondingly one works with the Ding--Iohara algebra, a slight extension of (the universal envelopping algebra of) the $q$W-algebra of type~$\lie{g}$~\cite{1106.4088}.}
When mass and Coulomb branch parameters are suitably quantized the equality can be proven using Dotsenko--Fateev integral representations of $qW_N$ conformal blocks~\cite{1105.0948,1309.1687,1403.3657,1506.04183,1602.01209} (also used in~\cite{1903.10817}).
See also~\cite{1308.2465,1309.4775}.

The 5d $\Nsusy=1$ quiver gauge theories admit realizations in terms of webs of $(p,q)$ fivebranes in IIB string theory~\cite{0906.0359}.  Applying S-duality exchanges the role of D5 and NS5~branes, thus equating $Z_{\inst}$ for a $\SU(N)^{M-1}$ linear quiver gauge theory to an $\SU(M)^{N-1}$ one (see~\cite{1412.8592} for a proof for $M=N=2$).  This 5d spectral duality (also called fiber-base duality~\cite{hep-th/9706110}) relates in general chiral blocks of different $qW_N$ theories~\cite{1112.5228,1606.03036}, it implies certain instances of 3d mirror symmetry~\cite{1712.08140}, and relations between spin chains~\cite{1905.09921}.
When reading the literature, one should keep in mind which of the two spectral duality frames is adopted, see~\cite{1612.07590} for a nice explanation.

The 4d case is retrieved as the limit $q\to 1$ with $t=q^{-\beta}$ and fixed $-\beta=b^2=\epsilon_1/\epsilon_2$ giving the 4d deformation parameters.
Other interesting limits than $q,t\to 1$ exist, especially taking $q$ and~$t$ a $k$-th roots of unity one obtains Nekrasov partition functions on $\CC^2/\ZZ_k$ \ac{ALE} space studied in~\cite{1111.2803,1211.2788,1308.2068,1408.4216,1409.3465,1512.01084,1705.03628}.  Another simplifying limit is the Hall-Littlewood limit $q\to 0$~\cite{1512.08016,1703.10990}.

An unrelated application of $Z_{\inst}$ and $qW_N$ conformal blocks is to construct solutions of $q$-Painlev\'e equations~\cite{1706.01940,1708.07479,1811.03285,1908.01278}, as in the 4d case.

\paragraph{Compact partition functions.}
Let us now glue instanton partition functions together.
While the partition function on~$S^4$ involves a pair of instanton contributions from the two fixed point of the supercharge squared, the partition function of 5d $\Nsusy=1$ theories on~$S^5$ combines three K-theoretic instanton partition functions because the supercharge has three fixed circles~\cite{1203.0371}.  Schematically,
\begin{equation}
  Z_{S^5} = \int \rmd a\,Z_{\pert} Z_{\inst,1} Z_{\inst,2} Z_{\inst,3} .
\end{equation}
The squashed $S^5$ has three axis lengths $\omega_1,\omega_2,\omega_3$ and here the different $Z_{\inst,i}$ are computed in the $\Omega$ background with parameters $(q,t)$ given by $(\frac{\omega_2}{\omega_1},\frac{\omega_3}{\omega_1})$, $(\frac{\omega_1}{\omega_2},\frac{\omega_3}{\omega_2})$, $(\frac{\omega_1}{\omega_3},\frac{\omega_2}{\omega_3})$, respectively.

The picture that emerges~\cite{1303.2626,1310.3841,1312.1294} is that there exists a $q$-deformed version of Toda \ac{CFT}, called $q$-Toda theory,\footnote{One should be careful that many papers talk about $q$-Liouville or $q$-Toda theory even when they only consider chiral blocks, which only involve the $q$-Virasoro and $qW_N$ symmetry algebras.}
that has $qW_N$ symmetry and whose correlators should match with $S^5$ partition functions.
The fact that three chiral factors need to be combined leads to a remarkable ``modular triple'' of $q$-Virasoro algebras~\cite{1710.07170} (for $N=2$), similar to the modular double combining $U_q(\lie{sl}(2))$ with $q=e^{2\pi ib^2}$ and $q=e^{2\pi i/b^2}$ in 2d \ac{CFT}.
A non-local Lagrangian for $q$-Liouville is proposed in~\cite{1710.07170}.

Half-\ac{BPS} operators with 3d $\Nsusy=2$ supersymmetry played an important early role right from the start.  On the squashed $S^5$ they can be inserted along three distinct~$S^3$ that intersect pairwise along~$S^1$.
The first explorations of the correspondence for $Z_{S^5}$ concerned the case of a single 3d operator in a simple 5d bulk theory, which can be obtained by Higgsing a larger 5d theory~\cite{1303.2626,1312.1294,1605.07029}.
Just as the analogous surface operators in the standard \ac{AGT} correspondence, these 3d operators correspond to degenerate $q$-Toda \ac{CFT} vertex operators.
They are useful to bootstrap structure constants of $q$-Toda, and show up in a Higgs branch localization expression of the instanton and $S^5$~partition functions~\cite{1711.06150,1807.11900,1812.11961}, again completely analogous to the 4d story~\cite{1508.07329,1612.04839}, albeit more technically involved.  A mathematical take on this is in~\cite{1911.02963}.

Codimension~$4$ operators of the 5d theory, specifically Wilson loops, are studied in~\cite{1907.03838}; they translate in $q$-Toda to stress tensor and higher-spin operator insertions.

\paragraph{Geometric setup.}

The correspondence is partially understood geometrically through 6d $(2,0)$ little string theories.  Contrarily to the \acp{SCFT}~$\Xg$, little string theories can only be reduced on zero-curvature surfaces, so the choice of Riemann surface and punctures is restricted.
When reducing on a cylinder with full punctures at the two ends, one would expect a 4d~reduction but the string winding modes give instead a 5d $\Nsusy=1$ theory on the T-dual circle times~$\RR^4$.
The correct limits to reproduce the 4d/2d \ac{AGT} correspondence and a dual version were discussed in~\cite{1701.03146,1912.09969}.

It is not clear at this point what $q$-Toda theory really is, in particular whether it is fundamentally a 2d theory that can only be placed on zero-curvature surfaces, as suggested by the little string construction, or whether it should be thought as a 1d theory in order to obtain an effective Lagrangian~\cite{1710.07170}.

\paragraph{Elliptic lift.}

Lifting one dimension up, 6d partition functions on $\RR^4\times T^2$ (twisted) and $S^5\times S^1$ (superconformal index)~\cite{1307.7660} are related to the elliptic deformation $(q,t)$-Toda: see~\cite{1309.4775,1507.00261,1511.00458,1511.00574,1512.06701,1603.05467,1604.08366,1607.08330,1608.02969,1608.04651,1611.07304,1701.08541,1703.04614,1711.07499,1712.08016,1801.04943}.

\subsection{\label{ssec:dim-index}Superconformal index and 2d \(q\)-\acsfont{YM}}

We now move on to partition functions of 4d $\Nsusy=2$ class~S theories on $M_4=S^3\times S^1$.

\paragraph{Supersymmetric index.}
See the review~\cite{1412.7131} for superconformal class~S theories and \cite{1608.02965}~for general 4d $\Nsusy=1$ theories.  We shall not write too much here, but we rather point to another course in this school~\cite{2006.13630}.
The \ac{AGT} relation to $q$-\ac{YM} is also surveyed briefly in~\cite{1608.02964}.

The $S^3\times S^1$ partition function is defined and computable for 4d $\Nsusy=1$ theories with an anomaly-free $\Lie{U}(1)$ R-symmetry.
Up to a factor involving the Casimir energy of the theory, expressible in terms of $a$ and~$c$ anomalies, the partition function coincides with the supersymmetric index, defined to be the Witten index of the theory quantized on $S^3\times\RR$.
Once refined by fugacities~$u_i$ for mutually commuting rotations, flavour, and R-symmetries (with charges~$K_i$), the index is written as
\begin{equation}
  \cI(u) = \Tr\biggl[(-1)^F e^{-\beta \tilde{H}} \prod_i u_i^{K_i}\biggr]
\end{equation}
where $(-1)^F$ counts bosonic and fermionic states with opposite signs, $\tilde{H}=\{Q,Q^\dagger\}$ for some supercharge~$Q$, and $\cI$ is $\beta$-independent thanks to cancellations between bosonic and fermionic states when $\tilde{H}\neq 0$.
This simplification means that $\cI(u)$ counts (with signs) short representations of the supersymmetry algebra.
Fugacities $u_i$ are encoded in the $S^3\times S^1$ partition function as holonomies around the~$S^1$ for background gauge fields coupled to the given symmetry: in particular, fugacities $(p,q)$ for two combinations of rotations and R-symmetries can be understood as a non-trivial fibration of~$S^3$ over~$S^1$.

The index formally does not depend on any continuous parameter beyond these:\footnote{In some contexts, this formal invariance of the index fails, which gives wall-crossing phenomena.} it is an \ac{RG} flow invariant and is independent of gauge couplings for instance, thus can be easily computed in any weakly-coupled Lagrangian description.  In this way $\cI(u)$ reduces to a simple signed count of local gauge-invariant operators built from the elementary fields in any given Lagrangian description (in other dimensions nonperturbative objects must be included).
Being an eminently computable \ac{RG} flow invariant makes the index a powerful window into nonperturbative physics of 4d $\Nsusy=1$ gauge theories, especially their \ac{IR} dualities.

Computing the index is much harder if we have no Lagrangian description, but part of the structure remains: if a theory $T$ is defined by gauging a common flavour symmetry~$G$ of two theories $T_1,T_2$, then the indices are related schematically as
\begin{equation}\label{index-gluing}
  \cI[T](a_1,a_2) = \int [\rmd z]_G \,\cI_{\text{vec}}(z)\,\cI[T_1](a_1,z) \,\cI[T_2](a_2,z) ,
\end{equation}
where we hid the $p,q$ dependence but kept explicit the fugacities $a_1$ and~$a_2$ for flavour symmetries of $T_1$ and~$T_2$ commuting with~$G$, which become flavour symmetries of~$T$.  The integral over the fugacity~$z$ for the symmetry~$G$ is done with a suitable measure~$\cI_{\text{vec}}$, which from the localization point of view is the vector multiplet one-loop determinant.
In fact, \eqref{index-gluing} gives a way to compute the index of a non-Lagrangian theory: embed it into a larger theory that is dual to a Lagrangian gauge theory, whose index is computable~\cite{1003.4244}.

\paragraph{Class~S.}
We now specialize to class~S theories, and specifically to superconformal ones.
Since the index cannot depend on gauge couplings, it only depends on the topology of the Riemann surface~$C$ and the type of punctures.
Thus, compared to the standard \ac{AGT} correspondence, the 2d \ac{CFT} side should be replaced by a \ac{TQFT}, as worked out in~\cite{0910.2225}.
Consider the theory $\TgCD$.
A flavour symmetry is associated to each puncture $z_i$, $i=1,\dots,n$, and we turn on corresponding fugacities~$a_i$.
For any pants decomposition of~$C$ we can express $\TgCD$ as the result of gauging flavour symmetries of a collection of tinkertoys (isolated \acp{SCFT}) associated to three-punctured spheres.
Through \eqref{index-gluing}, the index then reduces to an integral of superconformal indices of tinkertoys.
This precisely mimics the structure of correlators in a \ac{TQFT}:
\begin{equation}
  \cI\bigl[\TgCD\bigr](a_i) = \vev*{\cO_{D_1}(a_1)\dots\cO_{D_n}(a_n)}_{\text{some \ac{TQFT}}}
\end{equation}
for suitable operators~$\cO_D$ that depend on the type of puncture.

In analogy to Liouville \ac{CFT} bootstrap, which relied on using degenerate vertex operators that correspond in gauge theory to surface operators, one can bootstrap the index of all tinkertoy building blocks using surface operators~\cite{1207.3577}.  Adding a surface operator to the index corresponds to acting with a difference operator~$\Theta$ on fugacities associated to any one of the punctures, and by topological invariance it does not matter which puncture.  Expressing the result in an eigenbasis of~$\Theta$ labeled by representations~$\lambda$ of~$\lie{g}$ eventually gives
\begin{equation}
  \cI\bigl[\TgCD\bigr](a_i) = \sum_\lambda (C_\lambda)^{2g-2} \phi_\lambda^{D_1}(a_1)\dots\phi_\lambda^{D_n}(a_n)
\end{equation}
for some structure constants $C_\lambda(p,q,t)$ and wave functions $\phi_\lambda^D(p,q,t;a)$.\footnote{Here we introduced a fugacity~$t$ for the additional R-symmetry of 4d $\Nsusy=2$ theories.}  Wave functions for arbitrary punctures are related to those for full punctures by taking suitable residues in flavour fugacities~\cite{1203.5517}.  The sum may diverge if there are too few punctures or if they are too ``small'', signalling either that the given class~S theory does not exist or that the index is not sufficiently refined because there are additional flavour symmetries not associated to any of the punctures.

The wave functions can be computed order by order in $p,q,t$, but are not known in closed form.
In the Schur limit $q=t$ correlators are functions of~$q$ only ($p$-dependent terms are $Q$-exact), wavefunctions are proportional to Schur polynomials, and the corresponding \ac{TQFT} is $q$-deformed 2d \acl{YM} theory~\cite{1104.3850}.
In the more general Macdonald limit $p=0$, wavefunctions are essentially Macdonald polynomials in $q,t$ and the \ac{TQFT} must be deformed by changing the measure in the path integral of $q$-\ac{YM} theory~\cite{1110.3740}.
The Hall-Littlewood limit $p=q=0$ turns Macdonald polynomials to the easier Hall-Littlewood polynomials, which only depend on~$t$.
The Coulomb limit ($t,p\to 0$ with fixed $p/t$ and~$q$) is also interesting.
The 2d \ac{TQFT} description, which remains quite implicit for general $p,q,t$, was derived starting from the 6d theory~$\Xg$ in~\cite{1808.06744} (earlier derivations in~\cite{1206.5966,1210.2855,1505.06565} did not account for instanton corrections).

The correspondence is tested and extended in natural ways:
inserting Wilson--'t~Hooft loops at the poles of~$S^3$ and correspondingly loops in $q$-\ac{YM}~\cite{1201.5539,1205.0069,1603.02939,1701.04090},
inserting general surface operators~\cite{1303.4460,1401.3379,1407.4587,1606.01041,1905.02724},
replacing $S^3$ by the Lens space $L(p,1)=S^3/\ZZ_p$~\cite{1109.0283,1301.7486,1605.06528},
taking $C$ to have non-zero area~\cite{1207.3497},
generalizing to D-type gauge groups and non-simply-laced ones (using outer automorphism twists)~\cite{1212.0545,1212.1271}.
The relation with Hilbert series of instanton moduli spaces is explored in~\cite{1205.4722,1205.4741}.
The superconformal index of many \ac{AD} type theories is also known by now in the Macdonald limit~\cite{1505.05884,1505.06205,1509.05402,1509.06730,1705.07173}.
The key open question in this direction seems to be getting a handle on the full parameter space $(p,q,t)$ rather than its $p=0$ Macdonald slice.

\subsection{\label{ssec:dim-3d}3d/3d correspondence}

So far we have reduced the 6d $(2,0)$ theory of type~$\lie{g}$ with a partial topological twist along a Riemann surface.
Reducing it instead on $M_3\times C_3$, with a twist along a three-manifold~$C_3$, gives a 3d $\Nsusy=2$ gauge theory on~$M_3$.  One can add codimension~$2$ operators of the 6d theory to get analogues of punctures: boundary conditions along knots $K_1\subset C_3$ (which we leave implicit in our notation).  This defines a large class of 3d $\Nsusy=2$ gauge theories\footnote{This is not a standard notation; sometimes $\Theory(\lie{su}(N),C_3,K_1)$ is denoted $T_N[C_3\setminus K_1]$.}~$\Theory(\lie{g},C_3,K_1)$.
Their supersymmetric partition functions match $K_1\subset C_3$ partition functions of complexified Chern--Simons theory.
See~\cite{1412.7129,Pei:2016rmn,1608.02961} for reviews and \cite{1106.4550,1108.4389,1110.2115,1112.5179,1210.8393,1211.1986,1211.3730,1301.0192,1301.5902,1302.0015,1303.3709,1303.5278,1304.6721,1305.0937,1305.2891,1305.5451,1403.5215,1405.3663,Chung:2014xca,1409.0857,1410.1538,1501.01310,1503.04809,1510.03884,1510.05011,1512.07883,1602.05302,1603.01149,1604.02688,1610.09259,1701.06567,1702.05045,1706.06292,1803.00855,1803.04009,1804.02368,1806.03039,1808.02797,1809.10148,1905.01559,1907.03430,1909.05873,1909.11612,1910.01134,1910.14086,1911.08456,1911.10718,1912.13486} for works on this correspondence and its applications.

Natural building blocks for $C_3\setminus K_1$ are tetrahedra, and each triangulation of $C_3\setminus K_1$ yields an explicit gauge theory description of the 3d $\Nsusy=2$ theory as the \ac{IR} limit of an abelian Chern--Simons theory (and deformations by masses or other parameters).
More precisely, this description misses parts of the theory $\Theory(\lie{g},C_3,K_1)$, as pointed out in~\cite{1405.3663,Chung:2014xca}, and these additional branches are still under investigation (see e.g.~\cite{1911.08456}).
Similarly to how changing pants decompositions in the 4d/2d correspondence amounts to S-duality,
Pachner's 2-3 move for triangulated $3$-manifolds, which trades two neighboring tetrahedra for three tetrahedra covering the same part of the manifold, yields 3d $\Nsusy=2$ dualities.
Contrarily to S-duality, these are not all-scale dualities but only \ac{IR} dualities.

\paragraph{Statement of the correspondence.}
The 3d/3d \ac{AGT} correspondence was formulated in~\cite{1106.4550,1108.4389}, after several papers treating less general geometries: either reducing $\Theory(\lie{g},C_3,K_1)$ further on~$S^1$~\cite{1006.0977} (getting a 2d $\Nsusy=(2,2)$ theory), or taking $C_3$ to be a mapping cylinder or torus (Riemann surface fibered over an interval or circle)~\cite{1103.5748,1104.2589,1106.3066,1305.0937}.
The partition function of $\Theory(\lie{g},C_3,K_1)$ on certain manifolds~$M_3$ is equal to the partition function on $C_3\setminus K_1$ of Chern--Simons theory with a gauge group~$G_{\CC}$ whose Lie algebra is the complexification of~$\lie{g}$.  This, in turn, provides invariants of knots and of $3$-manifolds.
Complex Chern--Simons theory depends on levels~$(k,\sigma)$, one quantized $k\in\ZZ$ and one continuous $\sigma\in\RR\cup i\RR$, related to the choice of~$M_3$.
Its action is straightforward,
\begin{equation}
  S = \frac{k+i\sigma}{8\pi}\int_{C_3}\Tr\Bigl(\cA\wedge \rmd\cA+\tfrac{2}{3}\cA\wedge\cA\wedge\cA\Bigr)
  + \frac{k-i\sigma}{8\pi}\int_{C_3}\Tr\Bigl(\cAbar\wedge \rmd\cAbar+\tfrac{2}{3}\cAbar\wedge\cAbar\wedge\cAbar\Bigr) ,
\end{equation}
where $\cA=A+i\Phi$ is a complex gauge field.
However, defining Chern--Simons theory completely is subtle when the gauge group is noncompact~\cite{1001.2933}, and in fact the 3d/3d correspondence helps define it for complex gauge group~$G_{\CC}$~\cite{1409.0857,1608.02961,1710.04354,1811.06853}.
See also \cite{1509.00458,1511.05628} for a few applications of complexified Chern--Simons theory.

The squashed sphere partition function ($M_3=S^3_b$) corresponds to Chern--Simons at level $k=1$~\cite{1108.4389,1305.2891}.
The supersymmetric index ($M_3=S^2\times_q S^1$) corresponds to Chern--Simons at level $k=0$~\cite{1112.5179,1305.0291,1305.2429}.\footnote{I don't know if the differences between~\cite{1305.0291,1305.2429} have been resolved.}
The partition function on a squashed lens space $M_3=L(k,1)_b$ corresponds to a general Chern--Simons level~$k$~\cite{1409.0857,1710.04354}.
(These supersymmetric partition functions and more are reviewed in~\cite{1908.08875}.)
\begin{equation}
  \begin{aligned}
    Z_{S^2\times_q S^1}[\Theory(\lie{g},C_3,K_1)] & = Z_{C_3}\bigl[G_{\CC}\text{ at levels } \bigl(0,\sigma\bigr)] , \ q=e^{2\pi/\sigma} \\
    Z_{S^3_b}[\Theory(\lie{g},C_3,K_1)] & = Z_{C_3}\bigl[G_{\CC}\text{ at levels } \bigl(1,\tfrac{1-b^2}{1+b^2}\bigr)] \\
    Z_{L(k,1)_b}[\Theory(\lie{g},C_3,K_1)] & = Z_{C_3}\bigl[G_{\CC}\text{ at levels } \bigl(k,k\tfrac{1-b^2}{1+b^2}\bigr)] \\
  \end{aligned}
\end{equation}
All three can be decomposed into holomorphic blocks~\cite{1111.6905,1211.1986,1410.1538}, which are partition functions on the Omega background $\RR^2\times_q S^1$ or equivalently the cigar $D^2\times_q S^1$, and are wave functions on the Chern--Simons side.
A semi-classical version of this is that the set of supersymmetric vacua on $\RR^2\times S^1$ matches the space of flat $G_{\CC}$~connections on $C_3\setminus K_1$ with suitable boundary conditions along~$K_1$.

\paragraph{Boundaries and generalizations.}
When $C_3$ has 2d boundaries (on which the knot $K_1$~can end), the 3d $\Nsusy=2$ theory lives at the boundary of (and is coupled to) the 4d $\Nsusy=2$ class~S theory associated to $\del C_3$~\cite{1003.1112,1108.4389,1205.0069,1304.6721}.  In particular, when $C_3$ is a cobordism, namely $\del C_3$ consists of two disconnected components, it is more natural to think of the 3d $\Nsusy=2$ theory as a domain wall between the two corresponding 4d $\Nsusy=2$ class~S theories which are only coupled through their common 3d boundary.  The construction is thus functorial with respect to gluing.  One particularly simple example of the setup was described in~\cite{1003.1112}: consider $C_3=\RR\times C$ where the complex structure of~$C$ varies along the~$\RR$ direction; then on the gauge theory side we have the 4d $\Nsusy=2$ theory $\Theory(\lie{g},C)$ with a Janus domain wall defined by varying the 4d gauge couplings along one direction.
Further works in this direction include \cite{1405.3663,1702.05045}.

As known since Witten's~\cite{Witten:1988hf}, many knot invariants can be expressed as partition functions or other observables of gauge theories.  For a sample of references, see~\cite{1106.4789,1111.7035,1203.0667,1204.5975,1208.2282,1211.3730,1412.8455,1506.06695,1604.01158} and the review~\cite{1510.01795}, as well as calculations of knot invariants through the study of families of superpolynomials in~\cite{1303.2578,1304.1486,1310.2240,1310.7361,1310.7622,Sleptsov-2014tfa,1403.8087,1408.3076,Sleptsov:2014aba,1505.06221,1606.06015,Morozov:2016bcg,1712.03647,1902.04140,1906.09971}.

A different topological twist realizes homological invariants of knots and three-manifolds (monopole/\allowbreak Heegaard Floer and Khovanov--Rozansky homology) in terms of 3d $\Nsusy=2$ theory $\Theory(\lie{g},C_3,K_1)$ partially topologically twisted on a Riemann surface~\cite{1405.3663,1602.05302,1806.03039}.
Holographic calculations~\cite{1110.6400,1401.3595,1407.0403,1409.6206,1808.02797,1905.01559,1909.05873,1907.03430,1909.11612} probe or use the correspondence at large~$N$.
Dimensional reduction from the 4d $\Nsusy=2$ superconformal index to the 3d $\Nsusy=2$ sphere partition function translates to dimensional oxydation from 2d $q$-\ac{YM} to a hyperbolic manifold~\cite{1109.0283,1203.5792}.
The 3d/3d correspondence can also be refined using higher-form symmetries~\cite{1910.14086}.
Half-\ac{BPS} 2d $\Nsusy=(0,2)$ boundary conditions and domain walls of 3d $\Nsusy=2$ theories were studied in~\cite{1302.0015} and subsequent papers, and one offshoot is the 2d/4d correspondence~\cite{1306.4320} discussed next.

\subsection{\label{ssec:dim-other}Some more geometric setups}

\paragraph{2d/4d correspondence.}

Reducing the 6d $(2,0)$ theory on $T^2\times C_4$ with a partial topological twist along the four-manifold~$C_4$ gives 2d $\Nsusy=(0,2)$ supersymmetric gauge theories.
This setting has been somewhat less studied, owing to how the topology of four-manifolds is more complicated than for the 4d/2d and 3d/3d correspondences.
The relevant twist of the M5~brane action was constructed explicitly in~\cite{1311.3300,1406.4499}.

A dictionary \`a la \ac{AGT} is proposed in~\cite{1306.4320}: the Vafa--Witten partition function on~$C_4$~\cite{hep-th/9408074} is the superconformal index of the 2d $(0,2)$ theory, while 4d Kirby calculus translates to dualities of 2d $\Nsusy=(0,2)$ theories such as~\cite{1310.0818}.  Four-manifolds with a boundary $\del(C_4)$ correspond to domain walls between the 3d $\Nsusy=2$ theories associated to $\del(C_4)$ by the 3d/3d correspondence~\cite{1302.0015} (see also~\cite{1509.00466}).
Four-manifolds of the form $\CP^1\times C$ are considered in~\cite{1505.07110} and provide an analogue of class~S theories.
This can be enriched further by inserting defects of the 6d $(2,0)$ theory.
There are several variants: compactifying on $S^2\times C_4$~\cite{1604.03606}, generalizing to 6d $(1,0)$ theories~\cite{1610.00718}, and using a different twist to connect $4$-manifold invariants to 2d $\Nsusy=(0,2)$ chiral correlators~\cite{1705.01645,1806.02470}.  The abelian case is studied in detail in~\cite{1905.05173}.
See also~\cite{1707.01515,1811.07884}.

\paragraph{Less supersymmetry.}

One can learn properties of strongly-coupled 4d $\Nsusy=1$ theories, for instance analogues of Seiberg dualities, from supersymmetry-breaking deformations of 4d $\Nsusy=2$ theories and S-duality~\cite{0907.2625,0909.1327,1105.3215,1303.0836,1305.5250,1307.7703,1309.5299,1310.0467,1409.3077}.
These developments have led to finding 4d $\Nsusy=1$ Lagrangian descriptions for 4d $\Nsusy=2$ \ac{AD} theories that admit no 4d $\Nsusy=2$ Lagrangian description, as done for instance in~\cite{1505.05834,1606.05632,1607.04281,1609.08156,1610.05311,1707.04751,1707.05113,1710.06469,1802.05268,1806.08353,1808.00592,1809.04906,1906.05088,1910.09568,1912.09348} (see also~\cite{1807.02785} with more supersymmetry).  They also lead to new 4d $\Nsusy=2$ that may be ``minimal'' in the sense of having the smallest central charges $(a,c)$~\cite{1806.08353}.

Another approach to getting 4d $\Nsusy=1$ theories is to consider more general compactifications of 6d $\Nsusy=(2,0)$ \acp{SCFT} that amount to placing M5~branes on a complex curve inside a Calabi--Yau three-fold~\cite{1111.3402,1112.5487,1203.0303,1212.1467,1303.0836,1307.5877,1307.7104,1310.7943,1311.2945,1409.1908,1409.7668,1609.08156}.  Generalizations of \ac{SW} geometry appear to exist~\cite{1409.8306}.
A further reduction yields 3d $\Nsusy=2$ theories~\cite{1911.00956}.

A particularly natural path to lower supersymmetry arises from orbifolds of the M-theory setup.
The 6d $(1,0)$ theory of M5~branes at a $\ZZ_k$ singularity has very interesting reductions to 4d $\Nsusy=1$ theories called class~S$_k$~\cite{1503.05159,1503.06217,1504.05988,1504.08348,1505.05053,1508.00915,1512.06079,1606.01041,1606.01653,1609.01281,1702.04740,1703.00736,1712.01288,1804.00680,1812.04637}, see \cite{1709.02496,1802.00620,1806.07620} for M5~branes probing more general ADE singularities.  In principle this leads to an analogue of the \ac{AGT} correspondence, but the $Z_{S^4}$ partition function of $\Nsusy=1$ theories suffers some ambiguities, and the instanton partition function is not known (see however a very interesting proposal~\cite{1812.11188}).

Beyond these orbifolded M5~branes, there exists a zoo of 6d $(1,0)$ theories constructed from F-theory~\cite{1502.05405,1502.06594}, reviewed in~\cite{1805.06467,1903.10503}.
Compactifying them further on a torus gives 4d $\Nsusy=2$ supersymmetry, reproducing many class~S theories~\cite{1504.08348}.
The set-up has also been studied on a Riemann surface~\cite{1604.03560,1610.09178,1806.09196,1910.03603} or on a torus with fluxes turned on~\cite{1702.04740,1709.02496,1802.00620,1806.07620} to get 4d $\Nsusy=1$ theories, and with surface operators~\cite{1801.00960,1806.09196,1808.09509}.
The reduction from 6d to 5d is also interesting~\cite{1703.02981}.

Two other rich families of 4d $\Nsusy=1$ \acp{SCFT} are bipartite quivers~\cite{1211.5139,1404.3752}, and D3~branes probing orientifolds of toric singularities~\cite{1210.7799,1307.0466,1307.1701,1506.03090,1612.00853}.

\paragraph{Miscellaneous.}

By fine-tuning (and analytically continuing) parameters, one gets the \ac{AGT} correspondence for minimal models \cite{1404.7075,1404.7094,1504.01925,1507.03540,1507.05426,1509.01000,1912.04407} and a ``finite'' version of \ac{AGT} on the mathematical side~\cite{1008.3655,1107.5073,1609.04406}.

Class~S 4d $\Nsusy=2$ theories have also been localized on other geometries: $S^2\times S^2$ corresponds to Liouville gravity~\cite{1411.2762}, $S^2\times S^1\times I$ to complex Toda \ac{CFT}~\cite{1701.03298}, $S^2\times T^2$ in~\cite{1703.04614}.

Instead of reducing M5~branes on product geometries, reducing D3~branes yields a 2d/2d correspondence \cite{1412.8302,1508.00469,1511.09462,1605.05869}, while M2~branes give a 1d/2d correspondence \cite{1410.8180,1503.03906}.
A 3d/3d $\Nsusy=1$ correspondence was also proposed in~\cite{1804.02368}.

\section{\label{sec:con}Conclusions}

\apartehere
There is plenty more to be said about the \ac{AGT} correspondence.
Most obviously we have not placed this correspondence in the wider context of the \ac{BPS}/\ac{CFT} correspondence
between 4d $\Nsusy=2$ gauge theories and integrable models underlying \ac{SW} geometry.
We have also omitted the connections to refined topological strings and matrix models.
\startaparte
\begin{itemize}
\item \textbf{\ac{BPS}/\ac{CFT} correspondence:} \ac{SW}~curves of quite general 4d $\Nsusy=2$ theories are spectral curves of integrable systems.
\item \textbf{Topological strings} give $Z_{\inst}$ for theories obtained by reducing $(p,q)$ $5$-brane webs, or IIB~geometric engineering.
\item \textbf{Matrix models} with logarithmic potentials yield $Z_{\inst}$~of Lagrangian 4d $\Nsusy=2$ theories.
\end{itemize}
\stopaparte

\paragraph{Integrable systems.}

The \ac{SW} solutions of many 4d $\Nsusy=2$ theories can be realized as the spectral curve of known integrable systems~\cite{hep-th/9505035,hep-th/9509161,hep-th/9510101,hep-th/9511126}, such as the periodic Toda spin chain, Calogero-Moser, Ruijsenaars, sine-Gordon etc.
As a quite general example, for 4d $\Nsusy=2$ theories of class~S it is the Hitchin integrable system~\cite{0907.3987}.
Placing the 4d theory on the Omega background with $\epsilon_2=0$ (the Nekrasov--Shatashvili limit) corresponds to quantizing the integrable system, and turning on $\epsilon_2$~yields a further refinement~\cite{0908.4052}.
It is also understood how to include surface operators in these discussions.

Nekrasov has advocated for seeing these considerations as a \ac{BPS}/\ac{CFT} correspondence, reviewed in~\cite{1512.05388,1608.07272,1701.00189,1711.11011,1711.11582} (see also~\cite{1308.2465,1611.03478,1710.06970,1711.06150,1807.11900,1908.04394,1909.10352,1910.03247,1912.09969}), which relates supersymmetric gauge theories with $8$~supercharges (e.g.\ 4d $\Nsusy=2$) to integrable models and 2d \ac{CFT}.
Since this applies beyond class~S theories, one can consider the \ac{AGT} correspondence as merely an instance of it in which one can make further progress.

Another instance is the Bethe/gauge correspondence, which roughly speaking arises in the Nekrasov--Shatashvili limit.  In this limit, the Omega-deformed 4d $\Nsusy=2$ theories reduce to a 2d $\Nsusy=(2,2)$ theory whose properties match with those of quantum integrable systems.  For instance the twisted chiral ring of the 2d theory gives quantum Hamiltonian, supersymmetric vacua correspond to Bethe states, and the 2d twisted superpotential is the Yang--Yang function of the integrable system.  The limit was studied in~\cite{0901.4744,0901.4748,0908.4052,1005.4445,1006.4822,1103.4495,1103.5726,1104.3021,1202.5135,1205.3652,1207.0460,1212.6787,1310.0031,1312.4537,1312.6689,1401.4135,1504.08324,1506.01340,1511.02672,1705.09120,1804.04815,1911.01334} among others.

For further unsorted references regarding integrable models and class~S theories, see~\cite{0908.4052,0910.5670,0911.2396,0911.5721,1003.3964,He:2010zzc,1006.1214,1006.4505,1102.5403,1103.3919,1103.4843,1107.3756,1107.4234,1108.0300,1111.2892,Koroteev-2012ara,1204.0913,1206.6349,1307.1502,1209.3984,1210.3580,1302.0015,1303.0753,1303.2578,1303.4237,1304.0779,1305.5614,1306.4590,Popolitov:2013ria,1310.6022,1310.6958,1312.4537,1312.6382,1402.1626,1403.6454,1410.0698,Nekrasov:2014yra,1411.3313,1412.6081,Sciarappa-2015nvw,1505.07116,1507.00519,1512.02492,1601.05096,1601.08238,1602.02772,1603.00304,1604.03574,1606.08020,1612.07590,1701.03057,1708.06135,1709.04926,1710.01051,1711.07570,1711.07935,Bourgine:2017ahg,1712.02936,1801.04303,1801.08908,Fachechi:2018bhe,Poghosyan:2018vyl,1805.01308,1805.08617,1808.04034,1810.01970,1903.10372,1905.01513,1905.09921,1908.08030,1909.07990,1909.11100,1910.02606,1912.00870}.

\paragraph{Topological strings.}
Topological strings and their relation with the \ac{AGT} correspondence are reviewed in~\cite{1408.1240}.

For 4d $\Nsusy=2$ theories realized from IIB geometric engineering~\cite{hep-th/9604034,hep-th/9609239,hep-th/9706110} or as dimensional reductions of 5d $\Nsusy=1$ theories living on $(p,q)$ fivebrane webs, instanton partition functions can be expressed as partition functions $\Ztop$ of topological strings, as reviewed in~\cite{1412.7133}.
As advocated in~\cite{0909.2453} to explain the \ac{AGT} correspondence, $\Ztop$~can be further expressed in terms of Penner-like matrix models with logarithmic potentials~\cite{hep-th/0206255,hep-th/0208048,1010.4573,1012.3228,1106.4873,1107.5181,1207.1438}, which match with Dotsenko--Fateev integral representations of conformal blocks.  We return to these matrix models shortly.

The topological string partition function~$\Ztop$ is computed through the topological vertex formalism, developped in~\cite{hep-th/0212279,hep-th/0306032,hep-th/0310235,hep-th/0310272} in the unrefined case $\epsilon_1=-\epsilon_2$, and for general $\epsilon_1,\epsilon_2$ in two formulations in~\cite{hep-th/0305132,hep-th/0701156} and~\cite{hep-th/0502061,0805.0191}, whose equivalence is explained in~\cite{1112.6074} by realizing the refined topological vertex as an intertwiner of the \ac{DIM} algebra.  See also~\cite{1010.1210,1012.2147,1209.2425,1303.4237,1310.3854} for further tests and subtleties, \cite{1302.6993,1309.6688}~for a world-sheet perspective, \cite{1903.05905,1904.04766}~for a discussion of dualities that ensure that $\Ztop$~is independent of the so-called preferred direction, and \cite{1801.03916,1907.02382}~for a generalization beyond A-type quivers by introducing new topological vertices.
Sometimes, $\Ztop$~can instead be bootstrapped using holomorphic anomaly equations~\cite{1007.0263,1009.1126,1010.2635,1109.5728,1112.2718,1205.3652,1412.7133,1803.11222}, blowup equations~\cite{1711.09884}, or the quantum curve~\cite{1604.01690,1811.01978}.

The calculation in~\cite{1310.3841,1409.6313,1412.3395,1506.04183} of the partition function of $T_N$ theories, hence the three-point function of Toda \ac{CFT}, is particularly interesting.
Surface operators and their relation to geometric transition and qq-characters are discussed in~\cite{1004.2025,1007.2524,1008.0574,1010.1210,1103.1422,1608.02849,1705.03467,Bourgine:2017ahg,1801.04943,1910.10864}.
Other \ac{AGT} developments based on the refined topological vertex include~\cite{1004.2986,1010.4542,1201.6618,1212.0722,1305.7408,1410.3382,1412.4793,1504.01925,1510.01896,1511.02887,1603.01174,1604.01690,1609.07381,1702.07263,1809.00629,1811.03024,1906.06351}.

\paragraph{\label{par:symmetries}Symmetries and special polynomials.}
The renewed interest in refined topological strings in the context of~\ac{AGT} led to developping many families of special polynomials generalizing Jack and Macdonald polynomials, including Macdonald-Kerov functions, generalized Schur functions, elliptic generalized Macdonald polynomials and more: see~\cite{1404.5401,1409.4847,1606.04187,1607.00615,1612.09570,1706.02243,1712.10300,1810.00395,1812.03853,1903.07777,Ohkubo:2019atj,1906.09971,1907.02771,1907.05410,1908.05176,2001.05637} for recent developments.

These developments are based on various underlying symmetry algebras that generalize W-algebras~\cite{Feigin:1990pn,hep-th/9210010} and would deserve a review of their own by someone more qualified.
My survey of the literature suggests the following main players.  I found the references \cite{1202.2756,1404.5240} useful starting points.
\begin{itemize}
\item The (centrally-extended) \emph{elliptic Hall algebra~$\mathcal{E}_{\sigma,\sigmabar}$} introduced in~\cite{math/0505148} is an associative algebra generated by $u_{m,n}$ for $(m,n)\in\ZZ^2$ with $u_{0,0}=0$, and by commuting generators $\sigma,\sigmabar$.  These are subject to commutation relations expressing $[u_y,u_x]$ (for some $x,y\in\ZZ^2$) as a polynomial in $\sigma$, $\sigmabar$, and the generators~$u_z$ for $z\in\ZZ^2$ in the segment joining $x+y$ to the origin.
  The algebra may be thought as the stable limit~\cite{0802.4001} $\mathcal{E}_{\sigma,\sigmabar}=\ddot{\mathrm{SH}}_\infty$ of \emph{spherical \ac{DAHA}} $\ddot{\mathrm{SH}}_n$.  The latter, also called \emph{Cherednik algebras}, were introduced in~\cite{cherednik1992double} and reviewed in~\cite{math/0404307}.

  The algebra $\mathcal{E}_{\sigma,\sigmabar}$ admits an action of $\Lie{SL}(2,\ZZ)$ by automorphisms, induced by the action on~$\ZZ^2$.
  For any coprime $a,b$, the subalgebra generated by $u_{ma,mb}$, $m\in\ZZ$, forms a copy of the quantum group $U_q(\lie{gl}_1)$, and these copies are interchanged by the $\Lie{SL}(2,\ZZ)$ action.

  Let $\Mcal_{\Lie{U}(N)}$ be the moduli space of non-commutative $\Lie{U}(N)$ instantons on~$\CC^2$ (Gieseker framed moduli space).
  Its equivariant K-theory admits an action of~$\mathcal{E}_{\sigma,\sigmabar}$ \cite{0904.1679,0905.2555,1404.5240}.

\item The \emph{\ac{DIM} algebra} (discovered independently by Ding--Iohara~\cite{q-alg/9608002}, Miki~\cite{doi:10.1063/1.2823979}, and others~\cite{math/0505148,0904.1679}) is an associative algebra depending on complex parameters with $q_1q_2q_3=1$.  It is generated by $\psi_0^{-1}$ and $e_i,f_i,\psi_i$ for $i\in\ZZ$, with quadratic relations such as $[e_i,f_j]=\frac{1}{(1-q_1)(1-q_2)(1-q_3)} \bigl((\delta_{i+j>0}-\delta_{i+j<0})\psi_{i+j} + \delta_{i+j=0} (\psi_0-\psi_0^{-1})\bigr)$.

\item The \emph{quantum toroidal algebra $\ddot{U}_{q_1,q_2,q_3}(\lie{gl}_1)$}, also called \emph{quantum continuous~$\lie{gl}_\infty$}, is a quotient of \acs{DIM} by cubic Serre relations $[e_0,[e_1,e_{-1}]]=0$ and $[f_0,[f_1,f_{-1}]]=0$.  It appeared already in~\cite{doi:10.1063/1.2823979} and is explored in~\cite{1002.3100,1002.3113,1004.2575}.
It is also a specialization of the elliptic Hall algebra $\mathcal{E}_{\sigma,\sigmabar}$ at $\sigmabar=1$~\cite{1004.2575}, and as such, it acts on the aforementioned equivariant K-theory.

\item The \emph{affine Yangian of $\lie{gl}_1$}~\cite{1209.0429,1211.1287}, denoted $\ddot{Y}_{h_1,h_2,h_3}(\lie{gl}_1)$ for $h_1+h_2+h_3=0$ is generated by $e_i,f_i,\psi_i$ for $i\geq 0$ and relations deriving from those of $\ddot{U}_{q_1,q_2,q_3}(\lie{gl}_1)$.
  It is a stable limit of spherical degenerate \ac{DAHA}.
  The equivariant cohomology of $\Mcal_{\Lie{U}(N)}$ admits an action of $\ddot{Y}_{q_1,q_2,q_3}(\lie{gl}_1)$~\cite{1202.2756,1404.5240}.

\item The \emph{algebra denoted \SHc}~\cite{1202.2756,1211.1287} is isomorphic to the specialization of $\ddot{Y}_{h_1,h_2,h_3}(\lie{gl}_1)$ at $h_1=1$ \cite{1209.0429}.
  It has various constructions, such as the spherical degenerate \ac{DAHA} of $\Lie{GL}(\infty)$, or the spherical \ac{COHA} of the quiver with one vertex and one loop.

  Up to topological completions, \SHc\ at particular values of the parameters matches the universal envelopping algebra of the \emph{$W_N$~chiral algebra}~\cite{1202.2756}.
  This leads to an action of $W_N$~on the equivariant cohomology of the instanton moduli space~$\Mcal_{\Lie{U}(N)}$, which explains on the 4d gauge theory side the appearance of the $W_N$~symmetry of Toda \ac{CFT}.
\end{itemize}
The elliptic Hall algebra, \ac{DIM} algebra, and their numerous degeneration limits were used for \ac{AGT}-related applications in~\cite{1105.1667,1112.6074,1405.3141,1409.3465,1511.00458,1512.07178,1512.08016,1604.08366,1606.08020,1608.05351,1703.06084,1703.10759,1703.10990,1705.02941,1706.02243,1709.01954,1712.10300,1809.08861,1810.00301,1812.11961,1903.10372,1904.04766,Ohkubo:2019atj,1907.02382,1910.07997}.
As these algebras are intimately related and almost equivalent for physics applications, it is hard to disentangle which precise algebra is relevant to any given physics work.
The choice is often based on which generators of the algebra are the most physically meaningful: translating from one presentation of the algebras to another is highly involved, see for instance~\cite{1209.0429,1702.05100}.

Further algebras have also been considered.
\begin{itemize}
\item The quantum toroidal algebra $\ddot{U}_{q_1,q_2,q_3}(\lie{sl}_k)$ was introduced in~\cite{q-alg/9502013,q-alg/9506026,q-alg/9611030}.
  The relation between different presentations of $\ddot{U}_{q_1,q_2,q_3}(\lie{sl}_k)$ and its analogue $\ddot{Y}_{h_1,h_2,h_3}(\lie{sl}_k)$, and a further classical limit, were explored in~\cite{1504.01696,1512.09109,1603.08915,1605.01314}.
  I do not know if there is are analogues of the \ac{DIM} algebra or of the elliptic Hall algebra for this setting.
  This class of algebras is relevant to the \ac{AGT} correspondence in the presence of a $\ZZ_k$~orbifold~\cite{1610.04144,1906.01625,1912.13372}, see also perhaps~\cite{1705.03628}.

\item Just as the $W_N$~chiral algebra embeds into~\SHc, suitable specializations of  $\mathcal{E}_{\sigma,\sigmabar}$ contain~\cite{1002.2485} $q$W-algebras and quiver W-algebras, themselves explored further in~\cite{1509.07516,1512.08533,1607.05050,1608.04651,1612.07590,1703.00982,1705.00957,1705.04410,1710.02275,1711.07921,1712.07331,1712.08016,1804.06460,1805.00203,1806.02470,1810.08512,1810.10402,1811.03958,1905.03076,1905.03865,1907.06495,1908.04394,1910.00031,1910.00041,1910.03247,1910.10129,1912.09969,2001.05751}.

\item A general point of view on \ac{BPS} algebras is given by \acp{COHA}~\cite{Sala-2018coha,1810.10402,Zhao-2020Kha,1901.07641,1903.07253}.

\item The Skylanin algebra also appears in related literature.  It is a one-parameter deformation of the quantum group $U_q(\lie{sl}_2)$.
\end{itemize}

Based on these generalized symmetry algebras there exists an elliptic version of the refined topological vertex (an intertwiner of the elliptic \ac{DIM} algebra)~\cite{1603.00304,1712.10255,1805.12073,1905.00864} for use in the 6d lift of the \ac{AGT} correspondence, as well as a Macdonald refined topological vertex~\cite{1701.08541,1801.04943} and an analogue when the 4d spacetime is orbifolded~\cite{1906.01625}.

\paragraph{Matrix models.}
The \ac{AGT} correspondence (including its $q$-deformed version) can be explored by studying matrix models~\cite{0909.2453} since both instanton partition functions and conformal blocks are Penner-like matrix model integrals with logarithmic potentials.  See the reviews~\cite{1412.7124,1412.7132}.

On the 2d \ac{CFT} side the matrix model representation arises as Dotsenko--Fateev free-field representations of conformal blocks, which are available provided the sum of Liouville/Toda momenta in each three-punctured sphere is suitably quantized.
Internal momenta in the conformal block translate to choices of contours in the matrix model integral~\cite{1001.0563,1003.5752,1010.1734}.
Moving away from these quantized slices in parameter space requires analytic continuation, which is only completely under control in the $\lie{g}=\lie{su}(2)$ case since coefficients in various expansions are known to be rational functions of all parameters in this case.

The link between 4d gauge theory and matrix models shows up as an equality of $Z_{\inst}$ with the matrix model partition functions~\cite{0912.5476,1006.0828} (directly without going through the topological string~$\Ztop$), a matching of the \ac{SW} curve with the matrix model's spectral curve and of the \ac{SW} differential with the 1-point resolvent~\cite{0911.4244,0911.4797,0911.5721,1011.4491,1103.5470}.
Both this link and the one with 2d \ac{CFT} generalize to $b^2=\epsilon_1/\epsilon_2\neq-1$ in terms of $\beta$-deformed matrix models~\cite{0912.5476,1103.5470}, to 5d~\cite{0911.5337,0912.5476,1004.5122}, to asymptotically free theories~\cite{0911.4797,0912.2988,1011.3481,1209.6009,1303.5584,1909.09041}, to $\lie{su}(N)$ theories~\cite{0911.5337,1110.5255}, to quiver gauge theories~\cite{1009.5553} and generalized quivers~\cite{1011.5417} (higher genus~$C$), and to an orbifold of $\RR^4$~\cite{1105.6091}.

The relations are tested in various limits in~\cite{1003.2929,1004.2917,1005.5715,1104.2738,1112.3545,1209.6009} and proven in some cases~\cite{1309.1687,1403.3657}.
Since the Nekrasov--Shatashvili limit $\epsilon_2\to 0$ of $Z_{\inst}$ quantizes the integrable models underlying a 4d $\Nsusy=2$ theory's \ac{SW} solution, matrix models give useful information about integrable models, see for instance~\cite{1104.2589,1104.4016,1202.6029,1206.1696,1212.4972,1402.1626}.
Matrix models have also been studied for applications to wild punctures and \ac{AD} theories, especially in the classical limit $c\to\infty$ (Nekrasov--Shatashvili limit on the gauge theory side) in~\cite{1003.1049,1008.1861,1207.4480,1210.7925,1312.5535,1411.4453,1504.07910,1506.02421,1506.03561,1509.08164,1510.09060,Piatek:2016kij,1608.05027,1612.00348,1805.05057,1812.00811,1909.10770}.

In another direction, modular properties of (properly normalized) instanton partition functions under S-duality are studied in~\cite{1201.3633,1212.0722,1302.0686,1305.7408,1307.0773,1307.6648,1311.7069,1404.7378,1504.04360,1507.07476,1601.01827,Iqbal:2016pgq,1606.00179,1606.05324,1607.08327,1609.01189,1610.02000,1702.02833,1803.02320} through matrix model and other techniques.
Other works and reviews abound~\cite{1011.5629,1101.0676,1106.1539,1201.4595,1204.3953,Oota:2013qqa,1310.3566,1411.2602,1412.8592,1512.05785,1609.03681,1704.01517,1707.02443,1901.02811,1904.02088,1908.01278,1910.03261}, as well as PhD theses~\cite{Shakirov-2015sky,deCarmoVaz-2015wcf,Raman-2018igv}.

\paragraph{Other connected topics.}
We list disparate subjects that are connected in various ways with the \ac{AGT} correspondence.
Reference lists are both less complete and less properly filtered here than elsewhere in the review.

Holographic duals of the 6d $\Nsusy=(2,0)$ theory, of 4d $\Nsusy=2$ class~S theories, and in the presence of extended operators are explored in~\cite{0904.4466,0910.4234,1001.0906,1005.4527,1011.0154,1107.5763,1110.6400,1202.6613,1304.1643,1304.4954,1312.6687,1401.3595,1409.6206,1412.0489,1501.06072,Rota-2015bkn,1610.04144,1708.05052,1710.09479,1712.06596,1811.12375,1901.02888,1903.05095,1907.12561,1911.02185,1912.03791}.

Supersymmetric localization applies to many background geometries, and for a sample of interesting cases see~\cite{1308.1102,1404.0210,1405.2488,1412.7134,1507.05426,1508.04496,1603.02586,1608.02953,1702.01254,1712.01288,1801.01986,1803.06177,Fachechi:2018nrx,1812.06473,1812.11188} and reviews~\cite{1412.7134,1608.02953}.
Resurgence and Borel summability of various expansions of $Z_{\inst}$ (and of other exact results from supersymmetric localization) are studied in~\cite{1203.5061,1302.5138,1410.5834,1501.05671,1603.06207,1604.05520,1711.10799,1901.02076}.
These give some insight on how applicable resurgence techniques are in \ac{QFT}.

There are numerous interplays with other properties of 4d $\Nsusy=2$ theories.
\begin{itemize}
\item Topological anti-topological fusion ($tt^*$~equations)~\cite{0910.4963,1312.1008,1409.4212,1412.4793}.
\item The chiral algebra that appears as a protected subsector of 4d $\Nsusy=2$ theories~\cite{1312.5344,1408.6522,1411.3252,1505.05884,1506.00265,1509.00033,1509.05402,1511.01516,1511.07449,1602.01503,1603.00887,1604.02155,1606.08429,1610.05311,1612.01536,1612.02363,1612.06514,1612.08956,1701.08782,1704.01955,1706.01607,1706.03797,1708.05323,1710.04306,1710.06029,1710.06469,1711.07941,1805.08839,1807.04296,1810.03612,1811.01577,1811.03958,1811.11772,1812.06394,1812.07572,1901.07591,1902.02838,1903.07624,1903.11123,1904.00927,1904.02704,1904.09094,1906.07212,1909.04074,1910.02281,1910.04120,1911.05082,1911.05741,1912.00870,2001.08048} (see also~\cite{1404.1079,1412.0334,1601.05378,1712.09384,1805.00892,1912.01006} and references thereto).
\item Gauge/\ac{YBE} \cite{1011.3798,1203.5784,1504.04055,1504.05540,1606.01041,1611.07522,1701.05562,1709.00070,1803.00855} reviewed in~\cite{1808.04374}.  See also~\cite{1610.06925}.
\item The way class~S theories are built by combining building blocks through gauging suggests to introduce a notion of theory space~\cite{1304.0762,1704.02986}.
\item Conformal bootstrap: besides the constructions discussed in this review, another interesting method to find \acp{QFT}, specifically unitary \acp{CFT}, is the conformal bootstrap program started in~\cite{1203.6064} and applied to 4d $\Nsusy=2$ theories in~\cite{1412.7541}.
  Some \ac{AD} theories in particular are located at corners of the regions of parameter space allowed by the bootstrap.
\end{itemize}

The \ac{AGT} correspondence has increased the interest in several old questions about 2d~\ac{CFT}.
\begin{itemize}
\item Computing conformal blocks, correlators, and fusion matrices, either through recursion relations~\cite{0909.3412,0911.2353,0912.0504,1003.1049,1004.1841,1005.0216,1012.2974,1409.3537,1411.4222,1609.01189,Nemkov:2016qxf,1703.09805,1705.00629,1806.09563}, using holography in the large~$c$ limit~\cite{0909.3531,1109.6764,1309.4700,1311.2888,1502.07742,1504.05943,1508.04987,1510.00014,1510.06685,1511.05452,1512.03052,1601.06794,1602.04829,1609.00801,1706.07474,1707.09311,1709.03476,1712.08078,1807.07886,1910.04169,1911.01334}, or Chern--Simons theory~\cite{1801.08549}.
\item Studying variants of Toda \ac{CFT}, parafermionic Liouville \ac{CFT} etc.~\cite{1605.01933,1606.04328,1809.05568}.
\item Some (disputed) links to the fractional quantum Hall effect~\cite{1002.5017,1011.0154,1201.1903,1410.3575,1511.03372,1708.00419,1910.07369}.
\item Isomonodromy problems, as it is now known that conformal blocks (and hence Nekrasov partition functions), Fourier transformed with respect to internal momenta, give solutions to Painlev\'e equations arising in isomonodromy problems for Fuchsian connections~\cite{1008.4332,1011.0292,1104.3210,1202.2149,1206.5963,1207.0787,1207.6884,1302.1832,1302.5138,1303.0753,1307.0306,1307.4865,1308.4092,1309.4700,1309.7672,1401.6104,1403.1235,1405.1871,1406.3008,1505.00259,1505.02398,1506.06588,1507.08794,1508.04046,1601.05096,Ferrari:2016iro,1603.01174,1605.04554,1608.00958,1608.02568,1611.08971,1612.06235,1704.01517,1705.01869,1707.02443,1709.03476,1711.02063,1801.09608,1803.02320,1805.05057,1806.08344,1806.08650,Anselmo:2018zre,1811.01978,1811.11912,1901.02076,1901.10497,1902.06439,1905.03795,1906.10638,1909.07990}; likewise the chiral blocks of the $q$-deformed Virasoro algebra and $q$W-algebras give solutions of $q$-Painlev\'e equations~\cite{1706.01940,1708.07479,1811.03285}.
\end{itemize}
Incidentally, the Liouville \ac{CFT} has finally been defined mathematically from its path integral: see~\cite{1410.7318,1712.00829} and references therein.
Other mathematical references include the study of 6j symbols of (the modular double of) $U_q(\lie{sl}_2)$~\cite{1202.4698},
relations to the geometric Langlands correspondence or deformations thereof~\cite{1005.2846,1701.03146,1702.06499,1804.06460,1805.00203,Schweigert:2018zrz}.

\paragraph{Final thoughts.}

The construction of new theories by dimensionally reduction in various geometrical setups has proven very fruitful.  It has led to many new \aclp{QFT} that can be used as building blocks for yet more discoveries.
The large number of dualities uncovered in this way can be further enriched by considering extended operators in their various incarnations.
I hope that readers will participate in this exciting journey charting the space of \aclp{QFT}!

\mystarsec\section{Acknowledgments}

First, many thanks to Jaume Gomis for guiding me through basics of the \ac{AGT} correspondence starting a decade ago, and to my other coauthors on \ac{AGT}-related papers for sharing insights throughout the years: Alexander Gorsky, Alexey Milekhin, Yiwen Pan, Wolfger Peelaers, Nikita Sopenko and Gustavo Turiaci.
I wish to thank Antoine Bourget, Ioana Coman, Taro Kimura, Fabrizio Nieri, Wolfger Peelaers, Shlomo Razamat and Jaewon Song who helpfully provided references or comments, as well as Elli Pomoni and Alessandro Sfondrini for organizing \acs{YRISW} 2020.
Lastly, I am grateful to my wife and kids for their patience while I prepared and later revised the lectures.

\mystarsec\section{\label{sec:acro}Table of acronyms}

\setlength{\columnsep}{3em}
\begin{multicols}{2}\footnotesize
  \begin{acronym}[YRISW]
    \acro{ABJM}{Aharony--Bergman--Jafferis--Maldacena\acroextra{ (M2~brane worldvolume theory)}}
    \acro{AD}{Argyres--Douglas\acroextra{ (strongly coupled 4d $\Nsusy=2$ theories)}}
    \acro{ADHM}{Atiyah--Drinfeld--Hitchin--Manin\acroextra{ (construction of instantons)}}
    \acro{AGT}{Alday--Gaiotto--Tachikawa\acroextra{ (4d/2d correspondence)}}
    \acro{ALE}{asymptotically locally Euclidean\acroextra{ space (resolution of $\CC^2/\Gamma$)}}
    \acro{BMT}{Bonelli--Maruyoshi--Tanzini\acroextra{ (irregular states in 2d \ac{CFT})}}
    \acro{BPS}{Bogomol'nyi–Prasad–Sommerfield\acroextra{ (supersymmetric)}}
    \acro{CFT}{conformal field theory}
    \acrodefplural{CFT}{conformal field theories}
    \acro{COHA}{cohomological Hall algebra}
    \acro{DAHA}{double affine Hecke algebra}
    \acro{DIM}{Ding--Iohara--Miki\acroextra{ (algebra)}}
    \acro{DOZZ}{Dorn--Otto--Zamolodchikov--Zamolodchikov\acroextra{ (three-point function in Liouville \ac{CFT})}}
    \acro{DS}{Drinfeld--Sokolov\acroextra{ (reduction of W-algebras)}}
    \acro{FI}{Fayet--Iliopoulos\acroextra{ (parameter in supersymmetric action)}}
    \acro{GW}{Gukov--Witten\acroextra{ (surface defect)}}
    \acro{IR}{infra-red\acroextra{ (low energy/long distance)}}
    \acro{JK}{Jeffrey--Kirwan\acroextra{ (residue prescription)}}
    \acro{KK}{Kaluza--Klein\acroextra{ (reduction on circle)}}
    \acro{LMNS}{Losev--Moore--Nekrasov--Shatashvili\acroextra{ (formula for the instanton partition function)}}
    \acro{NS}{Nekrasov--Shatashvili\acroextra{ ($b\to 0$ limit, i.e., $\epsilon_1\to 0$)}}
    \acro{OPE}{operator product expansion}
    \acro{QFT}{quantum field theory}
    \acrodefplural{QFT}{quantum field theories}
    \acro{RG}{renormalization group}
    \acro{SCFT}{superconformal field theory}
    \acrodefplural{CFT}{superconformal field theories}
    \acro{SQCD}{super-QCD\acroextra{ (\ac{SYM} plus matter)}}
    \acro{SW}{Seiberg--Witten\acroextra{ (curve~$\Sigma$ and differential~$\lambda$ giving \ac{IR} description and prepotential~$F$ of 4d $\Nsusy=2$ theories)}}
    \acro{SYM}{super-Yang--Mills\acroextra{ (for us, 4d $\Nsusy=2$ vector multiplet)}}
    \acro{TQFT}{topological quantum field theory}
    \acrodefplural{TQFT}{topological quantum field theories}
    \acro{UV}{ultra-violet\acroextra{ (high energy/short distance)}}
    \acro{VEV}{vacuum expectation value}
    \acro{YBE}{Yang--Baxter equation}
    \acro{YM}{Yang--Mills\acroextra{ (non-supersymmetric)}}
    \acro{YRISW}{Young Researchers Integrability School and Workshop}
  \end{acronym}
\end{multicols}

\appendix
\section{\label{sec:special}Special functions}

In the main text we use the following special functions.
\begin{itemize}
\item The Gamma function $\Gamma(x)=\prod_{n\geq 0}^{\text{reg}}\frac{1}{x+n}$ has poles at $-\ZZ_{\geq 0}$ and obeys the shift formula $\Gamma(x+1)=x\Gamma(x)$.
\item The Barnes Gamma function $\Gamma_b(x)=\prod_{m,n\geq 0}^{\text{reg}}\frac{1}{x+mb+n/b}$ has poles at $-b\ZZ_{\geq 0}-b^{-1}Z_{\geq 0}$ and obeys the shift formula $\Gamma_b(x+b)/\Gamma_b(x)= \sqrt{2\pi}b^{xb-1/2}/\Gamma(xb)$.
\item The double-sine function $S_b(x)=\frac{\Gamma_b(x)}{\Gamma_b(b+1/b-x)}$ has poles at $-b\ZZ_{\geq 0}-b^{-1}Z_{\geq 0}$, zeros at $b\ZZ_{\geq 1}+b^{-1}Z_{\geq 1}$, and obeys the shift formula $S_b(x+b) / S_b(x)= 2\sin(\pi bx)$.
\item The Upsilon function $\Upsilon_b(x)=\frac{1}{\Gamma_b(x)\Gamma_b(b+1/b-x)}$ has zeros at $-b\ZZ_{\geq 0}-b^{-1}Z_{\geq 0}$ and $b\ZZ_{\geq 1}+b^{-1}Z_{\geq 1}$, and obeys the shift relation $\Upsilon_b(x+b) / \Upsilon_b(x) = b^{1-2bx} \Gamma(bx)/\Gamma(1-bx)$.
\end{itemize}
Note that $\Gamma_b,S_b,\Upsilon_b$ are invariant under $b\to 1/b$.

\setlength{\columnsep}{.5em} 
\begin{multicols}{2}
\printbibliography[heading=bibintoc,title={References}]
\end{multicols}

\end{document}